\RequirePackage{etex} 
\documentclass[12pt,oneside]{Thesis}
\usepackage{etex}
\usepackage{pst-node}
\usepackage{psfrag}
\usepackage{eso-pic,graphicx, import}
\usepackage{epsfig}
\usepackage{svg}
\usepackage{color, colortbl}
\usepackage{booktabs}
\usepackage{multirow}
\usepackage[utf8]{inputenc}
\usepackage{todonotes}
\usepackage{textcomp}
\usepackage[thinlines]{easytable}
\usepackage{float}
\usepackage[siunitx]{circuitikz}
\usepackage{braket}
\usetikzlibrary{arrows, decorations.markings}
\usepackage{timing-diagrams}
\usepackage{tikz-timing}
\usepackage{appendix}
\usepackage{epsfig}
\usepackage{algpseudocode}
\usepackage{psfrag}
\usepackage{url}
\usepackage[linesnumbered,ruled,vlined]{algorithm2e}
\usepackage{tfrupee}
\usepackage{longtable}
\usepackage{lscape}
\usepackage{siunitx,xspace}
\usepackage{hhline}
\usepackage{moresize}
\usepackage{amsmath}
\usepackage{makecell}
\usepackage{setspace}
\usepackage[justification=justified, font=small,labelfont=bf,labelsep=period]{caption}
\usepackage{array}
\usepackage{subfig}
\makeatletter
\renewcommand{\fnum@figure}{Figure \thefigure}
\makeatother
\usepackage{times}
\newcolumntype{?}{!{\vrule width 1.5pt}}
\DeclareUnicodeCharacter{00A0}{}
\DeclareUnicodeCharacter{00A0}{~}
\makeatletter
\def\Cline#1#2{\@Cline#1#2\@nil}
\def\@Cline#1-#2#3\@nil{%
	\omit
	\@multicnt#1%
	\advance\@multispan\m@ne
	\ifnum\@multicnt=\@ne\@firstofone{&\omit}\fi
	\@multicnt#2%
	\advance\@multicnt-#1%
	\advance\@multispan\@ne
	\leaders\hrule\@height#3\hfill
	\cr}
\makeatother

\usepackage[square, numbers, comma, sort&compress]{natbib} % Use the natbib reference package - read up on this to edit the reference style; if you want text (e.g. Smith et al., 2012) for the in-text references (instead of numbers), remove 'numbers' 
\hypersetup{urlcolor=blue, colorlinks=true} % Colors hyperlinks in blue - change to black if annoying
\title{\ttitle} % Defines the thesis title - don't touch this
\begin{document}
\sloppy
\frontmatter % Use roman page numbering style (i, ii, iii, iv...) for the pre-content pages

\setstretch{1.3} % Line spacing of 1.3

% Define the page headers using the FancyHdr package and set up for one-sided printing
\fancyhead{} % Clears all page headers and footers
\rhead{\thepage} % Sets the right side header to show the page number
\lhead{} % Clears the left side page header

\pagestyle{fancy} % Finally, use the "fancy" page style to implement the FancyHdr headers

\newcommand{\HRule}{\rule{\linewidth}{0.5mm}} % New command to make the lines in the title page

% PDF meta-data
\hypersetup{pdftitle={\ttitle}}
\hypersetup{pdfsubject=\subjectname}
\hypersetup{pdfauthor=\authornames}
\hypersetup{pdfkeywords=\keywordnames}

%----------------------------------------------------------------------------------------
%	TITLE PAGE
%----------------------------------------------------------------------------------------

%\begin{titlepage}
%\begin{center}
%
%\textsc{\LARGE \bf Turbo-Alternator Design Optimization Using Multi-objective Evolutionary Algorithms}\\[1.5cm] % University name
%\textsc{\Large \bf THESIS}\\[0.5cm] % Thesis type
%
%\HRule \\[0.4cm] % Horizontal line
%{\huge \bfseries \ttitle}\\[0.4cm] % Thesis title
%\HRule \\[1.5cm] % Horizontal line
% 
%\begin{minipage}{0.4\textwidth}
%\begin{flushleft} \large
%\emph{Author:}\\
%{\authornames} % Author name - remove the \href bracket to remove the link
%\end{flushleft}
%\end{minipage}
%\begin{minipage}{0.4\textwidth}
%\begin{flushright} \large
%\emph{Supervisor:} \\
%{\supname} % Supervisor name - remove the \href bracket to remove the link  
%\end{flushright}
%\end{minipage}\\[3cm]
%
%
%
%\large {Submitted in partial fulfillment \\
%	of the requirements for the degree of \\
%	 \bf DOCTOR OF PHILOSOPHY
%	}\\[1.9cm] % University requirement text
%{by}\\[0.4cm]
%\bf K. V. R. B. PRASAD\\[2cm] % Research group name and department name
% 
%\large {Under the Supervision of }\\[0.3cm] 
%
%\large {Dr. Pravin Singru }\\[0.7cm]  
% \includegraphics[scale=0.2]{logo.eps}\\[0.2 cm]
% 
%{\Large BIRLA INSTITUTE OF TECHNOLOGY AND SCIENCE
%	PILANI (RAJASTHAN) INDIA
%	}\\[0.5cm] 
%{\large \today}\\[4cm] % Date
% % University/department logo - uncomment to place it
% 
%\vfill
%\end{center}
%
%\end{titlepage}
\begin{titlepage}
	\begin{center}
		{\setstretch{2}
		\textsc{\Large \bf }
		%[1.1cm] % University name
		\textsc{\Large \bfseries ASPECTS OF VISIBLE SINGULARITIES IN GRAVITATIONAL COLLAPSE }\\[0.1cm] % Thesis type
		\textbf{THESIS}\\
	%	\HRule \\[0.4cm] % Horizontal line
		{\large  Submitted in partial fulfillment \\
			of the requirements for the degree of
			}\\ % Thesis title
	%	\HRule \\[1.5cm] % Horizontal line
		\textsc{\large \bfseries DOCTOR OF PHILOSOPHY}\\[0.3cm]
		{\Large by} \\
		\textsc{\large \bfseries Karim Mosani}\\[0.5cm]
		
		{\Large  Under the Supervision of	
		}\\
		\textsc{\large \bfseries \hspace{1.5cm} Prof. Gauranga C. Samanta (Supervisor)
		\newline
		Prof. Pankaj S. Joshi, and Dr. Mayank Goel (Co-Supervisors)}\\[1.1cm]
%	\begin{minipage}{0.4\textwidth}
%		\begin{flushleft} \large
%			\emph{Author:}\\
%			\href{mailto:pravinmane@goa.bits-pilani.ac.in}{\authornames} % Author name - remove the \href bracket to remove the link
%		\end{flushleft}
%	\end{minipage}
%	\begin{minipage}{0.4\textwidth}
%		\begin{flushright} \large
%			\emph{Supervisor:} \\
%			\href{mailto:pravinmane@goa.bits-pilani.ac.in}{\supname} % Supervisor name - remove the \href bracket to remove the link  
%		\end{flushright}
%	\end{minipage}\\[3cm]
		
		%\large \textit\\\deptname\\[2cm] % Research group name and department name
		
		%\includegraphics[scale=0.2]{logo.eps}\\[0.2 cm] % University/department logo - uncomment to place it
		\begin{figure}[htbp]
			\centering		
			\def\svgwidth{140\unitlength}
			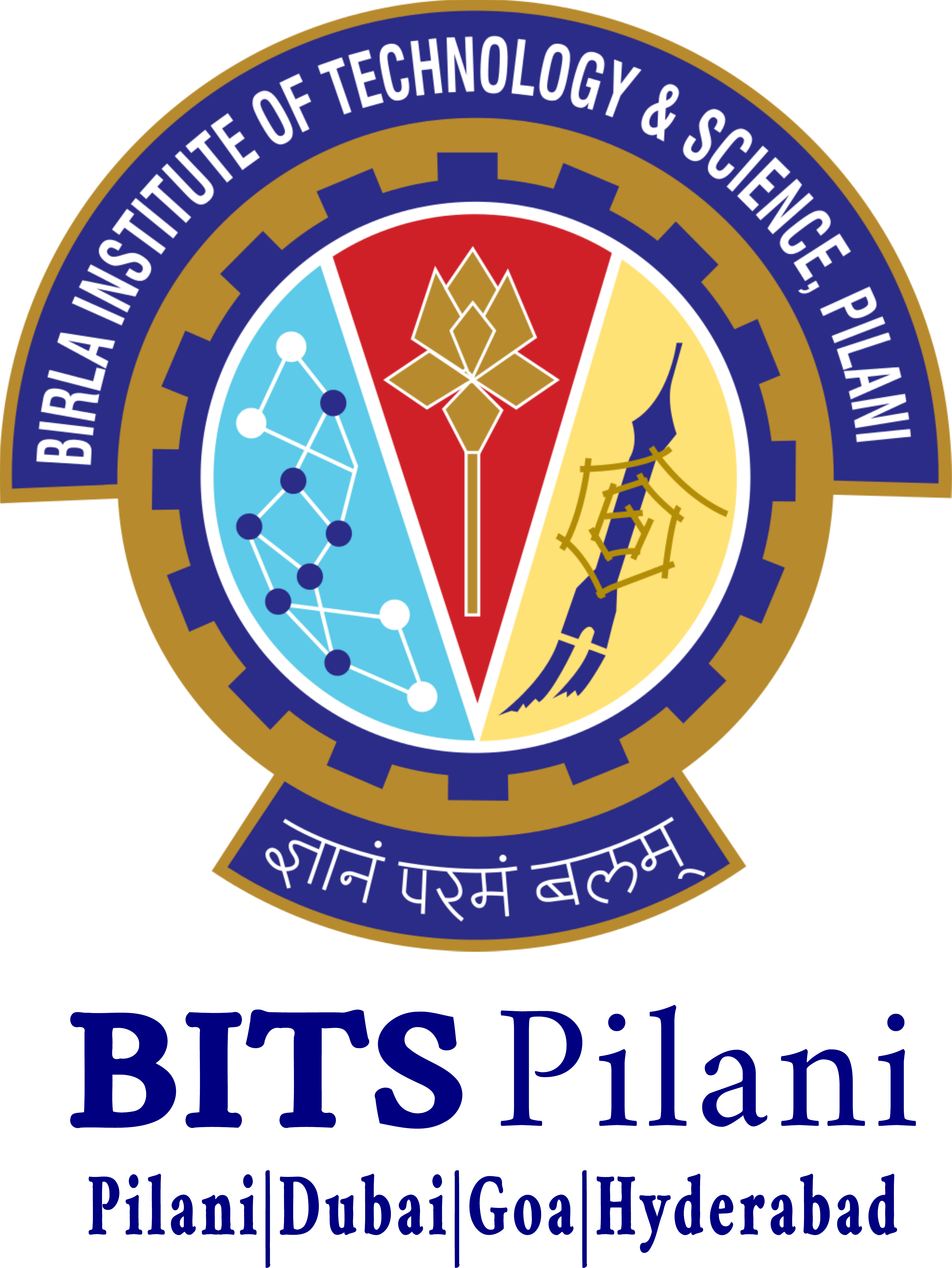		
		\end{figure}
		\vspace{0.3cm} 
		\textsc{\large \bfseries BIRLA INSTITUTE OF SCIENCE \& TECHNOLOGY, PILANI}\\
		{\large \bfseries 2022}\\ % Date
		%\includegraphics{Logo} % University/department logo - uncomment to place it
		
		%\vfill
	}
	\end{center}

\end{titlepage}
%----------------------------------------------------------------------------------------
%	DECLARATION PAGE
%	Your institution may give you a different text to place here
%----------------------------------------------------------------------------------------
\pagestyle{empty}
\begin{center}
	\textsc{\large \bfseries BIRLA INSTITUTE OF SCIENCE \& TECHNOLOGY, PILANI}\\ [2 cm]
	\textsc{\Large \bfseries CERTIFICATE\\ [0.5 cm]}
\end{center} 
{\setstretch{2}
	
 This is to certify that the project entitled  {\bf ``Aspects of visible singularities in gravitational collapse"} and submitted by  {\bfseries Karim Mosani} ID No   {\bfseries 2016PHXF0415G} for award of Doctor of Philosophy of the Institute embodies original work done by him under my supervision.}
 
 Signature of the Supervisor
 
 Name in capital letters : \textsc {\bfseries Gauranga C. Samanta}
 
 Designation : \textsc {\bfseries Associate Professor, FAKIR MOHAN UNIVERSITY, ODISHA}
 
 Date : 
 
 Signature of the Co-Supervisor
 
 Name in capital letters : \textsc {\bfseries Pankaj S. Joshi}
 
 Designation : \textsc {\bfseries Director and Professor, International Center for Cosmology, CHARUSAT, Gujarat}
 
 Date : 
 
  Signature of the Co-Supervisor
 
 Name in capital letters : \textsc {\bfseries Mayank Goel}
 
 Designation : \textsc {\bfseries Assistant Professor, BITS PILANI, K. K. Birla GOA CAMPUS}
 
 Date : 
 \clearpage
\Declaration{

\addtocontents{toc}{\vspace{1em}} % Add a gap in the Contents, for aesthetics

I, Karim Mosani, declare that this thesis titled, `Aspects of visible singularities in gravitational collapse' and the work presented in it are my own. I confirm that:
\begin{itemize} 
\item[\tiny{$\blacksquare$}] This work was done wholly or mainly while in candidature for a research degree at this University.
\item[\tiny{$\blacksquare$}] Where any part of this thesis has previously been submitted for a degree or any other qualification at this University or any other institution, this has been clearly stated.
\item[\tiny{$\blacksquare$}] Where I have consulted the published work of others, this is always clearly attributed.
\item[\tiny{$\blacksquare$}] Where I have quoted from the work of others, the source is always given. With the exception of such quotations, this thesis is entirely my own work.
\item[\tiny{$\blacksquare$}] I have acknowledged all main sources of help.
\item[\tiny{$\blacksquare$}] Where the thesis is based on work done by myself jointly with others, I have made clear exactly what was done by others and what I have contributed myself.\\
\end{itemize}
 
Signed:\\
\rule[1em]{25em}{0.5pt} % This prints a line for the signature
 
Date:\\
\rule[1em]{25em}{0.5pt} % This prints a line to write the date
}
\clearpage % Start a new page
%----------------------------------------------------------------------------------------
%	QUOTATION PAGE
%---------------------------------------------------------------------------------------- 

%\pagestyle{empty} % No headers or footers for the following pages
%
%\null\vfill % Add some space to move the quote down the page a bit
%
%\textit{``Thanks to my solid academic training, today I can write hundreds of words on virtually any topic without possessing a shred of information, which is how I got a good job in journalism."}
%
%\begin{flushright}
%Pravin Mane
%\end{flushright}
%
%\vfill\vfill\vfill\vfill\vfill\vfill\null % Add some space at the bottom to position the quote just right
%
%\clearpage % Start a new page

%----------------------------------------------------------------------------------------
%	ABSTRACT PAGE
%----------------------------------------------------------------------------------------

%\addtotoc{Abstract} % Add the "Abstract" page entry to the Contents
%
%\abstract{\addtocontents{toc}{\vspace{1em}} % Add a gap in the Contents, for aesthetics
%
%The Thesis Abstract is written here (and usually kept to just this page). The page is kept centered vertically so can expand into the blank space above the title too\ldots
%}
%
\clearpage % Start a new page

%%----------------------------------------------------------------------------------------
%	ACKNOWLEDGEMENTS
%%----------------------------------------------------------------------------------------

\setstretch{1.3} % Reset the line-spacing to 1.3 for body text (if it has changed)
%
%\acknowledgements{\addtocontents{toc}{\vspace{1em}} % Add a gap in the Contents, for aesthetics
%
%The acknowledgements and the people to thank go here, don't forget to include your project advisor\ldots
%}
%\clearpage % Start a new page

%----------------------------------------------------------------------------------------
%	LIST OF CONTENTS/FIGURES/TABLES PAGES
%----------------------------------------------------------------------------------------

\pagestyle{fancy} % The page style headers have been "empty" all this time, now use the "fancy" headers as defined before to bring them back

\chapter{Acknowledgements}

I owe my gratitude to Prof. Gauranga C. Samanta for guiding my research. He has supported me immensely at various points of my tenure here in BITS. Also, I thank him for giving me the book by Bernard Schutz, which he probably forgot to take back, and I forgot to give back. 

I express my gratitude to Prof. Pankaj S. Joshi for giving me insights to tackle problems and guide me. He has inspired me immensely. Discussions with him, including lunch meetings, have always been fruitful.   

I  thank Dr. Mayank Goel for helping me at a very crucial time last year.

I have learned many life lessons from Dr. Shilpa Gondhali. I am grateful to her for being so kind and helpful. I am also thankful to Prof. Prasanta Kumar Das for his valuable inputs, advice, and guidance. I appreciated his feedback during my progress seminar meetings. 

I thank Prof. Prasanna Kumar, Prof Tarkeshwar Singh, Prof. Amit Setia, and Prof. Bharat Deshpande for making the environment so research-friendly for Ph.D. students. I also acknowledge the Head of the Department (HoD), the Department Research Committee (DRC), and the Academic Graduate Studies and Research Division (AGSRD) for their cooperation.

During my stay at BITS, I met some wonderful people. They are Aditya, Arindam, Chitira, Dileep, Deepak, Harsha, Ishita, Jay Tushar, Malavika, Mona, Mubashshir, Nayan, Nilesh, Pabitra, Pritee, Ravi Pawar, Ram Singh, Sandhya, and Sayantan. 

I am thankful to Aditya, Arindam, and Sayantan, for inspiring me with their intellect. Sadly, Deepak left so early. He was one of my first friends in BITS. I thank him for being so kind. I thank Ishita, Tushar, Pritee, and Sandhya for being such wonderful friends. I enjoyed going out and having a fun time with them. I am indebted to Mubashshshshir for providing me with snacks during night time. He was the man I remembered when I felt hungry in the night during my stay inside BITS. In addition to this, I thank him for being so helpful.

During my stay at ICC, I met some wonderful people. They are Ashok, Aswathi, Chinmay, Dipanjan, Hari, Jun-Qi, Parth Joshi, Saikat, Samriddhi, Shailee, Vishwa, and Vitalii. 

I thank Ashok for being a very kind person. I am indebted to Aswathi, Vishwa, and Shailee to inspire me by questioning many things, having discussions without prejudices, and being patient while listening. I am grateful to Dr. Dipanjan Dey for working together and also for being a wonderful friend. I have enjoyed discussing gravity and brainstorming with him. I thank Samriddhi for being so generous, and a pleasant person. I admire the selfless attitude of Saikat and Hari and thank them for being my friends. I thank Saikat for introducing me to Teleparallel gravity. It was a good experience collaborating with Parth Joshi, Hari, and Vitalii.
%I thank my friends in BITS and ICC.

I thank all the members of the BITS administration, especially Ms. Niyata, Mr. Pratap, and Ms. Prasanthi, for their cooperation in all the administrative works.

I thank Robert Wald for writing such an excellent book on general relativity. 

Finally, I am grateful to my parents and my sister for supporting me throughout the journey. I am pleased that my family never forced me to abide by the norms imposed by the society. I thank them for having faith in me. It would not have been possible without their cooperation and understanding. 

I acknowledge the Council of Scientific and Industrial Research (CSIR, India, Ref: 09/919(0031)/2017-EMR-1) for funding my Ph.D. program.
\\

\hspace{13cm} \textbf{Karim Mosani}

%I would like to thank my friends in BITS and ICC $Names$ for constantly motivating me.

\lhead{\emph{Acknowledgements}}

\lhead{\emph{Contents}} % Set the left side page header to "Contents"
\tableofcontents % Write out the Table of Contents

\lhead{\emph{List of Figures}} % Set the left side page header to "List of Figures"
\listoffigures % Write out the List of Figures

\lhead{\emph{List of Tables}} % Set the left side page header to "List of Tables"
\listoftables % Write out the List of Tables

 % Write out the List of Figures
%----------------------------------------------------------------------------------------
%	ABBREVIATIONS
%----------------------------------------------------------------------------------------
 % Start a new page

\setstretch{1.5} % Set the line spacing to 1.5, this makes the following tables easier to read

\clearpage % Start a new page

%----------------------------------------------------------------------------------------
%	DEDICATION
%----------------------------------------------------------------------------------------

\setstretch{1.3} % Return the line spacing back to 1.3

\pagestyle{empty} % Page style needs to be empty for this page

\dedicatory{Dedicated To My Family} \clearpage

\addtocontents{toc}{\vspace{2em}} % Add a gap in the Contents, for aesthetics

%----------------------------------------------------------------------------------------
%	THESIS CONTENT - CHAPTERS
%----------------------------------------------------------------------------------------

\mainmatter % Begin numeric (1,2,3...) page numbering

\pagestyle{fancy} % Return the page headers back to the "fancy" style

% Include the chapters of the thesis as separate files from the Chapters folder
% Uncomment the lines as you write the chapters

% Chapter Template

\chapter{Introduction} % Main chapter title
What happens to a massive star at the end of its life cycle is one of the most important and intriguing questions in theoretical physics. An unhindered gravitational collapse of such a sufficiently massive star can give rise to a spacetime singularity. This thesis discusses the phenomenon of such unhindered gravitational collapse and the causal structure of the singularity, thus formed.
\label{Chapter1} % Change X to a consecutive number; for referencing this chapter elsewhere, use \ref{ChapterX}
 
\lhead{Chapter 1.\emph{Introduction}} % Change X to a consecutive number; this is for the header on each page - perhaps a shortened title

%----------------------------------------------------------------------------------------
\section{Gravitational collapse}\label{1.1}
%----------------------------------------------------------------------------------------

The contraction of a massive astrophysical or a cosmological body under its gravitational influence is called gravitational collapse. It is believed to be the reason for the formation of stars, galaxies, clusters, and in general, structure formations in the universe. The ultimate destiny of a star at the climax of its life is one of the fascinating questions in astrophysics and gravitational physics. The gradual collapse of overdense regions in the universe mostly made up of hydrogen and helium, gives birth to a star. A star, like our sun, harbors nuclear fusion reaction in its core. The pressure developed due to this nuclear fusion opposes the inward gravitational pull due to its mass, thereby reaching an equilibrium state. However, when the star's internal energy is exhausted and the nuclear fusion ceases to exist, the inward gravitational pull dominates its dynamics, and the star contracts in size, thereby becoming denser. It is possible that the contraction stops and reaches a new equilibrium state. Based on the mass of the collapsing star, the following are the possible end states:
\begin{enumerate}
    \item \textbf{White dwarf}:
    According to Pauli's exclusion principle, no two elementary particles with the same spin in a given volume can occupy the same energy state. Once an electron fills the lowest energy, the other electrons are forced into higher energy states, resulting in traveling at a progressively faster speed. This progression of speed by the electrons builds up a pressure called the electron degeneracy pressure. Suppose the electron degeneracy pressure opposes the gravitational collapse. After expelling some of its mass in a supernova, the collapsing cloud ends up in a white dwarf.

    \item \textbf{Neutron star}:
    If the elementary particles are neutrons, then the Paulis exclusion principle gives rise to neutron degeneracy pressure. Suppose the cloud is massive enough such that the electron degeneracy pressure is insufficient to oppose the gravitational contraction. In that case, the neutron degeneracy pressure can resist the gravitational collapse, giving rise to the core collapsing to a neutron star.

    \item \textbf{Unhindered gravitational collapse}:
    Chandrashekhar
 \cite{Chandrashekhar_31}
 and Landau 
    \cite{Landau_32}
suggested an upper limit to the mass of any astrophysical object such that any massive object (core progenitor) above this limit can not remain stable and fails to support itself against the inward gravitational pull. This leads to a continued contraction without any opposition.
\end{enumerate}
An unhindered gravitational collapse gives rise to a spacetime singularity as predicted by the singularity theorem 
\cite{Hawking_73}. 
A brief overview of the spacetime singularity is discussed in the next section.

\section{Spacetime singularities}\label{1.2}
Defining a spacetime singularity is not so straightforward. For example, the radial coordinate $r=2M$ in the Schwarzschild spacetime, where $M$ is the mass of the body generating this geometry, was once considered a singularity. However, later it was realized that one could get rid of such so-called ``singularity" by suitable coordinate transformation
\cite{Wald_84, Schutz_09}. 
It is, therefore, necessary to carefully define a spacetime singularity by keeping various aspects in mind. The following is the progression of its definition over more than eight decades:
\begin{enumerate}
    \item Consider the definition: The spacetime singularity is a `` place" where the curvature becomes unbounded. The issue with such definition is that we associate a ``place" to the singularity. In general relativity, the occurrence of an event (spacetime coordinates associated with it) makes sense only if the manifold (and a metric on it) is defined in the neighborhood of the spacetime coordinate of this event. Since a spacetime singularity is not considered a part of the manifold, its definition in terms of a ``place" is not very desirable and asks for a better explanation.
    
    \item To tackle this, one could now consider the set consisting of non-singular points, and add the singular points to this set, thereby giving a bigger set, and then define a topological structure on this entire bigger collection of singular as well as non-singular points. Doing so allows one to refer to the singularity as a place. However, a general definition based on the coordinate components does not work because the singularity property should be invariant under the coordinate transformation, which does not happen if we take this approach.
    
    \item The next thing we can do is to use the property of the curvature blow-up to define a spacetime singularity.  The curvature becomes unbounded as one approaches the singularity in the FLRW spacetime governing the universe, or the Schwarzschild spacetime. However, the curvature is represented by the Riemann curvature tensor $R^{a}_{bcd}$, which can be dependent on the choice of the coordinates. To give a coordinate independent definition, we can choose curvature invariant scalar quantities obtained by contracting the Riemann curvature tensor, such as the Ricci scalar $R$, the Kretschmann scalar $K=R_{abcd}R^{abcd}$, and analogous scalar terms, and see if these scalar terms diverge as one approaches the singularity. 
    
    The issue with such a way to define a spacetime singularity is the scenario where the scalar terms blow up to infinity only if one goes to infinity. We would want the blow-up of curvature terms in finite time to make physical sense. Such definition based on the blow-up of curvature scalars is hence not suitable.
    
    Another example that demotivates us to take this as a satisfactory definition is the formation of a conical singularity in the Minkowski spacetime by cutting a wedge $0<\phi<\phi_1$ (where $\phi$ is the azimuthal coordinate) and stitching (identifying) the points on $\phi=0$ with that of $\phi=\phi_1$. The Minkowski metric can then be naturally defined on the newly obtained manifold at every point except at $r=0$. Except at this point, the curvature tensor vanishes everywhere. Hence, even without any bad behavior of the curvature scalars, one can have a singularity. Therefore, describing the singularity only as a blow-up of some variable as one approaches it is not such a good idea. 
     
    \item Formation of ``holes" in the spacetime after removing the singularities can be proposed as a way to identify a spacetime singularity. These holes can be detected by the existence of inextendible geodesics\footnotemark, in at least one direction, with a finite affine parameter. We call such geodesics incomplete. However, one can show an example in which incompleteness does not happen for all types of geodesics, i.e., there is a hole for a timelike geodesic; however, a null / spacelike geodesic can easily be extended beyond this so-called hole.
    \cite{Wald_84}.
    \footnotetext{The geodesic is said to be future (past) inextendible if it has no future (past) endpoints in the manifold $\mathcal{M}$.}
    
    \item A later proposal of the definition of the spacetime singularity was made by Ellis and Schmidt
    \cite{Ellis_77}
    in 1977, as follows: If any object hits the singularity and is crushed to zero volume, then it is called a ``strong" singularity. 
    
    The mathematically precise statement given by Tipler 
    \cite{Tipler_77} 
    for such singularity is as follows: 

    \textit{Let $\mathcal{M}$ be a smooth manifold of four dimensions along with a smooth metric $g$ with Lorentz signature $(-,+,+,+)$ defined on it. For a causal geodesic $\gamma: [t_0,0) \rightarrow \mathcal{M}$, the volume element defined by wedge product of three independent Jacobi fields along $\gamma$, in a case $\gamma$ is a timelike geodesic (two independent Jacobi field in a case $\gamma$ is null geodesic), should approach to zero as $\lambda \rightarrow 0$, where $\lambda$ is the affine parameter along the geodesic.}

    We call such singularity ``Tipler" strong. Sufficient criterion for a singularity to be strong in this sense was provided by Clarke and Krolak 
    \cite{Clarke_85}.
    The criterion says that at least along one null geodesic, the following inequality needs to be satisfied:
   \begin{equation} \label{Krolak and Clarke criteria}
    \lim_{ \lambda \to 0} \lambda ^2R_{ij}K^iK^j>0.
    \end{equation}
    Here, $\lambda$ is the affine parameter along the null geodesic with $\lambda=0$ at the singularity, and $K^i$ are tangents of the null geodesics.
    
    \item The singularity as defined by Ellis, Shmidt, and Tipler 
    \cite{Ellis_77, Tipler_77} 
    involves vanishing of the volume element formed by independent Jacobi fields along the non-spacelike geodesic as it terminates in a strong singularity, rather than the behavior of individual Jacobi fields, as pointed out by Nolan 
    \cite{Nolan_99}
    in 1999.
    One can show examples of physically strong singularity wherein the volume element does not vanish and is classified as ``Tipler weak". These examples led Ori 
    \cite{Ori_99} 
    to redefine the physically strong singularity, which extends the class of strong singularity by including cases in which any of the Jacobi fields is unbounded
    \cite{Nolan_99, Ori_99}.
    Such singularities are termed as ``deformationally strong" singularities.
\end{enumerate}

The singularity theorems provided by Penrose and Hawking 
\cite{Hawking_73, Penrose_1965} 
prove that singularities could indeed form under very generic conditions in gravitational collapse as well as cosmology. These theorems suggest the existence of singularities in the universe.

It is worth noting that the singularity theorem can be interpreted in two different ways. One could interpret it as proof of the existence of the regime in which general relativity breaks down. According to this viewpoint, the existence of singularities cannot be accepted 
\cite{Wheeler_64, Hawking_79, Bergmann_80}.
Another viewpoint, proposed by Misner
\cite{Misner_1969},
says that the general relativistic predictions of the singularity and its properties should be taken into account as it may tell us about what one should expect from some modification in the general theory of gravity, which works in the regime of a strong field, e.g., a quantum theory of gravity
\cite{Burko_97}. 
The work presented in this thesis follows the latter viewpoint.

\section{Astrophysical horizons}\label{1.3}
A singularity formed due to an unhindered gravitational collapse may or may not be visible. By visible singularity, we mean that one can trace a non-spacelike geodesic initiating from the singularity to a non-singular region. Whether or not we can trace such geodesic depends on the formation of trapped surfaces around the singularity. 

Consider a setup where we have a congruence of null geodesics. Let us determine the evolution of the Jacobian vector (deviation vector) $\xi$, which connects two neighboring null geodesics with the affine parameter $\lambda$ as shown in Fig.(1.1). We know that 
\begin{equation}\label{relativefourvelocity}
    \xi^{\mu}_{;\nu}l^{\nu}=l^{\mu}_{;\nu}\xi^{\mu}.
\end{equation}
This is because the expressions on both sides of the equality in the above equation represents the same object, i.e. the relative four velocity. $ \xi^{\mu}_{;\nu}l^{\nu}$ reflects the parallel transport of the deviation vector along the null geodesic, and $l^{\mu}_{;\nu}\xi^{\mu}$ reflects the parallel transport of the tangent to the null geodesic along the deviation vector. Let us now define a tensor field
\begin{equation}\label{tenserfield}
    B_{\mu \nu}=l_{\mu;\nu}.
\end{equation}
Therefore we have
\begin{equation}\label{unabletoparalleltransport}
  \xi^{\mu}_{;\nu}l^{\nu}=B^{\mu}_{\nu}\xi^{\mu}.
\end{equation}
From the above equation, one can see that the tensor field $B^{\mu}_{\nu}$ tells us how the Jacobian vector $\xi^{\mu}$ is unable to be transported in a parallel manner along the null geodesic congruence.

Consider the tensor field 
\begin{equation}\label{transversemetric}
    h_{\mu \nu}=g_{\mu\nu}+l_{\mu} n_{\nu}+n_{\mu}l_{\nu}. 
\end{equation}
Here $n_{\mu}$ is a null vector field such that
\begin{equation}
     l^{\nu}n_{\nu}\neq 0.
\end{equation}
$ h_{\mu \nu}$ is a purely spatial component of the spacetime metric $g_{\mu\nu}$. We call it the transverse metric. One can see that $B_{\mu \nu}$ is orthogonal to $l_{\mu}$ but not orthogonal to $n^{\mu}$. Hence $B_{\mu \nu}$ has non-spatial contributions, which can be removed by twice contracting it by the the transverse metric and getting a new spatial tensor field as
\begin{equation}\label{contractB}
    \Tilde{B}_{\mu\nu}=h^{\alpha}_{\mu}h^{\beta}_{\nu}B_{\alpha \beta}.
\end{equation}
\begin{figure}\label{congruence}
\centering
\includegraphics[scale=0.4]{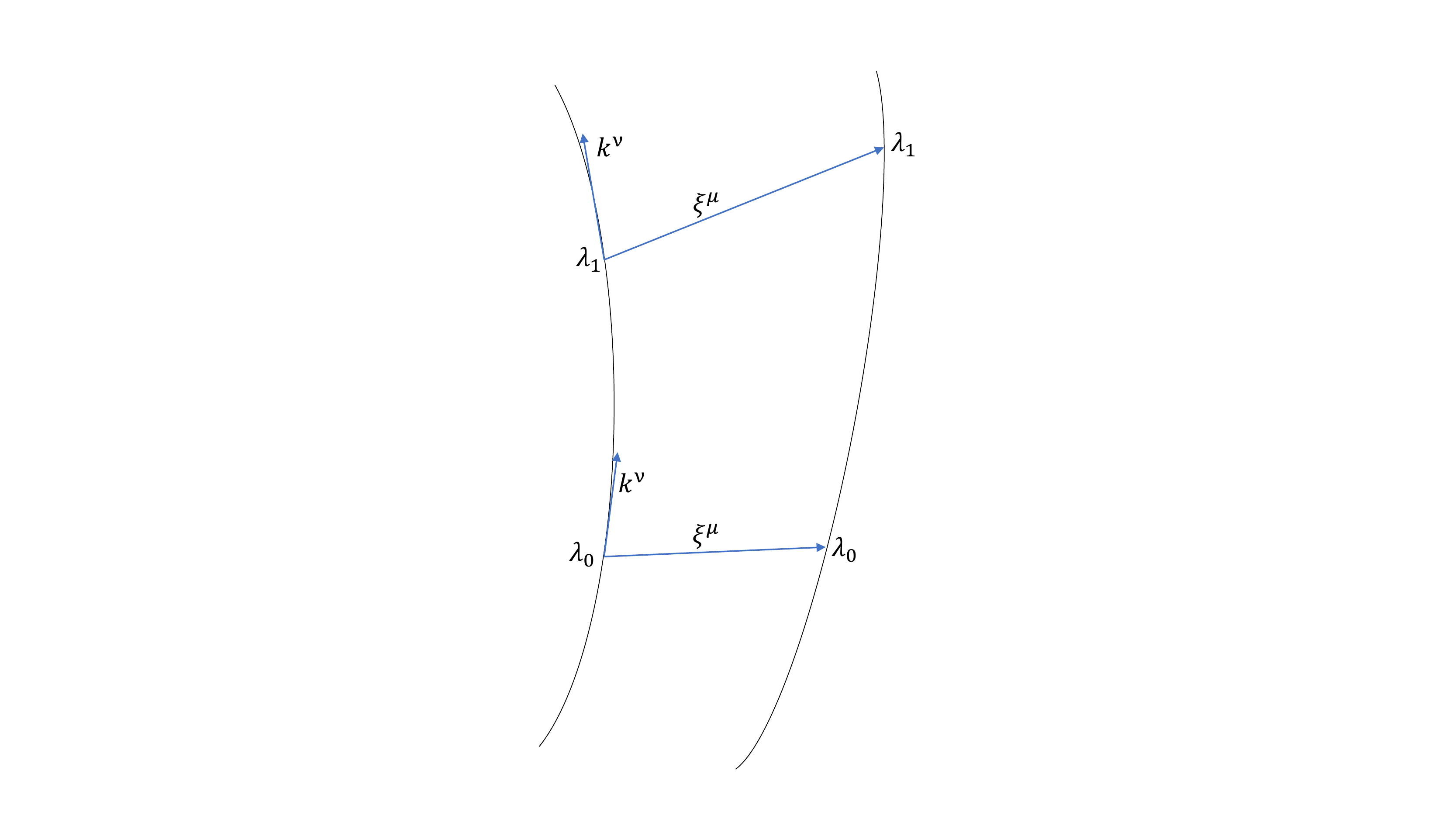}
\caption{Jacobian vector $\xi^{\mu}$ between two neighbouring null geodesics of the congruence, with affine parameter $\lambda$. $\kappa^{\nu}$ is the tangent to the geodesic.}
\end{figure}
This purely spatial tensor field can then be decomposed into minimal parts as
\begin{equation}
    \Tilde{B}_{\mu \nu}=\frac{1}{2}\theta h_{\mu\nu}+\sigma_{\mu\nu}+\omega_{\mu\nu}.
\end{equation}
As mentioned before and seen from Eq.(\ref{unabletoparalleltransport}), each component on the RHS of the above equation tells us how the null geodesic congruence evolves (The Jacobian vector fails to be parallelly transported along the null geodesic congruence if $B^{\mu}_{\nu}$ is not a null $3\times 3$ matrix). We discuss about each of these components as follows:
\begin{enumerate}
    \item  The first term contains the expansion scalar $\theta$ and is written as
    \begin{equation}\label{thetaB}
        \theta=\Tilde{B}^{\mu}_{\nu}.
    \end{equation}
    Physically, it is the fractional rate of change of the cross-section area of the null geodesic congruence. A positive value of $\theta$ hence means that this fractional change is positive, indicating that the null geodesic congruence is diverging outwards as we move forward with increasing affine parameter $\lambda$. Similarly, a negative value of $\theta$, likewise, indicates that the null geodesic congruence is converging inwards as we move forward with increasing affine parameter $\lambda$. 
    \item The second term is called the sheer parameter and is written as
    \begin{equation}
        \sigma_{\mu\nu}=\frac{1}{2}\left(\Tilde{B}_{\mu\nu}+\Tilde{B}_{\nu\mu}-\frac{1}{3}\theta h_{\mu\nu}\right).
    \end{equation}
    Physically, it indicates the sheering of the cross-section of the null geodesic congruence. $\sigma_{\mu\nu}$ does not contribute to the change in the area of this cross-section.
    
    \item The third term is called the rotation parameter and is written as
    \begin{equation}
        \omega_{\mu\nu}=\frac{1}{2}\left(\Tilde{B}_{\mu\nu}-\Tilde{B}_{\nu\mu}\right).
    \end{equation}
    Physically, it indicates the overall rotation of the cross section of the null geodesic congruence. $\omega_{\mu\nu}$ does not contribute to the change in the area and the shape of this cross section. 
\end{enumerate}
If we consider a sheer free and rotation free tensor field $\Tilde{B}$, then from Eq.(\ref{transversemetric}, \ref{contractB}, \ref{thetaB}), we get
\begin{equation}
    \theta=l^{\mu}_{;\mu},
\end{equation}
which can be rewritten in usable form as
\begin{equation}\label{thetausable}
    \theta=\left(g^{\mu \nu}-\frac{l^{\mu}n^{\nu}+l^{\nu}n^{\mu}}{l^{\alpha}n_{\alpha}}\right) l_{\nu;\mu}.
\end{equation}
Let us come back to the singularity formed due to unhindered gravitational collapse of of a massive cloud. Consider the outgoing null geodesic congruence from the singularity. Using Eq.(\ref{thetausable}) one can calculate the expansion scalar of this congruence. 

Trapped surfaces are those hypersurfaces on which $\theta<0$. The boundary of all the trapped surfaces is called the apparent horizon. On the apparent horizon, $\theta=0$. On the other side of the apparent horizon $\theta>0$. Hence one can get the evolution of the apparent horizon for a collapsing cloud by substituting $\theta=0$ in Eq.(\ref{thetausable}). 

The dynamics of the apparent horizon determine the causal structure of the singularity, at least locally. By locally, we mean that there could exist a situation in which the outgoing singular null geodesic has a positive expansion scalar at the center. Still, later in its evolution, the expansion scalar becomes negative, and hence, the geodesics fall back to the singularity. In such scenarios, the singularity is said to be only locally visible. The outgoing singular null geodesic cannot cross the boundary of the collapsing cloud, thereby making the singularity hidden from an asymptotic observer. A comoving observer inside the collapsing cloud can, however, observe the singularity. 

Visibility of the singularity in the global sense is determined by the behavior of the event horizon, which is an infinitely redshifted null surface, evolving from the center of the collapsing cloud towards its boundary and coinciding with the apparent horizon at the boundary. 

Consider a spherically symmetric collapsing cloud governed by spacetime metric with no non-diagonal terms, as follows:
\begin{equation}\label{metricnonondiagonal}
    ds^2=-e^{2\nu(t,r)}dt^2+e^{2\psi(t,r)}dr^2+R^2(t,r)d\Omega^2,
\end{equation}
where
\begin{equation}
   d\Omega^2= \left(d\theta^2+\sin^2{\theta}d\phi^2\right).
\end{equation}
%%%%%%%%%%%%%%%%%%%%%%%%%%%%%%%%%%%%%%%%%%%%%%%%%
The evolution of the event horizon for such a collapsing cloud is obtained from the above metric by substituting $ds^2=d\Omega^2=0$. We thus get a differential equation governing the outgoing null geodesic given by,
\begin{equation}
    \frac{dt}{dr}=e^{\psi-\nu}.
\end{equation}
The event horizon is then the solution of this first-order linear differential equation satisfying the initial condition $\theta(r_c)=0$. Here, $r_c$ is the largest comoving radius of the collapsing cloud, or the comoving radius of the boundary of the collapsing cloud, in other words. In the succeeding chapters, we will discuss further the role of event horizon in determining the causal structure of the singularity. Before that, we discuss the most simple model of the gravitational collapse in the following section.

\section{Oppenheimer-Snyder-Datt (OSD) Collapse}\label{1.4}
The first attempt to describe the process of gravitational collapse to a black hole was made B. Datt in 1938 \cite{Datt_1938}, and independently by J. Robert Oppenheimer and his student Hartland Snyder in 1939 \cite{Oppenheimer_1939}. The collapsing star was modeled by homogeneous dust: a zero pressured spherical matter cloud with uniform density. The spacetime governing the homogeneous spherical dust cloud is the Friedmann-Lemaître-Robertson-Walker(FLRW) metric in the comoving coordinates, given by:
\begin{equation}\label{OSDmetric}
    ds^2 = -dt^2 +\frac{a(t)^2}{1-k r^2} (dr^2 + r^2 d\Omega^2),
\end{equation}
where
\begin{equation}
    d\Omega = d\theta ^2 + \sin^{2}\theta d\phi ^{2}.
\end{equation}
Here, $k\in \{-1,0,1\}$. It represents the spatial curvature. The first of the Einsteins field equation gives the density $\rho=\rho(t)$ as
\begin{equation}\label{OSrho}
    \rho=\frac{F'}{r^2a^3},
\end{equation}
where
\begin{equation}\label{OSFmetric}
    F= r a \left(r^2 \dot a^2+kr^2\right)
\end{equation}
is called the Misner-Sharp mass function. It gives us the information about the mass of the cloud inside a spherical collapsing shell corresponding to the radial coordinate $r$, at time $t$.
The second Einstein's field equation relates the pressure of the collapsing cloud with the function of metric components and its derivatives as
\begin{equation}
    p= -\frac{\dot F}{r^3 a^2 \dot a}.
\end{equation}
However, since the collapsing matter cloud under consideration is pressureless, one can conclude that 
\begin{equation}
    \dot F=0,
\end{equation}
which means that $F=F(r)$ and does not depend on $t$. Thus, in the case of dust collapes, the mass inside a collapsing shell of fixed radial coordinate $r$ is conserved. Additionally, one can see from the expression of the density in Eq.(\ref{OSrho}) that $F'\sim O(r^2)$ so that $\rho$ is independent of $r$, and only evolves with the scale factor $a(t)$, thereby representing the density of a homogeneous collapsing cloud. Hence, we have
\begin{equation}
    F=F_0 r^3,
\end{equation}
where $F_0>0$. Now integrating Eq.(\ref{OSFmetric}) with respect to time, we get
\begin{equation}\label{OStminusts}
    t-t_s=-\frac{a^{\frac{3}{2}}}{\sqrt{F_0}}\mathcal{G}\left(\frac{k a}{F_0 r^2}\right).
\end{equation}
Here $\mathcal{G}(y)$ is defined as follows:
\begin{equation}\label{G}
    \begin{split}
        & \mathcal{G}(y)= \left(\frac{\text{arcsin}\sqrt{y}}{y^{\frac{3}{2}}}-\frac{\sqrt{1-y}}{y} \right) \hspace{0.5cm} \text{for} \hspace{0.5cm} 0<y<1, \\
        & \mathcal{G}(y)=\frac{2}{3} \hspace{0.5cm} \text{for} \hspace{0.5cm} y=0, \\
        & \mathcal{G}(y)= \left(\frac{-\text{arcsinh}\sqrt{-y}}{(-y)^{\frac{3}{2}}}-\frac{\sqrt{1-y}}{y} \right) \hspace{0.5cm} \text{for}  -\infty<y<0. 
    \end{split}
\end{equation}
Here $t_s$ is the constant of integration and gives us the time of formation of the singularity. It is expressed as
\begin{equation}\label{OSts}
    t_s=\frac{1}{\sqrt{F_0}}\mathcal{G}\left(\frac{k}{F_0}\right).
\end{equation}
From here, we can see that all the concentric spherical shells, each corresponding to the radial coordinate $r$, collapse simultaneously. 

We will see in sec.(\ref{1.6}), and the succeeding chapters that this is not always the case. To give a flavor here, if there is inhomogeneity involved in the density of the collapsing cloud such that the density decreases with increasing $r$, then the smaller shell collapses to a singularity before the larger shell. Hence, in general, we have a singularity curve $t_s(r)$, which gives us the time of the collapse of a shell of radial coordinate $r$ to a singularity.

The surrounding space-time is governed by the Schwarzschild metric, which, according to Birkhoff's theorem, is a unique static, spherically symmetric, vacuum, asymptotically flat spacetime. The line element is written as
\begin{equation}
    ds^2=-(1-\frac{r_s}{r})dt^2+(1-\frac{r_s}{r})^{-1}dr^2+r^2d\Omega^2.
\end{equation}
Here $r_s$ is called the Schwarzschild radius and is twice the total mass of the collapsing cloud in the units $c=G=1$.
\begin{figure}\label{OSspacetimeplot}
\centering
\includegraphics[scale=0.5]{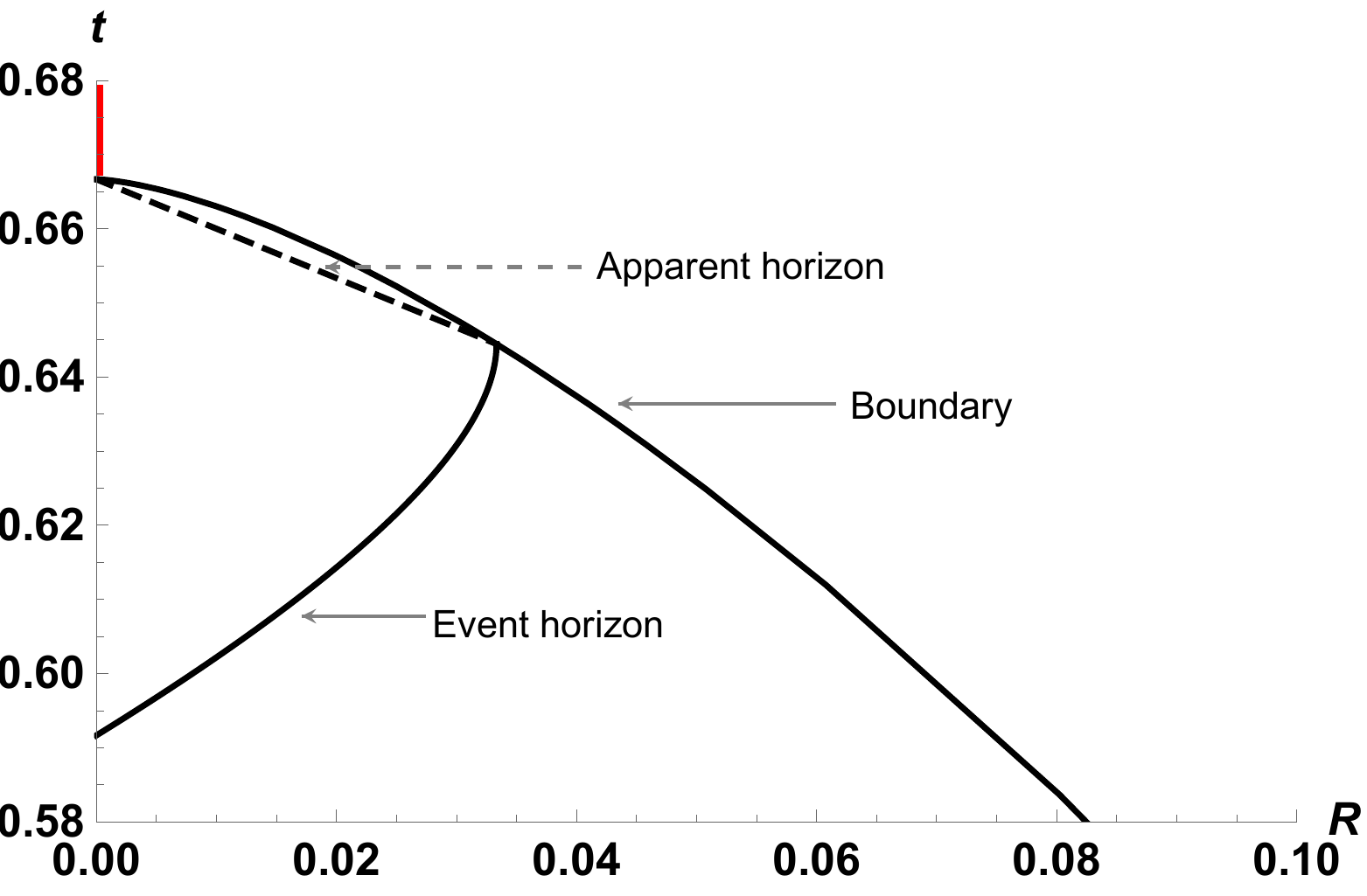}
\caption{Spacetime plot of the Oppenheimer-Snyder-Dutt collapse. Here, $R$ and $t$ are the physical radius of the collapsing cloud and the comoving time, respectively. The comoving radius of the boundary is $r_c=0.32813$. The Misner-Sharp mass function is $F=F_0 r^3$, where $F_0=1$. The collapse is marginally bound, i.e., $f=0$. The red segment on the vertical axis represents the worldline of the singularity.}
\end{figure}
%%%%%%%%%%%%%%%%%%%%%%%%%%%%%%%%%%%%%%%%%%%%%%%%%
%%%%%%%%%%%%%%%%%%%%%%%%%%%%%%%%%%%%%
Let us call the FLRW metric governing the spacetime of the collapsing cloud as the internal metric ($g_{\mu \nu}^{-}$), and the Schwarzschild metric governing the surrounding spacetime, as the external metric ($g_{\mu \nu}^{+}$). 

Suppose we desire to make the union of internal and external metrics a valid solution of Einstein's field equation. In that case, certain conditions need to be imposed on the otherwise free functions of the metric components. These are called junction conditions. 

The first junction condition dictates that the induced metric must be the same on both sides of the hypersurface partitioning the two regions governed by the internal and external metric, respectively. Maintaining this condition eliminates the singular term arising in $g_{\mu \nu, \gamma}$,
where $g_{\mu \nu}$ incorporates both the internal and the external metric such that 
\begin{equation}
    g_{\mu \nu}=\Theta(m)g_{\mu \nu}^{+}+\Theta(-m) g_{\mu \nu}^{-}.
\end{equation}
Here $\Theta(m)$ is called the Heaviside distribution function defined as
\begin{equation}
    \begin{split}
        \Theta(m) &= +1, \hspace{2.8cm} m>0,\\
                  &= 0, \hspace{3cm} m<0,\\
                  &=\textrm{indeterminite},\hspace{1cm} m=0.
    \end{split}
\end{equation}

The second junction condition dictates that the extrinsic curvature $K_{\mu \nu}$ should be the same on both sides of the hypersurface. Maintaining this condition eliminates the surface stress-energy tensor, thereby keeping the separating hypersurface non-singular. However, it should be noted that the violation of the second junction condition has an acceptable physical interpretation that the separating hypersurface is a thin shell with a non-vanishing stress-energy tensor. These junction conditions can put constraints on the otherwise free functions in the metric components.

%%%%%%%%%%%%%%%%%%%%%%%%%%%%%%%%%%%%%%%%%%%%%%%%%%%%%%%%%%%%%%%%%%%%%%%%%%%%%%%%%%%%%%%%%%%%%%%%%%%%
\begin{figure*}\label{OSdensityplot}
{\includegraphics[scale=0.315]{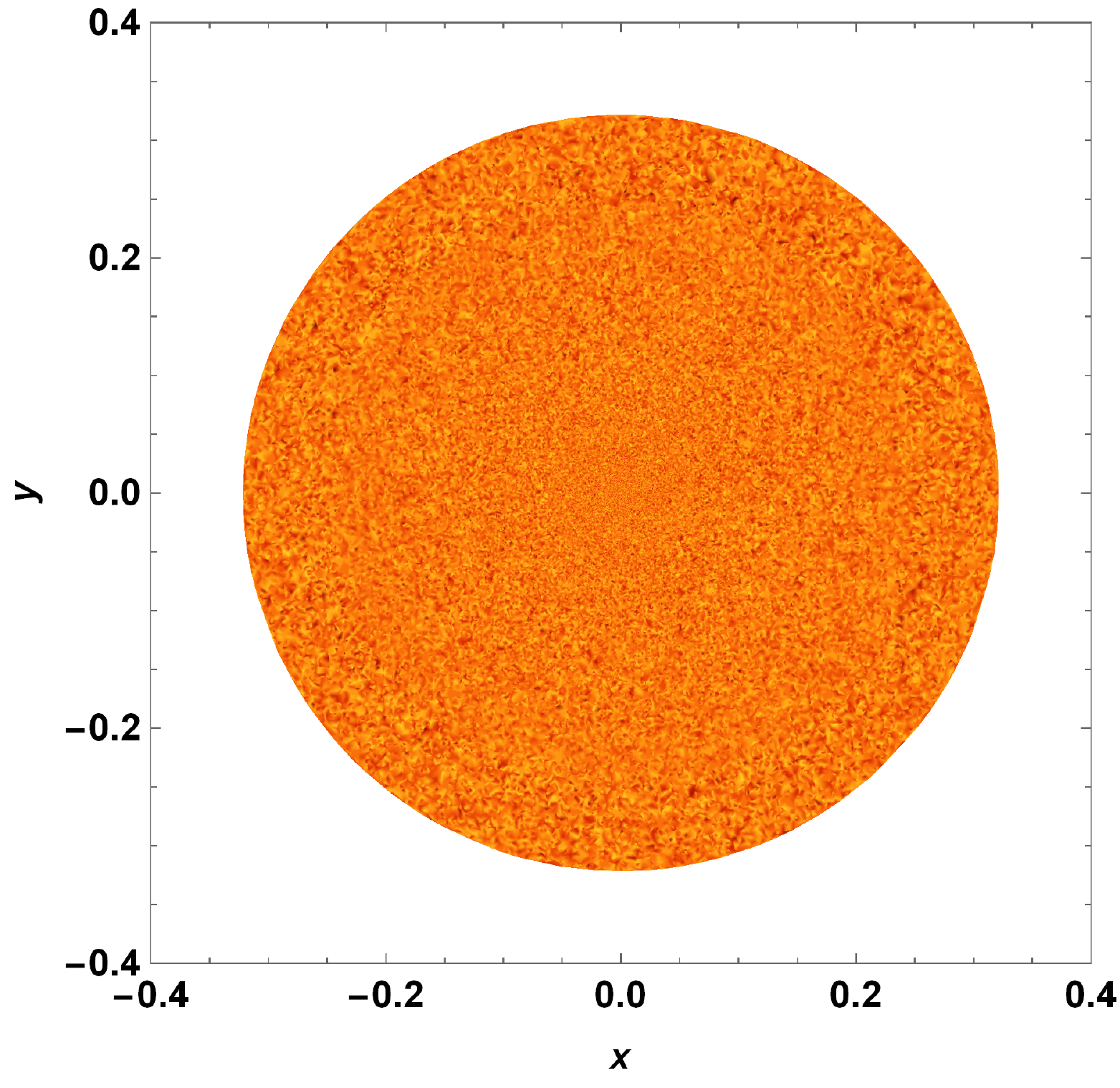}\hspace{0.5cm}\includegraphics[scale=0.315]{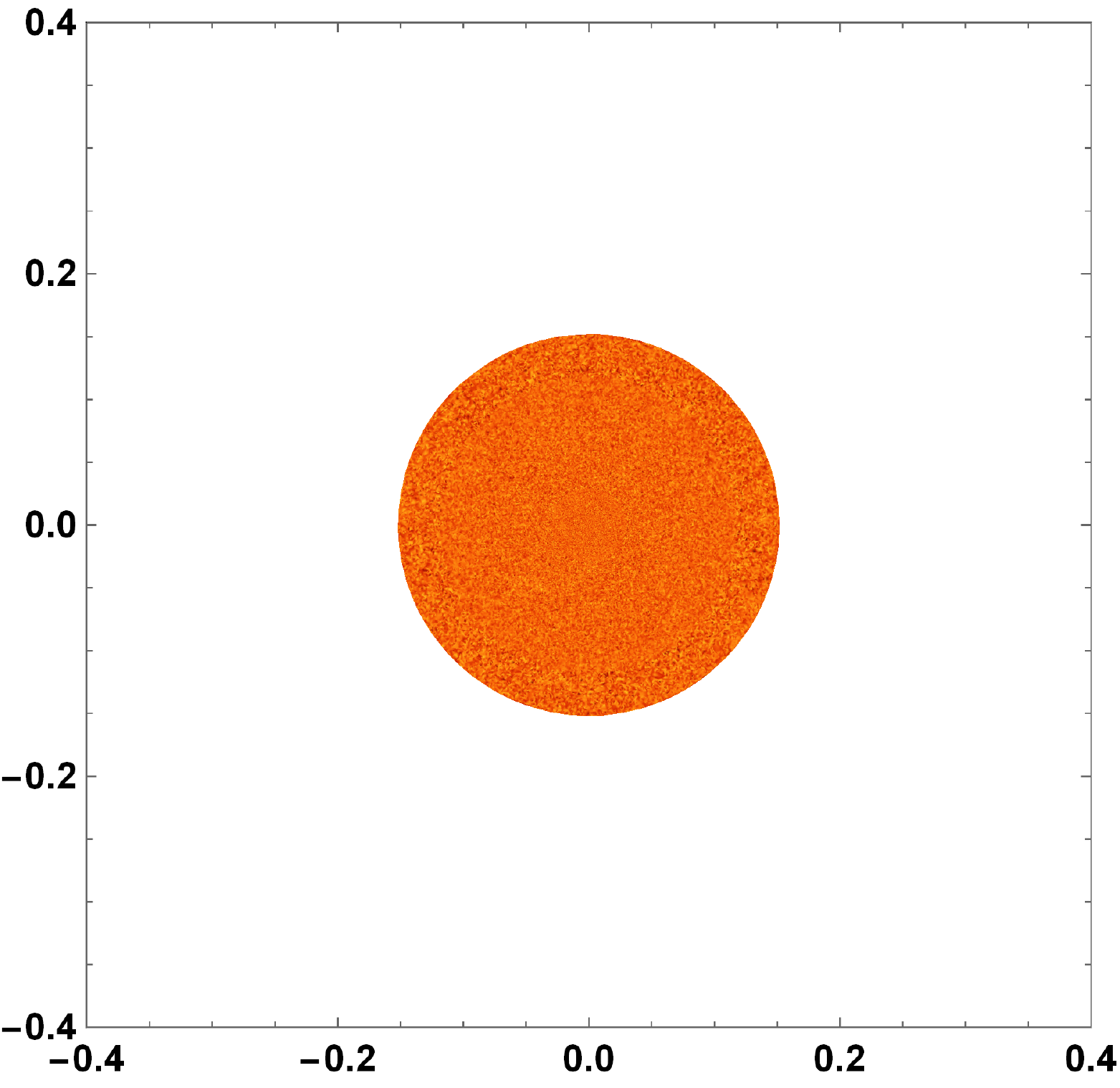}\hspace{0.5cm}\includegraphics[scale=0.315]{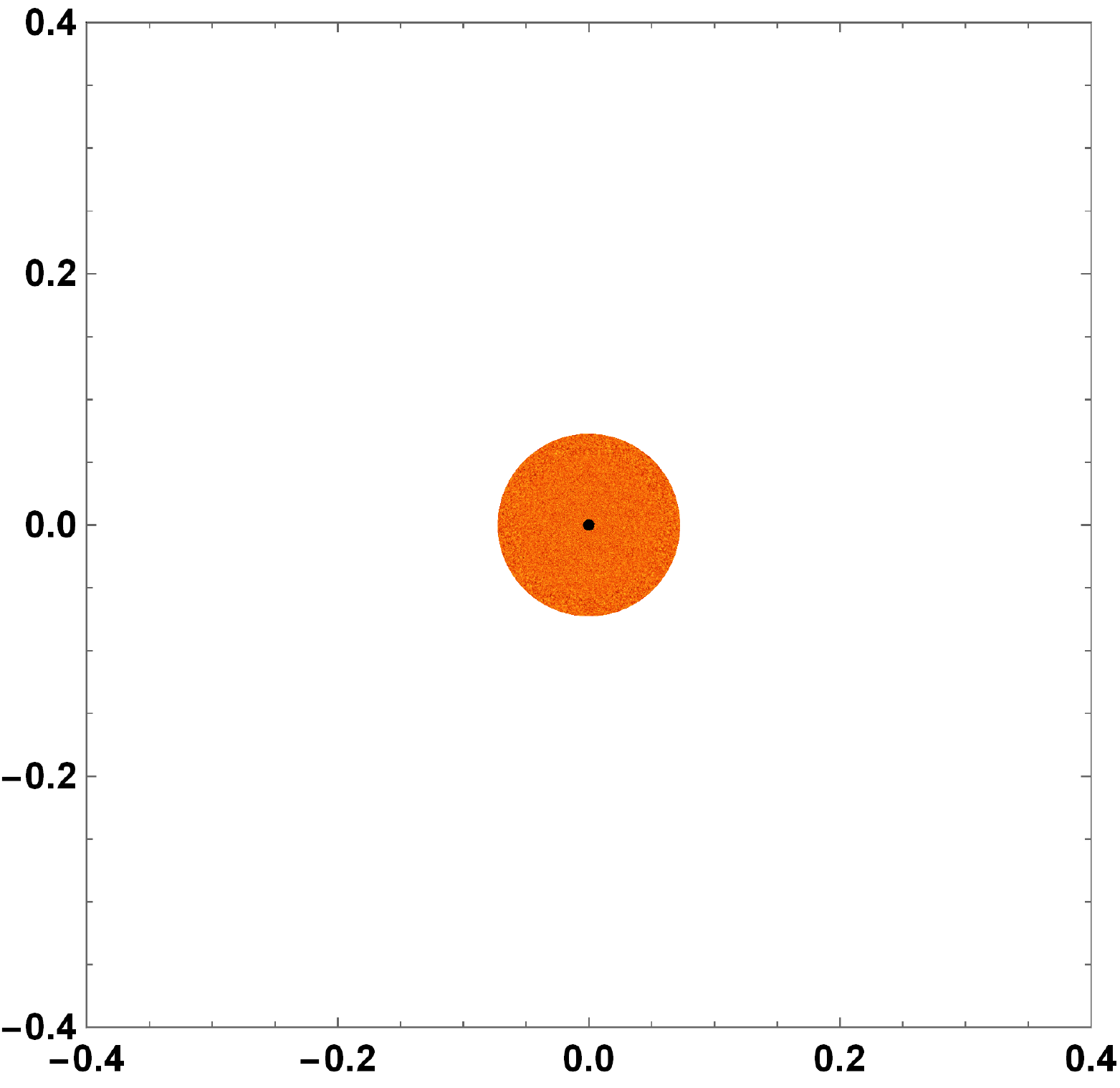}}\\
{\includegraphics[scale=0.315]{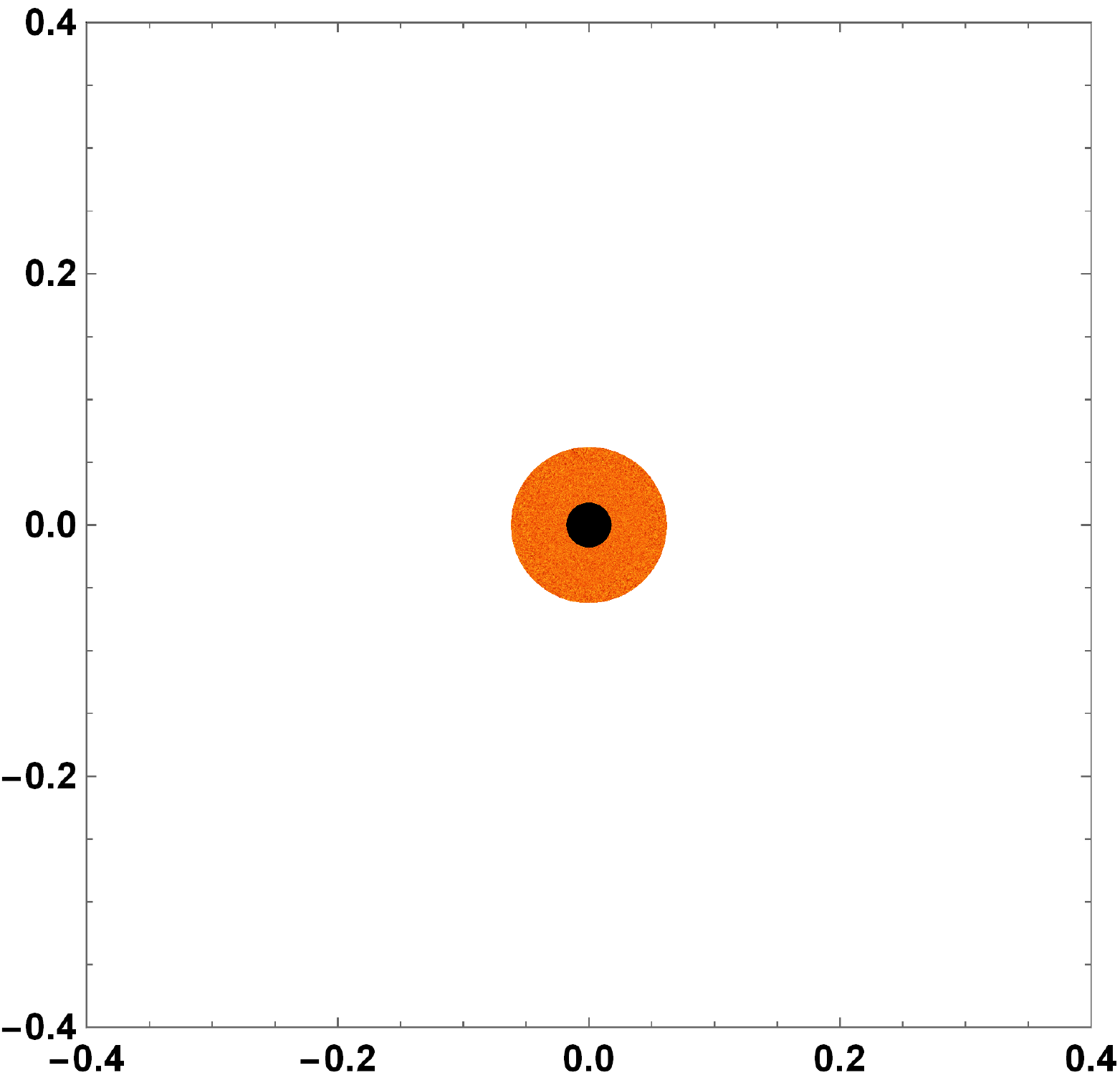}\hspace{0.5cm}\includegraphics[scale=0.315]{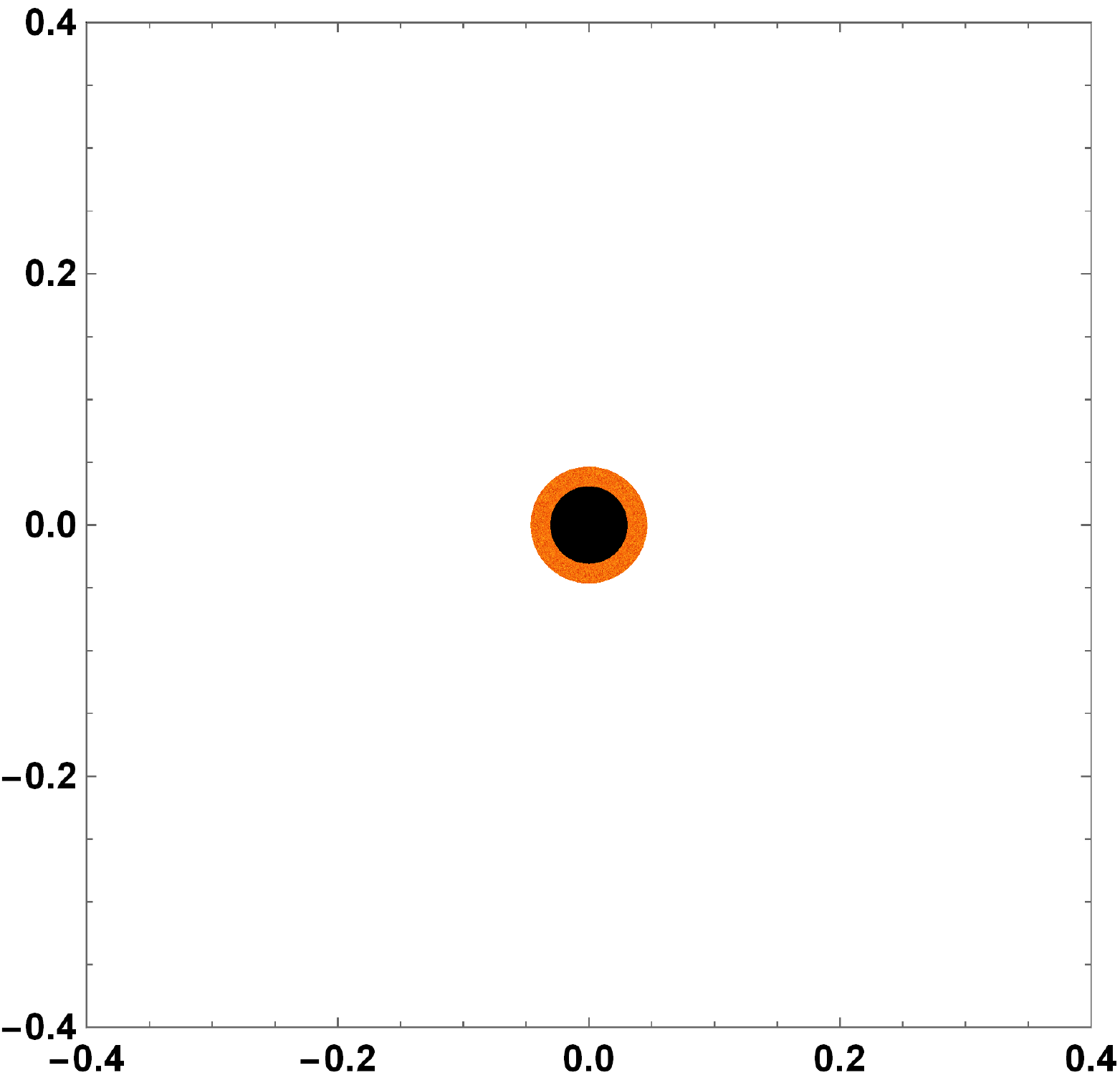}\hspace{0.5cm}\includegraphics[scale=0.315]{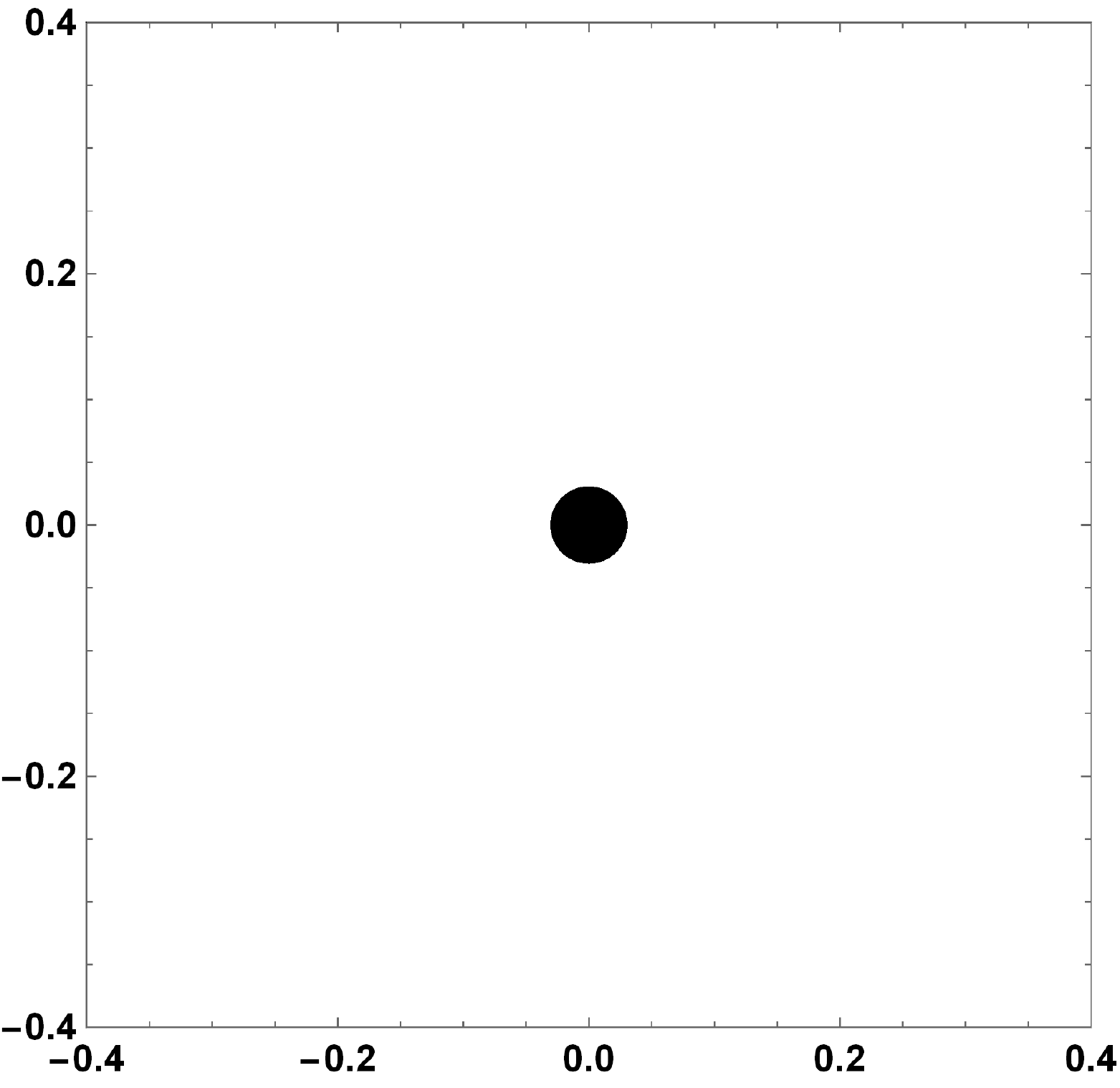}}
\caption{Density plot of the Oppenheimer-Snyder-Dutt collapse is depicted here. The evolution of the event horizon is depicted by the central black disk. The comoving radius of the boundary is $r_c=0.32813$. The Misner-Sharp mass function is $F=F_0 r^3$, where $F_0=1$. The collapse is marginally bound, i.e. $f=0$.}
\end{figure*}
%%%%%%%%%%%%%%%%%%%%%%%%%%%%%%%%%%%%%%%%%%%%%%%%%%
%%%%%%%%%%%%%%%%%%%%%%%%%%%%%%%%%%%%%%%%%%%%%%%%%%
%%%%%%%%%%%%%%%%%%%%%%%%%%%%%%%%%%%%%%%%%%%%%%%%%%%%%%%%%%%%%%%%%%%%%%%%%%%%%%%%%%%%%%%%%%%%%%%%%%%
The evolution of the apparent horizon in the case of OSD collapse can be found using Eq.(\ref{thetausable}) which in the metric under consideration (Eq.(\ref{OSDmetric})) becomes
\begin{equation}
    \theta=2\left(\frac{\dot a}{a}+\frac{\sqrt{1-k}}{r a}\right).
\end{equation}
Using Eq.(\ref{OSFmetric}), the expression of the expansion scalar of the outgoing singular null geodesic can be rewritten as
\begin{equation}
    \theta=\frac{2}{ra}\left(\sqrt{1-k}-\sqrt{\frac{F}{ra}-k}\right).
\end{equation}
Hence for $\theta$ to be zero, 
\begin{equation}\label{Fra}
    F=r a.
\end{equation}
For $k=0$, the scale factor of the collapsing homogeneous dust cloud is obtained using Eq.(\ref{OStminusts}) and Eq.(\ref{OSts}) as
\begin{equation}\label{OSa}
    a(t)=\left(1-\frac{3}{2}\sqrt{F_0}t\right)^{\frac{2}{3}}.
\end{equation}
Now, from Eq.(\ref{Fra}) and Eq.(\ref{OSa}), the evolution of the apparent horizon is obtained as
\begin{equation}\label{OSDah}
    t_{AH}(r)=\frac{2}{3}\left(\frac{1}{\sqrt{F_0}}-r^3F_0 \right).
\end{equation}
It can be seen from the above equation that the apparent horizon corresponding to the larger comoving radius forms earlier than that of the smaller comoving radius. This behavior is depicted in Fig.(1.2). Such, however, is not always the case. We will see in the succeeding chapters that introducing inhomogeneity in the collapsing cloud, such that the density is more towards the center and reduces as we move away from the center and towards the boundary, can change the dynamics of the apparent horizon such that the time of its formation is a monotone increasing function of the comoving radius. Hence, the apparent horizon evolves from the center towards the boundary of the collapsing cloud.  Such dynamics can make the singularity naked, at least to some observer inside the collapsing cloud.  

The evolution of the event horizon is obtained by substituting $ds^2=d\Omega^2=0$ in the metric mentioned in Eq.(\ref{OSDmetric}). We then get the following differential equation:
\begin{equation}
    \frac{dt}{dr}=\frac{ra}{\sqrt{1-kr^2}}.
\end{equation}
Here, $a$ is obtained using Eq.(\ref{OSa}). For $k=0$, the general solution of this differential equation is obtained as
\begin{equation}\label{OSDng}
    t_{NG}(r)=\frac{2}{3\sqrt{F_0}}+\frac{F_0 r^3}{12}+\frac{F_0 r^2 C}{16^{\frac{1}{3}}}+\frac{F_o r C^2}{4^{\frac{1}{3}}}+\frac{F_0 C^3}{3}.
\end{equation}
This equation represents the time evolution of the null geodesic from the center of the collapsing cloud. The center is non-singular in the OSD collapse. Here $C$ is a constant of integration. To obtain the evolution of the event horizon, we consider a particular curve that coincides with the apparent horizon at the boundary $r_c$ of the collapsing cloud. We thus obtain the value of $C$ for which 
\begin{equation}\label{OSDaheh}
    t_{AH}(r_c)=t_{NG}(r_c),
\end{equation}
Using Eq.(\ref{OSDah}), Eq.(\ref{OSDng}) and Eq.(\ref{OSDaheh}), the value of $C$ is obtained as
\begin{equation}\label{OSDC}
    C=-\frac{3 r_c}{4^{\frac{1}{3}}}.
\end{equation}
Substituting $C$ from Eq.(\ref{OSDC}) in Eq.(\ref{OSDng}), the event horizon curve $t_{EH}(r)$ is finally obtained as
\begin{equation}\label{OSDteh}
    t_{EH}(r)=\frac{2}{3\sqrt{F_0}}+\frac{F_0}{12}\left(r-3r_c\right)^3.
\end{equation}
It is clearly seen from Eq.(\ref{OSts}) and Eq.(\ref{OSDteh}) that at $r=0$, 
\begin{equation}
    t_{EH}(0)<t_s.
\end{equation}
Hence, in the OSD collapse, the event horizon at the center forms before forming the singularity. This dynamics is represented in Fig.(1.2). The event horizon is a one-way membrane since there is no causal connection from inside the event horizon to the outside universe. Hence the singularity formed in the OSD collapse cannot be seen from the outside universe. It is worth mentioning here that the necessary criteria for a singularity to be visible in the global sense is to have the event horizon form simultaneously with the singularity at $r=0$. 

This model of homogeneous pressureless collapsing cloud gave rise to the concept of ``\textit{Black hole}". It is, however, far from the reality due to its simplicity. E.g., a real star could have nonzero pressure, and the density is expected to be higher at the center and decrease outwards.  However, not much attention was paid in this direction until recently, mainly due to the complexities involved in dealing with the Einsteins' field equations, which are generally non-linear second-order partial differential equations. Sec.(\ref{1.6}) discusses recent developments in the study of gravitational collapse and formalism to deal with the complexities involved. But before that, let us give a brief overview of the cosmic censorship hypothesis.

\section{Cosmic censorship and violation of causality}\label{1.5}
Once the existence of singularities is assumed, the next step is to comprehend the nature of the singularity. One such property that needs to be investigated is the visibility of the singularity 
\cite{Singh_1999, Joshi_2000, Harada_2004, Joshi_07,  Goswami_2007}. 
It is known that the big bang singularity is visible in principle because we can see the null and timelike geodesics coming from it.   However, it was still unclear whether or not the singularities arising from gravitational collapse are necessarily censored completely from the outside universe by an event horizon. Penrose
\cite{Penrose_69}, 
in 1969, proposed what is now known as the cosmic censorship hypothesis, which is expressed in two forms: 
\begin{enumerate}
    \item The weak cosmic censorship hypothesis 
    \cite{Tipler_80}, suggest that a singularity can never be globally naked, i.e., it can never be visible to faraway asymptotic observers.  
    \item The strong cosmic censorship hypothesis 
    \cite{Geroch_79, Hawking_79(2), Penrose_79},
states that the singularity can never be locally naked (i.e., not even an observer comoving with the collapsing cloud can see the singularity).
\end{enumerate}

We now discuss an important motivation of the cosmic censorship hypothesis. The causal relationship between two events in a given spacetime is based on the spacetime metric $g_{\mu\nu}$. A timelike (non-spacelike)  worldline in the spacetime diagram such that its ending point is the same as its starting point is called a closed timelike (non-spacelike) curve. Such curves are counterintuitive because a person can then travel to their past, and hence, their existence is undesirable. Einsteins' equation, however, does not have a say in the existence/non-existence of global spacetime metrics, which contain closed non-spacelike curves. Some of the examples of the solutions of the Einsteins' field equation are the Godel metric
\cite{Godel_49}, 
and the Tipler cylinder
\cite{Tipler_76}.

To understand the relationship between cosmic censorship and causality violation, we require some definitions as follows
\cite{Hawking_73, Joshi_07}:

Consider a spacetime $(\mathcal{M},g)$. If $\exists$ a timelike curve which is future directed and smooth from $a$ to $b$, then we say that $a$ chronologically precedes $b$. It is denoted by $a<<b$. Similarly, if this future directed smooth curve is nonspacelike (timelike or null), instead of only timelike, then we say that $a$ causally precedes $b$. It is denoted by $a<b$. The chronological future $\mathcal{I}^+$, and the chronological past $\mathcal{I}^-$ are then defined respectively as
\begin{equation}
    \mathcal{I}^+(a)=\{b\in \mathcal{M}; a<<b\},
\end{equation}
and
\begin{equation}
    \mathcal{I}^-(a)=\{b\in \mathcal{M}; b<<a\}.
\end{equation}
Similarly, the causal future $\mathcal{J}^+$, and the causal past $\mathcal{J}^-$ are defined respectively as
\begin{equation}
    \mathcal{J}^+(a)=\{b\in \mathcal{M}; a<b\},
\end{equation}
and
\begin{equation}
    \mathcal{J}^-(a)=\{b\in \mathcal{M}; b<a\}.
\end{equation}
As discussed before, causality violation is not so desirable property of the spacetime $\mathcal{M}$. Various causality conditions are hence imposed on $\mathcal{M}$. These conditions are discussed as follows: 
\begin{itemize}
    \item \textbf{Chronological spacetime}: If the spacetime $\mathcal{M}$ does not permit closed timelike curve, then we call such spacetime as chronological The condition for $\mathcal{M}$ to be chronological is as follows:
\begin{equation}
    a\not\in \mathcal{I}^+(a), \hspace{1cm} \forall\textrm{ }a\in \mathcal{M}.
\end{equation}

\item \textbf{Causal spacetime}: If $\mathcal{M}$ does not permit closed nonspacelike curves, then we call such spacetime as causal. The required condition is as follows:
\begin{equation}\label{causalcondition}
    a\not\in \mathcal{J}^+(a), \hspace{1cm} \forall\textrm{ }a\in \mathcal{M}.
\end{equation} 

\item \textbf{Strongly causal spacetime}: Even if the above two conditions are satisfied, $\mathcal{M}$ can still almost violate the causality. In the spacetime depicted in Fig.(1.4), even though the condition mentioned in (\ref{causalcondition}) holds, still one can trace a nonspacelike curve from a point $a$ which passes through arbitrarily small neighborhood of $a$. To avoid such almost causal violation, the following condition can be imposed: $\forall$ $a \in \mathcal{M}$, and $\forall$ neighborhood $U$ of $a$, $\exists$ $V \subseteq U$, such that no causal curve through $a$ intersects $V$ more than once.

\begin{figure}\label{cylinder}
\centering
\includegraphics[scale=0.4]{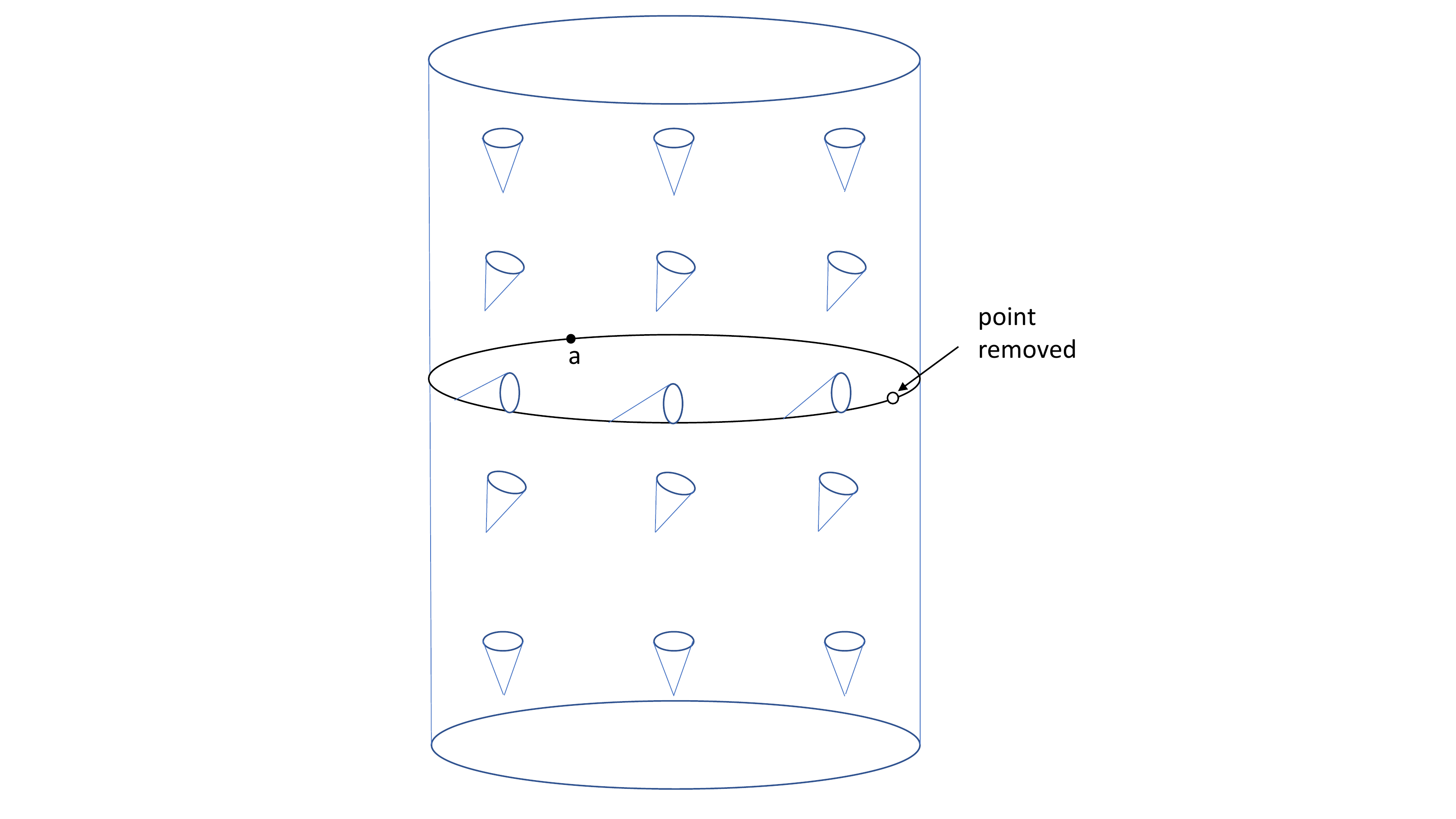}
\caption{A spacetime violating the strong causality condition. Here, a causal curve can be traced through $a$ which can come arbitrarily close to intersecting itself.}
\end{figure}

%This violation can be depicted by the metric given by
%\begin{equation}
%    ds^2=dt dx+t^2 dx^2.
%\end{equation}
%One can see that the null geodesics in this spacetime are given by vertical lines parallel to the time axis, and the solution of the differential equation
%\begin{equation}
%    \frac{dt}{dx}=-t^2,
%\end{equation}
%which is
%\begin{equation}
%    (x+k)t=1.
%\end{equation}
%As $t\to0$, $\frac{dt}{dx}\to 0$, hence one arm of the light cone tends to become parallel to the $x$ axis. \textcolor{red}{draw light cone. Show two distinct events p and q lying at t close to zero and x very large, and lying on the light cone. $q\in I^+(p)$. But $I^+(p)=I^+(q)$ hence, $q\in I^+(q)$. Also another example of cylinder in Wald depict. draw diagrams.}

%\begin{equation}
%    \mathcal{I}^+(a)=\mathcal{I}^+(b) \implies a=b,
%\end{equation}
%and
%\begin{equation}
%    \mathcal{I}^-(a)=\mathcal{I}^-(b) \implies a=b,
%\end{equation}

\item \textbf{Stably causal spacetime}: Even if the strong causality condition discussed above is satisfied, closed non-spacelike curves can come into existence if a small perturbation is given to the spacetime metric. This property seems to be a physical reality since we know that general relativity is, after all, a classical approximation of a more general quantum theory of gravity, and hence the uncertainty principle will not allow the spacetime metric to have the exact form at a given point. Thus, there should be stability in the spacetime metric in that all ``nearby" spacetime metrics should also have a similar property as far as the causality condition is concerned.

Giving a meaning to the word ``nearby" can be done in various ways. To do so, one has to define a topology on the set of all Lorentz metric. In $C^0$ topology, two metrics are said to be nearby if their actual values (the metric components) are nearby. In $C^k$ topology, two metrics are said to be nearby if the $k^{th}$ derivative of the metric components are nearby. In open topology, the metrics are said to be nearby if they are nearby everywhere. We will consider what is known as $C^0$-open topology. 

The job is now to define an ``open" set in $C^0$-open topology. First, let us consider the symmetric tensor space $T_a \mathcal{M}$ of type $\left(0,2\right)$ defined on each point $a\in\mathcal{M}$. The tensor bundle is defined as
\begin{equation}
    T\mathcal{M}=\cup_{a\in\mathcal{M}} T_a\mathcal{M}.
\end{equation}
A Lorentz metric $g$ on $\mathcal{M}$ is then a function 
\begin{equation}
    g:\mathcal{M}\to T\mathcal{M},
\end{equation}
which assigns an element of $T\mathcal{M}$ to each point on $\mathcal{M}$.  Consider an open set \footnotemark
\begin{equation}
    \mathcal{O}\subset T\mathcal{M}.
\end{equation}
\footnotetext{A subset $\mathcal{O}\subset T\mathcal{M}$ is \textit{open} if for any coordinate chart $\phi:\mathcal{U} \to \mathbb{R}^n$ on $\mathcal{M}$, the set $\tilde{\phi}(\mathcal{O}\cap T \mathcal{U})\subset \mathbb{R}^n \times \mathbb{R}^n$ is open, where $\tilde{\phi}:T\mathcal{U}\to \phi(\mathcal{U})\times \mathbb{R}^n$ is a homeomorphic map.}
 Let $\mathcal{S}(\mathcal{O})$ be a set of all the Lorentz metric of type $C^0$ such that $g(\mathcal{M})\subset \mathcal{O}$. In $C^0$-open topology of the $C^r$ Lorentz metric on $\mathcal{M}$, the open set is defined as the union of one of more sets of the form $\mathcal{S}$.
 
 A spacetime is said to be stably causal if the spacetime metric $g$ has an open neighborhood in $C^0$-open topology such that in any metric belonging to this neighborhood, no closed non-spacelike curve exists.

 \item \textbf{Globally hyperbolic spacetime}: Till now, the stably causal condition may be considered to be the most physically relevant of all the available conditions to maintain the causal regularity. However, a particular topology ($C^0$-open topology), out of the various possibilities, is chosen to define the notion of ``nearness" in this causality condition. One can equivalently consider $C^k$ topology on $T\mathcal{M}$ to give meaning to the word ``nearby". Therefore, there is no unique way to define the topology on $T\mathcal{M}$.
 
 %Nevertheless, it is possible to uniquely fix the overall topology of the spacetime if the spacetime is globally hyperbolic. 
 
 An alternative way to define the causality condition is as follows: Consider a spacetime $(\mathcal{M},g)$. We say that $(\mathcal{M},g)$ is globally hyperbolic if $\mathcal{M}$ is strongly causal, and the following set is compact and contained in $\mathcal{M}$ 
 \cite{Leray_52, Hawking_73}:
 \begin{equation}
     \mathcal{J}^+(a) \cap \mathcal{J}^-(b) \hspace{1cm} \forall \textrm{ } a,b \in \mathcal{M}. 
 \end{equation}
 In order to understand the significance of global hyperbolicity, consider the following set:
 \begin{equation}
     \mathcal{D}^+(A)=\{a\in \mathcal{M} \hspace{0.2cm}\vert \hspace{0.2cm} \textrm{every past inextendible \footnotemark causal curve from $a$ intersects } A\}
 \end{equation}
 \footnotetext{A curve $\gamma \subset \mathcal{M}$ is called past inextendible if it does not have a past end point in $\mathcal{M}$. Similarly, one can define a future inextendible curve.}
  defined on a closed, achronal \footnotemark,  
  \footnotetext{A subset $A \subset \mathcal{M}$ is called achronal if $\forall$ $a,b \in A$, $a \notin \mathcal{I}^+(b)$.}
  edgeless \footnotemark 
  \footnotetext{Edge of a closed achronal set $A$, denoted by $Edge(A)$ is a collection of points $a\in A$ such that every open neighborhood of $a$ contains $m \in \mathcal{I}^+(a)$ and $n \in \mathcal{I}^-(a)$ connected by a timelike curve not intersecting $A$. $A$ is edgeless if $Edge(A)= \emptyset$}
  set $A$ of $\mathcal{M}$. This set is called the future domain of dependence of $A$. It is also called the future Cauchy development of $A$. Past domain of dependence $\mathcal{D}^-(A)$ (or past Cauchy development) of $A$ is defined in a similar manner. The full domain of dependence (Cauchy development) is then defined as follows:
  \begin{equation}
      \mathcal{D}(A)=\mathcal{D}^+(A)\cup \mathcal{D}^-(A).
  \end{equation}
  A specific closed, achronal, edgeless set $S$ of $\mathcal{M}$ such that 
  \begin{equation}
      \mathcal{D}(S)=\mathcal{M}
  \end{equation}
  is called a Cauchy surface. A Cauchy surface, hence intersects exactly once with every inextendible causal curve $\gamma \in \mathcal{M}$. It can be shown that a non-empty, closed, achronal, edgeless set is a three dimensional, embedded, $C^0$ submanifold of $\mathcal{M}$ 
  \cite{Wald_84}.
  
  It was shown by Geroch
  \cite{Geroch_70}
  in 1970 that a globally hyperbolic spacetime $\mathcal{M}$ admits a Cauchy surface.
  Moreover, any two Cauchy surfaces are homeomorphic to each other. $\mathcal{M}$ therefore has a unique fixed topology since it is homeomorphic to $S\times \mathbb{R}$.
  
  If our universe is globally hyperbolic (i.e. if it admits a Cauchy surface $S$), one can predict the entire future, and trace the entire past of the universe. Conversely, if the universe does not admit a Cauchy surface, the power of predictability is lost. It can be shown the a globally hyperbolic spacetime is stably causal
  \cite{Wald_84}. 
  Hence, all the pathological features are taken care of, and it is the strongest causality condition of all.
 \end{itemize}
 
 Now we discuss why the property of global hyperbolicity implies the strong cosmic censorship. Let us assume that the strong cosmic censorship is not true. Then, at least locally naked (timelike or null) singularities exists. Therefore, there exists past inextendible timelike curve $\gamma$, ending at the singular point $b$. Consider a point $a$ such that $b$ lies in the causal future of $a$, i.e. $b\in\mathcal{J}^+(a)$. Now, consider a point $c\in \gamma$. $b$ also lies in the causal past of $c$, i.e. $b\in\mathcal{J}^-(c)$. Hence we have
 \begin{equation}
     b\in \mathcal{J}^+(a)\cap \mathcal{J}^-(c).
 \end{equation}
But since $b$ is a singular point, 
\begin{equation}
    b\notin \mathcal{M}.
\end{equation}
Therefore, 
\begin{equation}
    \mathcal{J}^+(a)\cap \mathcal{J}^-(c)\not\subset \mathcal{M},
\end{equation}
and hence, the spacetime is not globally hyperbolic \cite{Joshi_82, Joshi_07}. Therefore, if the strong cosmic censorship fails, then the causality property of the spacetime is violated.
\begin{figure}\label{nakedsingularitycauchysurface}
\centering
\includegraphics[scale=0.4]{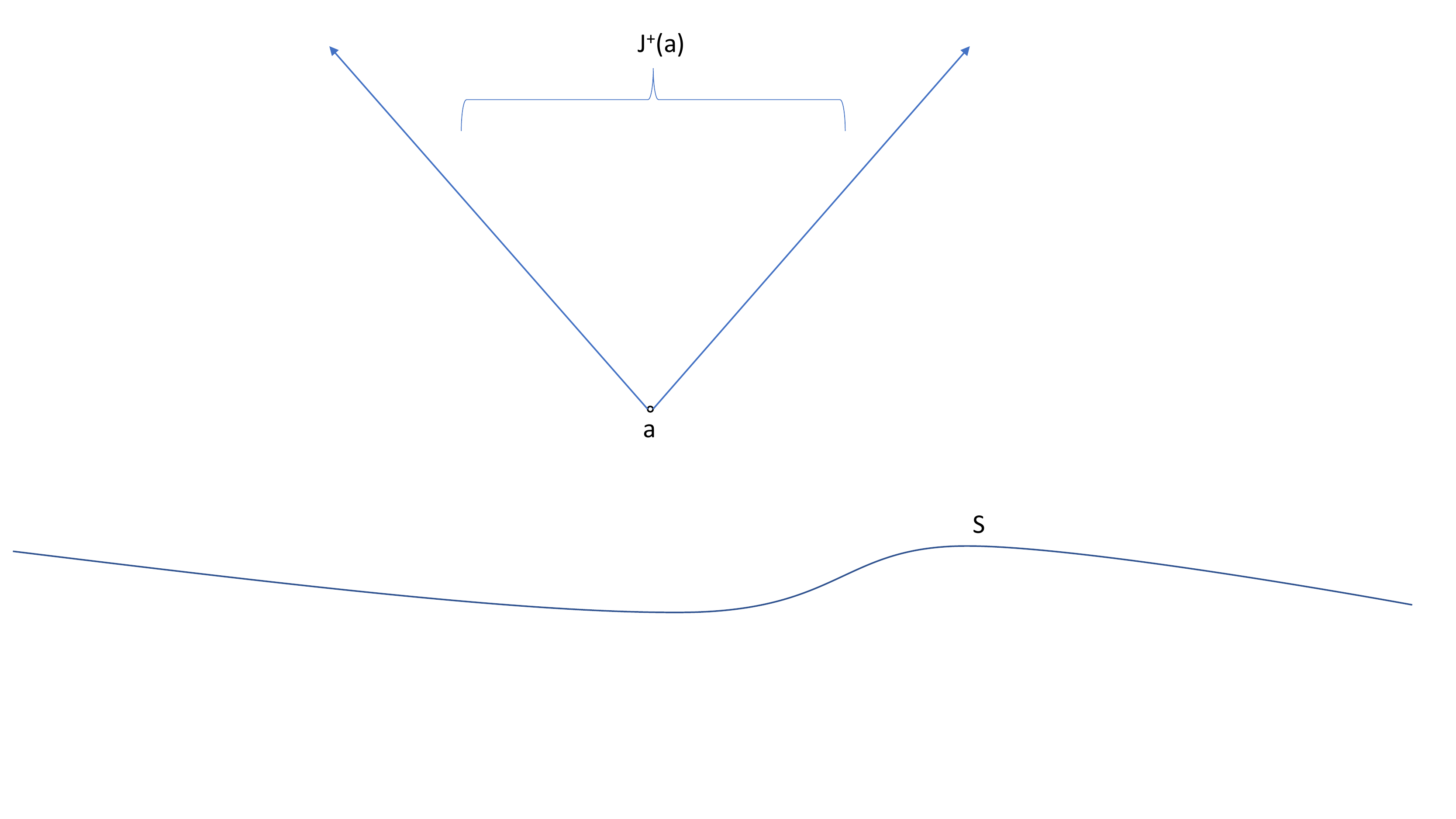}
\caption{There does not exist any closed, achronal, edgeless, spacelike three dimensional $C^0$ submainfold $S$ whose Cauchy development is the entire manifold $\mathcal{M}$, if there exists a naked singularity $a$.}
\end{figure}

An alternative way to realize the incompatibility of a globally hyperbolic spacetime with the presence of even a locally naked singularity is to argue the impossibility of the existence of a Cauchy surface in presence of a naked singularity. Consider a globally hyperbolic spacetime $(\mathcal{M},g)$ (refer Fig.(5.1)). Then $\mathcal{M}$ admits a Cauchy hypersurface $S$. Therefore, $D(S)=\mathcal{M}$. Now let us assume a naked singularity $a$.  One can see that $\mathcal{J}^+(a)\subset M$. However, $\mathcal{J}^+(a) \not\subset \mathcal{D}(S)$. More specifically, $\mathcal{J}^+(a) \cap \mathcal{D}(S)=\emptyset$. This is because $\forall$ $x\in \mathcal{J}^+(a)$, $\exists$ a past inextendible causal curve $\gamma$ ending at $a$. Therefore, $x\notin \mathcal{D}(S)$. Therefore, $\mathcal{D}(S)\neq \mathcal{M}$. This is a contradiction because $S$ is a Cauchy surface. Thus, one could say that global hyperbolicity implies strong cosmic censorship.

%Consider any closed, achronal, edgeless, non-empty set $S$ (refer Fig.(1.5)). As mentioned before, $S$ is therefore a three dimensional $C^0$ submanifold of $\mathcal{M}$. Let there exist a naked singularity denoted by $a$. One can see that $\mathcal{J}^+(a)\subset M$. However, $\mathcal{J}^+(a) \not\subset \mathcal{D}(S)$. More specifically, $\mathcal{J}^+(a) \cap \mathcal{D}(S)=\emptyset$. This is because $\forall$ $x\in \mathcal{J}^+(a)$, $\exists$ a past inextendible causal curve $\gamma$ ending at $a$. Therefore, $x\notin \mathcal{D}(S)$. 

%Therefore, $\mathcal{D}(S)\neq \mathcal{M}$. This is true for all $S$ with the abovementioned property. Hence, $\mathcal{M}$ does not admit any Cauchy surface, thereby the spacetime is not globally hyperbolic. Thus, one could say that global hyperbolicity implies strong cosmic censorship.   

One could refer to 
\cite{Bruhat_69, Israel_84, Krolak_1986, Christodoulou_86, Christodoulou_87, Christodoulou_91, Christodoulou_93, Christodoulou_94, Penrose_98, Ringstrom_13, Sbierski_2016} for further discussion on the development in the understanding of the cosmic censorship hypothesis.

\section{Visibility of the singularities}\label{1.6}
 Countering the cosmic censorship hypothesis, it was shown  by Eardley and Smarr
\cite{Eardley_79}
using numerical simulation of a collapsing star, and later by
Christodoulou
\cite{Christodoulou_84},
and Joshi and Dwivedi
\cite{Joshi_93},
analytically, that introducing inhomogeneity in the mass profile of the collapsing cloud could change the evolution of the apparent horizon, thereby possibly allowing non-spacelike geodesics to escape away from near the singularity without getting trapped. As such, the assumption of a homogeneous star is not very appropriate since it is expected that a star becomes denser as we move toward its center.  

The examples of the collapsing cloud ending up in at least locally visible singularity, as shown by 
\cite{Eardley_79, Christodoulou_84, Joshi_93}
has zero pressure, which may not seem desirable because, as mentioned before, a realistic collapsing cloud is expected to have nonzero pressure. Stronger counterexamples to the strong cosmic censorship are expected to be pressured clouds ending up in a locally naked singularity. At first, it was believed that naked singularities are unique to pressureless fluids. However, later on, various such counterexamples with nonzero pressures were depicted by exploiting symmetries, thereby simplifying the resulting field equations. We discuss some of them as follows: 
\begin{enumerate}
    \item Self-similar collapse
    \cite{Eardley_74}:
    In such a scenario, the Einsteins' field equations are invariant under scale transformation. The metric components $\nu$ and $\psi$ in Eq.(\ref{metricnonondiagonal}), and the scaling function $v=\frac{R}{r}$ can be completely written as a function of the similarity parameter defined by 
\begin{equation}
    X=\frac{t}{r}.
\end{equation}
Naked singularity formed due to self-similar collapse have been depicted by Ori and Piran 
\cite{Ori_87, Ori_88, Ori_90} 
wherein they considered a perfect fluid with a soft linear equation of state. Necessary conditions for the singularity to be naked in a marginally bound self-similar collapse were derived by Waugh, and Lake
\cite{Waugh_88, Waugh_89}.

\item Isentropic perfect fluid: It was shown by  Goswami and Joshi 
\cite{Goswami_04}
that the collapse of an isentropic perfect fluid, having isotropic pressure proportional to the density 
\begin{equation}
    p=\omega \rho, \hspace{1cm} \omega \in \left(-\frac{1}{3},1\right)
\end{equation}
can have the end state as locally visible for a suitable choice of initial data. Numerical results for such equation of state were bought forward by Harada
\cite{Harada_98},
which suggests that in the collapse of highly relativistic fluid with 
\begin{equation}
    \omega<0.1.
\end{equation}
it is possible to get a naked singularity formation from time-symmetric and spherically symmetric initial data.  This numerical result matches with that calculated by Ori and Piran
\cite{Ori_87, Ori_88, Ori_90}
where the initial data are not time-symmetric but imploding due to self-similarity assumption

\item Scalar field collapse: Scalar fields are fundamental matter fields obtained by adding suitable Lagrangian ($\mathcal{L}_{\phi}$) in the Einstein-Hilbert action as  
\begin{equation}
    S=\frac{1}{2}\int \left(R+ \mathcal{L}_{\phi} \right)\sqrt{-g}d^4x,
\end{equation}
where
\begin{equation}
    \mathcal{L}_{\phi}=-\frac{1}{2}g^{\mu\nu}\partial_{\mu}\phi\partial_{\nu}\phi -V(\phi)
\end{equation}
is the Lagrangian corresponding to a scalar field with potential $V(\phi)$.
Using variational principle with respect to the metric gives the energy momentum tensor corresponding to the scalar field as
\begin{equation}
    T^{(\phi)}_{\mu \nu}=-\frac{1}{\sqrt{-g}}\frac{\delta \left(\sqrt{-g}\mathcal{L}_{\phi}\right)}{\delta g^{\mu \nu}}.
\end{equation}
For the Friedmann-Lemaitre-Robertson-Walker (FLRW) metric governing the spacetime filled with the homogeneous scalar field, the density, and the isotropic pressure is expressed in terms of the time derivative of the scalar field and its potential as
\begin{equation}\label{rhopscalarfield}
    \rho= \frac{1}{2}\dot \phi^2+V(\phi), \hspace{2cm} p=\frac{1}{2}\dot \phi^2-V(\phi).
\end{equation}
It was shown by Goswami and Joshi
\cite{Goswami_04b}
that a class of scalar field model with nonzero potential can be constructed such that trapped surfaces do not form at the time of formation of the singularity, thereby exposing the singularity globally. It is worth noting here that the scalar field is homogeneous throughout the collapse, and all the shells, each corresponding to the radial coordinate $r$, collapse simultaneously to the singularity. It is hence proved that inhomogeneity is not a necessary criterion for a singularity to be visible.
\end{enumerate}

Soon after, in 2007, a general formalism was developed to check the local causal structure of an arbitrarily pressured cloud by Goswami and Joshi 
\cite{Goswami_2007}.
Joshi, Malafarina, and Saraykar
\cite{Joshi_12}
later used this formalism to give an example of a locally naked singularity formed due to the collapse of a nonzero pressured collapsing matter cloud.  Here we discuss a brief overview of the idea. 

Consider a spherically symmetric collapsing matter cloud governed by the spacetime metric with no non-diagonal terms, given by Eq.(\ref{metricnonondiagonal}). The matter cloud governed by this spacetime is of Type I, having the stress-energy tensor as
\begin{equation}
    T^{\mu}_{\nu}=\textrm{diag}(-\rho,p_r,p_{\theta},p_{\theta}),
\end{equation}
where $\rho$ is the density, and $p_r$ and $p_{\theta}$ are the radial and tangential pressures respectively. The Misner-Sharp mass function, which is defined as
\begin{equation}
    F=\int^{R}_{0}\rho(r,R) R^2 dR, 
\end{equation}
in the area-radial coordinate $(r,R)$ is expressed in terms of the metric components and its derivative (using the $(0,0)$ the component of the Einsteins' field equation) as
\begin{equation}\label{generalmsmf}
    F=R(1-G+H),
\end{equation}
where 
\begin{equation}
    G=\left(R'\right)^2 e^{-2\psi}, \hspace{1cm} and \hspace{1cm} H=(\dot R)^2 e^{-2\nu}.
\end{equation}
In case of zero isotropic pressure, the spacetime is governed by the Lemaitre-Tolman-Bondi metric given by
\begin{equation}
    ds^2=-dt^2+\frac{R'^2}{1+f}dr^2+R^2d\Omega^2,
\end{equation}
where $f=f(r)$ is called the velocity function. The Eq.(\ref{generalmsmf}) then reduces to 
\begin{equation}
    F=R\left(\dot R^2-f \right).
\end{equation}
The above equation can be integrated with respect to the time coordinate and the at least implicit relation between the physical radius $R$, the radial coordinate and the time coordinate can be obtained as 
\begin{equation}\label{Rfornonzerof1}
    R=\left(\frac{r^{\frac{3}{2}}G\left(-f r/F\right)-\sqrt{F}t }{G\left(-f R/F\right)}\right)^{\frac{2}{3}}.
\end{equation}
Once we obtain an explicit expression of the physical radius in terms of the time and radial coordinate, i.e., $R=R(t,r)$, we can then proceed to understand the evolution of the apparent horizon ($F=R$)  and the event horizon (solution of the differential equation $\frac{dt}{dr}=\frac{R'}{\sqrt{1+f}}$ satisfying $F=R$ at $r_c$). Eventually, using the knowledge of the evolution of both the horizons, one can check if the singularity is locally hidden, locally naked, but globally hidden or globally naked. Now, achieving the explicit expression of the physical radius is possible because of the simplicity achieved due to isotropically zero pressures in the collapsing cloud. However, when we consider nonzero pressure, integrating Eq.(\ref{generalmsmf}) is not such a straightforward task. 

Goswami and Joshi 
\cite{Goswami_2007}
in 2007 showed that the local causal structure of the singularity could be determined by the polarity of the coefficient $\chi_1$ of $r$ in the Taylor expansion of the singularity curve $t_s(r)$ (discussed in the sec.(\ref{1.4}) below Eq.(\ref{OSts})) around $r=0$. If $\chi_1>0$, then the singularity is locally naked. If negative, then locally hidden. In case $\chi_1=0$, we can check the polarity of the next nonzero coefficient $\chi_i$ of $r^i$ ($i\in \mathbb{N}$) in the Taylor expansion, and based on its polarity, one can determine the local causal structure of the singularity. 

Obtaining the value of the smallest nonzero $\chi_i$ may again require the knowledge of the dynamics of the collapse (or, in other words, the explicit expression of $R(t,r)$). One can, however, bypass this requirement if one considers a small suitable perturbation in the Misner-Sharp mass function corresponding to inhomogeneous dust. This small perturbation, in turn, gives rise to slight nonzero pressure. This perturbative approach was first made by Joshi, Malafarina, and Saraykar 
\cite{Joshi_12}
to show counterexamples of the strong cosmic censorship hypothesis formed from collapsing clouds having slight nonzero pressure. 

The next chapter will show that for a locally naked singularity, if the first nonzero (positive) coefficient in the Taylor expansion of the singularity curve is $\chi_1$ or $\chi_2$, then the singularity is weak in the sense of Tipler. One can, however, get Tipler strong locally naked singularity if $\chi_1=\chi_2=0$ and $\chi_3>0$, and for other such possibilities 
\cite{Mosani_20}.

A singularity that can emit non-spacelike geodesic is not physically relevant if it is not generic in nature. By genericity, we mean that the singularity property should not change drastically if one slightly changes the initial data producing it. The issue, however, is that genericity is not properly defined in general relativity. To get the idea, consider the following: Consider a set $X$ of all spacetime metrics. First, let us define the notion of distance on $X$ such that we say that the two spacetime metrics are nearby if the metric components are nearby. On the other hand, let us define the notion of distance on $X$ such that the two spacetime metrics are nearby if their first derivatives with respect to a certain coordinate of the metric components are nearby.  The problem lies in the fact that we can get different topologies in the former and latter cases. Hence, there is no properly established notion of the concept of genericity. Under this situation, one can approach the problem differently, each having answers unique to the approach undertaken. There have been studies to understand the genericity of naked singularity in alternative manners. Based on the dynamical system theory, the initial data leading to an outcome is said to be generic if it is open and dense \footnotemark in the set of entire initial data. Using this definition, Saraykar, and Ghate
\cite{Saraykar_99},
and later Sarve and Saraykar
\cite{Sarve_05}
showed that naked singularities formed due to gravitational collapse of Type-I matter field are non-generic.
\footnotetext{A subset $D \subset X$ of a topological space $X$ is said to be dense if every neighborhood of any $a\in X$ contains an element from $D$.}
However, if genericity is defined in such a way, then even a black hole formation is non-generic in nature. Let us now adopt more physical definitions in the sense of abundance as follows 
\cite{Joshi_11}:
\begin{enumerate}
    \item \textbf{Stability}: We say that the outcome of the gravitational collapse is stable under small perturbation in a certain subset $A$ of the entire initial data set $B$ if there exists a small neighborhood in $A$ giving rise to the same outcome.
    \item \textbf{Genericity}: We say that the outcome of the gravitational collapse is generic if the set of the initial data giving rise to the outcome has a nonzero measure in the entire initial data set.
\end{enumerate}
With these definitions, the outcomes in the class of scenarios discussed in 
\cite{Saraykar_99, Sarve_05}
(collapse of Type-I matter field, which is non-self similar and not a massless scalar field) becomes stable with respect to small changes in the initial data, and is also generic
\cite{Joshi_12}.

From the definitions above, we can see that one has to refer to a set of initial data to give the notion of stability. What this means is that a particular outcome (black hole or naked singularity) which is stable under small perturbation in a certain set of initial data (for, e.g., that of inhomogeneous dust), may not be stable under small perturbation in its superset (for, e.g., that of perfect fluid).

It was found by Christodoulou
\cite{Christodoulou_99}
that the cardinality of the linearly independent elements in the set of initial data giving rise to a naked singularity as an end state of a massless scalar field collapse is strictly less than that of the entire set of initial data. In other words, in the case of scalar field collapse with zero $\phi^2$ term in the scalar field Lagrangian, the set of initial data giving rise to naked singularity has positive codimension in the entire set of initial data (the whole set includes those initial data giving rise to a black hole, and those giving rise to a naked singularity). This outcome concludes that a massless scalar field collapses to a naked singularity that is non-generic. 

Furthermore, the dependence of the stability on the class of perturbations was demonstrated by Mena, Tavakol, and Joshi
\cite{Mena_2000}

It is worth noting here that in the case of OSD collapse, any small neighborhood in the set of initial data contains initial data, which leads to collapse evolution ending up in a naked singularity
\cite{Joshi_12(b)}
This work shows the instability of OSD collapse under small perturbation in pressure. 

%(One can refer to 
%\cite{Barve_1999, Jhingan_1996, Magli_1997, Magli_1998, Giambo_03, Harada_1999, Harada_2002, Joshi_11}. 
%for further examples of mass distributions ending up in a locally naked singularity.)

As far as the physical importance of a naked singularity is concerned, at par with the interpretation of the singularity theorem by Misner
\cite{Misner_1969}
as discussed in sec.(\ref{1.2}), let us consider a scenario: suppose we get some observational signatures from the extreme gravity region. These signatures can be obtained if a congruence of null geodesic can escape the region. Now, these signatures contain traces of a quantum theory of gravity. Suppose we already know the predictions of general relativity. In that case, the difference in these observational traces and the predictions of general relativity may tell us how to tune general relativity to give the predictions that match with the observational signatures. Hence, a naked singularity, more specifically, a globally naked singularity, acts as an astrophysical lab to test the laws of quantum gravity.

It should be emphasized that, even on a classical level, the general theory of relativity has certain limitations, which we discuss in the next section.

\section{Modified gravity}\label{1.7}

General relativity is a very successful theory and has been supported by various observational shreds of evidence like the measurement of the deflection angle of the light bent due to the curvature in the spacetime; which is predicted by the general relativity as twice that predicted by Newton's theory (and correspondingly the gravitational time delay)
\cite{Dyson_1920},
the perihelion precession of the mercury, the gravitational redshift
\cite{Schutz_09},
etc. However, general relativity may not give a complete picture of the working of the universe.

Recent observations of the High-Z Supernova Search Team 
\cite{Riess_98} 
and the type Ia supernova by the Supernova Cosmology Project 
\cite{Perlmutter_99}  
suggests that the universe is undergoing cosmic acceleration. In addition to this, the inflationary epoch in the early universe
\cite{Starobinsky_80, Kazanas_80, Guth_81, Sato_81}, 
which is needed to explain the flatness problem and the horizon problem,
requires a rapid accelerated expansion of the universe. 

These early and late time accelerated cosmic expansions are not at par with the outcome of general relativity, which predicts a positive deceleration parameter 
\cite{Liddle_03}, 
or in other words, the decelerated expansion of the universe provided the universe is filled with ordinary matter, the one which satisfies all the four energy conditions 
\cite{Hawking_73}. 
To see this, consider the FLRW metric in (\ref{OSDmetric}). This is assumed to govern the geometry of the universe on the large scale according to the cosmological principle, which states that the universe is homogeneous and isotropic on a large scale. The Friedmann's equation, which is considered one of the most important equations in cosmology can be obtained from general relativity, and is given by
\begin{equation}\label{friedmannsequation}
    \left(\frac{\dot a}{a}\right)^2=\frac{\rho}{3}-\frac{\Lambda}{a^2},
\end{equation}
where $\Lambda$ is the cosmological constant. Using the Bianchi Identity, we can find the equation of continuity as
\begin{equation}\label{equationofcontinuity}
    \dot \rho+\frac{3\dot a}{a}\left(\rho+p\right)=0,
\end{equation}
where $\rho$ and $p$ are the homogeneous density and isotropic pressure respectively. Differentiating Eq.(\ref{friedmannsequation}) and then using Eq.(\ref{equationofcontinuity}), we obtain the acceleration equation as follows:
\begin{equation}
    \frac{\ddot a}{a}=-\frac{1}{6}\left(\rho+3p \right).
\end{equation}
From the acceleration equation, we can see that if the dominating fluid in the universe obeys the strong energy condition, which is
\begin{equation}
    \rho+3p>0,
\end{equation}
then the second time derivative of the scale factor is negative, hence representing a decelerated expanding universe.

This mismatch of the observed cosmic acceleration with the predicted cosmic deceleration can be rectified if we drop the rigidity to choose the matter field to obey the energy conditions. The existence of dark energy models, which are exotic forms of matter fields not satisfying the energy conditions, have been considered to explain the cosmic acceleration, all the while maintaining the uprightness of general relativity. These models include the cosmological constant, quintessence, k-essence, phantom fluid, to name a few. A detailed discussion about the dark energy models can be found in 
\cite{Amendola_10}. Alternatively, one can also modify the Lagrangian of Einstein-Hilbert (EH) action, in which, instead of the Ricci scalar ($R$), curvature terms of different order can be added. Choosing a suitable Lagrangian can naturally give rise to cosmic acceleration
\cite{Nojiri_03, Nojiri_06, Nojiri_10, Samanta_16}
without compromising with the energy conditions.

Apart from the motivation to explain the cosmic acceleration, another motivation to modify gravity is that it has been found that general relativity is not renormalizable and hence can not be quantized 
\cite{Aharony_99, Shomer_07}. 
It was shown by Stelle 
\cite{Stelle_77} 
that actions with higher-order curvature terms can be renormalized. However, it had its shortcoming of unresolved unitarity problem.

We now give a brief overview of the $f(R)$ class of gravity theory, which is one of the type of modified gravity. The total action for $f(R)$ gravity is written as
\cite{Sotiriou_10}
\begin{equation}
    S=\frac{1}{2}\int f(R) \sqrt{-g}d^4x+ S_m,
\end{equation}
where $S_m$ is the matter Lagrangian. Using variational principle with respect to the metric gives the following field equation:
\begin{equation}
    G_{\mu \nu}=\frac{1}{F(R)}\left(T^{(m)}_{\mu \nu}+T^{(D)}_{\mu \nu}\right),
\end{equation}
where $F(R)=\frac{df(R)}{dR}$, and 
\begin{equation}
T^{(D)}_{\mu \nu}=\frac{(f-RF)}{2}g_{\mu \nu}+\nabla_{\mu}\nabla_{\nu}F-g_{\mu \nu}\Box F.    
\end{equation}
Here $\Box=g^{i j}\nabla_i \nabla_j$. In $f(R)$ theories of gravity, from the above field equation, even in the absence of matter field, the Einstein tensor can be non-zero, unlike in general relativity. One can interpret that $T^{(D)}_{\mu \nu}$  be considered as the energy-momentum tensor, which has a purely geometric origin.  However, it is not under obligation to obey the strong energy conditions. 

One such example of $f(R)$ gravity is the Starobinsky type gravity 
\cite{Starobinsky_80}, 
which was used to study curvature-driven inflationary scenarios. This model has the Lagrangian expressed as $R+\alpha R^2$, where $R$ is the Ricci scalar, and $\alpha>0$ is a constant 
\cite{Kofman_85}. 
In the strong gravity regime, the second-order curvature term $R^2$ can dominate and play a significant role.

In general, in the strong gravity region, higher-order curvature terms in action, if present, will dominate, and hence it may become essential to incorporate the modified action while investigating the neighborhood of the singularity formed due to gravitational collapse.

Various static as well as non-static solutions of the modified gravity models, have been studied before. Static spherically symmetric solutions in $f(R)$ model were studied by Capozziello, Nojiri, Odintsov, and Troisi
\cite{Capozziello_06}.
Blackhole solutions in these models were studied by various authors, for, eg. 
\cite{Aghmohammadi_10, Sebastiani_11, Bergliaffa_11}.
Spherically symmetric collapsing solutions were analyzed by Bamba, Nojiri, and Odintsov 
\cite{Bamba_11},
where the resolution of the singularity is discussed for suitable Lagrangian. However, the formation of naked singularities in $f(R)$ gravity has not been explored as much as the static solutions and the non-static ones who end up in a black hole.

 Naked singularity formation due to gravitational collapse of null dust in $f(R)$ gravity has been studied by Ghosh and Maharaj
\cite{Ghosh_12}.
Effect of additional matching condition, i.e., the smoothness of the Ricci scalar at the matching surface of the boundary of the collapsing cloud with the exterior Schwarzschild spacetime, in the collapse of spherically symmetric cloud has been explored by Goswami, Nzioki, Maharaj, and Ghosh
\cite{Goswami_14}
in the Starobinsky model, wherein they found that maintenance of the junction condition does not allow the collapse of a homogeneous dust cloud as a solution of the field equations, without the surface stress-energy tensor coming into the picture. More on the gravitational collapse of an LTB cloud in the Starobinsky gravity will be discussed in chapter $6$, where we show that evolution of the event horizon is affected if one goes from general relativity to $f(R)$ gravity, due to which the causal structure of the singularity may change.

\section{Summary of the work}\label{1.8}
The thesis is arranged as follows:
\begin{itemize}
    \item In chapter $2$, we investigate the locally naked singularity formed due to a spherically symmetric inhomogeneous collapsing cloud having nonzero isotropic pressure in terms of its strength. Sufficient condition provided by Clarke and Krolak for it to be Tipler strong has been used to restrict the parameters that represent the non-linear relation between the physical radius and the radial coordinate of the outgoing radial null geodesic at the singular center. Studying the end state of a collapsing cloud requires information about the collapse dynamics, which is unknown in a  general scenario. Hence we study small perturbations to the mass profile for inhomogeneous dust, which is possible using the formalism developed here. This perturbed mass profile, in turn, gives rise to nonzero pressure. We show the existence of a nonzero measure set of initial data giving rise to such strong curvature naked singularity.
    
    \item In chapter $3$, we investigate the local versus global visibility of a spacetime singularity formed due to the gravitational collapse of a spherically symmetric dust cloud having a nonzero velocity function. The conditions are investigated that ensure the global visibility of the singularity, in the sense that the outgoing null geodesics leave the boundary of the matter cloud in the future, whereas, in the past, these terminate at the singularity.  Explicit examples of this effect are constructed. We require that this must be a strong curvature singularity in the sense of Tipler to ensure the physical significance of the scenario considered. These examples may act as a counterexample to the weak cosmic censorship hypothesis. 
    
    \item In chapter $4$, we investigate the final state of gravitational collapse of a non-spherical and non-marginally bound dust cloud as modeled by the Szekeres spacetime. We show that a  directionally globally naked singularity can be formed in this case near the collapsing cloud boundary and not at its geometric center, as is typically the case for a spherical gravitational collapse. This is a strong curvature naked singularity in the sense of Tipler criterion on gravitational strength. The null geodesics escaping from the singularity would be less scattered in this case in certain directions since the singularity is close to the boundary of the cloud, as is the case in the current scenario. The physical implications are pointed out.
    
    \item In chapter $5$, we show the existence of a nonzero measured set of parameters: the total mass and the initial mean density of the collapsing cloud, giving rise to a physically strong globally visible singularity as the end state for a fixed velocity function. The existence of such a set indicates that such singularity is stable under small perturbation in the initial data causing its existence. This is true for marginally as well as non-marginally bound cases. The possibility of the presence of such suitable parameters in the astrophysical setup is then studied: $1)$ The singularities' requirements at the center of the M87 galaxy and at the center of our galaxy (SgrA*) to be globally visible are discussed in terms of the initial size of the collapsing cloud forming them, presuming that such singularities are formed due to gravitational collapse. $2)$ The requirement for the primordial singularities formed due to a collapsing configuration after getting detached from the background universe, at the time of matter-dominated era just after the time of matter-radiation equality, to be globally visible, is discussed. $3)$ The scenario of the collapse of a neutron star after reaching a critical mass, which is achieved by accreting the supernova ejecta expelled by its binary companion core progenitor, is considered. The primary aim of this chapter is to show that globally visible singularities can form in astrophysical setups under appropriate circumstances. 
    
    \item In chapter $6$, we investigate the global causal structure of the end state of a spherically symmetric marginally bound Lemaitre-Tolman-Bondi collapsing cloud (which is well studied in general relativity) in the framework of modified gravity having the generalized Lagrangian $R+\alpha R^2$ in the Einstein-Hilbert action. Here $R$ is the Ricci scalar, and $\alpha \geq 0$ is a constant. By fixing the functional form of the metric components of the LTB spacetime, using up the available degree of freedom, we realize that the matching surface of the interior and the exterior metric are different for different value of $\alpha$. This change in the matching surface can alter the causal property of the first central singularity. We depict this by showing a numerical example. Additionally, for a globally naked singularity to have physical relevance, a congruence of null geodesics should escape from such singularity to be visible to an asymptotic observer for infinite time. For this to happen, the first central singularity should be a nodal point. We here give a heuristic method to show that this singularity is a nodal point by considering the above class of theory of gravity, of which general relativity is a particular case.
    
    \item In chapter $7$, we end the thesis with the concluding remark.
    
\end{itemize}

% Chapter Template

\chapter{Strength of a visible singularity in gravitational collapse of a perfect fluid} % Main chapter title

\label{Chapter2} % Change X to a consecutive number; for referencing this chapter elsewhere, use \ref{ChapterX}

\lhead{Chapter 2. \emph{ }} % Change X to a consecutive number; this is for the header on each page - perhaps a shortened title

%----------------------------------------------------------------------------------------
%	SECTION 1
%----------------------------------------------------------------------------------------
Formation of a locally naked singularity due to the gravitational collapse of a Lemaitre-Tolman-Bondi (LTB) dust cloud, which has zero pressure, was shown to be possible by Joshi and Dwivedi 
\cite{Joshi_93}, 
under generic initial conditions. The significant role played by the inhomogeneity of the collapsing cloud was highlighted in such a phenomena. However, unlike the LTB dust cloud, a more realistic star does have non-zero pressure. Additionally, the matter is expected to behave like a perfect fluid at the center of the cloud as discussed in 
\cite{Goswami_2007}. 
The formalism developed to investigate the end state of a collapsing cloud having non-zero pressure suggests that the local nakedness or otherwise of the singularity thus formed, depends upon the polarity of the smallest non-zero component of the Taylor expansion of the singularity curve 
\cite{Joshi_07}. 
The positivity of such component implies that the tangent of the outgoing radial null geodesic (ORNG) is positive at the singularity, implying that the singularity is at least locally naked. 

Such locally naked singularities, however, may not be considered as evidence for the defiance of cosmic censorship if they are gravitationally weak. Any object hitting the singularity, if crushed to zero volume, is called ``strong" curvature singularity according to Tipler 
\cite{Tipler_77}.  
It was shown by Newman 
\cite{Newman_85} 
that naked singularities investigated by Eardley and Smarr 
\cite{Eardley_79} 
and Christodoulou 
\cite{Christodoulou_84}, 
formed in the classes of LTB cloud collapse are weak. Using the sufficient condition given by Clarke and Krolak 
\cite{Clarke_85} 
for a singularity to be strong in the sense of Tipler, naked singularity formation due to collapsing self-similar marginally bound singularity was studied by Waugh and Lake 
\cite{Waugh_88} 
and independently by Ori and Piran 
\cite{Ori_87, Ori_88}. 
Later, Joshi and Dwivedi investigated the naked singularity formed due to the collapse of LTB dust cloud under generic initial conditions for collapse, and derived the value of a certain parameter $\alpha$ for which the singularity is gravitationally strong 
\cite{Joshi_93}. 
The parameter $\alpha$ physically signifies the non-linear relation between the physical radius of the cloud and the radial coordinate of the ORNG at the singular center ($\alpha=1$ corresponds to a linear relation). The stability of such singularities against some perturbations in the initial data was later shown by Deshingkar, Joshi and Dwivedi 
\cite{Deshingkar_99}. 
Coming to the collapsing cloud having non-zero pressure, many models have been studied in which naked singularities are shown to arise, as discussed in the introduction. Here, we study the singularities formed due to a collapsing spherical cloud made up of a perfect fluid with non-zero pressure and derive an analogous criterion needed to be imposed on $\alpha$ for the singularity to be strong. Our basic purpose here is thus to examine and characterize the conditions that ensure that the naked singularities forming in collapse with non-zero pressure are strong curvature in nature.  

The chapter is arranged as follows: Einstein's field equations, regularity conditions, and the mathematical formalism to understand the final state of a collapsing spherically symmetric perfect fluid with arbitrary pressure are discussed in 2.1. The strength of singularity and related results are discussed in sec.(\ref{2.2}). An example showing the existence of a non-zero measured set of initial data giving rise to locally naked Tipler strong singularity is then illustrated and worked out in sec.(\ref{2.3}). Concluding discussions are given in sec.(\ref{2.4}). 
%The question about the strength of such locally naked singularity has not been answered yet.

\section{Collapse formalism}\label{2.1}
The collapse of a spherically symmetric cloud made up of perfect fluid is governed by three functions $\nu(t,r)$, $\psi (t,r)$ and $R(t,r)$, and the metric is expressed as:
%The metric representing the collapse of spherically symmetric cloud of perfect fluid, having energy-momentum tensor $T^{\mu}_{\nu}=diag(-\rho,p,p,p)$, is given by 
\begin{equation} \label{metric}
    ds^2=-e^{2\nu (t,r)}dt^2+e^{2\psi(t,r)}dr^2+R^2(t,r)d\Omega ^2
\end{equation}
in the comoving coordinates $t$ and $r$. The stress-energy tensor for a general type I matter field, more specifically a perfect fluid, has non-diagonal terms as zero, and diagonal terms as
\begin{equation}
    T^t_t=-\rho, T^r_r=T^{\theta}_{\theta}=T^{\phi}_{\phi}=p.
\end{equation} 
Here the $\rho$ is the energy density and $p$ is the isotropic pressure of the collapsing cloud. The matter field under consideration is assumed to be satisfying the weak energy condition thereby restricting the components of stress-energy tensor in the following way:
\begin{equation}
 \rho\geq 0, \rho+p\geq 0.   
\end{equation}
In the units of $8\pi G=c=1$, the Einstien's field equations relates the metric functions $\nu(t,r)$, $\psi (t,r)$ and $R(t,r)$ with the components of stress-energy tensor in the following way:
\begin{eqnarray}
      \rho &=\frac{F'}{R^2R'},\label{efe11} \\
     p &=-\frac{\dot F}{R^2\dot R}, \label{efe21}\\
     \nu ' &=-\frac{p'}{\rho +p}, \label{efe3}\\
     2 \dot R' &= R'\frac{\dot G}{G}+ \dot R\frac{H'}{H},\label{efe4}
\end{eqnarray}
where,
\begin{equation} \label{gh}
    G(t,r)=e^{-2\psi}R'^2; \hspace{0.5cm} H(t,r)=e^{-2 \nu}\dot R^2.
\end{equation}
The superscript dot and prime are the notations used for partial derivative with respect to time and radial coordinates respectively. Here $F$ is the Misner-Sharp mass function given by
\begin{equation}\label{msmf}
     F=R(1-G+H).
\end{equation}
It physically signifies the mass of the cloud inside a shell of radius $r$ at time $t$. It can be expressed as $F=r^3\mathcal{M}$, where $\mathcal{M}$ is such that it maintains regularity. By regularity, we mean that $\mathcal{M}$ is a suitably differentiable function which does not blow up or vanish as $r \to 0$. Doing so ensures that the energy density at the regular center does not blow up before the formation of central shell focusing singularity. Another regularity condition needed to be fulfilled by the collapsing matter field to be well behaved is the absence of cusp in the energy density at the center which is taken care by the equation
\begin{equation}
    \mathcal{M}'(t,0)=0.
\end{equation}
The physical radius of the cloud is represented by the component of metric, $R(t,r)$. For different shells to avoid crossing each other, $R$ has to follow the inequality $R'>0$. To get a collapsing solution of Einstein's field equations, we have to restrict $\dot R (t,r)$ to be less than zero. This indicates that given a shell of radial coordinate $r$, the corresponding physical radius $R$ decreases as time passes until it becomes a singularity, i.e. $R(t,r)=0$. It is to be noted that $R(t,r)$ vanishes also at the regular center, i.e. at $r=0$. This means that vanishing $R(t,r)$ does not necessarily imply the formation of a singularity. The representation of the distinction in both the cases can be achieved by expressing $R$ as
\begin{equation}\label{R}
    R(t,r)=r v(t,r),
\end{equation}
where $v(t,r)$ can be viewed as a scale factor. The scaling freedom accessible for $r$ can be used to define $R(t_i,r)=r$, where $t_i$ is the initial epoch. This allows us to write the following:
\begin{equation}
   v(t_i,r)=1; \hspace{0.5cm} v(t_s(r),r)=0; \hspace{0.5cm} \dot v<0,
\end{equation}
where $t_s(r)$ is called the singularity curve which gives the time of formation of singularity due to collapsing shell having radial coordinate $r$. Now it can be said that this shell collapses to form a singularity if $v(t_s,r)=0$, thereby distinguishing the case from a regular center. An additional benefit of introducing the scale factor $v$, also known alternatively as the scaling function, is the freedom to study the collapse formalism in the transformed $(r,v)$ coordinates, instead of $(t,r)$ coordinates, which will be apparent in the forthcoming approach.

Let us now recall briefly the formalism developed earlier 
\cite{Joshi_93(2), Joshi_07} 
to study the end state of the collapse. We start with defining an appropriately differentiable function $A(r,v)$ as follows:
\begin{equation}\label{av}
    A_{,v}=\nu ' \frac{r}{R'}.
\end{equation}
Eq.(\ref{efe3}), after integrating, can be used to express $G$ in terms of $A(r,v)$ as
\begin{equation} \label{G2}
    G(r,v)=b(r)e^{2A(r,v)},
\end{equation}
where the integration constant $b(r)$ is related to the velocity with which the matter shell falls in. It can be expressed near the regular center as
\begin{equation}
    b(r)=1+r^2b_0(r).
\end{equation}
$b_0(r)$ is interpreted in analogy with the Lemaitre Tolman Bondi (LTB) dust model in which $b_0<0$ means bounded, $b_0>0$ means unbounded and $b_0=0$ means marginally bound dust collapse. The equation of motion can be found using Eq.(\ref{msmf}) as 
\begin{equation} \label{eom}
    \sqrt{v}\dot v=-e^{\nu} \sqrt{\mathcal{M}+\frac{v(be^{2A}-1)}{r^2}}.
\end{equation}
This can be integrated to achieve the time curve $t(r,v)$ as follows:
\begin{equation}\label{timecurve1}
    t(r,v)=t_i+\int_v^1 \frac{e^{-\nu}}{\sqrt{\frac{\mathcal{M}}{v}+\frac{be^{2A}-1}{r^2}}}dv.
\end{equation}
The time curve dictates the time required for a collapsing shell of radial coordinate $r$ to arrive at an event $v$. This could now be used to get the singularity curve,
\begin{equation}\label{singularitycurve1}
    t_s(r)=t(r,0)=t_i+\int_0^1 \frac{e^{-\nu}}{\sqrt{\frac{\mathcal{M}}{v}+\frac{be^{2A}-1}{r^2}}}dv
\end{equation}
which tells us the time required for a shell of radial coordinate $r$ to collapse to a singularity. Near the center, the time curve can be Taylor expanded  around $r=0$ as 
\begin{equation}\label{taylorsinglaritycurve}
    t(r,v)=t(0,v)+r\chi_1(v)+r^2\chi_2(v)+r^3\chi_3(v)+O(r^4),
\end{equation}
where
\begin{equation}\label{chii}
    \chi_i(v)=\frac{1}{i!}\frac{d^i t}{dr^i}\bigg |_{r=0}.
\end{equation}
For a singularity to be at least locally naked, there have to be families of timelike or null geodesics leaving the singularity. If the trapped surfaces in the neighborhood around the center are formed before the formation of the singularity, the geodesics will not be able to escape, thereby giving a black hole as the end product. The existence or otherwise of such escaping  geodesics can be investigated by considering the equation for outgoing radial null geodesics (ORNG) as follows: 
\begin{equation}\label{efong}
    \frac{dt}{dr}=e^{\psi-\nu}.
\end{equation}
If these geodesics were to be incomplete in the past at the singularity, $R \to 0$ as $t \to t_s$ (or $v\to 0$) along these curves, that ensures a visible singularity. The above equation can be expressed using chain rule in terms of $R$ and $u=r^{\alpha}$, where $\alpha>1$,  as
\begin{equation}
    \frac{dR}{du}=\frac{1}{\alpha}\frac{R'}{r^{\alpha-1}}\left (1+ \frac{\dot R}{R'}e^{\psi-\nu} \right ). 
\end{equation}
which can be rewritten as
\begin{equation} \label{dRbydu}
    \frac{dR}{du}=\frac{1}{\alpha} \left ( \frac{R}{u}+ \frac{\sqrt{v}v'r^{\frac{5-3\alpha}{2}}}{\sqrt{\frac{R}{u}}} \right ) \left (\frac{1-\frac{F}{R}}{\sqrt{G}(\sqrt{G}+\sqrt{H})} \right ).
\end{equation}
Here we have used the relation obtained from Eq.(\ref{msmf}). Along constant $v$ surface, $dv=v'dr+\dot v dt=0$, and hence, $\sqrt{v}v'$, appearing in the above equation, could be obtained from Eq.(\ref{eom})as
\begin{equation}\label{sqrtv'}
    \sqrt{v}v'=e^{\psi}\sqrt{e^{2A}vb_0+vh+\mathcal{M}},
\end{equation}
where
\begin{equation}
    h(r,v)=\frac{e^{2A}-1}{r^2}.
\end{equation}
For a singularity to be naked (at least locally), the tangent to the future directed radially null geodesic, which ceases at the singularity in the past, should have $\frac{dR}{du}>0$ at the singularity in the $(R,u)$ plane 
\cite{Joshi_93(2)}. 
Also, it should be finite. L'Hospital's rule then gives us
\begin{equation} \label{x0}
    X_0= \lim_{(R,u)\to (0,0)} \frac{R}{u}=\frac{dR}{du}.
\end{equation}
The mass profile $\mathcal{M}$ near the center can be Taylor expanded around $r=0$ as
\begin{equation}\label{temp}
    \mathcal{M}(r,v)=M_0(v)+M_2(v)r^2+M_3(v)r^3+M_4(v)r^4+o(r^5).
\end{equation}
At the limit $(r,v)\to (0,0)$ we obtain
\begin{equation}\label{limsqrtvv'}
\lim_{(r,v)\to 0}\sqrt{v}v'=  \big (\chi_1(0) +2r\chi_2(0)+3r^2\chi_3(0) 
 +4r^3\chi_4(0)+ o(r^4)\big )  \sqrt{M_0(0)}.
\end{equation}
Substituting for $\sqrt{v}v'$ from Eq.(\ref{limsqrtvv'}) in the limiting case of Eq.(\ref{dRbydu}) along with using Eq.(\ref{taylorsinglaritycurve}-\ref{chii})  and Eq.(\ref{x0}) gives
\begin{equation}\label{X02}
X_{0}^{\frac{3}{2}}=  \lim_{r\to 0}\frac{1}{\alpha-1}\big ( \chi_1(0) +2r\chi_2(0)+3r^2\chi_3(0) +4r^3\chi_4(0)+o(r^4)\big ) \sqrt{M_0(0)}r^{\frac{5-3\alpha}{2}}.
\end{equation}
It can be seen from the above equation that the problem of determining the local nakedness of the singularity is reduced to determining the polarity of $X_0$. Eq.(\ref{X02}) depicts the relation between the tangent of ORNG at singularity $X_0$ and the components $\chi_i$ of the Taylor expansion of the singularity curve. Here, a specific value of $\alpha$ is chosen so that $X_0 \neq 0$. For instance, if $\chi_1 \neq 0$ then $\alpha=5/3$ has to be chosen, and Eq.(\ref{X02}) is reduced to
\begin{equation}
    X_0^{\frac{3}{2}}=\frac{3}{2}\chi_1(0)\sqrt{M_0(0)}
\end{equation}
at the limit $r\to 0$. This implies that polarity of $\chi_1(0)$ is  the deciding factor for the local visibility or otherwise of the singularity. 

Another possible value which $\alpha$ can take is $\alpha=7/3$, for which the deciding factor is $\chi_2$ as seen in the following specific form of Eq.(\ref{X02}) as follows:
\begin{equation}
    X_0^{\frac{3}{2}}=\frac{3}{2}\chi_2(0)\sqrt{M_0(0)}.
\end{equation}
Here, $\chi_1(0)$ should be of the order of $r$ in order to avoid the blowing up of $X_0$, hence $\chi_1(0)$ has to be zero in the limit $r\to 0$. Generally, $\alpha$ is restricted to the following values so that $X_0 \neq 0$:
\begin{equation}\label{alpha}
  \alpha \in \left\{\frac{2n+3}{3};\hspace{0.2cm} n\in \mathbb{N} \right\}.
\end{equation}
Additionally, near $(r,v)\to (0,0)$,  we should have 
\begin{equation}\label{orderofchi}
  \chi_i(v) \sim O\left(r^{\frac{3\alpha-1}{2}-i}\right), \hspace{0.5cm} \forall \ i<\frac{3}{2}(\alpha-1).
\end{equation}
This ensures that $\chi_i(0)=0$ and thereby preventing $X_0$ from blowing up. Whether or not these values of $\alpha$ in (\ref{alpha}) corresponds to a singularity which is strong, in the sense of Tipler, is investigated in the next section.

\section{Strength of Singularities}\label{2.2}
The tangents of the outgoing timelike or null geodesic from a singularity formed due to gravitational collapse of an inhomogeneous spherically symmetric perfect fluid with non-zero pressure are as follows:

\begin{equation} \label{tangent equation}
    \begin{split}
& K^t =\frac{dt}{d\lambda}=\frac{\mathcal{P}}{R}, \\ 
& K^r =\frac{dr}{d\lambda}=\frac{\sqrt{G}}{RR'}\sqrt{\mathcal{P}^2\frac{\dot R^2}{H}-l^2+BR^2},\\ 
& K^{\theta^2}+\sin ^2\theta K^{\phi^2} =\frac{l^2}{R^4}.\\
\end{split}
\end{equation}

Here the value of $B$ denotes the type of geodesics such that for null geodesic $B=0$ and for timelike geodesic $B=-1$. Also, $l$ is called the impact parameter which vanishes for radial geodesics. The function $\mathcal{P}(t,r)$ satisfies the following geodesic equation:
\begin{equation}\label{geodesic equation}
\begin{split}
& \frac{d\mathcal{P}}{d\lambda}-\frac{\mathcal{P}^2}{R}\left(\frac{\dot R}{R} -\frac{ \dot H}{2H}+\frac{\ddot R}{\dot R} -\frac{\dot R'}{R'}+\frac{\dot G}{2G} \right)-\frac{\mathcal{P}\sqrt{G}}{R}\sqrt{\frac{\mathcal{P}^2\dot R^2}{H}-l^2+BR^2} \\
&\left (\frac{1}{R}+\frac{1}{R'}\left( \frac{2\dot R'}{\dot R}-\frac{H'}{H} \right)\right) +\frac{H}{\dot R}\left(\frac{l^2}{R}\left(\frac{1}{R}+\frac{\dot G}{2G \dot R}-\frac{\dot R'}{\dot R R'}\right)+\frac{BR}{\dot R}\left(\frac{ \dot R'}{R'}-\frac{\dot G }{2G}\right)\right)=0.
\end{split}
\end{equation}
For radial null geodesic, close to $\lambda=0$, i.e. near the singularity, using L'Hospital's rule in the above equation gives us the expression of $\mathcal{P}$ as follows:
\begin{equation} \label{P expression}
    \mathcal{P}= \lim_{r\to 0}\frac{R}{\lambda}\Bigg (\frac{\dot R}{R}-\frac{\dot H}{2H}+\frac{\ddot R}{\dot R}+\frac{\dot G}{2G}-\frac{\dot R'}{R'}+\sqrt{\frac{G}{H}}\Bigg (\frac{\dot R}{R} 
     -\frac{H'\dot R}{HR'}+\frac{2\dot R'}{R'}\Bigg )\Bigg )^{-1}.
\end{equation}
The sufficient condition for a singularity to be strong in the sense of Tipler 
\cite{Tipler_77}, 
provided by Clarke and Krolak 
\cite{Clarke_85}, 
is that at least along one null geodesic with the affine parameter $\lambda$, with $\lambda=0$ at the singularity, the inequality (\ref{Krolak and Clarke criteria}) should be satisfied. Eq.(\ref{efe11}) and Eq.(\ref{efe21}), gives
\begin{equation}\label{psi}
R_{ij}K^iK^j =  \frac{1}{2R^2}\Big ( \frac{\dot R}{H R'}\left(F'\dot R-3\dot FR'\right)(K^t)^2  +\frac{R'}{G\dot R}\left(F'\dot R+\dot FR'\right)(K^r)^2 \Big ).
\end{equation}
Substituting for the tangents to the radial null geodesic from Eq.(\ref{tangent equation}), we get
\begin{equation}\label{clarkkrolak2}
     \lim_{ \lambda \to 0}\lambda^2R_{ij}K^iK^j = 3 \lim_{ \lambda \to 0}\left(\frac{\lambda \sqrt{F}\mathcal{P}\dot R}{R^2\sqrt{rR'}\sqrt{H}}\right)^2.
\end{equation}
Here, we have used the following limiting values arising from the regularity conditions:
\begin{equation}\label{etazeta}
 \lim_{r\to 0}\frac{rF_{,r}}{F}=3,\hspace{0.5cm} \lim_{v\to 0}\frac{vF_{,v}}{F}=0.
\end{equation}
The particular case of LTB collapse reduces the expression Eq.(\ref{clarkkrolak2}) to 
\begin{equation}
      \lim_{ \lambda \to 0}\lambda^2R_{ij}K^iK^j = 3 \lim_{ \lambda \to 0}\left(\frac{\lambda \sqrt{F}\mathcal{P}}{R^2\sqrt{rR'}}\right)^2,
\end{equation}
which agrees with the result obtained in 
\cite{Joshi_93}. The above equation is obtained by substituting $H=\dot R^2$ in Eq.(\ref{clarkkrolak2}).
Using Eq.(\ref{P expression}) and Eq.(\ref{clarkkrolak2}), the Clarke and Krolak's criteria is restated as
\begin{equation} \label{clarkkrolak}
\begin{split}
    & \lim_{(r,v)\to (0,0)} \left(\frac{F'}{R'}-\frac{\dot F}{\dot R}\right)\Bigg ( \sqrt{G}\left(1-\frac{H'R}{HR'}+\frac{2R\dot R'}{\dot R R'} \right) \\
    & +\sqrt{H}\left(1-\frac{\dot H R}{2H \dot R}+\frac{\ddot R R }{\dot R^2}+\frac{R\dot G}{2\dot R G}- \frac{R\dot R'}{\dot R R'}\right)\Bigg )^{-2}>0,
\end{split}    
\end{equation}
which should hold at least along one null geodesic which is past incomplete at the singularity, for the singularity to be Tipler strong. $H$ can be expressed using Eq.(\ref{gh}) and Eq.(\ref{eom}) as
\begin{equation}\label{H}
    H(r,v)=\frac{\mathcal{M}r^2}{v}+be^{2A}-1.
\end{equation}
Differentiating Eq.(\ref{H}) with respect to $r$ can lead to 
\begin{equation}\label{H'}
    \lim_{(r,v)\to(0,0)}\frac{H'}{H}=\lim_{r\to 0}\frac{1}{r}+\frac{M_{,r}}{M}.
\end{equation}
Differentiating $H$ in Eq.(\ref{gh}) with respect to $t$ gives the following equation:
\begin{equation}\label{hdot}
    \frac{\ddot R R}{\dot R^2}-\frac{\dot H R}{2H\dot R}=\nu_{,v}v.
\end{equation}
Differentiating $G$ in Eq.(\ref{gh}) with respect to $t$ and using Eq.(\ref{G2}) gives the following equation:
\begin{equation}\label{gdot}
    \frac{r\dot v'}{\dot v}=2vA_{2,v}r^2+2\psi_{,v}v.
\end{equation}
In the $(r,v)$ coordinate, we have
\begin{equation}\label{F'/R'-Fdot/Rdot}
  \lim_{(t,r)\to(t_s,0)}\frac{F'}{R'}-\frac{\dot F}{\dot R}  =  \lim_{(r,v)\to(0,0)}\frac{1}{2}\left(\frac{F_{,r}}{v}-\frac{F_{,v}}{r} \right)
   =\frac{3}{2}\frac{\mathcal{M}(0,0)}{X_0} \lim_{r\to 0}  r^{3-\alpha}.
\end{equation}
Using Eq.(\ref{H'}, \ref{hdot}, \ref{gdot}, \ref{F'/R'-Fdot/Rdot}) in Eq.(\ref{clarkkrolak}) we obtain the condition of Clarke and Krolak as
\begin{equation} \label{final clarke and krolak criteria}
     \lim_{(r,v)\to (0,0)}  \Bigg (\sqrt{\frac{\lvert X_0 \rvert}{\mathcal{M}(0,0)} }r^{\left(\frac{\alpha-3}{2}\right)}\bigg (\frac{1}{2}-\frac{M_{,r}r}{2M}+2\psi_{,v}v  + 2vA_{2,v}r^2\bigg ) +1 +\nu_{,v}v-\psi_{,v}v \Bigg )^{-2}>0
\end{equation}
at least along one null geodesic. The above inequality can be satisfied only if 
%either $\zeta_0<\eta_0\leq 3 \leq \alpha$ or $\zeta_0<\eta_0 \leq \alpha <3$. 
%However, it can be shown that for regularity condition to be satisfied, $\zeta_0=0$ and $\eta_0\geq 3$. This implies that
\begin{equation}\label{alphageq3}
    \alpha\geq 3,
\end{equation}
for if $\alpha<3$, then the denominator on the left hand side of the inequality (\ref{final clarke and krolak criteria}) will blow up in the limit $(r,v)\to (0,0)$, thereby not satisfying the inequality anymore.

From Eq.(\ref{orderofchi}) and Eq.(\ref{alphageq3}) it can be seen that in order to maintain the finiteness of $X_0$, $\chi_1$ and $\chi _2$ should be of the order of at least $r^2$ and $r$ respectively, implying that 
\begin{equation}
\chi_1(0)=\chi_2(0)=0.
\end{equation}
%This is only the necessary condition for a singularity to be strong and locally naked. 

It is to be noted that $\alpha$ can take values as follows:
\begin{equation}
    \alpha \in \left\{\frac{2n+1}{3};\hspace{0.2cm} n\geq 4; \hspace{0.2cm} n\in \mathbb{N} \right\}.
\end{equation}
We now carry out a case study for one such value of $\alpha$  in the next section.

\section{Collapse Endstates}\label{2.3}
If $\alpha=3$, the equation for tangent of the null geodesic at the singularity for $r=0$ follows from Eq.(\ref{X02}) as 
\begin{equation}\label{X0chi3}
      X_{0}^{\frac{3}{2}}=\lim_{r\to 0}\frac{3}{2}\sqrt{M_0(0)}\chi_3(0).
\end{equation}
Polarity of $\chi_3$ then determines the polarity of $X_0$ which in turn determines the nakedness or otherwise of the Tipler strong singularity. Substituting for density and pressure of the cloud from Eq.(\ref{efe11}) and Eq.(\ref{efe21}) in Eq.(\ref{efe3}) gives us 
\begin{equation}\label{nudash}
    \nu '=\frac{\mathcal{M},_{vr}v+\left(\mathcal{M},_{vv}v-2\mathcal{M},_{v} \right)w}{\left(3\mathcal{M}+r\mathcal{M},_r-\mathcal{M}_v v\right)v}R'.
\end{equation}
Here, $v'$, which is the partial derivative of $v$ in $(t,r)$ coordinate has been expressed as a function $w(r,v)$ in the $(r,v)$ coordinate.  One could use the above equation in Eq.(\ref{av}) for obtaining the integral expression of $A(r,v)$ as
\begin{equation}\label{iearv}
    A(r,v)=\int_v^1 \frac{\mathcal{M},_{vr}v+\left(\mathcal{M},_{vv}v-2\mathcal{M},_{v} \right)w}{\left(3\mathcal{M}+r\mathcal{M},_r-\mathcal{M}_v v\right)v} rdv.
\end{equation}
Also regularity condition demand that $A \simeq r^2 $. Hence one can Taylor expand it around $r=0$ as 
\begin{equation}\label{tearv}
    A(r,v)=A_2(v)r^2+A_3(v)r^3+...
\end{equation}
where the components $A_i(v)$, $i\geq 2$, can be obtained using Eq.(\ref{iearv}) as follows:
\begin{equation}\label{A2}
    A_2(v)=\int_v^1 \frac{2M_{2,v}+\left(M_{0,vv}-\frac{2M_{0,v}}{v} \right)w,_{r}}{3M_0-M_{0,v}v} dv,
\end{equation}
\begin{equation}\label{A3}
    A_3(v)=\int_v^1 \frac{6M_{3,v}+\left(M_{0,vv}-\frac{2M_{0,v}}{v} \right)w,_{rr}}{3M_0-M_{0,v}v}dv,
\end{equation}
\begin{equation}\label{A4}
    \begin{split}
 A_4(v)= & \int_v^1\frac{1}{\left(3M_0-M_{0,v}v\right)^2}\Bigg (2M_{2,v}\left(vM_{2,v}-5M_2\right)+4M_{4,v}\left(3M_{0}-vM_{0,v}\right) +\\  & w,_r \bigg ( M_0\big (3M_{2,vv} -\frac{6M_{2,v}}{v}\big )+M_{0,v}\left(\frac{10M_2}{v}-vM_{2,vv}\right)+M_{0,vv}\left(v M_{2,v}-5M_2\right)\bigg )\Bigg ) \\ &+w,_{rrr}\frac{\left(M_{0,vv}v-2M_{0,v}\right)}{6v\left(3M_0-M_{0,v}v\right)} dv,
    \end{split}
\end{equation}
and
\begin{equation}\label{A5}
  \begin{split}
A_5(v)= & \int_v^1\frac{1}{(3M_0-M_{0,v}v)^2} \Bigg (-12M_3M_{2,v}-15M_2M_{3,v}+5v M_{2,v}M_{3,v} + w,_{r} \Big (-\frac{6M_0M_{3,v}}{v} \\ 
& +M_3\left(\frac{12M_{0,v}}{v}-6M_{0,vv}\right)+M_{0,vv}M_{3,v}v+3M_0M_{3,vv}-M_{0,v}M_{3,vv}v \Big ) \\ 
&+w,_{rr}\Bigg (5M_2\left(\frac{M_{0,v}}{v}-\frac{M_{0,vv}}{2}\right) \\ 
& +3M_0\left(-\frac{M_{2,v}}{v}+\frac{M_{2,vv}}{2}\right)+\frac{v}{2}\left(M_{2,v}M_{0,vv}-M_{0,v}M_{2,vv}\right)\Bigg )\Bigg) +\\ 
& w,_{rrrr} \frac{\left(-\frac{M_{0,v}}{12v}+\frac{M_{0,vv}}{24}\right)}{3M_0-M_{0,v}v} dv.    
\end{split}
\end{equation}
These components of Taylor expansion of $A(r,v)$ around $r=0$ are then used to determine $\chi_3$ by differentiating the singularity curve thrice. We also have,
\begin{equation}\label{timecurvecomponent}
    \frac{be^{2A}-1}{r^2}=\sum_{i=0}^{\infty} ((i+2)A_{i+2}+b_{0i})r^i
\end{equation}
 near the center. Here, $b_{0i}$ are the coefficients of $r^i$ in the Taylor expansion of $b_0(r)$ around $r=0$. Substituting from Eq.(\ref{timecurvecomponent}) in Eq.(\ref{singularitycurve1}) along with using Eq.(\ref{taylorsinglaritycurve}) and Eq.(\ref{chii}), we obtain the expression of $\chi_3$ as
\begin{equation}\label{chi3}
\begin{split}
    \chi_3= &  \int_v^1 \frac{3A_3+b_{01}}{\left(\frac{M_0}{v}+2A_2+b_{00}\right)^{\frac{3}{2}}}\left(\frac{g_2}{2}-\frac{5}{16}\left(\frac{3A_3+b_{01}}{\frac{M_0}{v}+2A_2+b_{00}}\right)^2+\frac{3}{4}\left(\frac{\frac{M_2}{v}+4A_4+b_{02}}{\frac{M_0}{v}+2A_2+b_{00}}\right)\right) \\
    & - \frac{1}{2}\frac{\left(\frac{M_3}{v}+5A_5+b_{03}\right)}{\left(\frac{M_0}{v}+2A_2+b_{00}\right)^{\frac{3}{2}}}dv.
\end{split}
\end{equation}
Here, $g_2=\frac{1}{2}A_{2,v}v.$
As is apparent from Eq.(\ref{X0chi3}), polarity of $\chi_3$ is the deciding factor for local visibility of Tipler strong singularity. The expressions for $A_2$, $A_3$, $A_4$ and $A_5$ can be obtained from Eq.(\ref{A2}-\ref{A5}) for a given mass profile $\mathcal{M}(r,v)$, which is then substituted in Eq.(\ref{chi3}). However, while calculating the $A_i$s, we also require the derivatives of $w(r,v)$ with respect to $r$, which is not known in general. Nevertheless, for a well-chosen mass profile such that the components non-minimally coupled with the derivatives of $w(r,v)$ in the integral expressions for $A_i$ vanish, we could bypass the requirement of the information of the collapse dynamics. Since, there is no mention of equation of state here, we have total five field equations in six unknown parameters namely $p$, $\rho$, $\nu$, $\psi$, $R$ and $F$, i.e. two matter variables, three metric tensor components and the Misner-Sharp mass function. Therefore, there is one degree of freedom left, thereby allowing us to specify the evolution of mass profile $\mathcal{M}$. The idea is to give a small perturbation to the mass profile corresponding to inhomogeneous dust which upto fourth order is expressed close to the center as 
\begin{equation}\label{massprofile}
    \mathcal{M}(r,v)=m_0+m_2r^2+m_3r^3+m_4r^4,
\end{equation}
where $m_0$, $m_2$, $m_3$ and $m_4$ are constants. The perturbation term $\delta(v)$ is then coupled minimally to the fourth order component of $\mathcal{M}$. The reason for this form of perturbation is to vanish the terms involving the derivative of $w(r,v)$ in the expression of $A_i$.
One such example of a perturbed mass profile is as follows:
\begin{equation}\label{perturbedmassprofile}
   \mathcal{M}(r,v)=m_0+m_2r^2+m_3r^3+m_4r^4+\delta(v) r^4.
\end{equation}
This mass profile can give rise to non-zero pressure near the center. 

Now, let us consider the mass profile Eq.(\ref{massprofile}), with  $m_0=1$, $m_2=-0.1$, $m_3=0$ and $m_4=-0.1$. Let us give a fourth order perturbation, $\delta (v)=-0.1 (1-v^2)$. This perturbed mass profile corresponds to a perfect fluid with non-zero pressure associated with it. Fixing $b_{00}=-0.5$ and $b_{01}=-0.1$, a non-zero measured set of initial data $(b_{02},b_{03})$ satisfying the inequality 
$$
9.4685 b_{02}+48.4614 b_{03}<1
$$
is obtained for which $\chi_3>0$, and hence the end state of the collapse for such initial data is a locally visible Tipler strong singularity.

\section{Concluding remarks}\label{2.4}

Some concluding points and open concerns are discussed below:
\begin{enumerate}

\item The necessary criterion for a central shell-focusing singularity formed due to gravitational collapse of a spherically symmetric inhomogeneous perfect fluid with non-zero pressure to be visible is that the relation between the physical radius and the radial coordinate of ORNG should be of the form
\begin{equation*}
        R=X_0r^{\alpha}, \hspace{0.5cm} X_0>0,
    \end{equation*}
where $\alpha$ is restricted to the values given by the set 
\begin{equation*}
    \alpha \in \left\{\frac{2n+3}{3};\hspace{0.2cm} n\in \mathbb{N} \right\}.
\end{equation*}

\item For this singularity to be strong in the sense of Tipler, set of possible values of $\alpha$ is further refined as follows:
\begin{equation*}
     \alpha \in \left\{\frac{2n+3}{3};\hspace{0.2cm} n\geq 3; \hspace{0.2cm} n\in \mathbb{N} \right\}.
\end{equation*}
 This restriction on $\alpha$ concludes that the locally naked singularities in \cite{Joshi_12}
 and 
 \cite{Goswami_2007} 
 are not Tipler strong because of the fact that $\alpha$ was chosen to be $\frac{5}{3}$ and $\frac{7}{3}$. 
 
\item While investigating the end state, the requirement of pre-knowledge of the dynamics of the collapse, $v(t,r)$, causes a hindrance to proceed further to determine the visibility of the singularity, as observed in Eq.(\ref{A2}-\ref{chi3}). Nevertheless, due to a degree of freedom available with us, we have freedom of choice of fixing an unknown function. In our case, this unknown function is the mass profile of the fluid. By wisely choosing the mass profile, the requirement of the knowledge of  $v(t,r)$ could be bypassed. To achieve this, we have given a perturbation to the mass profile for dust in such a way that the components non-minimally coupled with the derivative terms of the scaling function in Eq.(\ref{A2}-\ref{A5}) vanish. One way to obtain such a mass profile is to add a perturbed term of order four in $r$. For an example of such a form of a mass profile, there indeed exists a non-zero measured set in the $(b_{02},b_{03})$ plane for which the end state after the collapse is a Tipler strong locally visible singularity. Existence of such a set of initial data guarantees that the naked singularities forming due to perfect fluid collapse are stable against any perturbation in the initial data from which the collapse begins.  

\item This acts as a counter-example to at least the strong cosmic censorship hypothesis which does not allow the existence of such locally visible singularity. It is to be noted that the matter fluid formed due to such perturbed term satisfies the weak energy condition and has non-zero pressure $p=-\frac{\delta,_{v}}{X_0^2}$, not restricted to any equation of state. Non-zero pressure in the collapsing cloud arises because of the time dependence property of the perturbed mass profile. Hence, we have shown that there exists collapsing cloud having certain mass profile with non-zero pressure which collapses to form a Tipler strong singularity which is locally visible. Also, since the collapsing cloud is scale independent, if its size is very large, an observer sufficiently close to the singularity will be able to detect the singularity even if it is only locally naked. Hence, even a locally naked singularity is a serious defiance of the cosmic censorship. 
\item In 
\cite{Joshi_93}, 
in the case of inhomogeneous collapsing dust, it has been shown that $\alpha\leq 3 $ for a singularity to be naked ($X_0>0$). This puts a further restriction on $\alpha$, fixing it to $\alpha=3$ for a singularity to be Tipler strong and locally visible. In our case study, we have shown that $\alpha=3$ indeed gives Tipler strong locally visible singularity formed due to collapsing perfect fluid cloud with non-zero pressure. Whether or not $\alpha>3$ gives a naked singularity is yet to be studied.

\item Throughout this chapter, we have considered the possibility of strong singularities which are locally naked. Whether or not they are globally naked is still unknown. The existence of such singularities would be a big blow to the weak cosmic censorship. 
%\item quantum gravity
\end{enumerate}
 
% Chapter Template

\chapter{Globally visible singularity in gravitational collapse of nonmarginally bound dust} % Main chapter title

\label{Chapter3} % Change X to a consecutive number; for referencing this chapter elsewhere, use \ref{ChapterX}

\lhead{\emph {Chapter 3. }} % Change X to a consecutive number; this is for the header on each page - perhaps a shortened title

%----------------------------------------------------------------------------------------
%	SECTION 1
%----------------------------------------------------------------------------------------
In the introduction, and in the previous chapter, we have seen that the local causal structure of the end state of the collapse has been studied wherein possibilities of locally naked singularities have been depicted. Nevertheless, the globally naked singularities, rather than the locally naked singularities, may have more observational significance. Some of the work dealing with global visibility can be found in \cite{Deshingkar_98, Harada_04, Miyamoto_13, Jhingan_14, Kong_14, Ortiz_15}. 
Deshingkar, Jhingan and Joshi 
\cite{Deshingkar_98} 
depicted some examples of mass functions giving rise to a globally visible singularity where the mass profile is a function of only $r$, i.e., the fluid under consideration was dust. The collapse, in this case, is considered to be marginally bound. Later, Jhingan and Kaushik
\cite{Jhingan_14}
used a certain transformation of coordinates to put a restriction on the mass profile of a marginally bound collapsing dust to ensure global visibility of the singularity thus formed. 
On the contrary, Miyamoto, Jhingan and Harada 
\cite{Miyamoto_13} 
investigated some stellar models (density distribution and total mass as the parameters) influenced by marginally stable configurations of neutron stars for various equations of state
\cite{Shapiro_83}
and realized that for such configurations, the outgoing null geodesic, if at all it exists, gets trapped inside the event horizon, thereby making the singularity globally invisible. Suggestions in support of the validity of weak cosmic censorship have also been discussed by Wald 
\cite{Wald_84} 
and Hod 
\cite{Hod_08}.

It is to be noted that the strength of singularities formed due to the depicted mass functions in 
\cite{Deshingkar_98, Miyamoto_13, Jhingan_14} 
was not investigated. 

Our basic purpose here is to examine the global causal structure of a singularity, keeping in mind the maintenance of its strength in the sense of Tipler
\cite{Tipler_77}, 
to ensure the physical relevance of the scenario considered. Also, marginally bound collapse is a very special case which corresponds to a very specific dynamics of the collapse, as we will see in the next sections. Considering such a collapsing scenario makes it easy to integrate one of Einstein's field equations. However, the generality is lost by doing so. Hence, we take into consideration here the nonmarginally bound collapse which incorporates all the possible dynamics of the collapse  (except one corresponding to marginally bound) depending on the functional form taken by the velocity function and permitted by the  Einstein's field equations, thereby widening our scope of understanding the gravitational collapse and its end state to a more general scenario.

The chapter is arranged as follows: In 3.1, Einstein's field equations corresponding to an inhomogeneous collapsing dust cloud is discussed. In 3.2, the possibility of global visibility of singularities formed due to bound dust collapse is discussed.  In 3.3, the strength of such globally visible singularity in the sense of Tipler is discussed. We end the chapter with the concluding remarks and stating a few open concerns in 3.4. 

\section{Lemaitre-Tolman-Bondi spacetime}\label{3.1}
The Lemaitre-Tolman-Bondi metric 
\cite{Lemaitre_33, Tolman_34, Bondi_47} 
is a spherically symmetric metric governing the spacetime of collapsing dust clouds. It is given by
\begin{equation}\label{ltb}
    ds^2=-dt^2+\frac{R'^2}{1+f}dr^2+R^2d\Omega^2
\end{equation}
in the comoving coordinates $t$ and $r$. We consider here a \textit{type I} matter field
\cite{Hawking_73}.
In such a matter field, the energy-momentum tensor has nondiagonal entries as zero in a comoving coordinate system. One of the eigenvalues $\rho$ represents the energydensity as measured by a comoving observer at a point $p$. All observed fields with nonzero rest mass can be classified under type I matter field. The corresponding energy-momentum tensor along with vanishing pressure is given by
\begin{equation}
    T^{\mu \nu}=\rho U^{\mu} U^{\nu},
\end{equation}
where $U^{\mu}$, $U^{\nu}$ are the components of the four-velocity.
Einstein's field equations give us the expression of density and pressure, and the information about the dynamics of the collapse as
\begin{equation}\label{efe0}
    \rho=\frac{F'^2}{R^2R'},
\end{equation}
\begin{equation}\label{efe1}
    p=-\frac{\dot F}{R^2 \dot R},
\end{equation}
and
\begin{equation} \label{efe2}
    \dot R^2=\frac{F}{R}+f
\end{equation}
respectively. The superscripts dot and prime denote the partial derivative with respect to $t$ and $r$, respectively. Here, $F$ and $f$ are, respectively, called the Misner-Sharp mass function and the velocity function. The Misner-Sharp mass function in case of dust is a function of $r$ only and independent of $t$. This can be seen from Eq.(\ref{efe1}) which tells us that $\dot F=0$ since $p=0$ in case of dust. $F$ tells us about the mass of the collapsing cloud inside a shell of radial coordinate $r$ at time $t$. For zero pressure, this mass is conserved inside a fixed radial shell. For the collapsing matter field to be well behaved at the initial time and at the center of the cloud, certain regularity conditions need to be maintained. The metric functions should by $\mathcal{C}^2$ differentiable everywhere according to the obligations of the Einstein's field equations. The Misner-Sharp mass function should have the following expression:
\begin{equation}\label{regularitycondition}
    F(r)=r^3 M(r).
\end{equation}
Here, $M>0$ and is a regular, at least $\mathcal{C}^2$ function, having a finite value at the limit of approach to the center, and is called the mass profile of the collapsing cloud. In the case $F$ goes as $r^2$ or lower power, it could be seen from the Einstein's field equation (\ref{efe0}) that the density blows up at the center at the initial epoch itself, which is undesirable. Additionally, in order to avoid cusp in the energy density, the function space of $M$ is further restricted to follow the condition 
\begin{equation}
    M'(0)=0.
\end{equation}
The positivity of energy density is achieved by restricting $F'>0$ and $R'>0$. This is maintained by restricting the mass profile as follows:
\begin{equation}
    3M+rM'>0.
\end{equation}
Energy density can also be positive when $F'<0$ and $R'<0$. However, in such a case, the mass profile becomes negative near the center, which is not allowed.

A collapsing solution of Einstein's field equation is obtained by restricting the physical radius as $\dot R<0$. This means that a particular shell of fixed radial coordinate collapses to form a singularity when $R=0$ for this shell. However, $R$ vanishes also at the regular center. Both these cases can be differentiated by expressing the physical radius as
\begin{equation}
    R(t,r)=r v(t,r).
\end{equation}
Now, a shell of radial coordinate $r$ is said to form a singularity a time $t_s$ when $v(t_s,r)=0$. Rescaling of the physical radius is done using the coordinate freedom such that
\begin{equation}\label{scaling freedom}
    R(t_i,r)=r,
\end{equation} 
where $t_i$ is the initial time. This can be rewritten as $v(t_i,r)=1$.

The polarity of $f$ classifies the spacetime in three different categories: bound (elliptic), marginally bound (flat) and unbound (hyperbolic) collapse, corresponding to the restrictions $f<0$, $f=0$ and $f>0$ respectively. Rewriting Eq.(\ref{efe2}) as $\dot v=\sqrt{M+f/r^2}$ demands that the velocity function should have the form
\begin{equation}
    f=r^2b_0(r)
\end{equation}
as a regularity requirement. Here, $b_{0}(r)$ is a sufficiently differentiable function.

Equation (\ref{efe2}) can be integrated to get 
\begin{equation}\label{timecurve}
    t-t_s(r)=-\frac{R^{\frac{3}{2}}\mathcal{G}(-fR/F)}{\sqrt{F}}.
\end{equation}
Here $\mathcal{G}(y)$ is defined as in Eq.(\ref{G})
The constant of integration in Eq.(\ref{timecurve}) can be obtained using Eq.(\ref{scaling freedom}) as
\begin{equation}\label{singularitycurve}
     t_s(r)=\frac{r^{3/2}\mathcal{G}(-fr/F)}{\sqrt{F}}.
\end{equation}
This is called the singularity curve. It gives us the information about the time at which a shell of radial coordinate $r$ collapses to form a singularity $R=0$. 

A singularity can be only locally naked if the null geodesic can escape from the neighborhood of the singularity but later in its path, comes across the trapped surfaces, and falls back to the singularity. The boundary of all trapped surfaces is called the apparent horizon. The evolution of the apparent horizon is determined by equating the physical radius with the Misner-Sharp mass function as $R=F(r)$. This, along with Eqs.(\ref{timecurve}) (\ref{singularitycurve}) gives us the time of formation of the apparent horizon as a function of radial coordinate as
\begin{equation}\label{ahcurve}
    t_{AH}(r)=\frac{r^{3/2}\mathcal{G}(-rf/F)}{\sqrt{F}}-F \mathcal{G}(-f).
\end{equation}
It is also called the apparent horizon curve. 
%Various examples of mass profiles and velocity functions have been shown to give rise to a singularity such that
$F(0)=0$ implies $t_s(0)=t_{AH}(0)$, thereby creating a possibility for nonspacelike geodesic to have a positive tangent at $r=0$. Such singularities are at least locally naked. The geodesics may later get trapped, thereby keeping the weak cosmic censorship intact. However, it is also possible that the singular geodesic avoids getting trapped by the trapped surfaces and reaches the boundary of the collapsing cloud unhindered. Such singularities are studied in detail in the next section.

\section{Global visibility}\label{3.2}
The singularities which are only locally visible may not be of much observational significance. This is because, in such a case, an observer outside the event horizon will not be able to receive any signal escaping from the neighborhood of the singularity. For this reason, it is of extreme importance to investigate whether or not there exists a globally visible singularity. 

For a singularity to be globally visible, null geodesics originating from the neighborhood of the singularity should not only avoid getting trapped by the trapped surfaces but also reach the boundary of the star before the event horizon. It turns out that for globally visible singularity, the latter always implies the former. This is because, at $r=r_c$, the apparent horizon coincides with the event horizon. We know that the evolution of the event horizon of the collapsing cloud is the same as the evolution of the null geodesic along with the condition that at the boundary of the cloud $r_c$, the following equality should be satisfied:
\begin{equation}\label{icforeh}
    F(r_c)=R(t,r_c).
\end{equation}
Now, the event horizon cannot start forming after the initiation of the formation of trapped surfaces (or its boundary, i.e., the apparent horizon). This is because any null geodesic, more specifically outgoing null geodesic, forming inside the apparent horizon, will have a negative tangent and fall back into the singularity. The evolution of EH can be thought of as the evolution of the last outgoing radial null geodesic escaping the center without getting trapped and falling back to the singularity. The equation of the null geodesic is given by
\begin{equation}\label{NG}
    \frac{dt}{dr}=\frac{R'}{\sqrt{1+f}}.
\end{equation}
In the case of inhomogeneous dust, at $r=0$, the time of formation of AH is the same as the time of formation of central singularity, as seen from Eq.(\ref{singularitycurve}) and Eq.(\ref{ahcurve}) and the fact that $F$ vanishes. Hence, it can be concluded that the  EH starts forming either before or during the formation of the  singularity due to the collapse of the central shell i.e.
\begin{equation}
t_{EH}(0)\leq t_s(0).    
\end{equation} 
Here, $t_{EH}(r)$ is the event horizon curve which is the solution of Eq.(\ref{NG}) with the condition given by Eq.(\ref{icforeh}).
 
We now define a small neighborhood around the time of formation of central singularity such that the null geodesic escaping the center at a time belonging to this neighborhood (an interval around $t_s(0)$ rather than a single point $t_s(0)$) will be termed singular. This neighborhood should have a size of the order of Planck time.

One may question the choice of the size of this neighborhood as a magnitude influenced by quantum theory, even when general relativity is assumed to be fundamental. Let us recall that, as mentioned in the Introduction (1.2); we interpret the singularity theorem as proposed by Misner
\cite{Misner_1969}.
Hence, we will determine the result obtained by the general relativistic approach in the strong gravity regime, which may help us to predict what we must expect from a quantum theory of gravity.

If we can trace a singular null geodesic (SNG) reaching the boundary before the event horizon, then we have 
\begin{equation}
 t_{SNG}(r_c)<t_{EH}(r_c).   
\end{equation}
Now, we know that $R$ is a monotone decreasing function of $t$ since $\dot R<0$. Hence we have
\begin{equation}
R( t_{SNG}(r_c),r_c)>R(t_{EH}(r_c),r_c).
\end{equation}
However, we know that $R(t_{EH}(r_c),r_c)=F(r_c)$ from Eq.(\ref{icforeh}). Hence for a singular null geodesic reaching the boundary before the event horizon, the following inequality should be satisfied:
\begin{equation}
    R(t,r_c)>F(r_c).
\end{equation}
Geometrically, the above inequality gives a positive value of the expansion parameter for outgoing null geodesic congruence $\Theta_l$, at $r=r_c$, which is expressed in terms of physical radius, Misner-Sharp mass function and velocity function as follows:
\begin{equation}
   \Theta_l= \frac{2}{R}\left( \sqrt{1+f}-\sqrt{\frac{F}{R}+f} \right).
\end{equation}
This specifies the divergent nature of these outgoing null geodesic congruences at $r=r_c$.

Now, the expression of $R$ in terms of the comoving coordinates $t$ and $r$ is obtained from Eqs.(\ref{timecurve}) and (\ref{singularitycurve}) as 
\begin{equation}\label{Rfornonzerof}
    R=\left(\frac{r^{\frac{3}{2}}\mathcal{G}\left(-f r/F\right)-\sqrt{F}t }{\mathcal{G}\left(-f R/F\right)}\right)^{\frac{2}{3}}.
\end{equation}
In the case of marginally bound collapse, this is reduced to 
\begin{equation}\label{R(t,r)mb}
    R=\left(r^{\frac{3}{2}}-\frac{3}{2}\sqrt{F}t \right)^{\frac{2}{3}}.
\end{equation}
However, in the case of nonmarginally bound collapse, we use the Taylor expanded expression for the function $\mathcal{G}(y)$ given by Eq.(\ref{G})  around $y=0$ for $0<y\leq 1$ as
\begin{equation}\label{taylorexpandedG}
    \mathcal{G}(y)=\frac{2}{3}+\frac{1}{5}y+\frac{3}{28}y^2+o(y^3).
    \end{equation}
This  can then be used in Eq.(\ref{Rfornonzerof}) to write $R$ explicitly as
\begin{equation}\label{R(t,r)}
    R(t,r)=\frac{5F}{2f}\left(1-\sqrt{1- o(y_1^3)-\frac{4f}{5F}\left(r^{\frac{3}{2}}\left(1-\frac{3fr}{10F}+o(y_2^2)\right)-\frac{3}{2}\sqrt{F}t\right)^{\frac{2}{3}}}\right),
\end{equation}
for nonvanishing velocity function, i.e. $f\neq 0$. 
%Here, we have considered the expansion of $\mathcal{G}$ from Eq.(\ref{taylorexpandedG}) in Eq.(\ref{R(t,r)})) only up to first order. 
Here, 
\begin{equation}
    y_1=-\frac{fR}{F}, \hspace{1cm} y_2=-\frac{fr}{F}.
\end{equation}
Ignoring higher order, i.e. $o(y_1^3)$ and $o(y_2^2)$ in Eq.(\ref{R(t,r)}) is equivalent to considering the expansion of $\mathcal{G}$ from Eq.(\ref{taylorexpandedG}) only up to first order. Hence, large value of the ratio $\frac{fR}{F}$ may not give a good approximation. Therefore, in our investigation, we make sure to keep this ratio small by considering positive velocity function having small deviation from zero. 
%%%%%%%%%%%%%%%%%%%%%%%%%%%%%%%%%%%%%%%%%%%%%%%%%%%%%%%%%%%%%%%%%%%%%%%%%%%%%%%%%%%
%%%%%%%%%%%%%%%%%%%%%%%%%%%%%%%%%%%%%%%%%%%%%%%%%%
\begin{figure*}
{\includegraphics[scale=0.5]{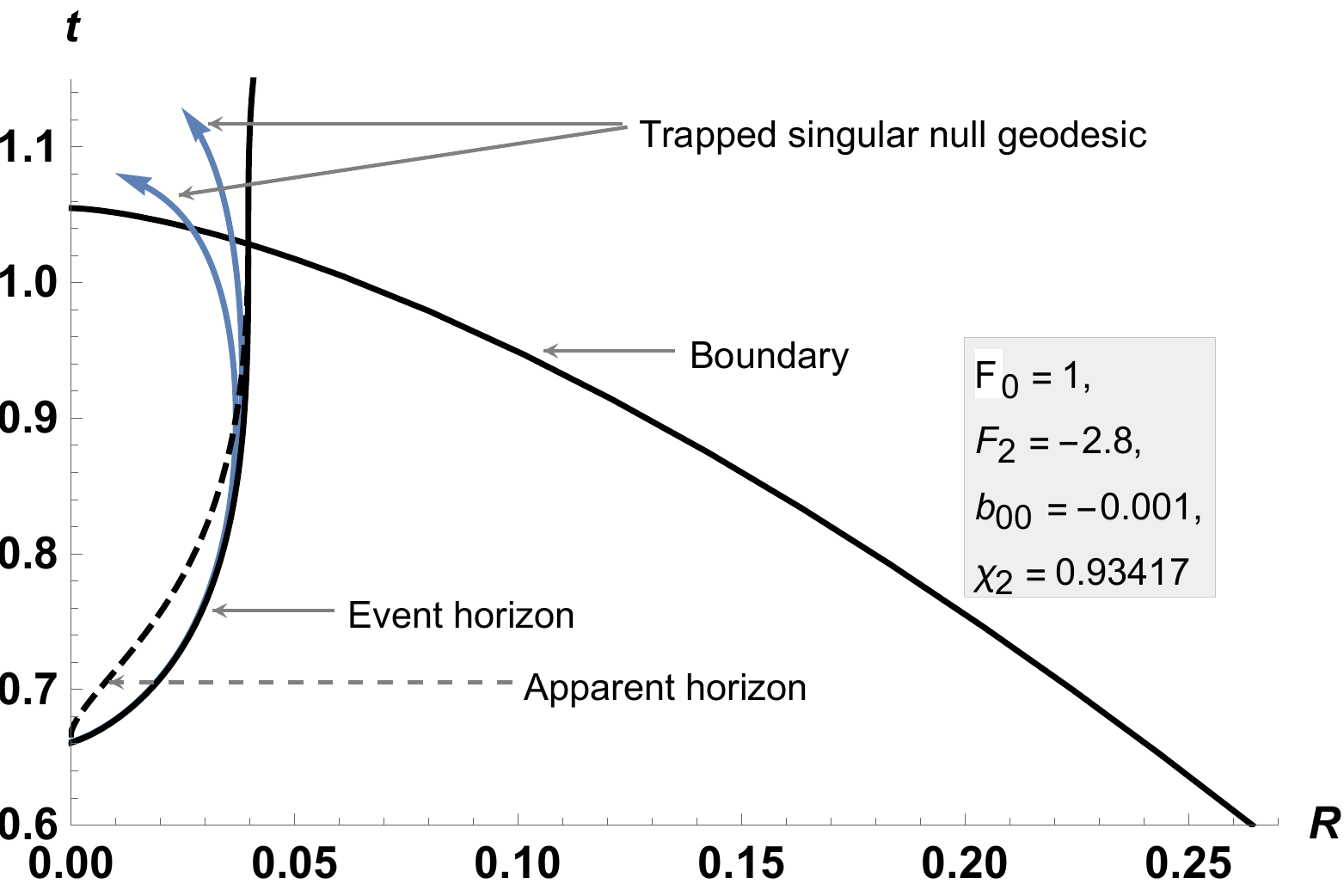}
\hspace{0.2cm}\includegraphics[scale=0.5]{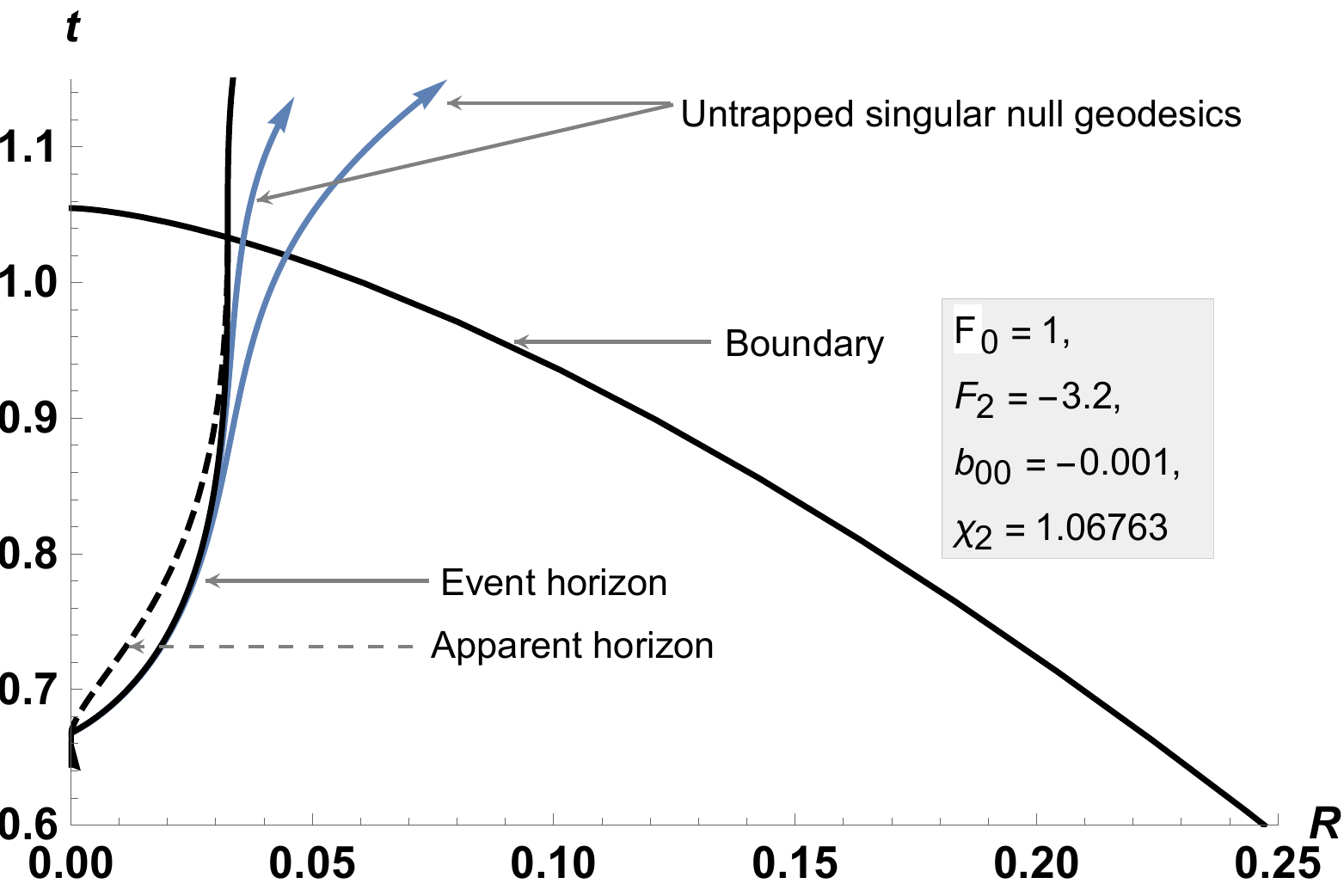}}
\caption{Causal structure of a singularity formed as an end state of a bound (elliptic) collapsing dust cloud. Apparent horizon, event horizon, and singular null geodesics are represented by dashed black curves, solid black curves, and
solid blue curves, respectively. (a) The evolution of the event horizon starts from the center before the formation of the central singularity. Singular null geodesics, if at all, can escape the singularity gets trapped later and falls back in, making the singularity only locally naked. (b) The evolution of the event horizon starts during the formation of the central singularity. Singular null geodesics can escape and reach the faraway observer. Here, $\frac{fR}{F}\sim10^{-3}$ initially, and reduces in magnitude thereafter, in both these cases. Higher-order terms: $o(y_1^3)$ and $o(y_1^2)$, arising in Eq.(\ref{R(t,r)}) are neglected.} 
\label{weakglobal}
\end{figure*}
%%%%%%%%%%%%%%%%%%%%%%%%%%%%%%%%%%%%%%%%%%%%%%%%%%%%%%%%%%%%%%%%%%%%%%%%%%%%%%%%%%%%%%%%%%%%%%%%%%%
Deshingkar, Jhingan, and Joshi 
\cite{Deshingkar_98} 
studied the global causal structure of the end state of marginally bound collapse, wherein three different mass distributions were considered. These mass distributions had first, second, and third-order inhomogeneity terms, respectively, in the initial density. (Here, $nth$ order inhomogeneity term means the initial density profile is of the form $\rho(r)=\rho_0+\rho_n r^n$, where $\rho_0$ and $\rho_n$ are constants. Also, the corresponding Misner-Sharp mass function is of the form $F=F_0r^3+F_{n+3} r^{n+3}$ where $F_0$ and $F_{n+3}$ are constants). The general result obtained was that a higher magnitude of the inhomogeneity term corresponded to the end state as a globally visible singularity. Here, we analyze the global behavior of the singularity formed by bound collapse and for a mass function and the velocity function given by
\begin{equation}
 F=F_0r^3+F_2r^5,\hspace{1cm} f=b_{00}r^2
\end{equation}
The boundary of the cloud is found such that the density smoothly matches to zero there. Hence, the boundary is given by
\begin{equation}
    r_c=\sqrt{-\frac{3F_0}{5F_2}}.
\end{equation}
 This is a second-order inhomogeneity in the mass function.  As seen in Fig.(\ref{weakglobal}), the singularity is at least locally naked for chosen values of $F_0$ and $F_2$. However, in the left panel of Fig.(\ref{weakglobal}) the event horizon starts forming before the formation of the central singularity, thereby making the singularity globally hidden. The singular geodesic can escape the singularity but later gets trapped and falls back. Now, increasing the magnitude of the inhomogeneity term $F_2$, as seen in the right panel of Fig.(\ref{weakglobal}) affects the evolution of the event horizon in such a way that its time of formation is delayed and now overlaps with the time of formation of a central singularity. A null geodesic with the property $F(r_c)<R(t,r_c)$ can be traced with the criteria that the difference between the time of escape of the null geodesic from the center and the time of formation of the central singularity can be reduced as much as we desire. In such a case, the singularity is considered as globally visible.

It should be noted that even if such globally visible singularities exist, it should not create a problem for the cosmic censorship if such singularities are gravitationally weak. We discuss this in more detail in the following section.

\section{Strength of the singularity}\label{3.3}

%%%%%%%%%%%%%%%%%%%%%%%%%%%%%%%%%%%%%%%%%%%%%%%%%%%%%%%%%%%%%%%%%%%%%%%%%%%%%%%%%%%%%%%%%%%%%%%%%%%%
\begin{figure*}
{\includegraphics[scale=0.5]{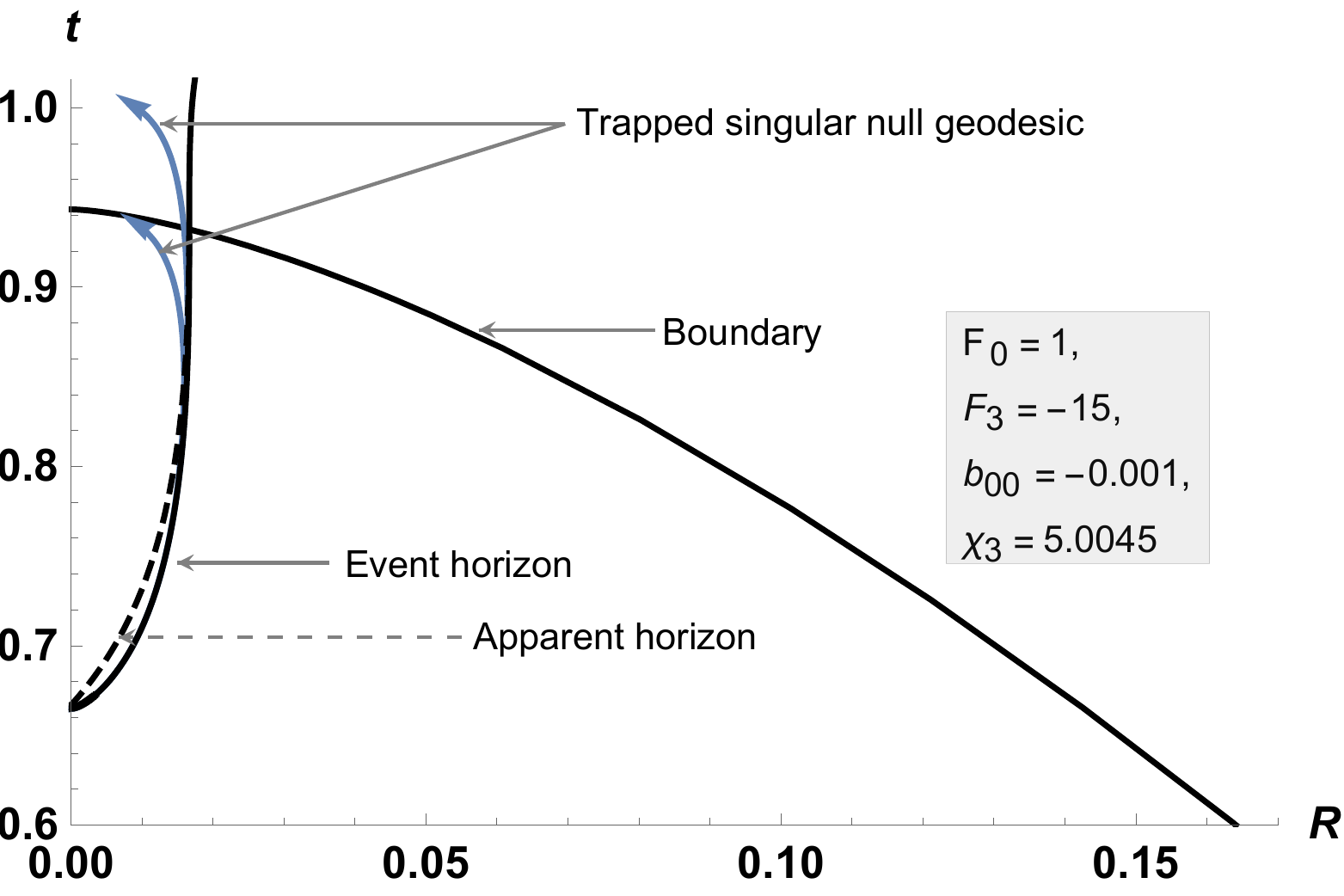}\hspace{0.2cm}\includegraphics[scale=0.5]{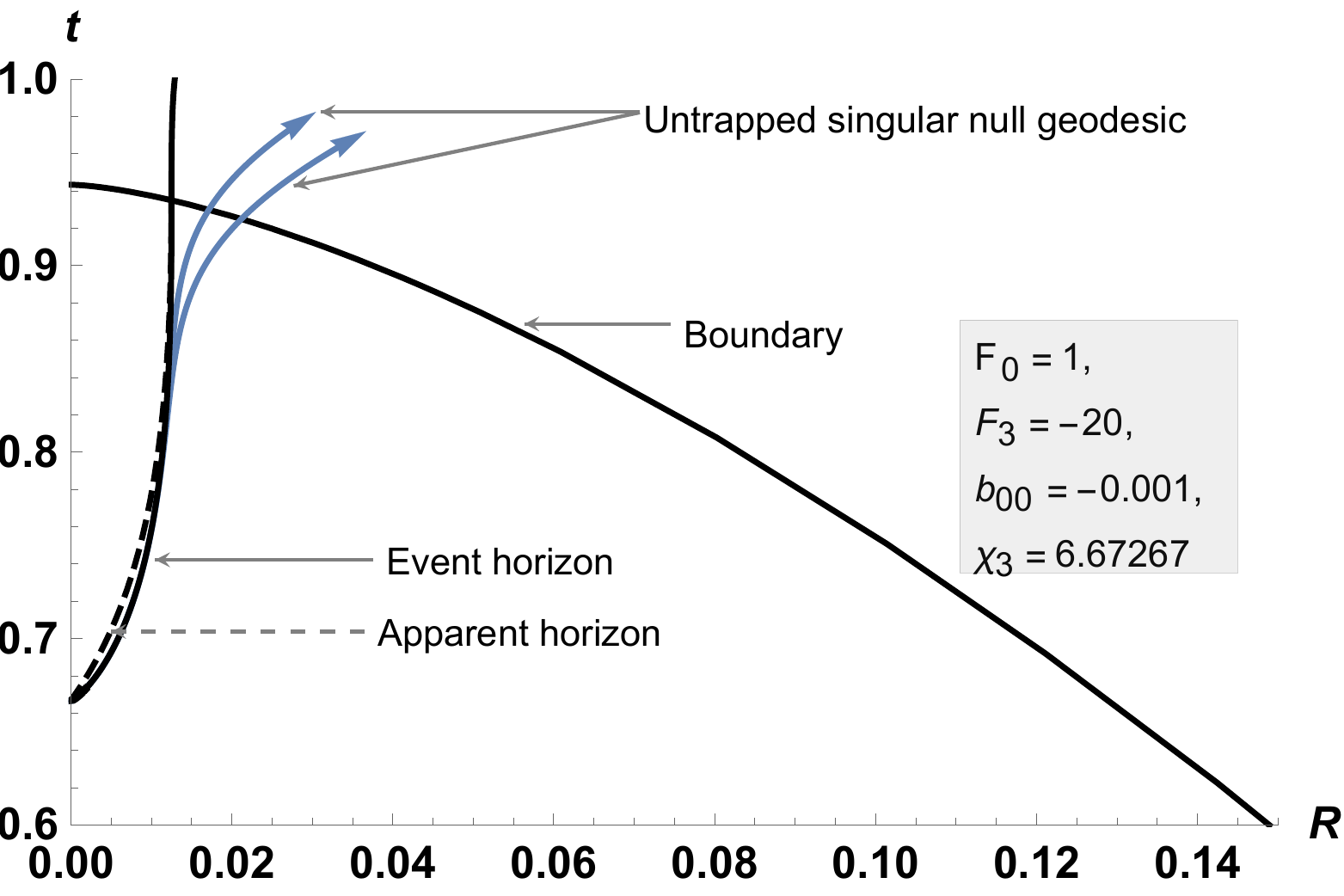}}
\caption{Causal structure of a Tipler strong singularity formed as an end state of a bound (elliptic) collapsing dust cloud. Apparent horizon, event horizon, and singular null geodesics are represented by dashed black curves, solid black curves, and
solid blue curves, respectively. $\chi_1$ = $\chi_2$ = 0 and  $\chi_3>0$. (a) The evolution of the event horizon starts from the center before the formation of the central singularity. Singular null geodesics, if at all, can escape the singularity gets trapped later and falls back in, making the singularity only locally naked. (b) The evolution of the event horizon starts during the formation of the central singularity. Singular null geodesics can escape and reach the faraway observer. Here, $\frac{fR}{F}\sim10^{-3}$ initially, and reduces in magnitude thereafter, in both these cases. Higher-order  terms: $o(y_1^3)$ and $o(y_1^2)$, arising in Eq.(\ref{R(t,r)}) are neglected.}
\label{strongglobalbound}
\end{figure*}
%%%%%%%%%%%%%%%%%%%%%%%%%%%%%%%%%%%%%%%%%%%%%%%%%%%%%%%%%%%%%%%%%%%%%%%%%%%%%%%%%%%%%%%%%%%%%%%%%%%%
%%%%%%%%%%%%%%%%%%%%%%%%%%%%%%%%%%%%%%%%%%%%%%%%%%%%%%%%%%%%%%%%%%%%%%%%%%%%%%%%%%%%%%%%%%%%%%%%%%%%

\begin{figure}
\centering
\includegraphics[scale=0.5]{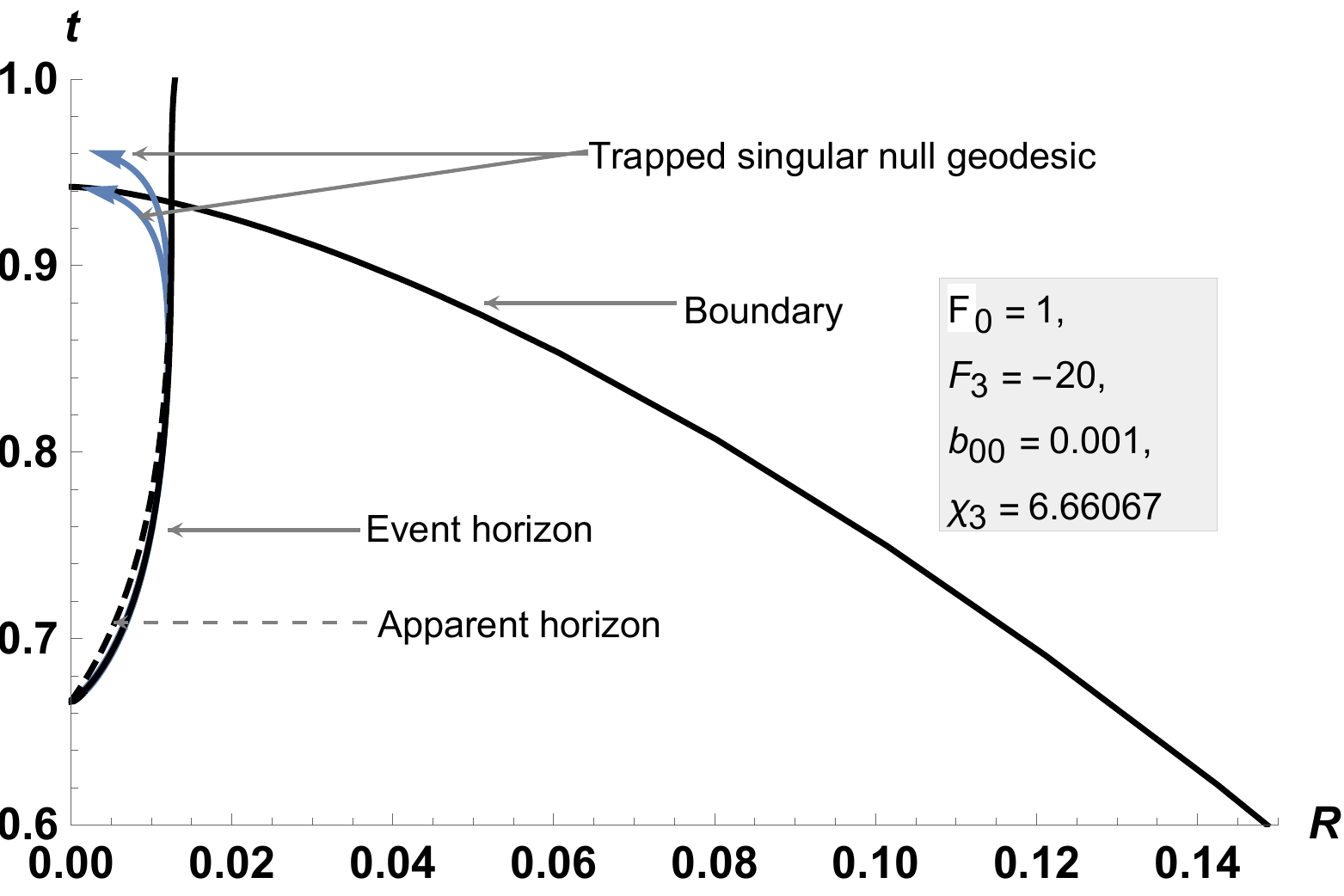}
\caption{Causal structure of a Tipler strong singularity formed as an end state of an unbound (hyperbolic) collapsing dust cloud. Apparent horizon, event horizon, and singular null geodesics are represented by dashed black curves, solid black curves, and solid blue curves, respectively.    $\chi_1=\chi_2=0$ and $\chi_3>0$. The mass profile, which ends as a globally visible singularity in bound case (see Fig.(\ref{strongglobalbound})), ends as a globally hidden singularity in unbound case. Here, $\frac{fR}{F}\sim10^{-3}$ initially, and reduces in magnitude thereafter. Higher-order  terms: $o(y_1^3)$ and $o(y_1^2)$, arising in Eq.(\ref{R(t,r)}) are neglected.}
\label{strongglobalunbound}
\end{figure}
%%%%%%%%%%%%%%%%%%%%%%%%%%%%%%%%%%%%%%%%%%%%%%%%%
%%%%%%%%%%%%%%%%%%%%%%%%%%%%%%%%%%%%%

To maintain the strength of the singularity in the sense of Tipler, Clarke, and Krolak has given a necessary and sufficient condition which needs to be satisfied, as mentioned in the previous chapter \eqref{Krolak and Clarke criteria}. We can use this criterion to put a restriction on a particular parameter signifying the nonlinear relation between the physical radius and the tangent of the outgoing radial null geodesic at the singular center.  The time curve can be Taylor expanded around the center $r=0$ as in Eq.(\ref{taylorsinglaritycurve}).

For a singularity to be at least locally visible, the tangent of the future directed radial null geodesic from the singularity at $r\to 0$ should be positive. In the $(R,u)$ frame, where $u=r^{\alpha}$ with $\alpha>1$, this tangent is written as $X_0=\lim_{r\to 0}\frac{dR}{du}$.   The relation between the tangent of outgoing radial null geodesic at the singularity and the components $\chi_i$ of the Taylor expansion of the time curve at $v=0$ is depicted in Eq.(\ref{X02}). To ensure the positivity of $X_0$, the first nonzero $\chi_i$ should be positive.

Now, it is known that Eq.(\ref{Krolak and Clarke criteria}) can be satisfied only if  $\alpha \geq 3$. Also, the necessary criterion for the singularity to be at least locally naked is given by $\alpha \leq 3$. Hence, the necessary criterion for a singularity to be strong and locally naked is given by 
\cite{Joshi_93} 
\begin{equation}\label{alpha=3}
    \alpha=3.
\end{equation}

%%%%%%%%%%%%%%%%%%%%%%%%%%%%%%%%%%%%%%%%%%%%%%%%%%%%%%%%%%%%%%%%%%%%%%%%%%%%%%%%%%%%%%%%%%%%%%%%%%%%
\begin{figure*}
{\includegraphics[scale=0.315]{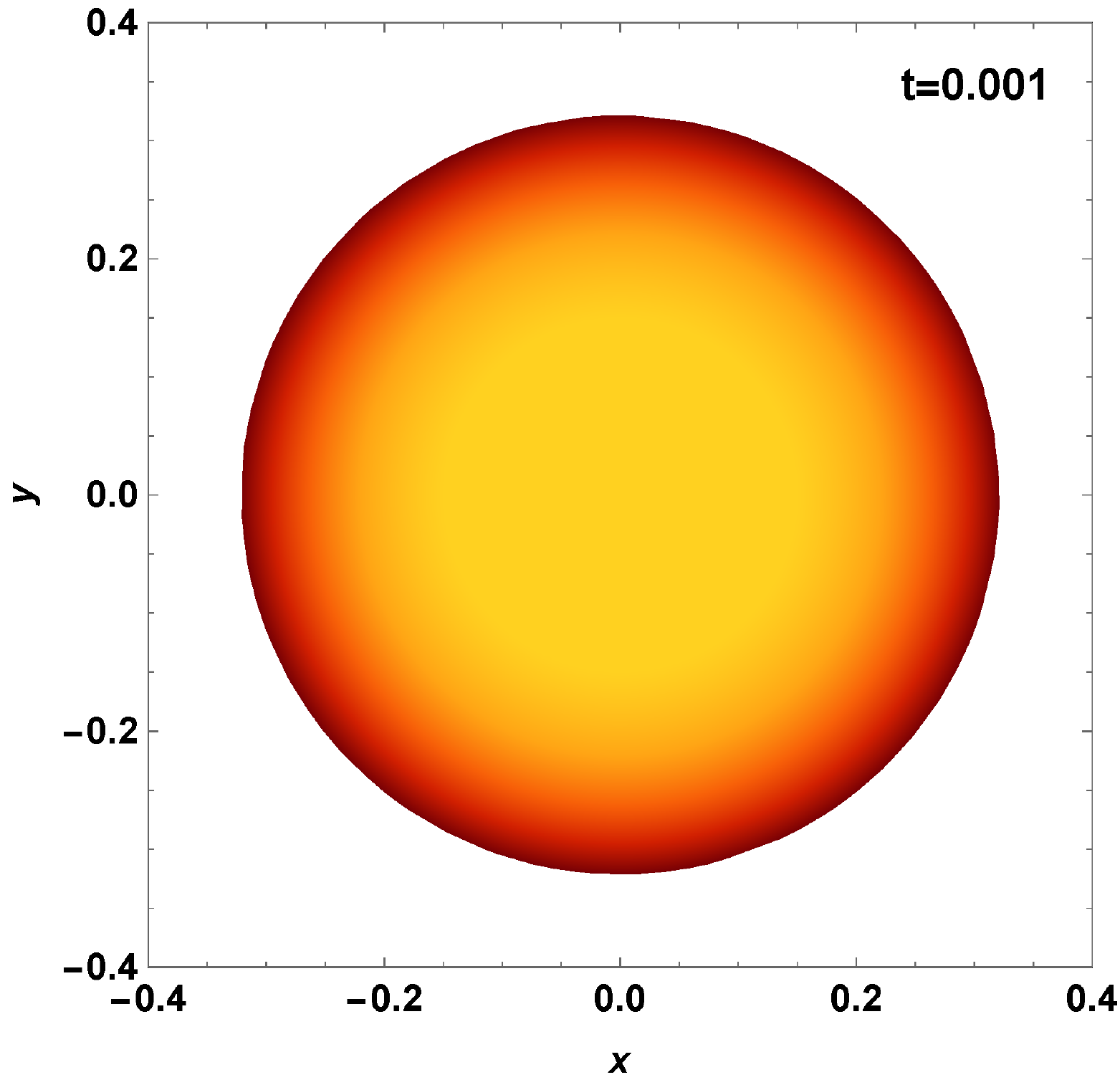}\includegraphics[scale=0.315]{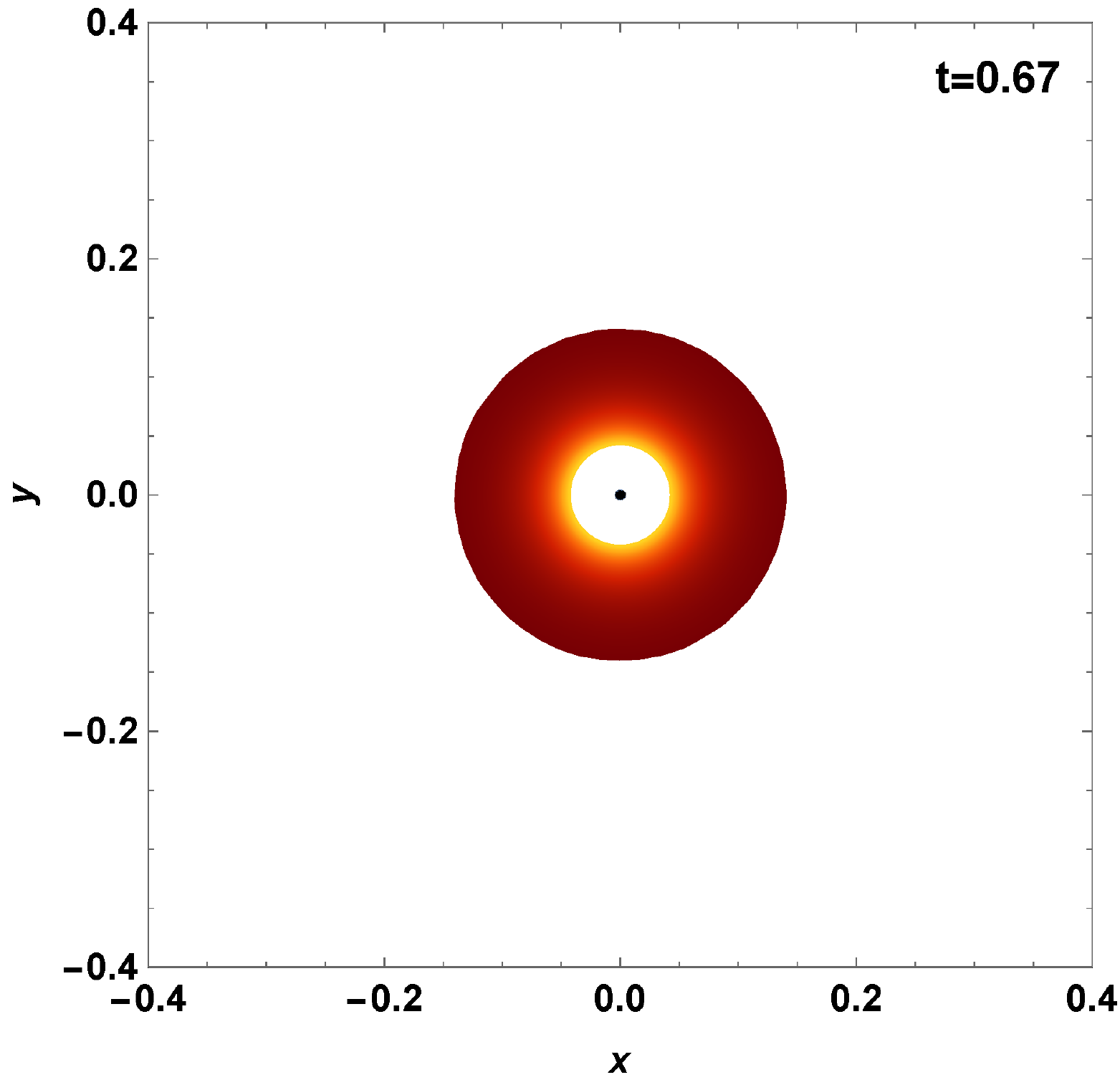}\includegraphics[scale=0.315]{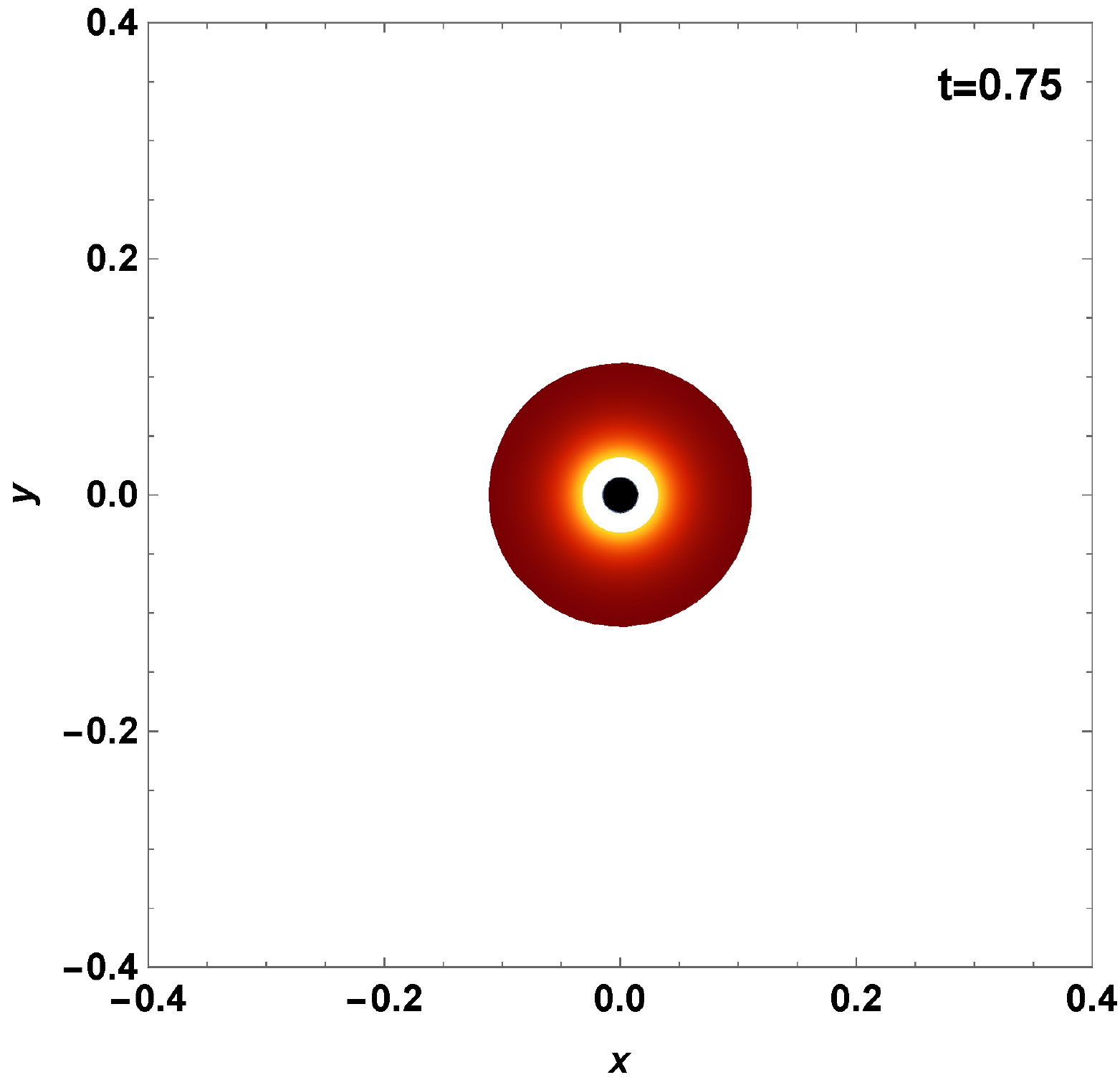}}
{\includegraphics[scale=0.315]{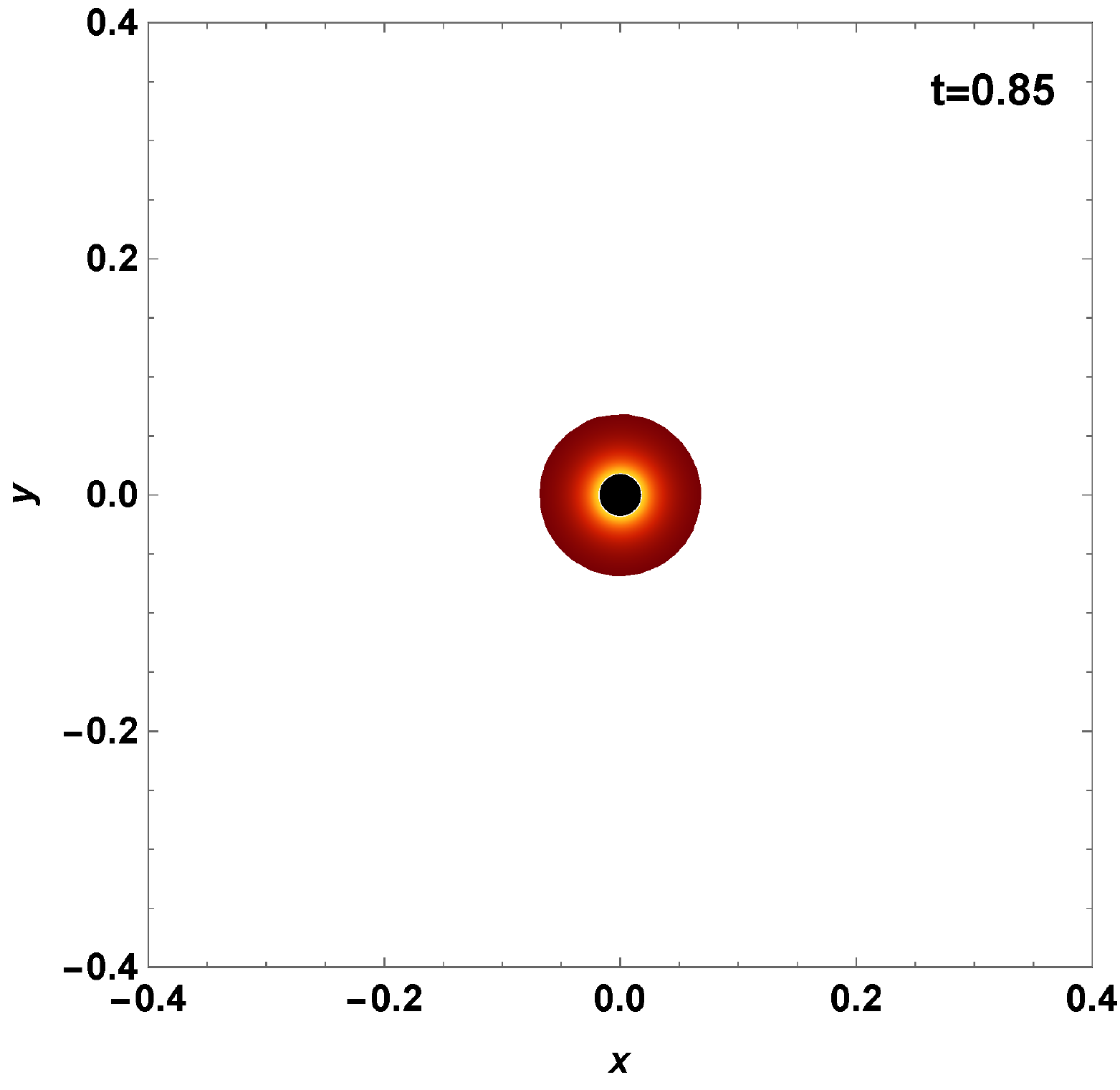}\includegraphics[scale=0.315]{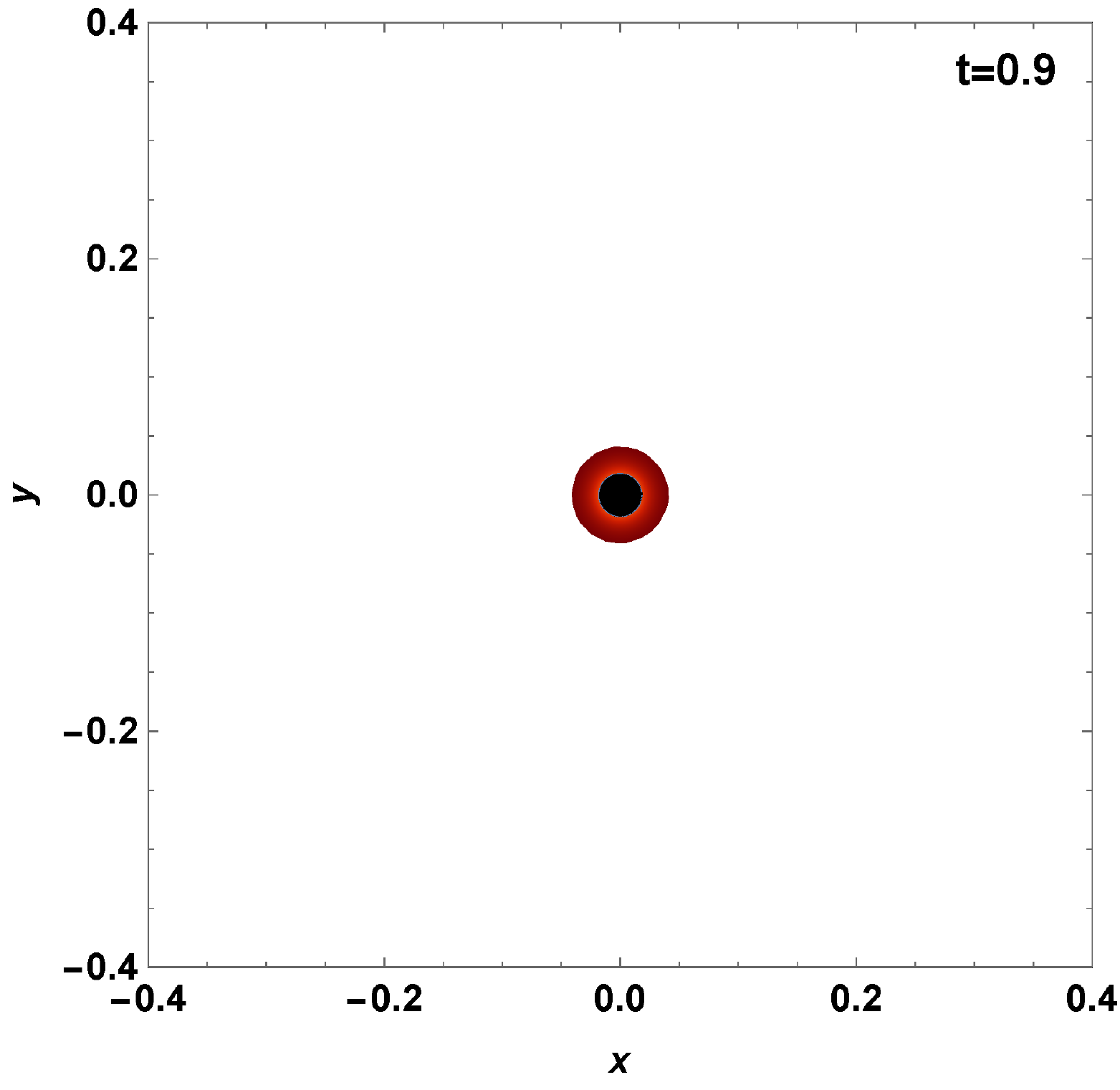}\includegraphics[scale=0.315]{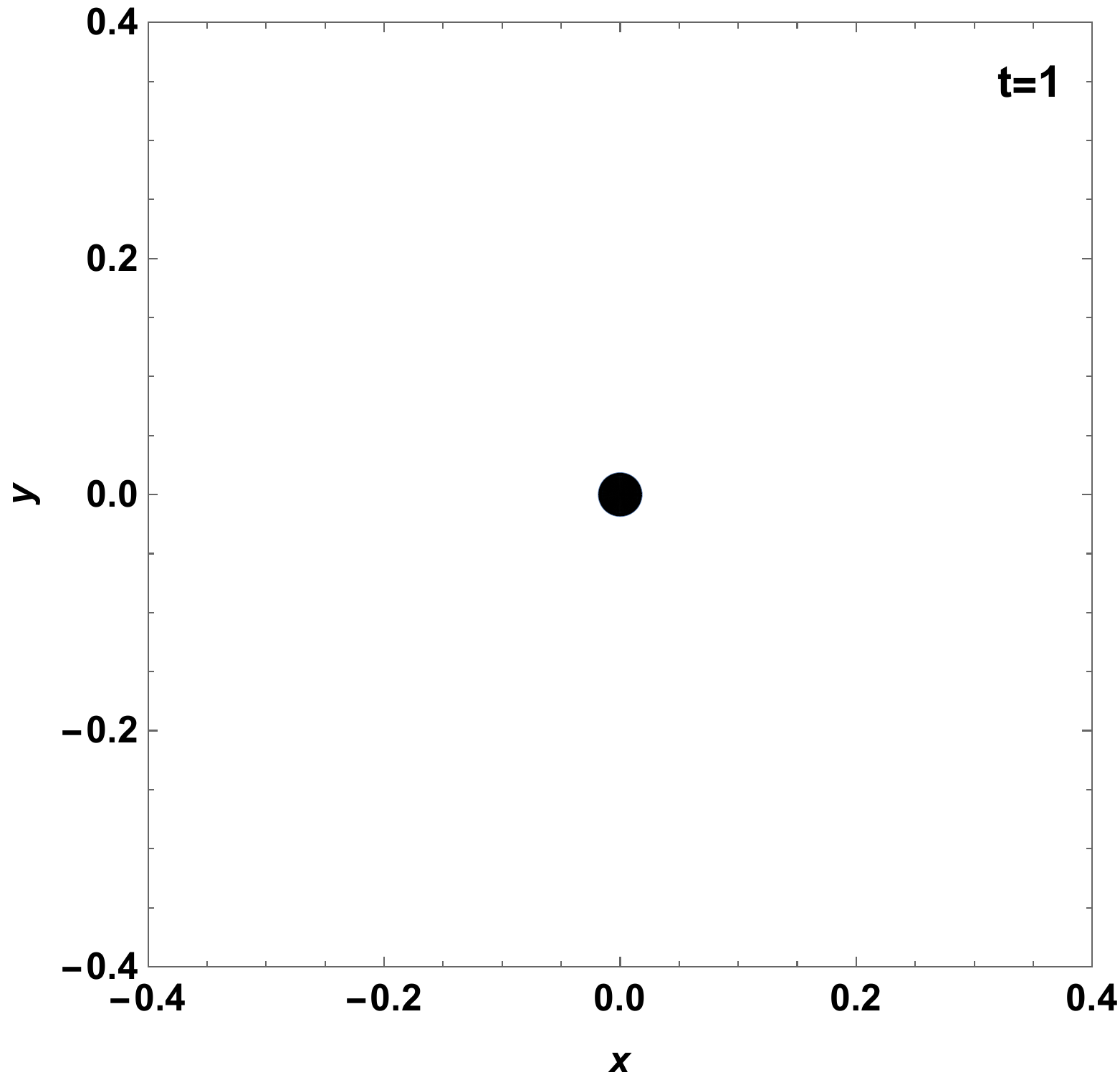}}
\caption{Evolution of the collapsing star and the global causal structure is depicted here. $F_0=1$, $F_3=-15$ and $F_i=0$ for $i \neq 1,3$. $b_{00}=-0.001$ and $b_{0j}=0$ for $j \neq 0$. $\frac{fR}{F}\sim10^{-3}$ initially and reduces in magnitude thereafter. Higher-order  terms: $o(y_1^3)$ and $o(y_1^2)$, arising in Eq.(\ref{R(t,r)}) are neglected. The singularity is Tipler strong with $\chi_1=\chi_2=0$ and $\chi_3\neq 0$.  The solid black disk represents the event horizon which increases in size with time. No singular geodesic can escape and reach the boundary.}
\label{globallyhiddendensityfull}
\end{figure*}
%%%%%%%%%%%%%%%%%%%%%%%%%%%%%%%%%%%%%%%%%%%%%%%%%%%%%%%%%%%%%%%%%%%%%%%%%%%%%%%%%%%%%%%%%%%%%%%%%%%%
\begin{figure*}
{\includegraphics[scale=0.315]{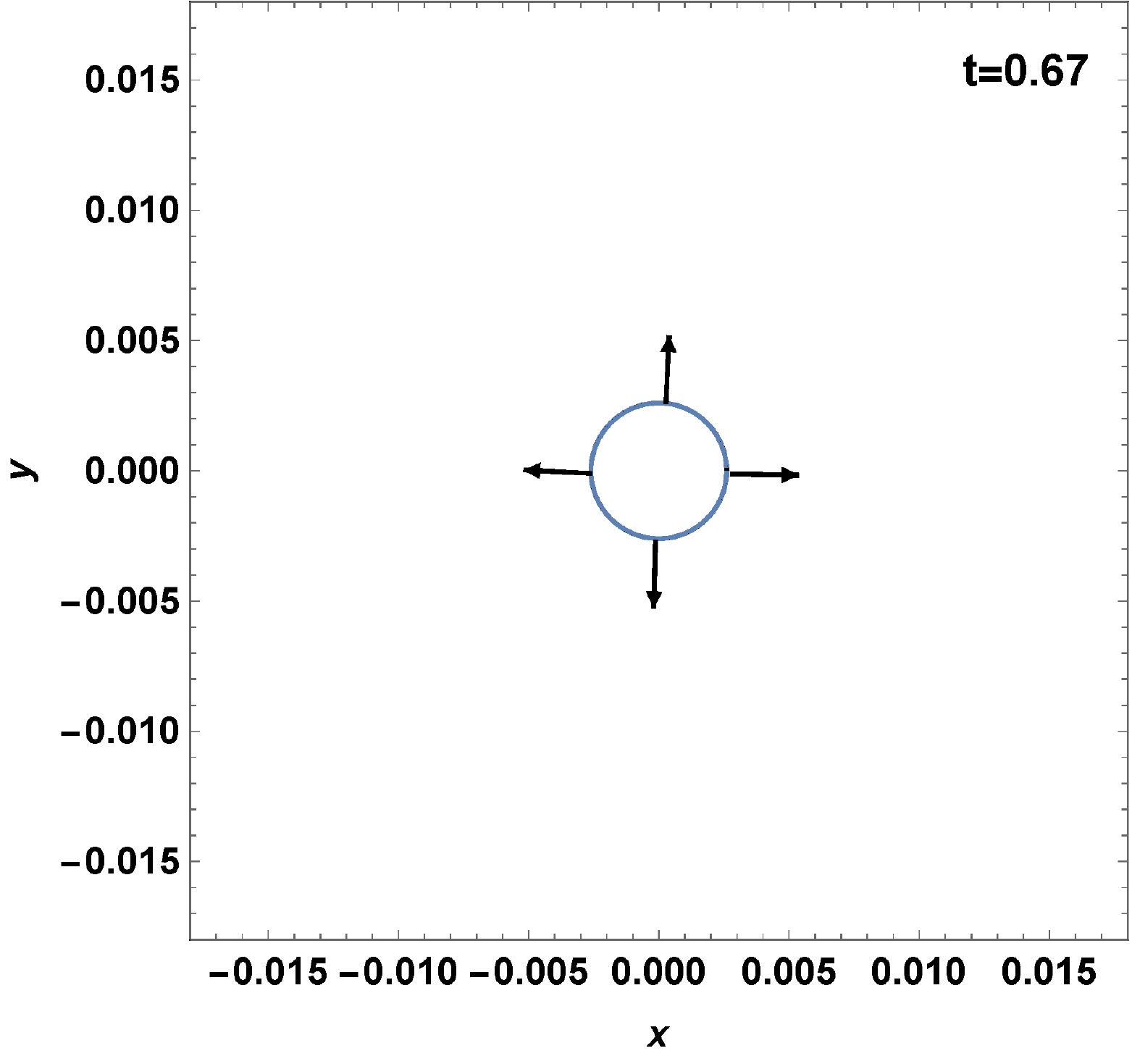}\includegraphics[scale=0.315]{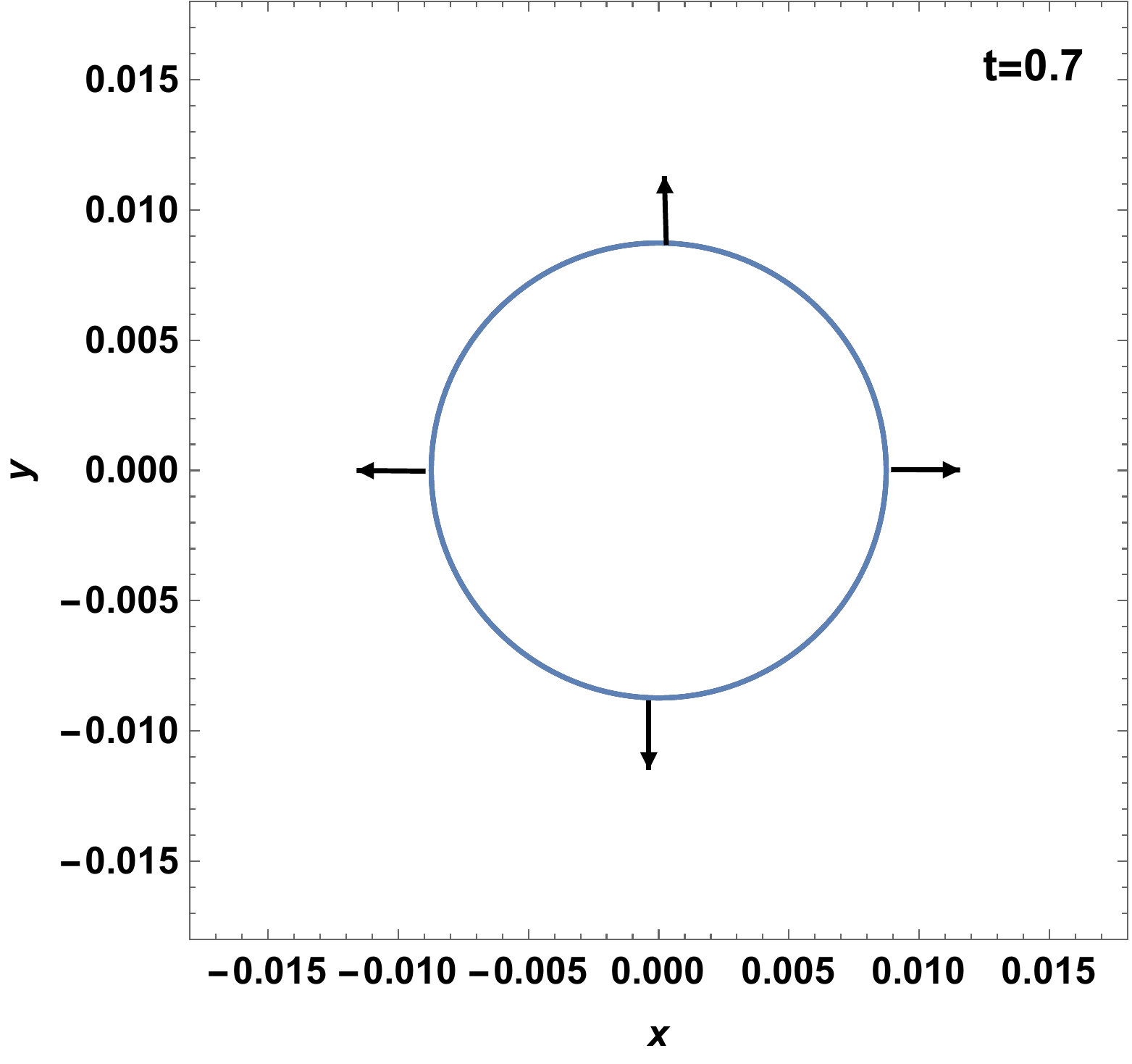}\includegraphics[scale=0.315]{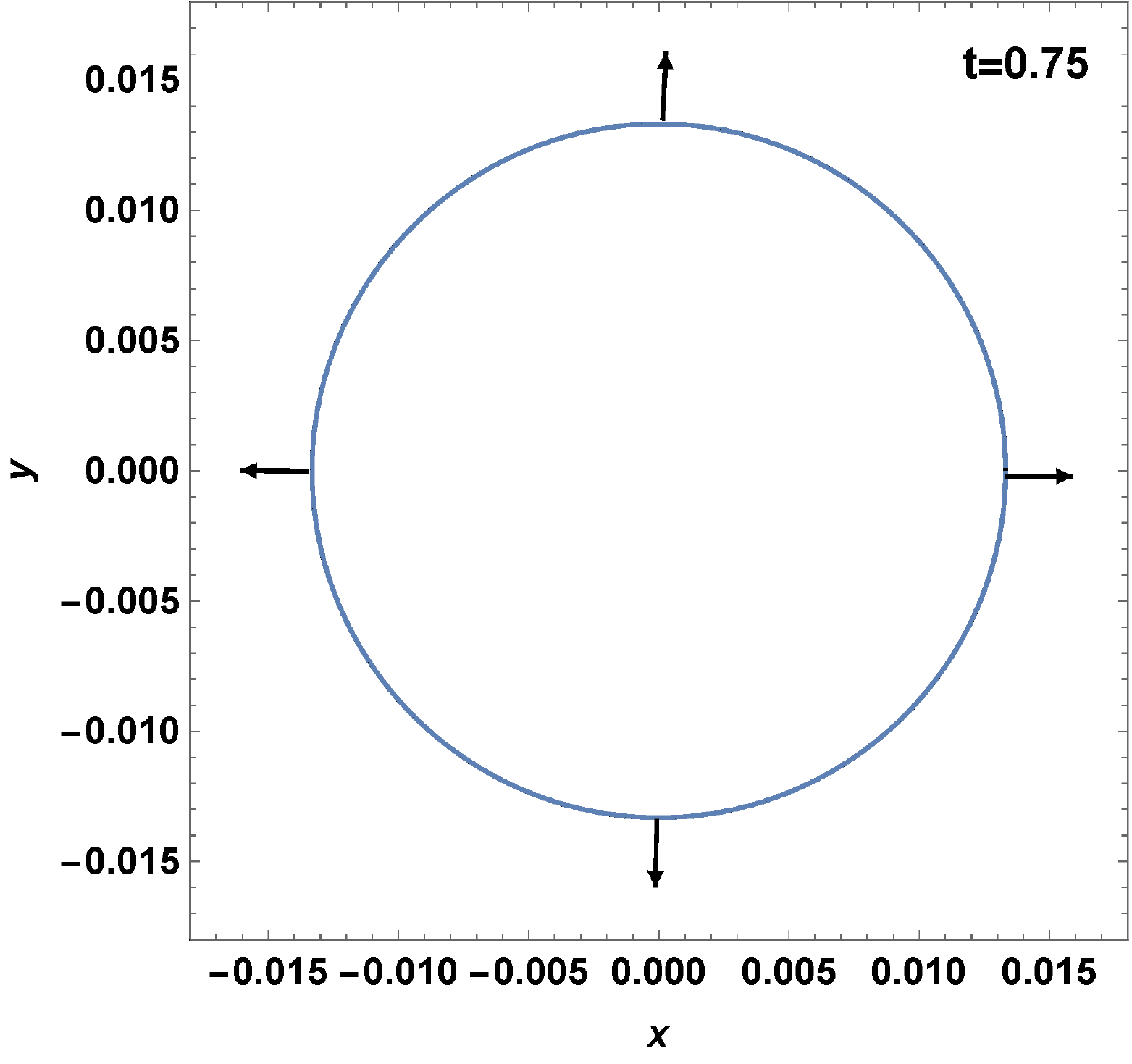}}
{\includegraphics[scale=0.315]{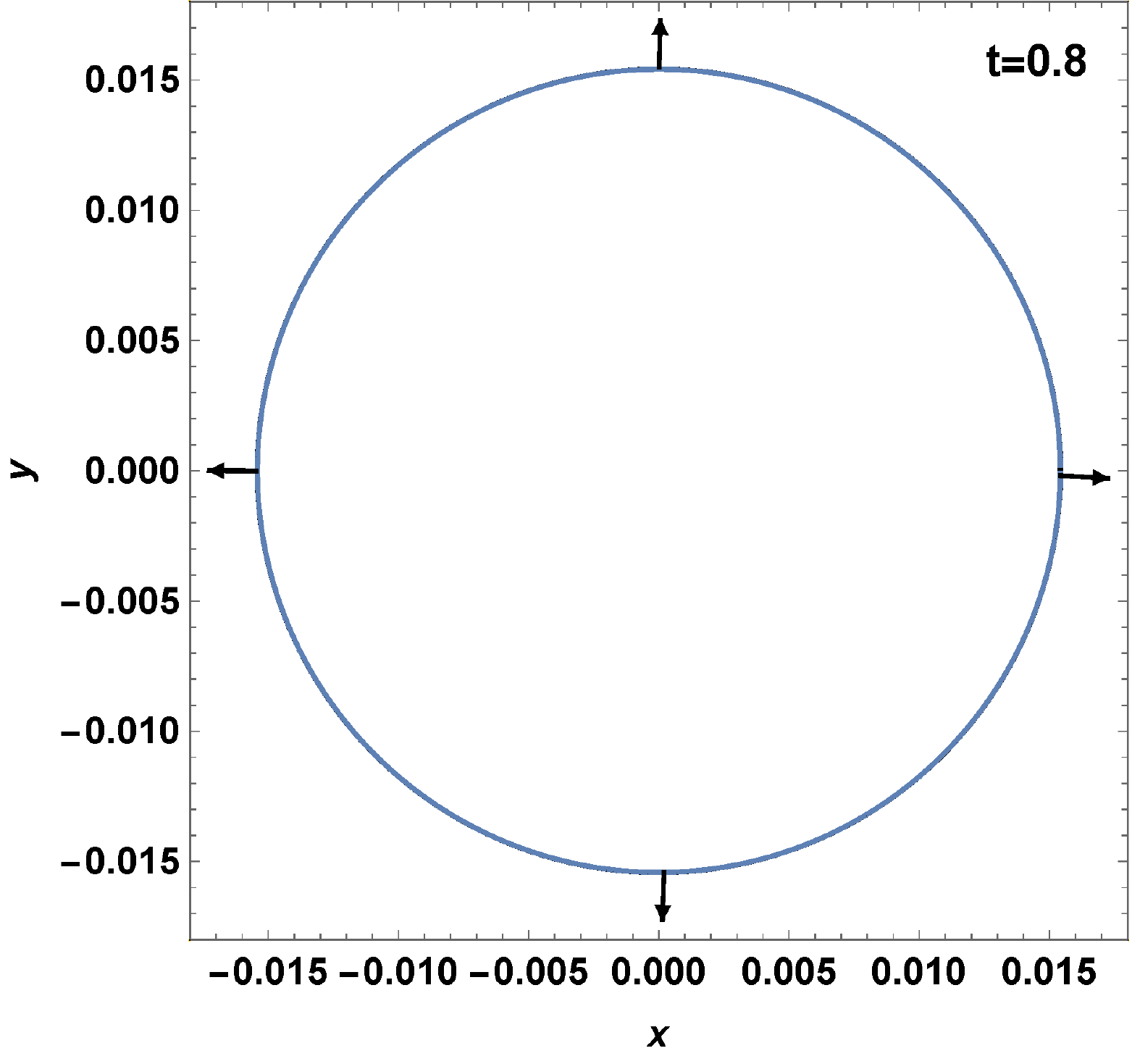}\includegraphics[scale=0.315]{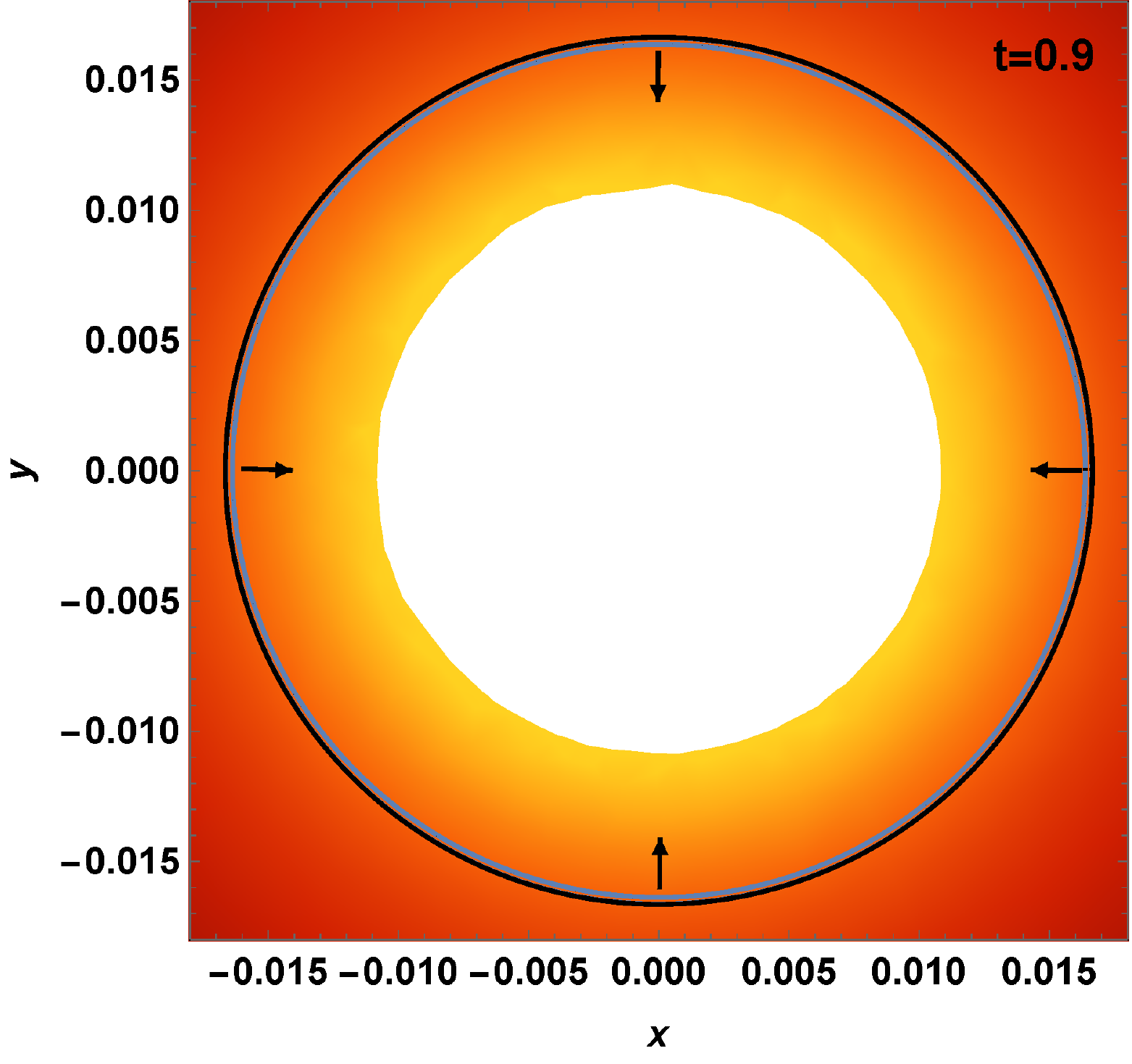}\includegraphics[scale=0.315]{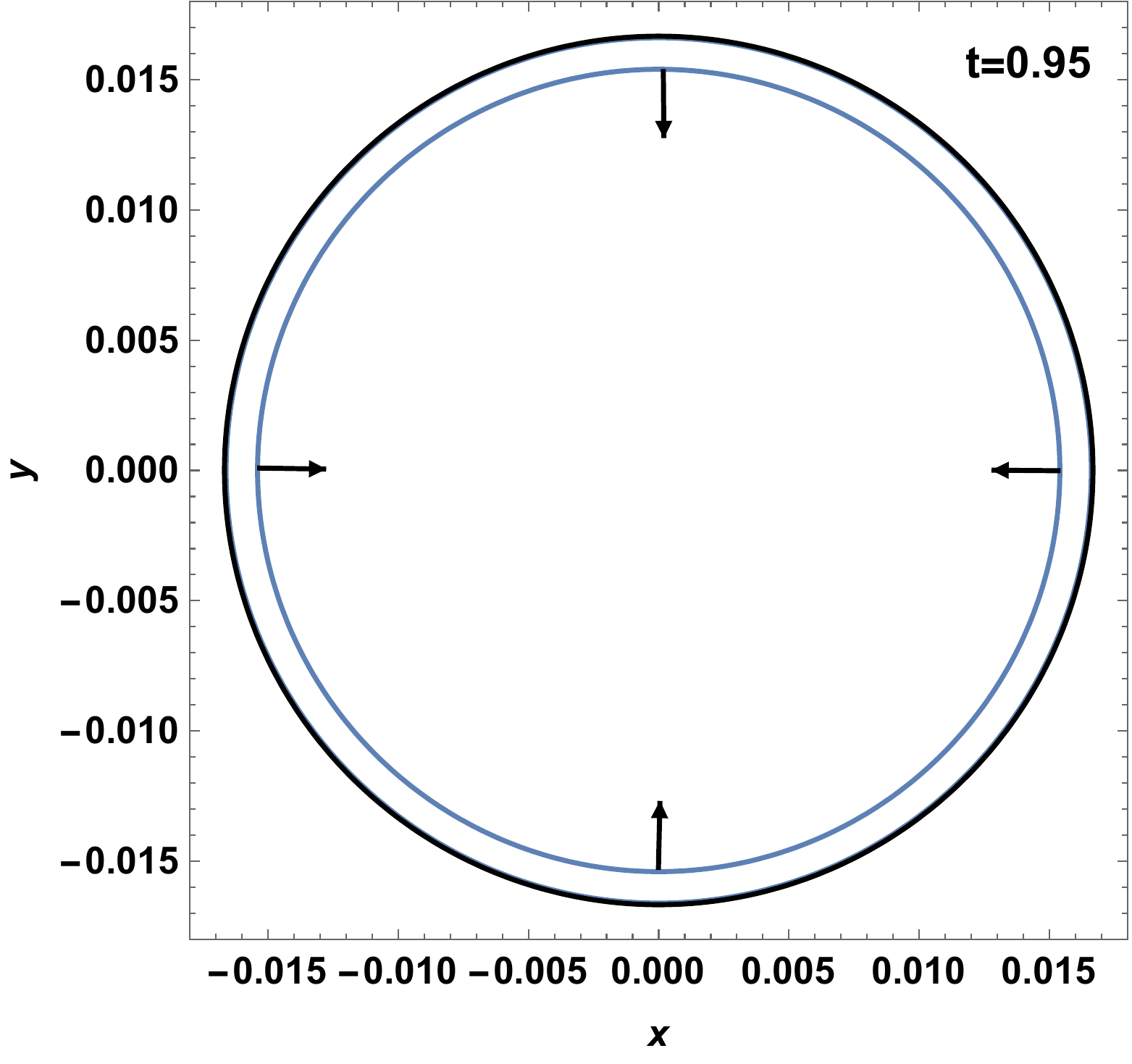}}
\caption{Local causal structure is depicted here. $F_0=1$, $F_3=-15$ and $F_i=0$ for $i \neq 1,3$. $b_{00}=-0.001$ and $b_{0j}=0$ for $j \neq 0$. $\frac{fR}{F}\sim10^{-3}$ initially and reduces in magnitude thereafter. Higher-order  terms: $o(y_1^3)$ and $o(y_1^2)$, arising in Eq.(\ref{R(t,r)}) are neglected. The singularity is Tipler strong with $\chi_1=\chi_2=0$ and $\chi_3\neq 0$. Behavior of singular outgoing radial null geodesic wave front is represented by blue color. Event horizon is represented by black colored circle.}
\label{globallyhiddendensity}
\end{figure*}
%%%%%%%%%%%%%%%%%%%%%%%%%%%%%%%%%%%%%%%%%%%%%%%%%%
%%%%%%%%%%%%%%%%%%%%%%%%%%%%%%%%%%%%%%%%%%%%%%%%%%
%%%%%%%%%%%%%%%%%%%%%%%%%%%%%%%%%%%%%%%%%%%%%%%%%%%%%%%%%%%%%%%%%%%%%%%%%%%%%%%%%%%%%%%%%%%%%%%%%%%
\begin{figure*}
{\includegraphics[scale=0.315]{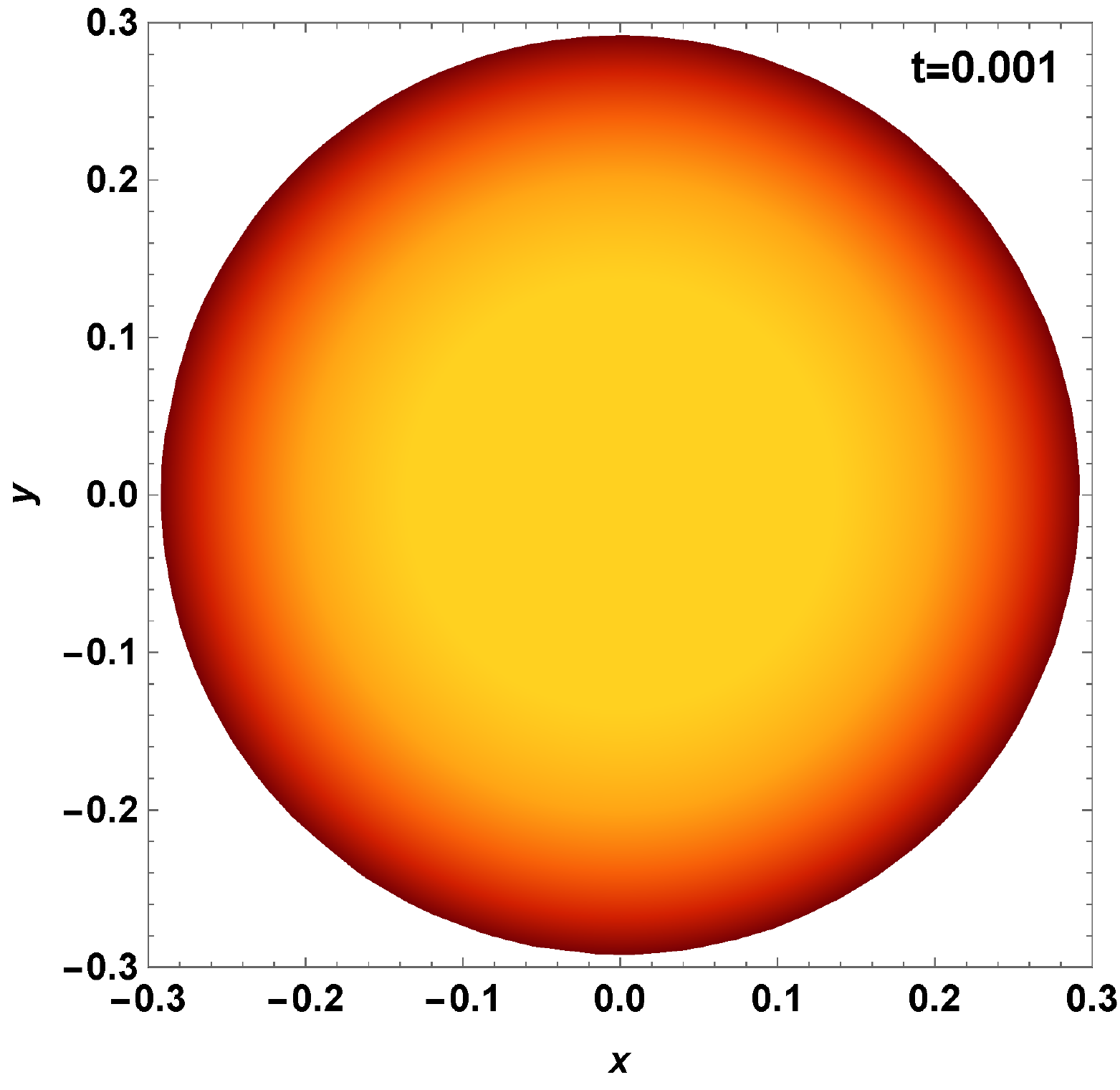}\includegraphics[scale=0.315]{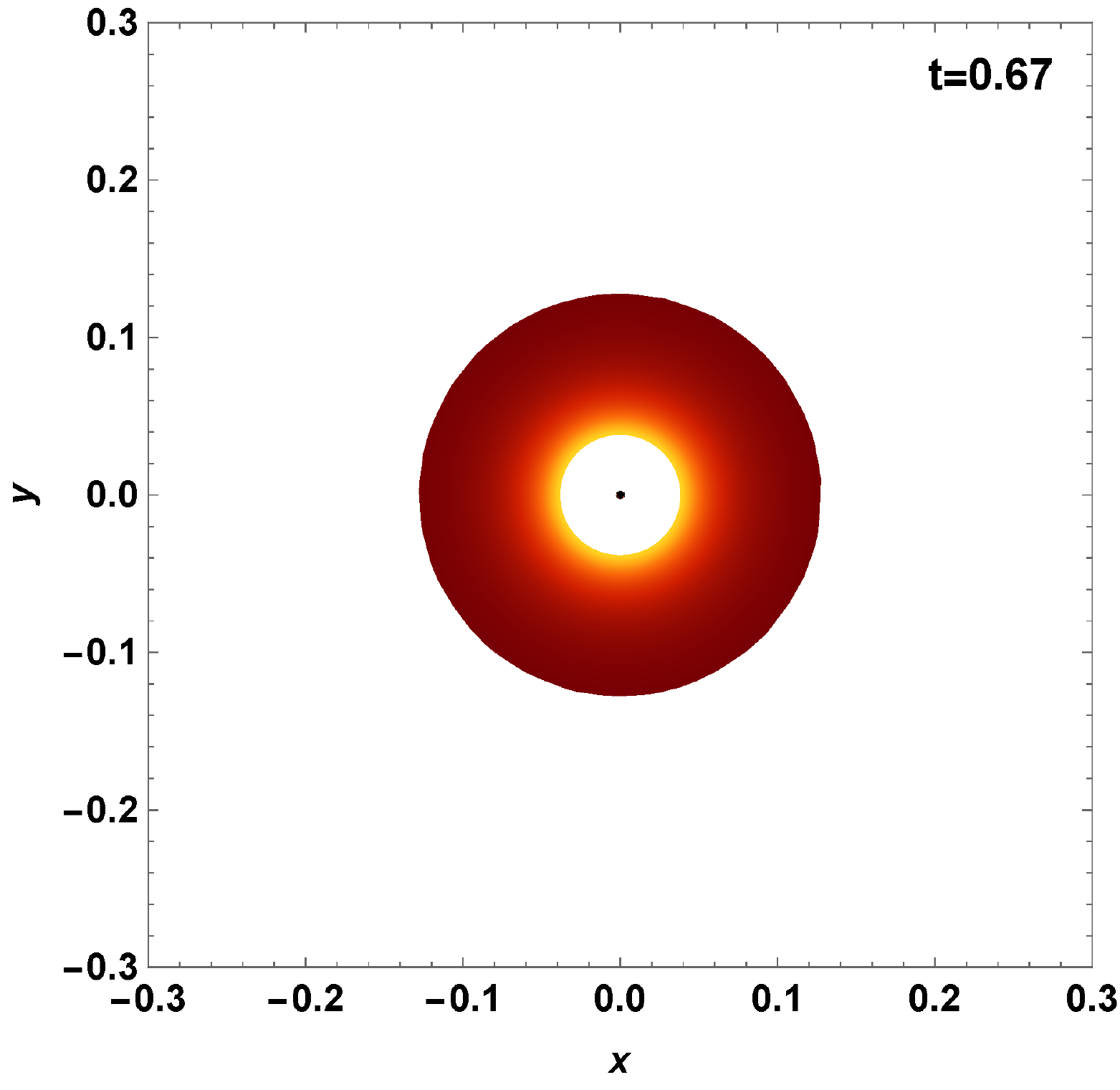}\includegraphics[scale=0.315]{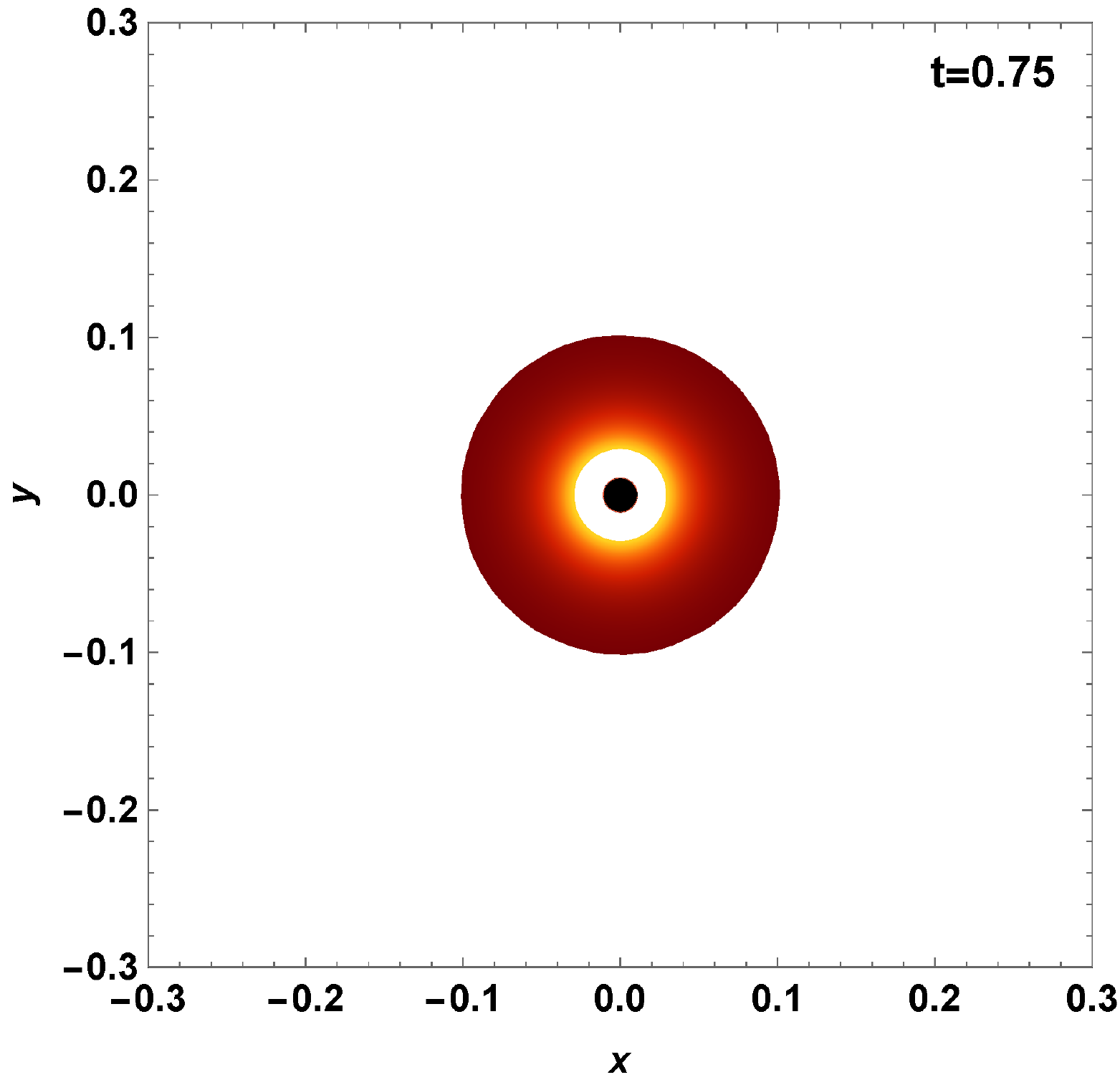}}
{\includegraphics[scale=0.315]{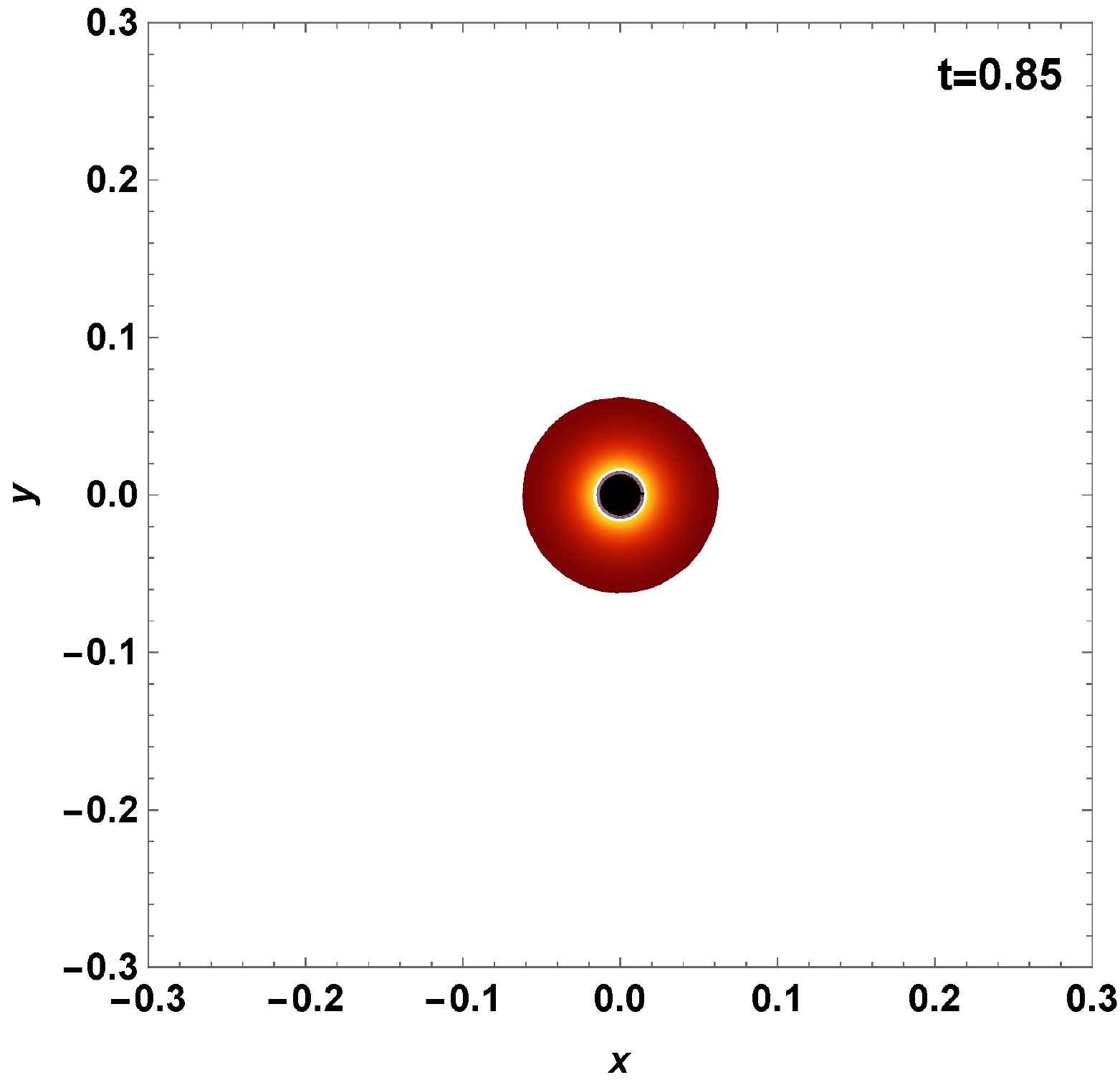}\includegraphics[scale=0.315]{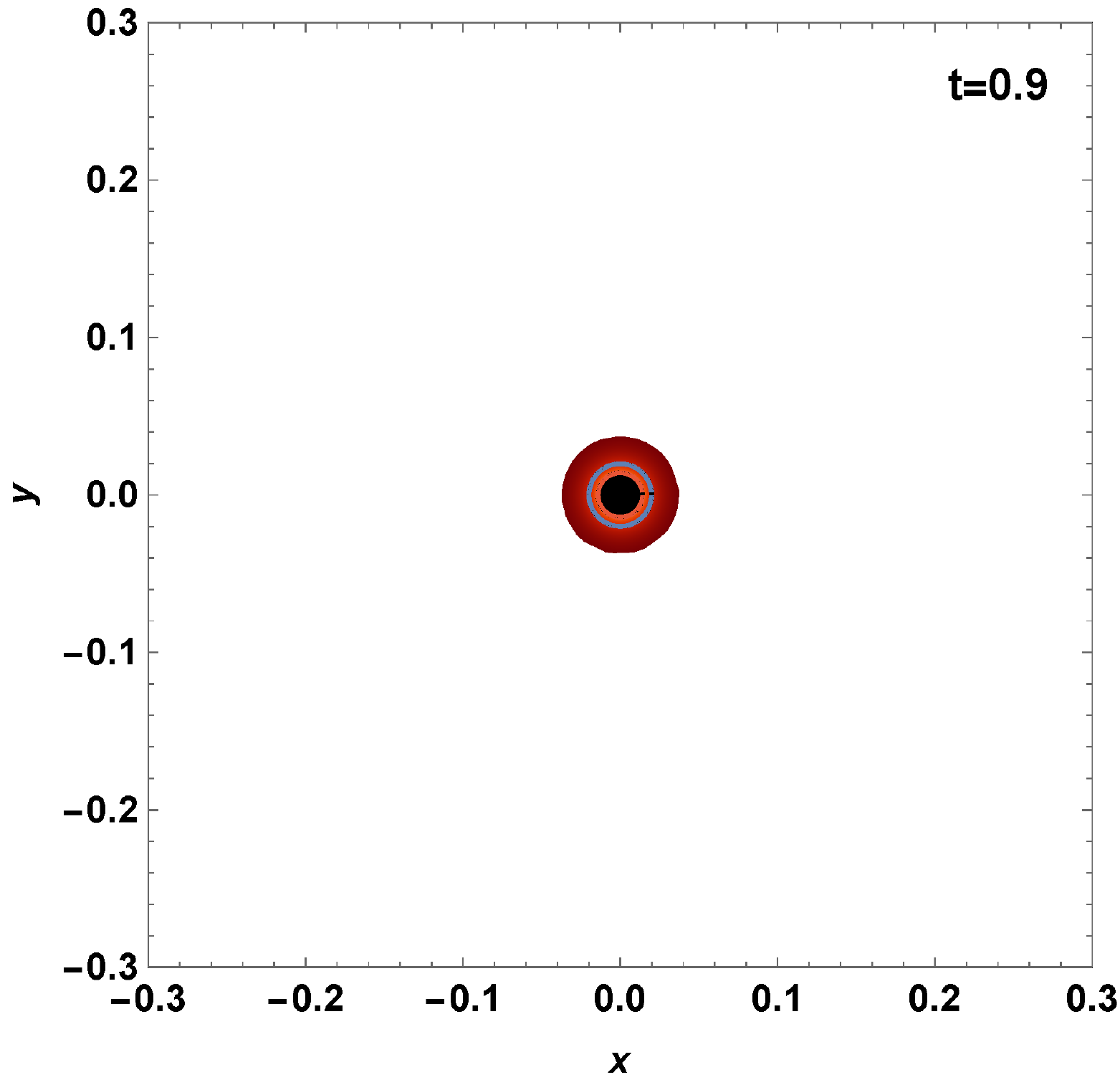}\includegraphics[scale=0.315]{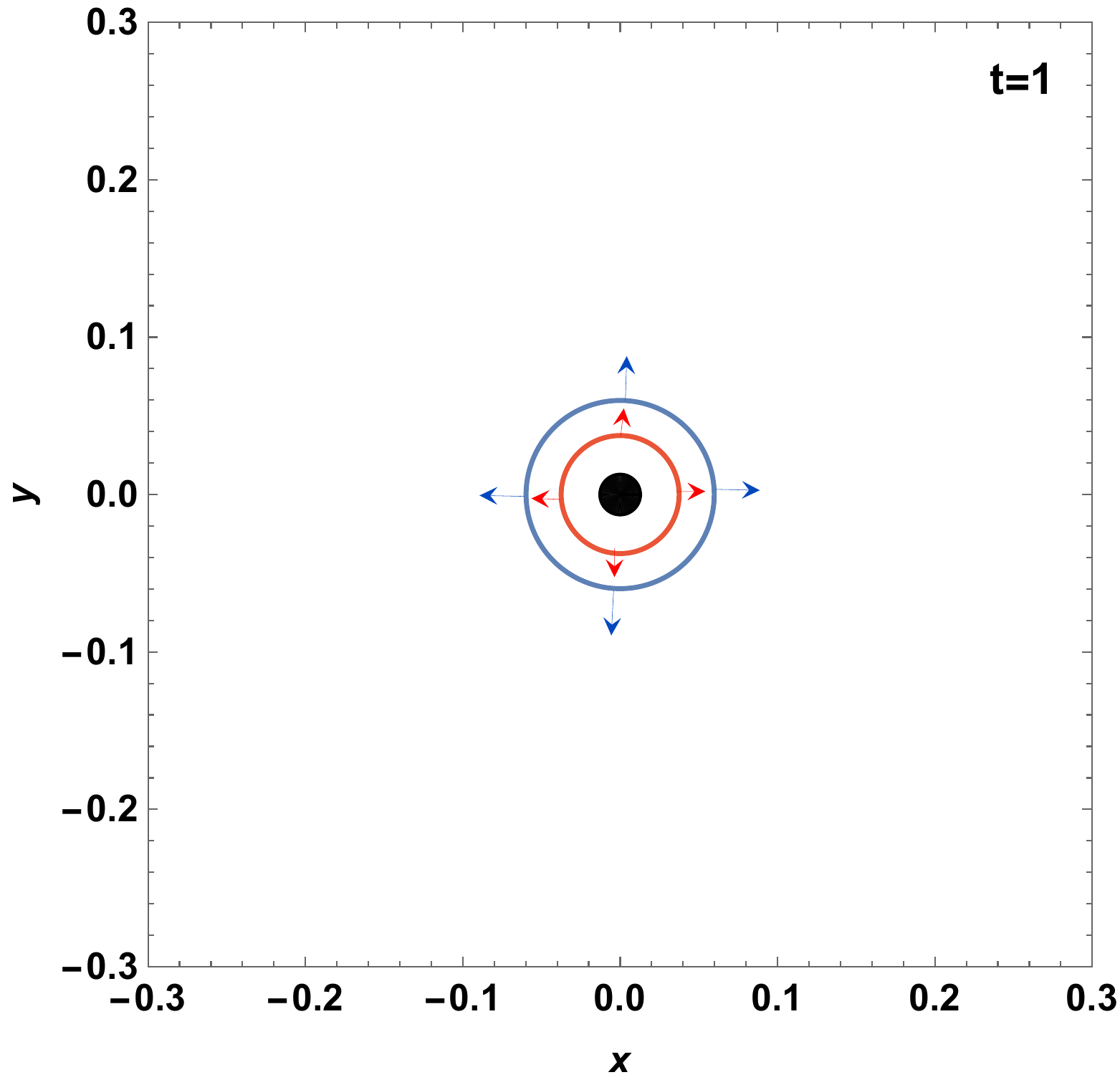}}
\caption{Evolution of the collapsing star and the global causal structure is depicted here. $F_0=1$, $F_3=-20$ and $F_i=0$ for $i \neq 1,3$. $b_{00}=-0.001$ and $b_{0j}=0$ for $j \neq 0$. $\frac{fR}{F}\sim10^{-3}$ initially and reduces in magnitude thereafter. Higher-order  terms: $o(y_1^3)$ and $o(y_1^2)$, arising in Eq.(\ref{R(t,r)}) are neglected. The singularity is Tipler strong with $\chi_1=\chi_2=0$ and $\chi_3\neq 0$. The solid black disk represents the event horizon which increases in size with time. Escaping singular null geodesic wave fronts are represented by red and blue circles which increases with time.}
\label{globallyvisibledensity}
\end{figure*}
%%%%%%%%%%%%%%%%%%%%%%%%%%%%%%%%%%%%%%%%%%%%%%%%%%%%%%%%%%%%%%%%%%%%%%%%%%%%%%%%%%%%%%%%%%%%%%%%%%%
However, for $\alpha=3$, if at all $\chi_1$ or $\chi_2$ is/are nonzero, then $X_0$ blows up. Hence, we will have to make sure that $\chi_1$ and $\chi_2$ should be zero. More specifically, $\chi_1$ and $\chi_2$ should be of order at least $r^3$ and $r^2$, respectively, to avoid blowing up of $X_0$ . The integral expression of $\chi_1$, $\chi_2$ and $\chi_3$ is as follows 
\cite{Joshi_12, Mosani_20}:
\begin{equation}\label{chi1}
    \chi_1(v)=-\frac{1}{2}\int^{1}_{v} \frac{\frac{M_1}{v}+b_{01}}{\left(\frac{M_0}{v}+b_{00}\right)^{\frac{3}{2}}}dv, 
\end{equation}
\begin{equation}\label{chi2}
    \chi_2(v)=\int^{1}_{v}\left[\frac{3}{8}\frac{\left(\frac{M_1}{v}+b_{01}\right)^2}{\left(\frac{M_0}{v}+b_{00}\right)^{\frac{5}{2}}}-\frac{1}{2}\frac{\frac{M_2}{v}+b_{02}}{\left(\frac{M_0}{v}+b_{00}\right)^{\frac{3}{2}}}\right] dv
\end{equation}
and
\begin{equation}\label{chi3(2)}
     \chi_3=  \int_v^1 \frac{b_{01}}{\left(\frac{M_0}{v}+b_{00}\right)^{\frac{3}{2}}}\left(-\frac{5}{16}\left(\frac{b_{01}}{\frac{M_0}{v}+b_{00}}\right)^2 + \frac{3}{4}\left(\frac{\frac{M_2}{v}+b_{02}}{\frac{M_0}{v}+b_{00}}\right)\right)-\frac{1}{2}\frac{\left(\frac{M_3}{v}+b_{03}\right)}{\left(\frac{M_0}{v}+b_{00}\right)^{\frac{3}{2}}}dv.
\end{equation}

%%%%%%%%%%%%%%%%%%%%%%%%%%%%%%%%%%%%%%%%%%%%%%%%%%%%
\begin{table}
\centering
\begin{tabular}{lccccc}
\hline
$b_{00}$  & $F_3$&  $t_{EH}(0)$\\
\hline
    $10^{-1}$     & -1  & 0.146174    \\
    
   $10^{-1}$       & -5   & 0.586922    \\
   
   $10^{-1}$    & -20  & 0.646240   \\

   $10^{-1}$     & -50  &  0.646667  \\

 $10^{-1}$      & -100  & 0.646667 \\

 $10^{-1}$      & -200  & 0.646667   \\

 $10^{-2}$     & -1  & 0.170522    \\

 $10^{-2}$      & -5   & 0.609225   \\

  $10^{-2}$       & -20  & 0.664501   \\

  $10^{-2}$       & -50  & 0.664667   \\

 $10^{-2}$      & -100  & 0.664667   \\

  $10^{-2}$       & -200  & 0.664668    \\
\hline
\end{tabular}
\quad
\begin{tabular}{lccccc}
\hline
$b_{00}$  & $F_3$&  $t_{EH}(0)$\\
\hline
$10^{-3}$       & -1  & 0.172916    \\

$10^{-3}$      & -5   & 0.611443   \\

$10^{-3}$     & -20  & 0.666319   \\
 
$10^{-3}$     & -50  & 0.666467   \\

$10^{-3}$     & -100  & 0.666467   \\

$10^{-3}$  & -200  & 0.666468   \\

$10^{-4}$     & -1  & 0.173141   \\

$10^{-4}$     & -5   & 0.611663   \\

$10^{-4}$     & -20  & 0.666500   \\

$10^{-4}$     & -50  & 0.666647   \\

$10^{-4}$     & -100  & 0.666648    \\

$10^{-4}$     & -200  & 0.666647    \\
\hline 
\end{tabular}
 \caption{Here, mass function and velocity function of Eq.(\ref{mfvf}) are considered. $F_0=1$ and $t_s(0)=\frac{2}{3}$. The collapse is unbound (hyperbolic). The singularity thus formed is strong and globally hidden since $t_{EH}(0)<\frac{2}{3}$. $t_{EH}(0)$ is achieved by numerical approximation rounded up to six decimal digits.}
\label{tab1}
\end{table}
%%%%%%%%%%%%%%%%%%%%%%%%%%%%%%%%%%%%%%%%%%%%%%%%%%%%

Here $M_i$ are the components nonminimally coupled to $r^i$ in the Taylor expansion of $M$ around $r=0$. $M$ is the mass profile, having relation with the Misner-Sharp mass function, as shown in Eq.(\ref{regularitycondition}). Also $b_{0i}$ in Eqs.(\ref{chi1})-(\ref{chi3(2)}) are the components nonminimally coupled with $r^{i}$ in the Taylor expansion of the velocity profile $b_0(r)$ around the center $r=0$. Regularity condition dictates that $f(r)=r^2b_0(r)$ 
The mass profile and the velocity profile together determine the polarity of $\chi_3$. For positive $\chi_3$, we have a strong at least locally naked singularity provided $\chi_1$ and $\chi_2$ vanish at $v=0$. Such analysis was not done in 
\cite{Deshingkar_98} 
in the case of marginally bound collapse for various mass functions considered therein. One such example of mass function and velocity function for which $\chi_1$ and $\chi_2$ vanish is given as follows:
\begin{equation}\label{mfvf}
    F=F_0 r^3+F_3 r^6, \hspace{1cm} f=b_{00}r^2.
\end{equation}
The boundary of the cloud is found such that the density smoothly matches to zero there. Hence, the boundary is given by
\begin{equation}
    r_c=\left(-\frac{F_0}{2F_3}\right)^{\frac{1}{3}}.
\end{equation}
Similar to the previous mass function, this mass function, along with a positive velocity, also gives at least a locally naked singularity for chosen values of $F_0$ and $F_3$. $\chi_3>0$ in this case. However, in the left panel of  Fig.(\ref{strongglobalbound}), outgoing singular radial null geodesics having positive tangent at the center later gets trapped and falls back to the singularity. Increasing the magnitude of the inhomogeneity term, $F_3$, alters the evolution of the event horizon in such a way that its initiation now coincides with the time of formation of singularity due to collapsing central shell, thereby allowing singular null geodesics to escape and reach the faraway observer, as observed in the right panel of Fig.(\ref{strongglobalbound}). 

In the case of a marginally bound collapse of dust, third-order inhomogeneity in the mass profile can give globally naked singularity for a wide range of $F_3<0$ 
\cite{Deshingkar_98}. 
It can be seen from Eqs.(\ref{chi1})-(\ref{chi3(2)}) that such singularity is Tipler strong.

In the case of unbound collapse, we consider a velocity function to have a positive value. It is depicted in Fig.(\ref{strongglobalunbound}) that the mass function giving rise to the globally naked singularity as the end state of bound collapse gives a globally hidden singularity as the end state of unbound collapse having velocity function with the same magnitude but opposite polarity. Furthermore, it is observed in Tab.(\ref{tab1}) that at least so long as the mass function and the velocity function are of the form Eq.(\ref{mfvf}) along with $b_{00}>0$, a wide range of coefficients in such mass and velocity function give a globally hidden singularity as the end state.

In Fig.(\ref{globallyhiddendensityfull}), dynamics of the collapse of the fluid are shown for a particular mass function such that the outgoing radial null geodesics get trapped, and there is no causal connection between the singular region and the outside observer. The singularity thus obtained is, however, locally naked, as seen in Fig.(\ref{globallyhiddendensity}). Fig.(\ref{globallyvisibledensity}) depicts the evolution of the density profile, event horizon, and singular geodesics escaping the boundary of the cloud without getting trapped by any trapped surfaces. A different value of the mass function is considered here. The magnitude of the inhomogeneity term in the Misner-Sharp mass function is more in this case.  An asymptotic observer may observe the wave fronts of the escaped singular null geodesic highly redshifted. Null geodesic escaping from closer to the singularity will be more redshifted. The light traveling from more close to the singularity is also traveling closer to the event horizon. One could deduce that more redshifted the light is, more significant it is, in respect of holding traces of the quantum gravity. All the evolutions are in the comoving frame. Fig.(\ref{globallyhiddendensityfull}-\ref{globallyvisibledensity}) help in visualizing the evolution of the collapsing cloud along with the evolution of the event horizon and null trajectories. They also depict the dynamics of density variation of the collapsing cloud due to inhomogeneous mass distribution, bright light indicating denser.

\section{Concluding remarks}\label{3.4}
Some concluding remarks and open concerns are mentioned as follows:
\begin{enumerate}
    \item End state of a marginally bound collapse has been studied in 
    \cite{Deshingkar_98}. 
    Considering $f=0$ eases the integration of Eq.(\ref{efe2}) to obtain the expression of $R$ as in Eq.(\ref{R(t,r)mb}).  However, such a scenario is a very particular case  corresponding to a very specific dynamics of the collapse, as mentioned in the Introduction, with a scaling function expressed as 
    \begin{equation}
        v(t,r)=\left(1-\frac{3}{2}\frac{\sqrt{F}t}{r^{\frac{3}{2}}}\right)^{\frac{2}{3}},
    \end{equation}
    which is obtained from Eq.(\ref{R(t,r)mb}). Here we consider a nonmarginally bound collapse of the inhomogeneous dust cloud and study the causal structure of the singularity formed as the end state. Investigating the nonmarginally bound gravitational collapse increases our scope of understanding the gravitational collapse to a more general scenario. It is worth mentioning that such a general scenario also encapsulates a very important case wherein, initially, all the fluid elements are at rest, i.e., $\dot R(0,r)=0$. This is obtained by considering the velocity function as 
    \begin{equation}
        f=-\frac{F}{r}
    \end{equation}
    which is obtained by substituting $\dot R(0,r)=0$ in Eq.(\ref{efe2}). Such momentarily static initial condition is motivated from the idea that collapse to a singularity begins when some dynamical instability sets in, as discussed in 
    \cite{Miyamoto_13, Brown_94}.
    
    \item Unless the globally naked singularity is physically strong, it should not be taken as a serious counterexample to the weak cosmic censorship. It is important to note that the strength of the singularity as defined by Tipler 
    \cite{Tipler_77} 
    involves vanishing of the volume element formed by three independent Jacobi fields along the timelike geodesic as it terminates in a strong singularity, rather than the behavior of individual Jacobi fields, as pointed out by Nolan 
    \cite{Nolan_99}.
    One can show examples of physically strong singularity wherein the volume element does not vanish and hence are classified as ``Tipler weak". This led Ori 
    \cite{Ori_99} 
    to redefine the physically strong singularity which extends the class of strong singularity by including cases in which any of the Jacobi fields is unbounded.  
    \cite{Nolan_99, Ori_99}.
    Such singularities are termed as ``deformationally strong" singularity. However, here we have taken an interest in singularities wherein the volume element defined by independent Jacobi fields vanishes (Tipler strong). We have proved the existence of such Tipler strong singularities that are globally naked and formed due to bound gravitational collapse.
    
    \item In deriving the explicit expression of the physical radius in terms of $t$ and $r$ in Eq.(\ref{R(t,r)}) for nonvanishing velocity function, only the first component of the Taylor expansion of $\mathcal{G}$ is used from Eq.(\ref{taylorexpandedG}). Hence the accuracy of our further analysis will get affected for large values of the term $\frac{fR}{F}$. To minimize the error, small values of the magnitude of the velocity function are considered. For larger values, higher-order terms in the expansion of $\mathcal{G}$ from Eq.(\ref{taylorexpandedG}) will have to be taken into account. Once the explicit expression of the physical radius is achieved, one can study the dynamics of the event horizon, apparent horizon, and singular radial null geodesics to investigate the global causal structure of the singularity.

   \item It is the event horizon, which evolves like an outgoing radial null geodesic, which starts from the singularity satisfying the equality of the physical radius and the Misner-Sharp mass function at the boundary of the collapsing fluid. Hence, any outgoing radial null geodesic with the property that $F<R$ at $r=r_c$ has to start from the center at a time before the formation of the singularity. However, this time difference between the escape of the light and the formation of the singularity can be reduced as much as desired. For such a null geodesic to be singular, it should escape from the region, which is in a small neighborhood of the singularity. This small neighborhood should have a measure of the order of Planck length. Only then will such untrapped null geodesic be considered significant and will be expected to contain traces relevant to deepen our understanding of how gravity works in the quantum regime.
    
    \item In terms of observational significance,  if at all there exists a globally visible singularity, it may be difficult to distinguish between singular and nonsingular geodesics escaping such singularity and received by a telescope. However, light wave front, which is more redshifted, is expected to come from the region, which is more close to the singularity as compared to the wave front, which is less redshifted. 
    
    \item Consider Eq.(\ref{mfvf}) with negative $F_3$ and positive $b_{00}$. This corresponds to the unbound collapse of fluid with third-order inhomogeneity in mass profile. It is found that as far as such mass and velocity functions are considered,  we may have $t_{EH}(0)<t_s(0)$, which means that globally visible singularity may not be achieved. This argument is supported by data in Tab.(\ref{tab1}). So far, no concrete statement about the global visibility of a strong singularity formed due to unbound collapse of dust can be made, and further investigation is needed. It may be possible that for some other combination of mass function and velocity function (unbound), the collapse ends in a globally visible singularity. This will be investigated in more detail in our future work.
    
    \item A very important concern is that our analysis is restricted to the end state of a collapsing dust cloud, i.e., the pressure of the collapsing fluid is considered to be zero. The effect on the global causal structure of the singularity in the presence of pressure is unknown. To understand the behavior of singular null geodesic numerically requires information about the explicit expression of the physical radius. However, this is difficult to obtain when the Misner-Sharp mass function varies with time, which is the case when there is nonzero pressure. Investigating the global visibility of a Tipler strong singularity formed due to the collapse of a cloud having such time-varying Misner-Sharp mass function will be a significant step toward understanding the cosmic censorship.  
\end{enumerate}

% Chapter Template

\chapter{Globally visible singularity in gravitational collapse of non-spherical dust} % Main chapter title

\label{Chapter4} % Change X to a consecutive number; for referencing this chapter elsewhere, use \ref{ChapterX}

\lhead{\emph{Chapter 4. }} % Change X to a consecutive number; this is for the header on each page - perhaps a shortened title

%----------------------------------------------------------------------------------------
%	SECTION 1
%----------------------------------------------------------------------------------------
A spherically symmetric collapsing cloud forms a singularity at the center which may be hidden in a black hole or it is a visible naked singularity. A relevant question here is, even if the singularity is naked, whether it would be visible to faraway observers in the universe in physical reality. For example, in spherical collapse models, even when the singularity is globally naked allowing the timelike and null geodesics to escape away from the collapsing cloud, the singularity occurs only at the center of the collapsing matter cloud. Such a singularity, even when it is visible faraway in principle, may not be able perhaps to radiate away energy, being embedded within very high density regions of the collapsing star. Thus the physical implications of such a naked singularity would need a more detailed investigation.

Even if the singularity of collapse is not hidden to faraway observers, one could argue that the information about the extremely strong gravity region carried by the outgoing null geodesics can be opaque or get distorted  because of the scattering of the null geodesic due to the collapsing matter surrounding the singularity. The spherical symmetry is, however, a strict presupossition and analysing more general solutions of the Einstiens field equation would be of considerable interest. In other words, it is important to ask whether the same scenario persits in collapse models that are not exactly spherical, or which represent small perturbations from spherical symmetry.

Existence of strong (in the sense of Tipler 
\cite{Tipler_77, Clarke_85}), 
naked singularities formed due to quasi-spherical collapsing clouds governed by what is known as the Szekeres spacetimes
\cite{Szekeres_75}
was studied previously by  Joshi and Krolak 
\cite{Joshi_96}. 
It was shown that the criteria for such singularity to be  strong naked is the same as in the case of spherical symmetry. Globally visibile singularity was shown to exist in collapse of such non-spherical marginally bound cloud by Deshingkar \textit{et.al.} 
\cite{Deshingkar_98}. 

What we need to inquire and examine is whether the causal structure of the naked singularity remains the same, i.e. embedded in the interior of the matter cloud, or whether it exibits other novel causal features. In particular, 
the distortion of the information can be reduced or avoided if the singularity is formed near the boundary of the collapsing cloud. Closer the singularity is to the boundary, lesser will be the scattering of the outgoing singular null geodesics in certain directions. We show here that such collapsing spacetime solutions, which can end up in a singularity which is not at the geometric center, can be obtained from the Szekeres solution 
\cite{Szekeres_75}.

Marginally bound collapse is again a special case, so to offer generality we investigate here a non-marginally bound non-spherical gravitational collapse.
We show that for a suitable choice of free functons, arising due to available degrees of freedom in the Einstein's field equations, one could achieve a scenario wherein the globally visible singularity is formed not at the geometric center but closer to the boundary of the collapsing cloud. 

It thus turns out that introducing asphericity in collapse can radically alter the causal structure of the naked singularity, in that its visibility can be greatly enhanced. This indicates that small perturbations from spherical symmetry are important to consider in order to examine the physical implications of naked singularities. The chapter is arranged as follows: In the next sec.(\ref{4.1}), we discuss the non-spherical Szekeres spacetime and derive the required equations using the Einstein's field equations. In sec.(\ref{4.2}), we discuss the visibility aspects and the strength of the singularity.  We end the chapter with the concluding remarks and stating a few open concerns in 4.3. 

\section{Szekeres spacetime and the Einstein's equations}\label{4.1}
Using the units in which $c=8\pi G=1$, the general Szekeres metric in the comoving coordinates is given by,
%\begin{equation}
 %   ds^2=-dt^2+M^{2}dr^2+N^{2}(dx^2+dy^2).
%\end{equation}
%The energy-momentum tensor of \textit{type I} matter field with vanishing pressure is written as $T^{\mu \nu}=\rho U^{\mu} U^{\nu}$, where $U^{\mu}$ are the components of the four velocity. We now consider a pair of conjugate coordinates defined as $\zeta=x+i y$ and $\Bar{\zeta}=x-i y$. The metric can then be rewritten in the $(t,r,\zeta,\Bar{\zeta})$ coordinates as
\begin{equation}\label{szekeresmetric}
    ds^2=-dt^2+M^{2}(t,r,\zeta,\Bar{\zeta})dr^2+N^{2}(t,r,\zeta,\Bar{\zeta})d\zeta d\Bar{\zeta},
\end{equation}
where  
\begin{equation}
    \zeta=x+i y \hspace{1cm}\textrm{and}\hspace{1cm} \Bar{\zeta}=x-i y
\end{equation}
 is a pair of conjugate coordinates. Also, we have
 \begin{equation}\label{Nszekeres}
     N=\frac{S(t,r)}{Q(r,\zeta,\Bar{\zeta})}
 \end{equation}
 and
 \begin{equation}\label{Mszekeres}
     M=\frac{Q N'}{\sqrt{1+f(r)}},
 \end{equation}
 where $f>-1$ is the velocity function. $f$ greater than, equal to, and less than zero corresponds to bound, marginally bound, and unbound collapse respectively. Additionally, $N'\neq 0$. Here the subscripts prime and dot denotes the partial derivative with respect to $r$ and $t$ respectively. Also,
\begin{equation}\label{Qszekeres}
    Q=a(r)\zeta \Bar{\zeta}+B(r)\zeta+\Bar{B}(r)\Bar{\zeta}+c(r),
\end{equation}
where $a$ and $c$ are real, and $B$ is a complex function having the relation
\begin{equation}
    ac-B\Bar{B}=\delta/4, \hspace{1cm}\textrm{where} \hspace{1cm}  \delta=0,\pm 1.
\end{equation}
The energy-momentum tensor of \textit{type I} matter field with vanishing pressure is written as $T^{\mu \nu}=\rho U^{\mu} U^{\nu}$, where $U^{\mu}$ are the components of the four velocity. Now, our aim is to write the last term of Eq.(\ref{szekeresmetric}) in terms of the line element of two sphere, i.e. $d\Omega^2$. Let 
\begin{equation}
    \zeta=\zeta(\theta,\phi), \hspace{1cm} \textrm{and} \hspace{1cm} \Bar{\zeta}=\Bar{\zeta}(\theta,\phi)
\end{equation}
The infinitesimal increments $d\zeta$ and $d\Bar{\zeta}$ are then written using the chain rule as
\begin{equation}
    d\zeta=\frac{\partial \zeta}{\partial \theta} d\theta+\frac{\partial \zeta}{\partial \phi}d\phi,
\end{equation}
and
\begin{equation}
    d\Bar{\zeta}=\frac{\partial \Bar{\zeta}}{\partial \theta} d\theta+\frac{\partial \Bar{\zeta}}{\partial \phi}d\phi.
\end{equation}
We therefore have
\begin{equation}
     d\zeta d\Bar{\zeta}=\frac{\partial \zeta}{\partial \theta}\frac{\partial \Bar{\zeta}}{\partial \theta} d\theta^2+ \left(\frac{\partial \zeta}{\partial \theta}\frac{\partial \Bar{\zeta}}{\partial \phi}+\frac{\partial \zeta}{\partial \phi}\frac{\partial \Bar{\zeta}}{\partial \theta}     \right)d\theta d\phi + \frac{\partial \zeta}{\partial \phi}\frac{\partial \Bar{\zeta}}{\partial \phi} d\phi^2.
\end{equation}
Let us choose
\begin{equation}
 \zeta(\theta,\phi)=\zeta_1 e^{i \phi}, \hspace{1cm} \textrm{and} \hspace{1cm} \Bar{\zeta}(\theta,\phi)=\zeta_1 e^{-i\phi}.   
\end{equation}
Such choice leads to the following:
\begin{equation}
    N^2 d\zeta d\Bar{\zeta}=N^2 \left(\zeta_1,_{\theta}^2 d\theta^2+\zeta_1 ^2 d\phi^2\right).
\end{equation}
%%%%%%%%%%%%%%%%%%%%%%%%%%%%%%%%%%%%%%%%%%%%%%%%%%%%%%%%%%%

%Choosing $\zeta=f_1(\theta)e^{i \phi}$ gives us
%\begin{equation}
%   d\zeta d\Bar{\zeta}=\left \vert \frac{\partial f_1}{\partial \theta}\right \vert^2d\theta^2+\left \vert \frac{\partial f_1}{\partial \phi} \right \vert^2 d\phi^2.
%\end{equation}
Choosing  
\begin{equation}
    \zeta=\tan(\theta/2)e^{i \phi} \hspace{1cm}\textrm{and}\hspace{1cm} B=0,
\end{equation}
 one can express $N^2d\zeta d\Bar{\zeta}$ as 
\begin{equation}\label{Nsquareszekeres}
    N^2d\zeta d\Bar{\zeta}=
    R^2(t,r,\theta) \left(d\theta^2 +\sin^2{\theta}d\phi^2\right),
\end{equation}
where
\begin{equation}\label{Rszekeres}
    R(t,r,\theta)=\frac{S(t,r)\sec^2(\frac{\theta}{2})}{2(a(r)\tan^2(\frac{\theta}{2})+c(r))}
\end{equation}
is the physical radius of the collapsing cloud, which tells us the distance of a point $(t,r,\theta, \phi)$ from its center of mass. Note that $R$ is symmetric with respect to change in $\phi$. The density and pressure of the collapsing cloud, obtained from the Einstein's field equation, are expressed respectively as
\begin{equation}\label{rhoaspherical}
    \rho(t,r,\theta)=\frac{QA'-3AQ'}{S^2(Q S'-S Q')},
\end{equation}
and
\begin{equation}\label{paspherical}
  p(t,r)=-\frac{\dot A}{S^2 \dot S},
\end{equation}
where
\begin{equation} \label{Aaspherical}
    A=S(\dot S ^2-f).
\end{equation}
Assumption of zero pressure corresponds to $A$ being a function of only $r$, as is apparent from Eq.(\ref{paspherical}). 
%It should be noted that the density can also be written as $\rho(t,r,\theta)=F'/(R^2R')$, where $F$ is twice the mass inside a collapsing shell of constant radial coordinate $r$ at time $t$. 
Eq.(\ref{Aaspherical}) can be integrated to obtain
\begin{equation}\label{timecurveaspherical}
    t-t_s(r)=-\frac{S^{\frac{3}{2}}\mathcal{G}(-fS/A)}{\sqrt{A}},
\end{equation}
where $\mathcal{G}(y)$ is defined as in Eq.(\ref{G}). Here, $t_s(r)$, is the time at which a collapsing shell of fixed radial coordinate $r$ becomes singular, and is given by
\begin{equation}\label{singularitycurveaspherical}
    t_s(r)=\frac{S_0(r)^{\frac{3}{2}}}{\sqrt{A}}\mathcal{G}\left(-\frac{f S_0(r)}{A}\right),
\end{equation}
where $S_0(r)=S(0,r)$. Consider the comoving radius corresponding to the boundary of the collapsing cloud $r_c$. In our aspherical collapsing model, the density vanishes at the boundary, hence $r_c=r_c(\theta)$. Therefore, from the above equation, $t_s(r_c)=t_s(r_c(\theta))$. Hence, the boundary of the cloud does not collapse to the singularity simultaneousy, but rather falls in the singularity at different times along different directions. 

\section{Visibility and strength of the singularity}\label{4.2}
%%%%%%%%%%%%%%%%%%%%%%%%%%%%%%%%%%%%%%%%%%%%%%%%%%%%%%%%%%%%%%%%%%%%%%%%%%%%%%%%%%%%%%%%%%%%%%%%%%%%
\begin{figure}
\centering
{\includegraphics[scale=0.45]{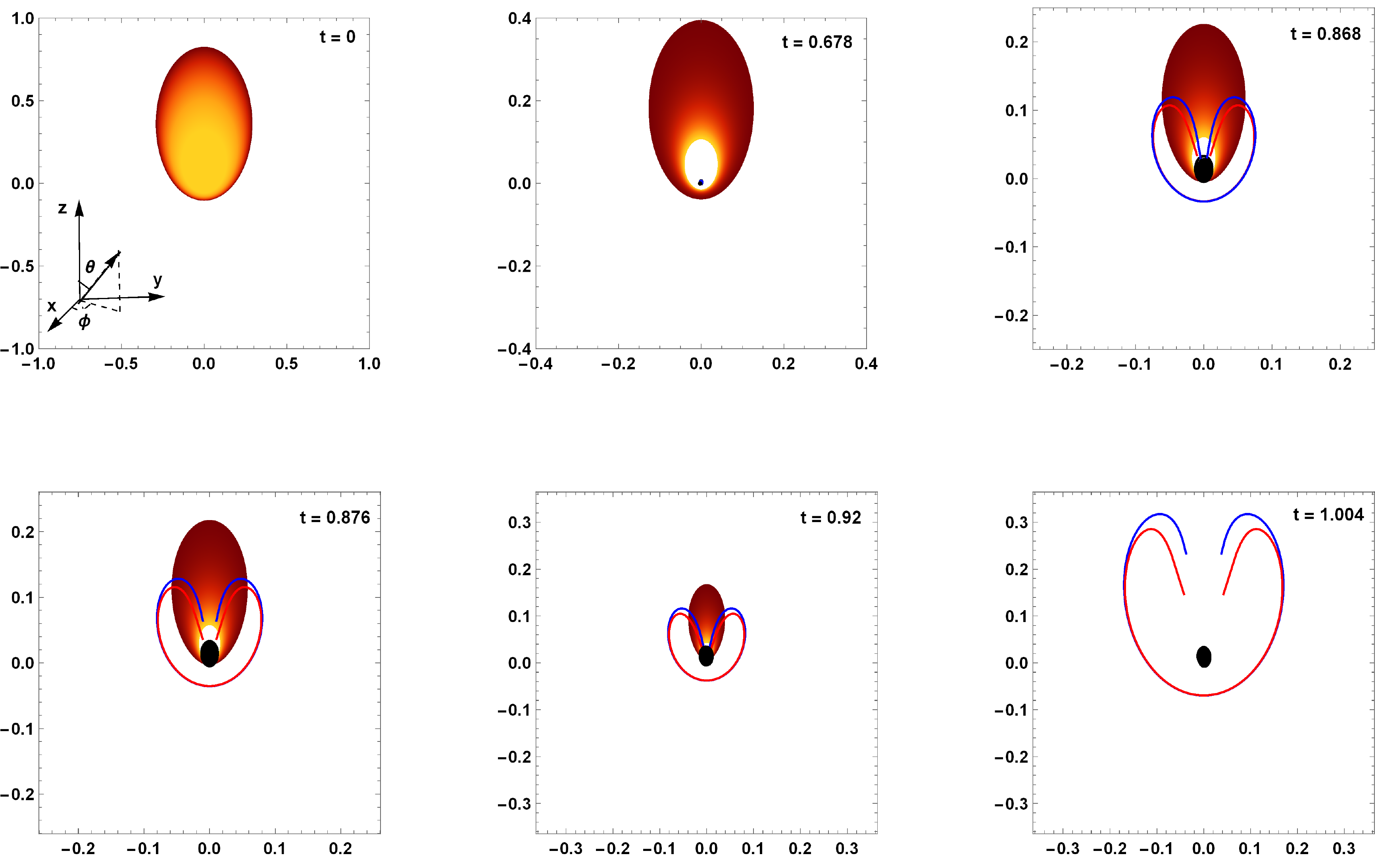}}
%\hspace{0.2cm}
%\subfigure[]
\caption{Cross-section along fixed $\phi=\pi/2$ of the evolution of the collapsing aspherical dust cloud, and the global causal structure of the singularity thus formed, is depicted. The figures are symmetric with change in $\phi$. Here, $A=r^3-20r^6$, $f=-10^{-3}r^2$ and $Q=\text{exp}(r) \tan^2{\left (\frac{\theta}{2}\right)}+\frac{\text{exp}(-r)}{4}$. The singularity forms at $t=2/3$. Evolution of event horizon depends on $\theta$. The event horizon touches the boundary of the collapsing cloud first at $\theta=\pi$ at $t=0.868$. The aspherical wavefronts of singular null geodesics (red and blue colored) escape the singularity, thereby making it globally visible. However, part of these outgoing wavefronts which lies in the neighbourhood of $\theta=0$ gets trapped by the trapped surfaces and falls back to the singularity, hence making it only directionally globally naked in the neighbourhood around $\theta=\pi$.   }
\label{densityplotaspherical}
\end{figure}
%%%%%%%%%%%%%%%%%%%%%%%%%%%%%%%%%%%%%%%%%%%%%%%%%%%%%%%%%%%%%%%%%%%%%%%%%%%%%%%%%%%%%%%%%%%%%%%%%%%
 %%%%%%%%%%%%%%%%%%%%%%%%%%%%%%%%%%%%%%%%%%
%%%%%%%%%%%%%%%%%%%%%%%%%%%%%%%%%%%%%%%%%%%%%%%%%%%%%%%%%%%%%%%%%%%%%%%%%%%%%%%%%%%%%%%%%%%%%%%%%%%%
\begin{figure}
\centering
{\includegraphics[scale=0.45]{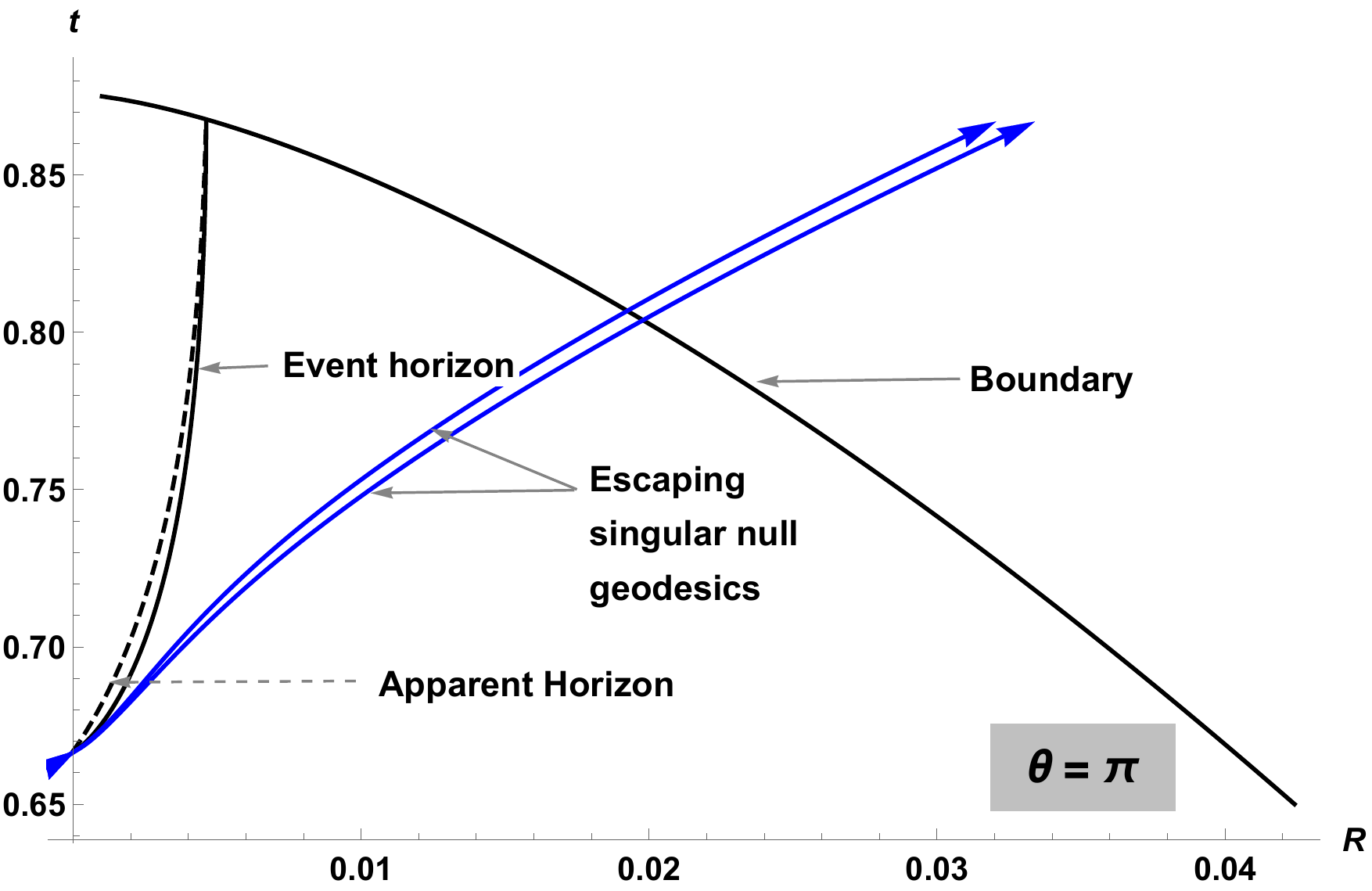}\includegraphics[scale=0.45]{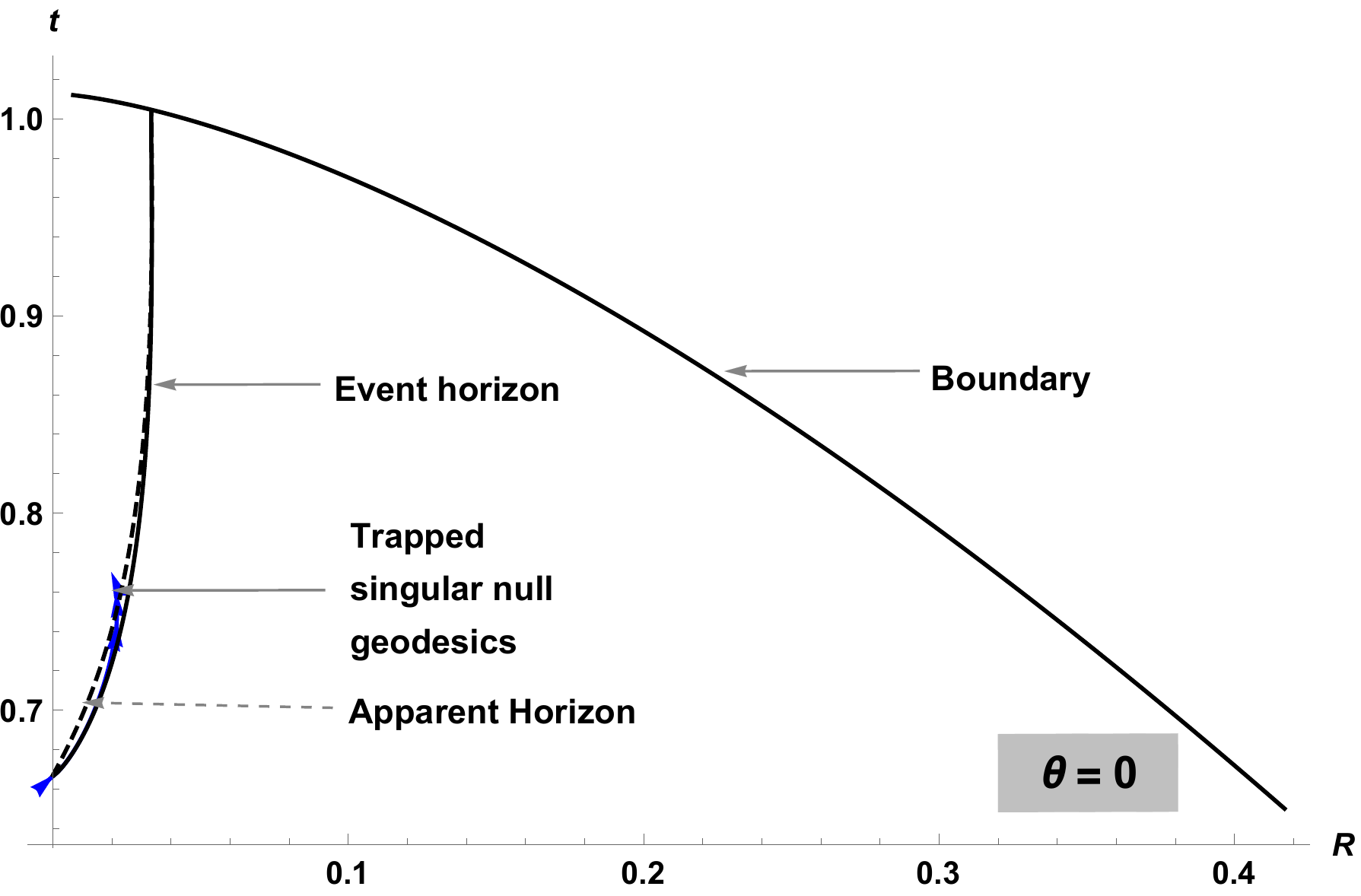}}
%\hspace{0.2cm}
%\subfigure[]
\caption{Causal structure of the singularity formed as the end state of a bound (elliptic) collapsing aspherical dust cloud along different inclination angles $\theta=\pi$ and $\theta=0$.  The apparent horizon, event
horizon, and singular null geodesics are represented by dashed black curves, solid black curves, and solid blue curves, respectively. Here, $A=r^3-20r^6$, $f=-10^{-3}r^2$ and $Q=\text{exp}(r) \tan^2{\left (\frac{\theta}{2}\right)}+\frac{\text{exp}(-r)}{4}$. Escaping singular null geodesics can reach the boundary of the collapsing cloud at $\theta=\pi$, and these get trapped and fall back to the singularity at $\theta=0$. The singularity is thus directionally globally visible.}
\label{ehnullaspherical}
\end{figure}
%%%%%%%%%%%%%%%%%%%%%%%%%%%%%%%%%%%%%%%%%%%%%%%%%%%%%%%%%%%%%%%%%%%%%%%%%%%%%%%%%%%%%%%%%%%%%%%%%%%
The apparent horizon, which is the boundary of trapped surfaces forming inside the collapsing cloud, is represented by vanishing $\Theta_l$, where
\begin{equation}\label{thetal}
    \Theta_l=h^{\mu \nu}l_{\nu;\mu}=\left (g^{\mu \nu}+\frac{l^{\mu}n^{\nu}+l^{\nu}n^{\mu}}{-l^{\alpha}n_{\alpha}}\right)l_{\nu;\mu}
\end{equation}
is the expansion scalar of outgoing null geodesic congruence (Refer Eq.(\ref{thetausable}) in the Introduction). Here, $h^{\mu \nu}$ is the transverse metric, and $l^{\alpha}$ and $n^{\alpha}$ are the tangents of the outgoing and incoming null geodesics respectively. For the metric shown in Eq.(\ref{szekeresmetric}), and using the expressions given in Eq.(\ref{Nszekeres}, \ref{Mszekeres}, \ref{Qszekeres}, \ref{Nsquareszekeres}, \ref{Rszekeres}), we can find $e^{\mu}_{\alpha}$, where 
\begin{equation}
    e^{\mu}_{\alpha}=\frac{\partial x^{\mu}}{\partial y^{\alpha}}
\end{equation}
is the orthonormal basis. Here, $y^{\alpha}$ are the coordinate in local inertial frame, and $x^{\mu}$ are the coordinates in global non-inertial frame. We hence obtain:
\begin{equation}
\begin{split}
     & e^{\mu}_{0}=\left(1,0,0,0\right),\\
     & e^{\mu}_{1}=\left(0,\frac{\sqrt{1+f}}{\left(R'-\frac{R Q'}{Q}\right)},0,0\right),\\
     & e^{\mu}_{2}=\left(0,0,\frac{Q}{R \zeta_1,_{\theta}},0\right),\\
     & e^{\mu}_{3}=\left(0,0,0,\frac{Q}{R \zeta_1,_{\theta}\sin{\theta}}\right).
\end{split}
\end{equation}
The tangents of the outgoing and incoming null geodesics are then expressed respectively as
\begin{equation}\label{lmu}
    l^{\mu}=e^{\mu}_{0}+e^{\mu}_{1}=\left(1,\frac{\sqrt{1+f}}{\left(R'-\frac{R Q'}{Q}\right)},0,0\right)
\end{equation}
and
\begin{equation}\label{nmu}
     n^{\mu}=e^{\mu}_{0}-e^{\mu}_{1}=\left(1,-\frac{\sqrt{1+f}}{\left(R'-\frac{R Q'}{Q}\right)},0,0\right).
\end{equation}
Using Eq.(\ref{lmu}) and Eq.(\ref{nmu}) in the Eq.(\ref{thetal}) and substituting $\theta_{l}=0$, we  obtain from that along the apparent horizon curve, 
\begin{equation}
    A(t,r)=S(r).
\end{equation}
Using this along with Eq.(\ref{timecurve}), we obtain the apparent horizon curve as
\begin{equation}
    t_{AH}(r)= t_s(r)-A\mathcal{G}(-f).
\end{equation}
From this equation, it can be seen that $t_s(0)=t_{AH}(0)$, since $A(0)=0$, as demanded by the regularity conditions 
\cite{Joshi_11}. 
%[HOW SPECIAL OR GENERIC ARE THESE VALUES CHOSEN? IN OTHER WORDS, IS THERE A PARAMETER RANGE FOR WHICH SAME OR SIMILAR RESULTS ARE OBTAINED, OR THIS IS JUST A SINGLE EXAMPLE?] 

The event horizon is a null surface and its evolution is described by the solution of the nul geodesic equation
\begin{equation}
\frac{dt}{dr}=M(t,r,\theta),    
\end{equation}
 satisfying the condition, $A=S$ at $r=r_c$. For a singularity to be visible globally in a certain constant $\theta$ direction, the event horizon at the center of mass should form not before, but at the time of formation of the singularity obtained due to the collapsing shell corresponding to $r=0$. If the event horizon at the center of mass forms before the singularity, in a given direction, even if the singular null geodesic escapes, it later gets trapped and falls back in, before reaching the collapsing boundary.
 %Additionally, the singularity should be a nodal point. This ensures that entire family of null geodesics escape from the singularity $(t_s(0),0)$ in the $(t,R)$ plane with fixed $\theta$ coordinate, thereby exposing the singularity globally, for infinite time.

As seen in Fig.(\ref{densityplotaspherical}), for suitable functions $A$, $f$ and $Q$ (freedom of choice possible due to the availability of three degrees of freedom in the given set of Einstein's equations), satisfying regularity conditions 
\cite{Joshi_11}, 
a  directional globally visible singularity forms at $(0,0)$, which is away from the geometric center of the collapsing cloud, and is located near its boundary. Directional visibility is attributed to the fact that the event horizon starts evolving from the center at different times along different $\theta$ (also see Fig.(\ref{ehnullaspherical})). The null geodesics should escape from the region sufficiently close to the singularity to contain signatures of the strong gravity regions (mimicing possibly the quantum gravity effects); i.e. the difference in time of escape of the null singularity at $r=0$ and the time of formation of the singularity at $r=0$ should be of the order of the Planck time. The outermost aspherical wavefront of the null geodesic satisfying this criterion lies partly inside and partly outside the aspherical event horizon surface, intersecting it such that the locus of the points of intersection is a closed curve. This closed curve subtends a solid angle at the center $(0,0)$ such that the part of escaping singular null geodesic wavefront lying inside this solid angle gets trapped and the part outside escapes, thereby making the singular region visible, only if the asymptotic observer lies outside the region subtended by this solid angle.

It is worth mentioning that the aspherical collapse example which we have shown is not a small perturbation on the LTB metric. We know that in the case of spherical symmetry, there exists a non-zero measured set of initial parameters giving rise to such globally visible singularity  
\cite{Joshi_11}.
Similarly, one can show that there exists a non-zero measured set of initial parameters $A$ and $f$ giving rise to a directional globally visible singularity which is offcentric and close to the boundary of the collapsing cloud.

%[A RELEVANT QUESTION HERE IS, CAN THESE BE REGARDED AS "SMALL PERTURBATIONS" FROM SPHERICAL SYMMETRY? IS IT A SMALL CHANGE IN VALUES FROM THE LTB METRIC?] 
This directonally visible singularity is physically strong in the sense that at least along one null geodesic with affine parameter $\lambda$, with $\lambda=0$ at the singularity, the inequality (\ref{Krolak and Clarke criteria})
\begin{equation}
\lim_{\lambda \to 0} \lambda^2R_{ij}K^iK^j>0
\end{equation}  
holds. Here $R_{ij}$ is the Ricci tensor, and $K^{\alpha}$ is the tangent of the outgoing null geodesics from the singularity. This condition ensures that the volume element formed by independent Jacobi fields along the geodesic vanishes as the geodesic terminates at the singularity, as discussed in the Introduction. 

\section{Concluding remarks}\label{4.3}
Some concluding remarks and open concerns are mentioned as follows:
\begin{enumerate}
    \item Once the matter cloud falls inside the event horizon, the exterior spacetime is aspherical, static, vacuum, and asymptotically flat. It was shown by
    \cite{Cruz_70, Price_71} 
    using perturbation theory, that for a small deformation from spherical symmetry, a non-rotating, collapsing body radiates away the deformatons in the form of gravitational waves and assumes the shape of minimum curvature (as visualized by Misner 
    \cite{Thorne_70}).
    Hence, the end state is possibly a spherical event horizon. For highly deformed spacetime, this may or may not be true 
    \cite{Thorne_70}.
    
     \item The singular null geodesics escaping radially in the neighbourhood around $\theta=\pi$, as shown in Fig.(\ref{densityplotaspherical}), are less scattered due to interaction with the collapsing matter, than those in other directions. These less scattered geodesics  in a less distorted form may contain traces of gravity theory that governs the strong gravity regime. One might wonder if there is a scenario, a solution of the Einstein's field equation, wherein the singularity forms `at' the boundary (and not close to it) so that the original form of the escaping singular null geodesic is maintained with no distortion at all.
\end{enumerate}

% Chapter Template

\chapter{Globally visible singularity in an astrophysical setup} % Main chapter title

\label{Chapter5} % Change X to a consecutive number; for referencing this chapter elsewhere, use \ref{ChapterX}

\lhead{\emph{Chapter 5. }} % Change X to a consecutive number; this is for the header on each page - perhaps a shortened title

%----------------------------------------------------------------------------------------
%	SECTION 1
%----------------------------------------------------------------------------------------

Observations about compact objects such as the shadow of the M87 galactic center by the Event Horizon Telescope 
\cite{Akiyama_19}, 
earlier observations such as  Sagittarius A* (Sgr A*) at our galactic center 
\cite{Schodel_02, Ghez_08}, 
and similar compact objects at the center of other galaxies 
\cite{Kormendy_13} 
may hint that singularities do exist in our universe. Some theoretical predictions motivate us to consider the existence of a naked singularity. One of them is that shadows are not unique to black holes, but even naked singularities under certain circumstances can cast a shadow 
\cite{Shaikh_18}. 
Apart from this, the precession of the orbit of stars around the compact object at the galactic center can also put light on the causal structure of the central singularity. It has been found that  while the precession of particle orbit in the external Schwarzschild spacetime is in the direction of the particle motion, the naked singularity spacetimes like the Joshi-Malafarnia-Narayan (JMN) spacetime
\cite{Joshi_11(2)}
and the Janis-Newman-Winicour (JNW) spacetime
\cite{Janis_68}
can cause the orbiting particle to precess in the opposite direction of the particle's motion under certain conditions 
\cite{Bambhaniya_20, Joshi_20}.
The S2 star orbiting around the center of our galaxy may also exhibit such behavior
\cite{Dey_19}.
Other observational signatures like the property of the accretion disk  
\cite{Gyulchev_19, Gyulchev_20, Shaikh_19}, 
the Einstein ring formed due to gravitational lensing  
\cite{Dey_13, Dey_13(2), Bhattacharya_20}, 
the property of the shadow 
\cite{Ortiz_15, Abdikamalov_19, Joshi_19, Dey_20, Dey_21, Bambhaniya_21}, 
and the  gravitational redshift of the photons coming from the center of the collapsing cloud 
\cite{Ortiz_14, Ortiz_15(2)}, 
can help in distinguishing between a black hole and a naked singularity. On the other hand, 
\cite{Kong_14} 
has discussed about the qualitative indifference between the radiation escaping from the collapsing dust cloud forming a naked singularity and that forming a black hole, thereby highlighting the extreme challenge to observationally distinguish the birth of a naked singularity from that of a black hole.

Mathematically, it is possible to achieve a globally naked singularity. However, whether suitable configurations required for its existence are possible astrophysically is our matter of concern, and we will discuss it in this chapter. The chapter is arranged as follows: Sec.(\ref{5.1}) consists of the mathematical formalism of the gravitational collapse of a spherically symmetric dust cloud and the collapse criteria to end in a globally visible singularity. In sec.(\ref{5.2}), the astrophysically reasonable parameters, which can give a globally visible singularity, are discussed. The collapse of the cloud possibly forming the central singularity at the galactic center,  the collapse of the matter cloud in the primordial time, and the collapse of an accreting neutron star are investigated. We end the paper with the concluding remarks and open concerns in sec.(\ref{5.3}). 

\section{Gravitational collapse formalism}\label{5.1}

We consider the gravitational collapse of a spherically symmetric cloud of dust. The spacetime of such dust cloud is described by Lemaitre-Tolman-Bondi metric
\cite{Lemaitre_33, Tolman_34, Bondi_47}, 
which is given by
\begin{equation}
    ds^2=-c^2dt^2+\frac{R'}{1+f}dr^2+R^2d\Omega^2.
\end{equation}
Here $R=R(t,r)$ is the physical radius of the collapsing cloud and $f(r)$ is called the velocity function. The energy-momentum tensor in the comoving frame is given by
\begin{equation}
    T^{\mu\nu}=\rho(t,r) U^{\mu}U^{\nu}.
\end{equation}
where $\rho(t,r)$ is the energy density of the cloud  $U^{\mu}$ is the four velocity. Using the Einstein's field equations, we get the following relation:
\begin{equation}
    \frac{F'}{R^2R'}=8\pi \rho
\end{equation}
and
\begin{equation}\label{dotF}
    \dot F=0,
\end{equation}
where
\begin{equation}\label{friedmann}
    F=\frac{c^2}{G}\left(\frac{R \dot R^2}{c^2}-fR\right).
\end{equation}
The superscripts dot and prime denotes the partial derivative with respect to time and radial coordinate respectively. The term $F$ is the Misner-Sharp mass function 
\cite{Misner_1969}. 
%In case of a collapsing dust, the Misner-Sharp mass function is a conserved quantity for a given radial coordinate and hence is a function of only $r$ as seen in Eq.(\ref{dotF}). 
Integrating Eq.(\ref{friedmann}), gives
\begin{equation}\label{t-tsr}
    t-t_s(r)=-\frac{R^{\frac{3}{2}}}{\sqrt{G}\sqrt{F}}\mathcal{G}\left(-\frac{c^2fR}{GF}\right),
\end{equation}
where $\mathcal{G}(y)$  is given in Eq.(\ref{G})
Now, we can rescale the physical radius using the coordinate freedom such that
\begin{equation}
    R(0,r)=r.
\end{equation}
This along with Eq.(\ref{t-tsr}) gives
\begin{equation}\label{tsr}
    t_s(r)=\frac{r^{\frac{3}{2}}}{\sqrt{G}\sqrt{F}}\mathcal{G}\left(-\frac{c^2fr}{GF}\right),
\end{equation}
where $t_s(r)$ is the time taken by a shell of radial coordinate $r$ to collapse to a singularity. From Eq.(\ref{t-tsr}), Eq.(\ref{tsr}) and Eq.(\ref{G}), one can write $R$ explicitly as a function of $r$ and $t$ as
\begin{equation}\label{rnmb}
     R(t,r)_{\textrm{NMB}}=\frac{5GF}{2c^2f}\bigg (1-\bigg (1-\frac{4fc^2}{5FG}\bigg (r^{\frac{3}{2}}-\frac{3c^2f}{10GF}r^{\frac{5}{2}}-\frac{3}{2}\sqrt{G}\sqrt{F}t\bigg )^{\frac{2}{3}}\bigg )^{\frac{1}{2}}\bigg),
\end{equation}
for $f\neq 0$. The subscript ``NMB" stands for non-marginally bound collapse. Here, we have used the Taylor expansion representation of $\mathcal{G}(y)$ up to first order, which is given by
\begin{equation}\label{taylorexpansionofG}
    \mathcal{G}(y)=\frac{2}{3}+\frac{y}{5}+o(y^2), \hspace{1cm} 0<y<1.
\end{equation}
Eq.(\ref{rnmb}) is a good approximation only if the velocity function is such that $|f|<<\frac{G F}{c^2 r}$. Velocity functions with higher magnitude require higher orders to be considered in the Taylor expansion of $\mathcal{G}(y)$.
For the case of marginally bound (MB) collapse, $R$ is expressed as
\begin{equation}\label{rmb}
    R(t,r)_{\textrm{MB}}=\left(r^{\frac{3}{2}}-\frac{3}{2}\sqrt{G}\sqrt{F}t\right)^{\frac{2}{3}}.
\end{equation}
The condition for the apparent horizon curve can be obtained from the following equality:
\begin{equation}
    g^{\mu\nu}R,_{\mu}R,_{\nu}=0.
\end{equation}
This gives the following relation needed to be satisfied for the apparent horizon curve:
\begin{equation}
    \frac{\dot R^2}{c^2}=1+f.
\end{equation}
The above equation along with Eq.(\ref{friedmann}) gives
\begin{equation} 
    F=\frac{c^2}{G}R.
\end{equation}
The event horizon evolves like a null geodesic wavefront and meets the apparent horizon at the boundary of the collapsing cloud $r_c$. Hence, the event horizon is described by the solution of the differential equation:
\begin{equation}\label{ngde}
    \frac{dt}{dr}=\frac{1}{c}\frac{R'}{\sqrt{1+f}},
\end{equation}
satisfying the condition:
\begin{equation}\label{ngdeic}
    F(r_c)=\frac{c^2}{G}R(t,r_c).
\end{equation}
We consider the Misner-Sharp mass function to be of the form
\begin{equation}\label{strongmassfunction}
    F=F_0\left(\frac{r}{r_c}\right)^3+F_3\left(\frac{r}{r_c}\right)^6.
\end{equation}

For $F_3<0$, the singularity formed at the end of the collapse is locally naked. In the case of non-marginally bound collapse, the strength of the singularity and its local causal property depends on the velocity function in addition to the mass function
\cite{Mosani_20(2)} 
(Please refer to the Appendix for further discussion on the local visibility of the singularity).    

For the density to decrease as we move away from the center and smoothly vanish at the boundary of the collapsing cloud ($\rho(0,r_c)=0$), $F$ has to be rewritten as 
\begin{equation}\label{msmf5}
    F(r)=F_0\left(\frac{r}{r_c}\right)^3\left(1-\frac{1}{2}\left(\frac{r}{r_c}\right)^3\right).
\end{equation}
It should be noted that in the case of marginally bound pressureless collapsing cloud with density vanishing smoothly at its boundary, the above expression is the only possible mass function which corresponds to a Tipler strong singularity as its end state (See the Appendix).

Substituting Eq.(\ref{rnmb}) and Eq.(\ref{msmf5}) in Eq.(\ref{ngdeic}), we get \begin{equation}\label{tehrc}
    t_{\textrm{EH}_{\textrm{NMB}}}(r_c)=\frac{2\sqrt{2}}{3\sqrt{G}\sqrt{F_0}}r_C^{\frac{3}{2}}-\frac{2\sqrt{2}}{5}\frac{c^2f(r_c)}{\left(G F_0\right)^{\frac{3}{2}}}-\frac{G F_0}{3c^3}\left(1-\frac{f(r_c)}{5}\right)^{\frac{3}{2}}
\end{equation}
for $f\neq 0$. For $f=0$, substituting Eq.(\ref{rmb}) and Eq.(\ref{msmf5}) in Eq.(\ref{ngdeic}), we get
\begin{equation}\label{tehrc2}
    t_{\textrm{EH}_{\textrm{MB}}}(r_c)=\frac{2\sqrt{2}}{3\sqrt{F_0}\sqrt{G}}\left(r_c^{\frac{3}{2}}-\left(\frac{GF_0}{2c^2}\right)^{\frac{3}{2}}\right).
\end{equation}
%%%%%%%%%%%%%%%%%%%%%%%%%%%%%%%%%%%%%%%%%%%%%%%%%%%%%%%%%%%%%%%%%%%%%%%%%%%%%%%%%%%%%%%%%%%%%%%%%%%%
\begin{figure*}\label{fig1..}
{\includegraphics[scale=0.5]{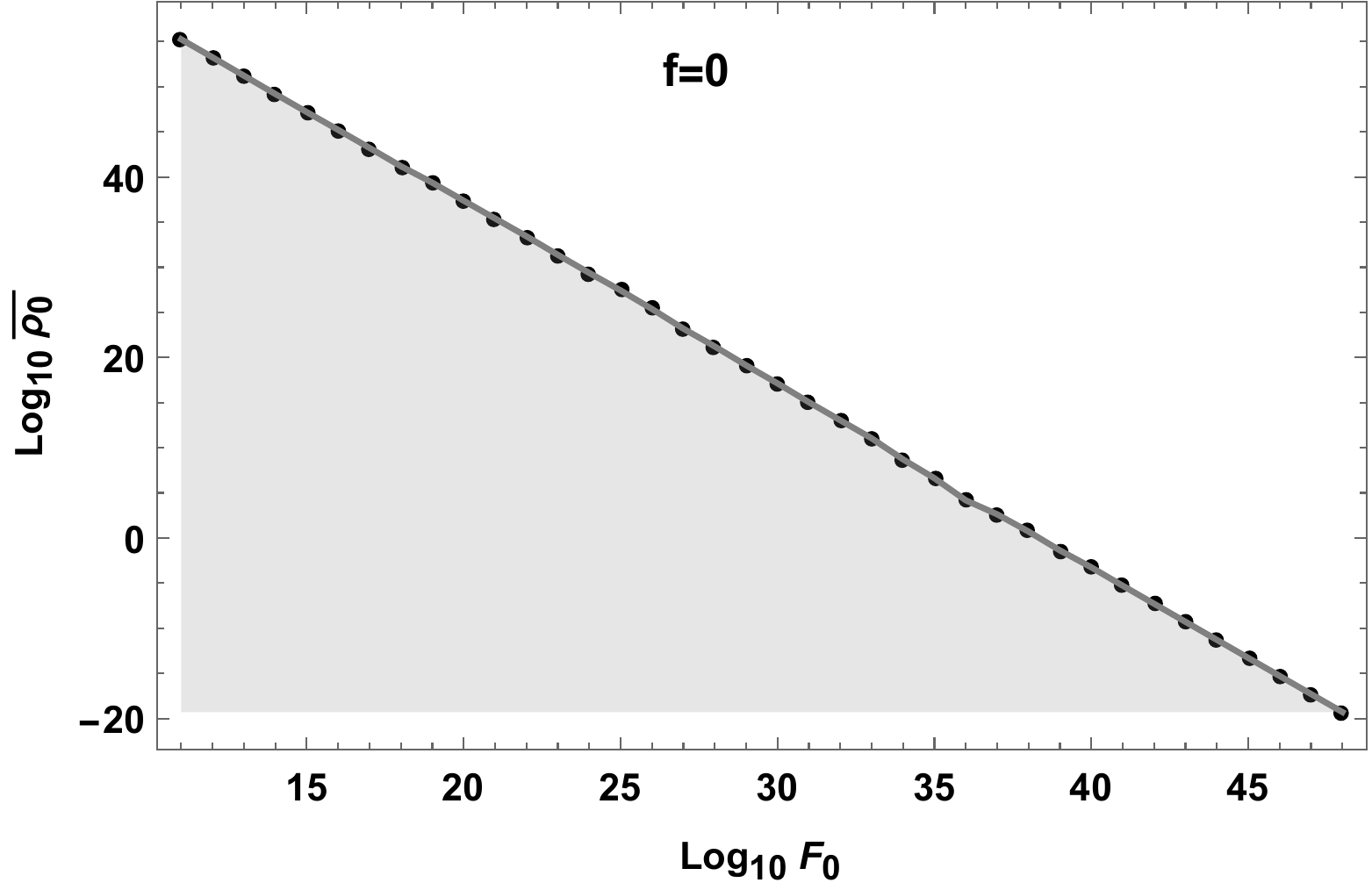}\includegraphics[scale=0.5]{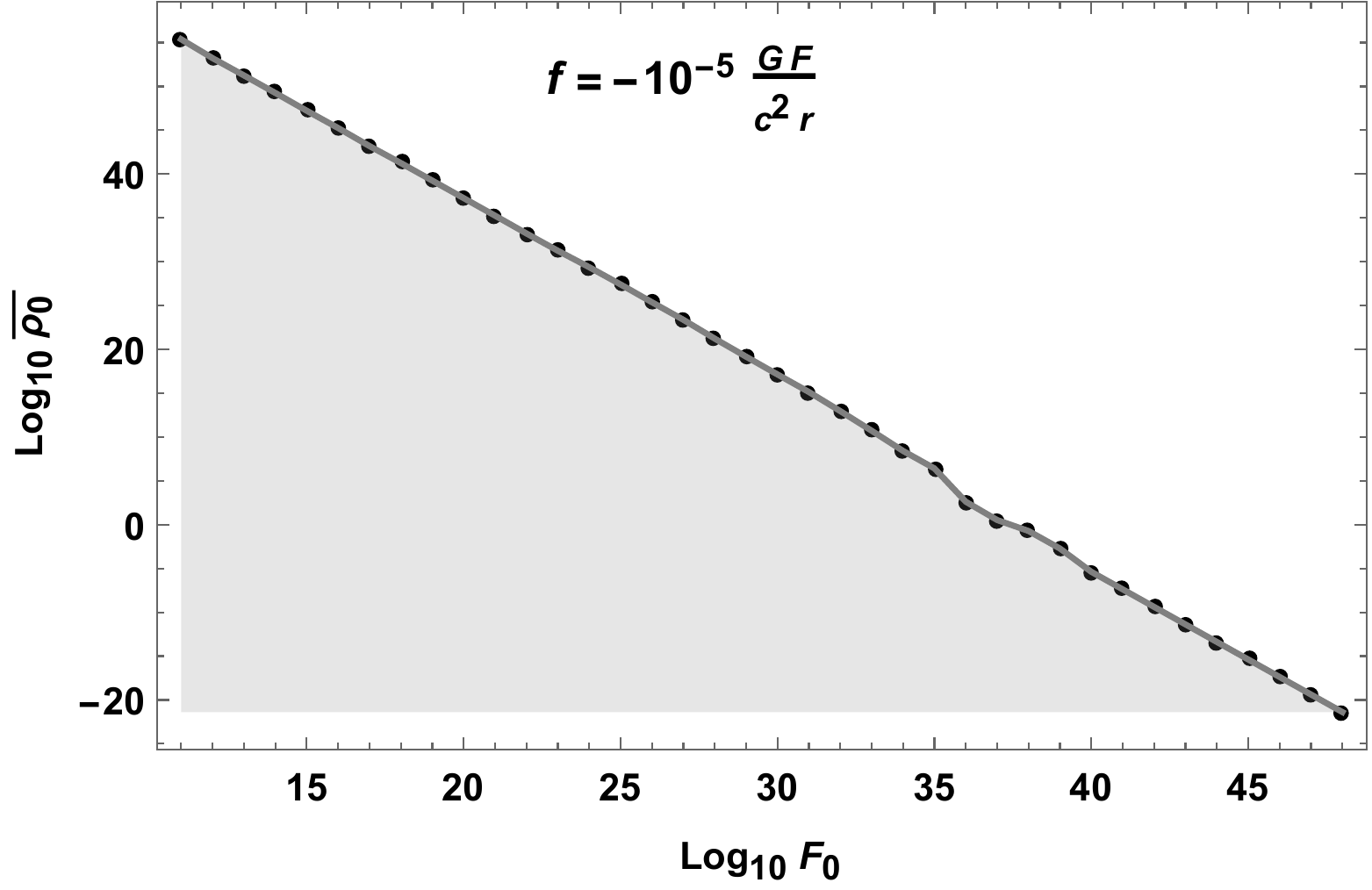}}
\caption{Non-zero measured set of initial data giving rise to a globally visible singularity (shaded region) and globally hidden singularity (unshaded region) as an end state of a marginally bound collapse (a) and a non-marginally bound collapse (b) of a pressureless inhomogeneous dust cloud having the mass function as in Eq.(\ref{msmf5}). Here, $\Bar{\rho_0}$ and $\frac{F_0}{4}$ are the initial mean density and the total mass of the collapsing cloud having units $\textrm{kg}/\textrm{m}^3$ and kg respectively.  }
\label{astrophysicalregionplot}
\end{figure*}
%%%%%%%%%%%%%%%%%%%%%%%%%%%%%%%%%%%%%%%%%%%%%%%%%%%%%%%%%%%%%%%%%%%%%%%%%%%%%%%%%%%%%%%%%%%%%%%%%%%
The event horizon curve $t_{EH}(r)$ is now the solution of the differential Eq.(\ref{ngde}) satisfying Eq.(\ref{tehrc}) in case of non-marginally bound collapse (Eq.(\ref{tehrc2}) in case of marginally bound case). At the time of formation of the central singularity, the apparent horizon (the boundary of all trapped surfaces) starts forming from the center. It evolves in the outward direction for an inhomogeneous mass function as in Eq.(\ref{msmf5}) 
\cite{Joshi_93(2), Mosani_20(2)}. 
Hence, any null geodesic leaving the center after the formation of the singularity, due to the collapse of the central shell, gets trapped by trapped surfaces. Therefore, the event horizon forms either before or during the formation of the central singularity. If it forms at the center before the formation of this central singularity, any null geodesic escaping from the center after $t_{EH}(0)$, having a positive tangent at $r=0$, will fall back to the singularity, thereby making the singularity globally hidden. Hence, the necessary criteria for a singularity to be globally visible is as follows: 
\begin{equation}\label{ts0equalteh0}
t_s(0)=t_{\textrm{EH}}(0). 
\end{equation}
We note here that the singularity has to be a nodal point apart from satisfying the above equality. This is to make sure that an entire family of null geodesic escapes from the point $(t_s(0),0)$ in the $(t,R)$ plane so that the central singularity remains visible to an asymptotic observer for an infinite time 
\cite{Joshi_93(2)}. 
These escaping null geodesics are solutions of the differential equation (\ref{ngde}) starting from the point $(t_s(0),0)$ in the $(t,R)$ plane, and reaching the boundary $r_c$ before the event horizon, for which it has to satisfy the following inequality:
\begin{equation}
    F(r_c)<\frac{c^2}{G}R(t,r_c).
\end{equation}
We now use Eq.(\ref{ts0equalteh0}) to plot Fig.(\ref{astrophysicalregionplot}) on the basis of which we study the astrophysical relevance of globally visible singularity in the following section.

\section{Astrophysical relevance}\label{5.2}
It is evident that for a pressureless marginally bound collapsing cloud, with smoothly vanishing density at its boundary, ending up in a Tipler strong singularity, the global visibility or otherwise of the singularity depends on two parameters: the total mass $\frac{F_0}{4}$ and the initial size $r_c$ of the collapsing cloud at the beginning of its collapse (or equivalently $\frac{F_0}{4}$ and its initial mean density $\Bar{\rho_0}=\frac{3F_0}{8\pi r_{c}^{3}}$). This is because, in the case, $f=0$, the equality or otherwise of the singularity curve Eq.(\ref{tsr}) and the event horizon curve (which is a solution of the differential Eq.(\ref{ngde}) satisfying the condition Eq.(\ref{ngdeic})), which decides the global visibility of the singularity, depends on only $F_0$ and $r_c$. A similar situation may or may not occur in the case of a non-marginally bound collapse.  It is possible to show some examples of the mass profile, values of $F_0$, and $\Bar{\rho_0}$, which gives rise to the globally visible singularity as an end state. However, do such values arise in nature? Fig. (1) is a plot of $\textrm{log}_{10}F_0$ v/s $\textrm{log}_{10}\Bar{\rho_0}$ (for marginally bound case as well as non marginally bound case) in which the dotted line is the cut off for $\Bar{\rho_0}$ for a given $F_0$ having the property that any $\Bar{\rho_0}$ less than this cut off gives a globally visible singularity for a given $F_0$. In the case of non-marginally bound collapse (Fig(\ref{astrophysicalregionplot})), the velocity function is taken such that mass and size remain the only variables that control the global causal property of the singularity. Also, it is small enough to avoid error due to the approximation of Taylor expansion of Eq.(\ref{taylorexpansionofG}) up to only the first order. Here, we study three types of astrophysical scenarios giving rise to a singularity and investigate if they can be observed by an asymptotic observer. 

\subsection{The galactic center}
We investigate the causal structure of the central singularity of SgrA*, which is a compact object at the center of our galaxy. The mass of this object as calculated by 
\cite{Gillessen_09}
is $4.31 \times 10^6 M_{\odot}$, and its Schwarzschild radius is $\sim 1.26 \times 10^{10}$ m. For the SgrA* singularity to be globally visible, the collapsing spherical cloud forming such singularity should have the initial radius greater than equal to $\sim 3.33\times 10^{11}$ m (in case of a marginally bound case), which translates to roughly at least $26.42$ times the size of its Schwarzschild radius. A lesser initial radius will make the singularity globally hidden. This corresponds to having the initial mean density less than $55.42 \textrm{kg}/\textrm{m}^3$. Fig.(2) is the visual representation of both the possible outcomes.

Similarly, the compact object at the center of the M87 galaxy has mass $6.6 \times 10^9 M_{\odot}$ as calculated by the Event Horizon Telescope collaboration
\cite{Akiyama_19, Akiyama_19(2)},
and a Schwarzschild radius $\sim 1.9 \times 10^{13}$ m. For the central singularity to be globally visible, the collapsing spherical cloud forming such singularity should have the initial radius greater than or equal to $\sim 5.14\times 10^{14}$ m (in case of marginally bound collapse), which roughly translates to at least $27.05$ times its Schwarzschild radius. A lesser initial radius will make the singularity globally hidden. This corresponds to having the initial mean density less than $2.31 \times 10^{-5} \textrm{kg}/\textrm{m}^3$. Fig.(3) is the visual representation of both the possible outcomes.
%%%%%%%%%%%%%%%%%%%%%%%%%%%%%%%%%%%%%%%%%%%%%%%%%%%%%%%%%%%%%%%%%%%%%%%%%%%%%%%%%%%%%
%%%%%%%%%%%%%%%%%%%%%%%%%%%%%%%%%%%%%%%%%%%%%%%%%%%%%%%%%%%%%%%%%%%%%%%%%%%%%%%%%%%%%%%%%%%%%%%%%%%%
\begin{figure}
{\includegraphics[scale=0.5]{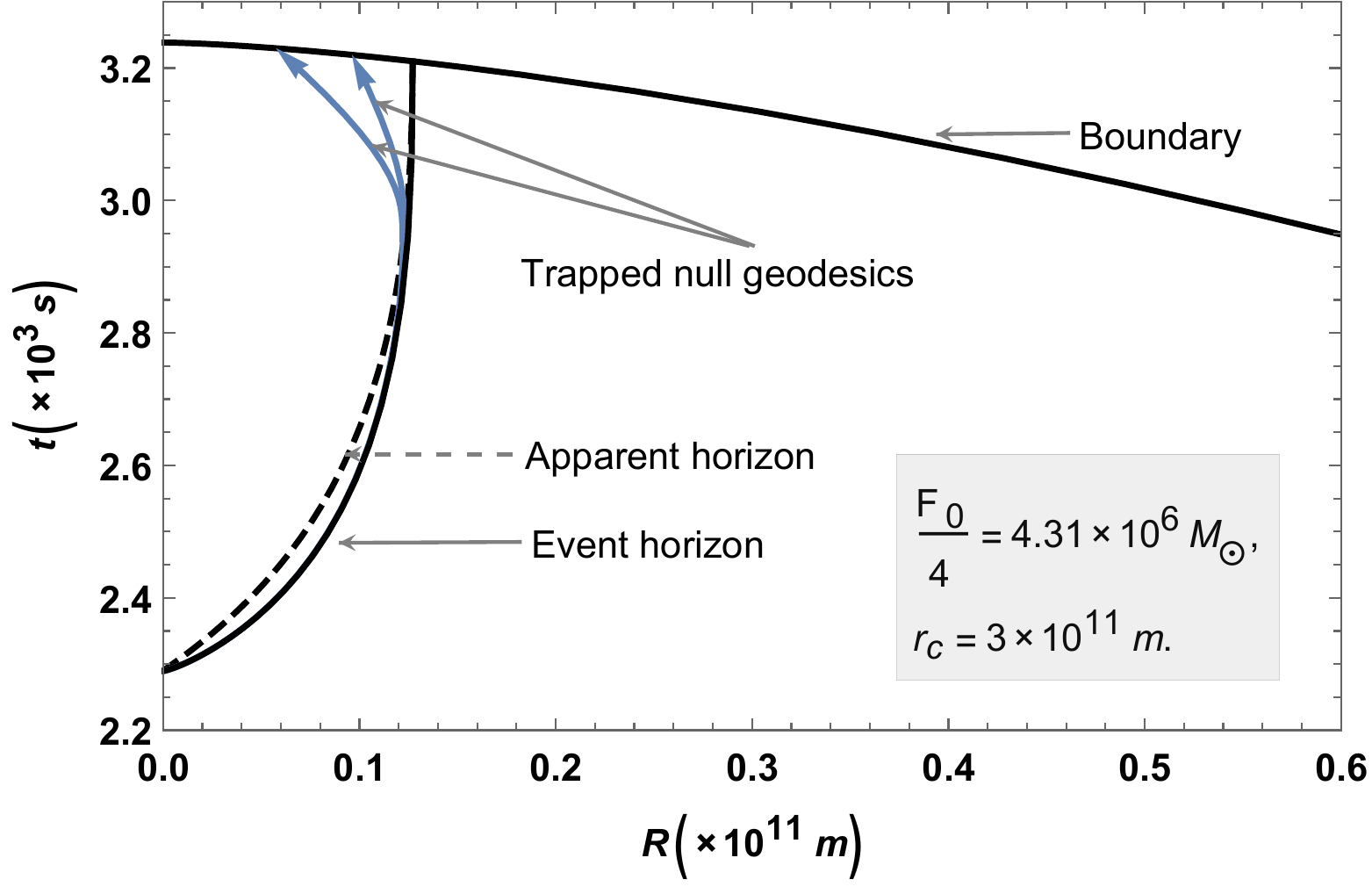}
\includegraphics[scale=0.5]{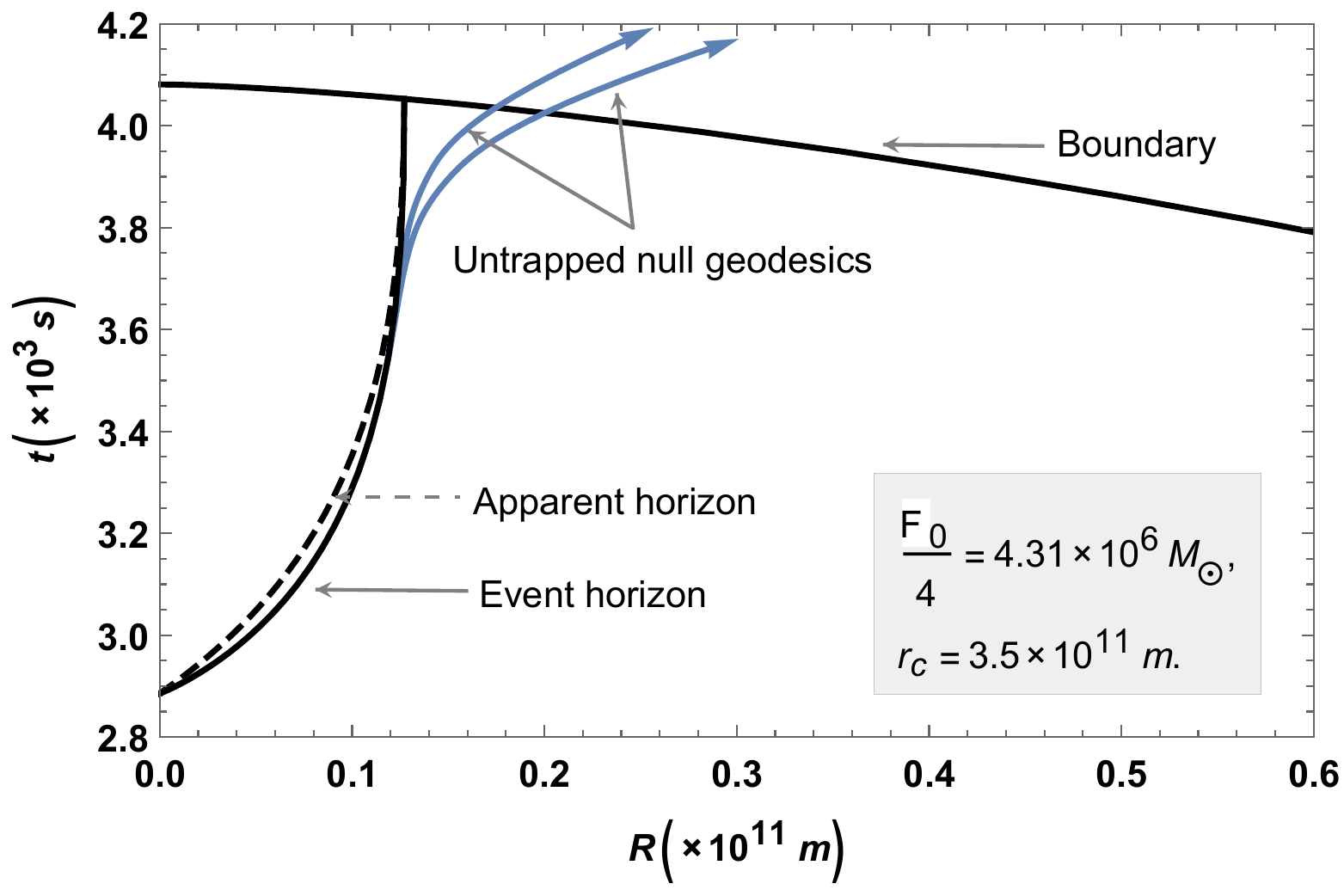}}
\caption{Causal structure of space-time singularity formed at the center of the milky way galaxy (the compact object SgrA*). The initial unknown radius of the collapsing core forming such singularity decides its end state. $r_c\leq 3.32\times 10^{11} \textrm{m}$ and $r_c \geq 3.33\times 10^{11} \textrm{m}$ ends up globally hidden and a globally visible singularity respectively. Collapsing core is assumed to be spherical, pressureless and marginally bound. Mass function is chosen such that the singularity is Tipler strong.}
\label{sgrA}
\end{figure}
%%%%%%%%%%%%%%%%%%%%%%%%%%%%%%%%%%%%%%%%%%%%%%%%%%%%%%%%%%%%%%%%%%%%%%%%%%%%%%%%%%%%%%%%%%%%%%%%%%%
%%%%%%%%%%%%%%%%%%%%%%%%%%%%%%%%%%%%%%%%%%%%%%%%%%%%%%%%%%%%%%%%%%%%%%%%%%%%%%%%%%%%%%%%%%%%%%%%%%%%%
\begin{figure}
{\includegraphics[scale=0.5]{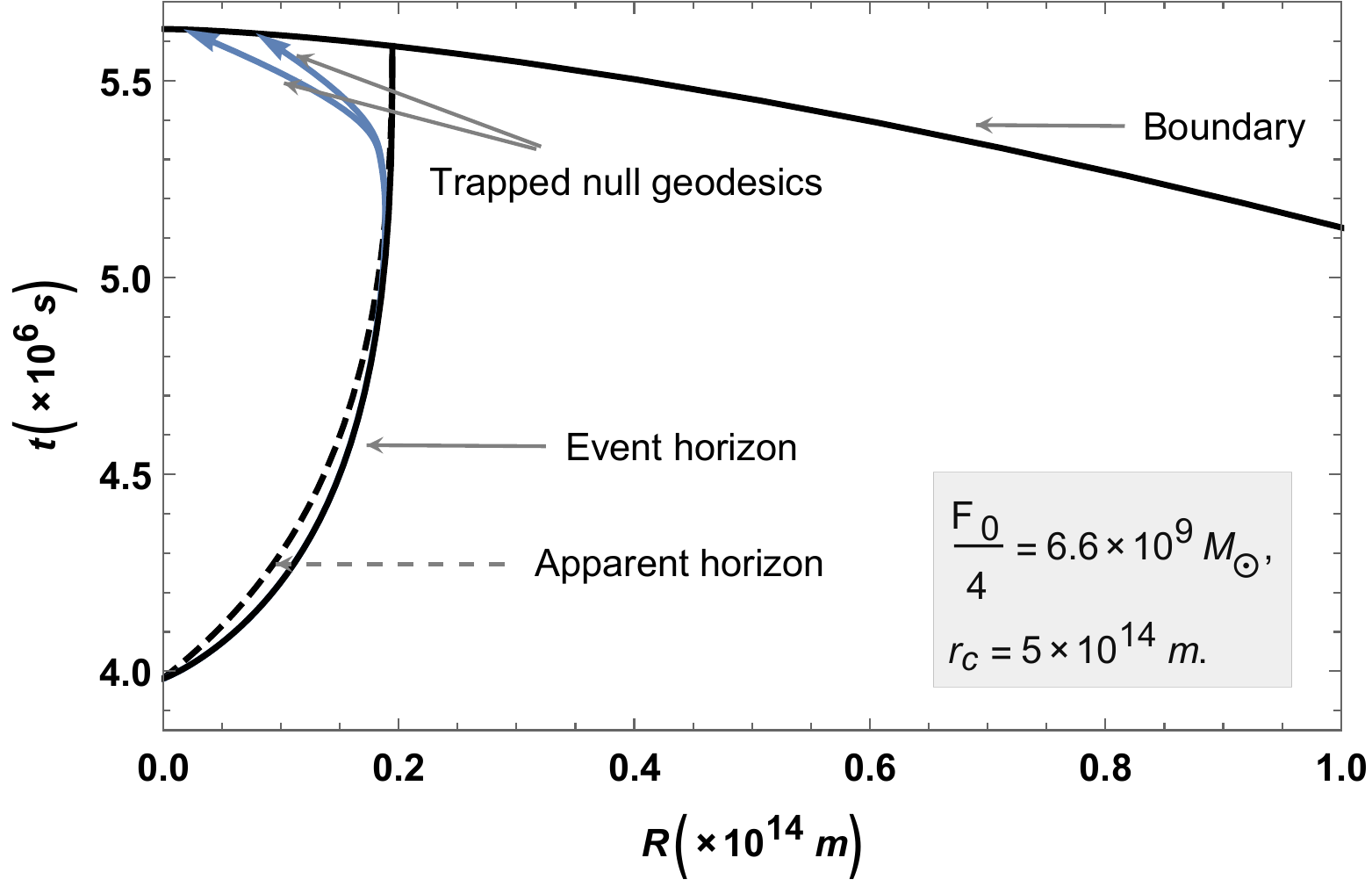}\includegraphics[scale=0.5]{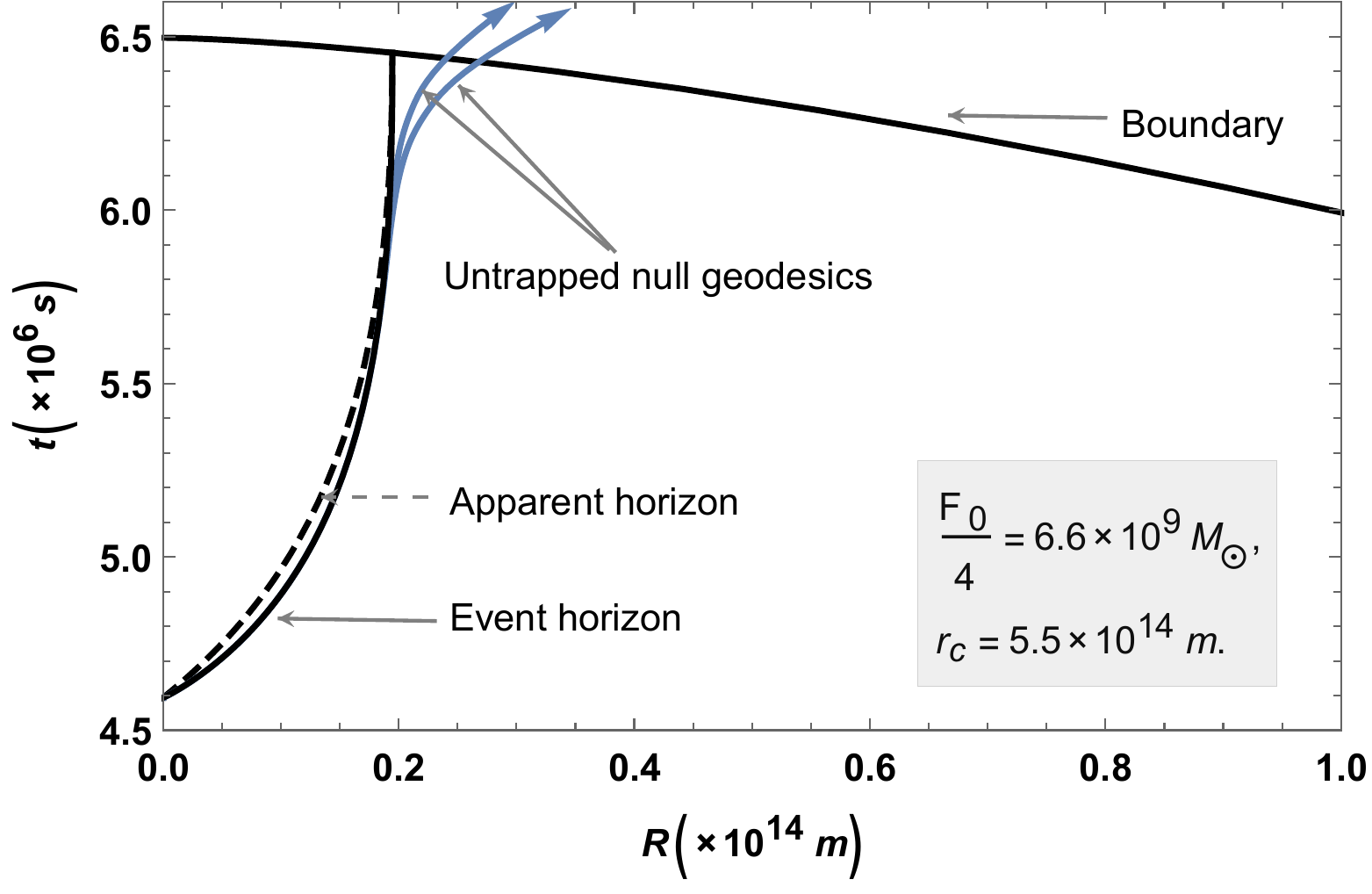}}
\caption{Causal structure of space-time singularity formed at the center of the M87 galaxy. The initial unknown radius of the collapsing core forming such singularity decides its end state. $r_c\leq 5.13\times 10^{14} \textrm{m}$ and $r_c \geq 5.14\times 10^{14} \textrm{m}$ ends up in a globally hidden and a globally visible singularity respectively. Collapsing core is assumed to be spherical, pressureless and marginally bound. Mass function is chosen such that the singularity is Tipler strong.}
\label{M87}
\end{figure}
%%%%%%%%%%%%%%%%%%%%%%%%%%%%%%%%%%%%%%%%%%%%%%%%%%%%%%%%%%%%%%%%%%%%%%%%%%%%%%%%%%%%%%%%%%%%%%%%%%%
\subsection{Singularities having primordial origin}
Another astrophysically motivated entity that we consider here is the primordial singularities. Primordial fluctuations in the very early universe could give rise to such singularities
\cite{Zeldovich_67, Hawking_71}. 
For a primordial singularity to form due to gravitational collapse, we investigate the collapse of matter in the era just after the matter-radiation equality, i.e., the start of the matter-dominated era. The reason for considering this time as the initiation of the collapse is that the temperature during this epoch drops down to $9000$ K, and the matter configuration can now collapse. We know that the epoch of matter-radiation equality occurred at redshift $z\sim 3000$
\cite{Frieman_08}. 
For a $\Lambda\textrm{CDM}$ model, the deceleration to acceleration transition of the universe is calculated to begin at $z\sim 0.72$
\cite{Farooq_17}. 
The relation between the density of the universe at these epochs depends on $z$ at both these epochs as follows:
\begin{equation}\label{rhoz}
   \frac{\rho_{\textrm{MRE}}}{\rho_{\textrm{DA}}}=\left (\frac{1+z_{\textrm{MRE}}}{1+z_{\textrm{DA}}}\right )^{3}. 
\end{equation}
This is because the era bounded below and above by the time of matter-radiation equality and the time of deceleration to acceleration transition, respectively, has matter as the dominant fluid, and the density of the universe is dependent on the linear equation of state parameter $\omega$ of the dominating fluid, as follows
\cite{Liddle_03}:
\begin{equation}
    \rho(\omega) \propto \left(1+z\right)^{3(1+\omega)}.
\end{equation}
Hence for $\omega=0$ (dust domination), Eq.(\ref{rhoz}) holds. If we consider the dark energy candidate driving the present accelerated expansion as the cosmological constant ($\omega=-1$), its density $\rho_{\Lambda}$ remains constant in time. Therefore, we can say that
\begin{equation}\label{rhodalambda}
    \rho_{\textrm{DA}}=\rho_{\Lambda}.
\end{equation}
 Using the above equality in Eq.(\ref{rhoz}), one can obtain the density of the universe at the time of matter-radiation equality as
 \begin{equation}
    \rho_{\textrm{MRE}}= 3.092 \times 10^{-17} \textrm{kg}/\textrm{m}^3.
 \end{equation}
Here, we have used the present Hubble parameter, $H_0=67.4 \textrm{km s}^{-1}\textrm{Mpc}^{-1}$, and the density parameter of $\Lambda$, $\Omega_{\Lambda}=0.686$, from the Planck 2018 results
\cite{Planck_18}
to get the value of $\rho_{\Lambda}= 5.82298 \times 10^{-27} \textrm{kg}/\textrm{m}^3$ using the relation $\rho_{\Lambda}=\frac{3H_0^2\Omega_{\Lambda}}{8\pi G}$
\cite{Liddle_03}.
The overdensity configuration, formed due to primordial fluctuation just after the epoch of matter-radiation-equality, detaches itself from the background universe and starts contracting if its density contrast $\delta \rho/ \rho$ at the Hubble horizon crossing time  is between zero and one
\cite{Carr_75, Khlopov_80, Harada_13}. 
Hence, at the start of the collapse, one can assume the order of the initial mean density of the collapsing configuration, which we denote by $\rho_{\textrm{CONFIG}}$, same as the order of the density of the background universe, i.e. $10^{-17} \, \textrm{kg}/\textrm{m}^3$. Therefore, we have 
\begin{equation}
    \rho_{\textrm{CONFIG}}< \Bar{\rho_0} \left (\frac{F_0}{4} \right )
\end{equation}
for all configuration having $\frac{F_0}{4}< 10^{45} \,\textrm{kg}$ as is evident from Fig.(1) in both marginally bound as well as non-marginally bound case for very small velocity function. Hence, we can state that any configuration with $0<\frac{\delta \rho}{\rho}<1$, and having mass less than $10^{45}\, \textrm{kg}$ collapses to give a globally visible singularity if the collapse initiates just after the time $t_{\textrm{MRE}}$. 
%It is worth mentioning that the overdense region collapses even if the density contrast is of the order greater than one, in which case, the result may vary accordingly. 

\subsection{Collapse of a matter accreting neutron star}
%Typically, a star, after exhausting its nuclear fuel, undergoes a supernova explosion, and the central core collapses to form either a white dwarf, a neutron star, or an infinitely dense singularity, based on its initial mass. 
We now consider a close binary system scenario, one of which is a star undergoing supernova explosion and another is a neutron star, as investigated by 
\cite{Rueda_12}.
We give a brief overview of this scenario.

The material expelled from the core progenitor is accreted by the companion neutron star and can get captured if it falls in the region at a distance less than
\begin{equation}
    R_{cap}=\frac{2GM_{NS}}{v_{rel}^{2}}
\end{equation} from the center of the neutron star. Here, $M_{NS}$ is the mass of the neutron star, and $v_{rel}$ is the velocity of the ejected particle relative to the neutron star orbital motion, and is given by 
\begin{equation}
    v_{rel}=\sqrt{v^{2}_{orb}+v^{2}_{ej}},
\end{equation} 
where 
\begin{equation}
v_{orb}=\sqrt{\frac{G(M_{prog}+M_{NS})}{a}}.
\end{equation}
Here, $v_{ej}$ is the ejecta velocity, which at the start of the supernova explosion has a value of the order of $10^7$ $m s^{-1}$, $v_{orb}$ is the orbit velocity of the neutron star, $M_{prog}$ is the mass of the progenitor, and $a$ is the binary separation. We neglect the neutron star magnetic field's effect by considering the magnetospheric radius $R_{m}$ comparatively smaller than $R_{cap}$. This is possible for a suitably high rate of accretion since 
\begin{equation}
    R_{m}=\frac{B^2R_{NS}^6}{(\dot M \sqrt{2G M_{NS}})^{\frac{2}{7}}},
\end{equation}
where $\dot M=\frac{dM}{dt}$ is the matter accretion rate of the supernova ejecta by the neutron star, and $B$, $M_{NS}$ and $R_{NS}$ are the magnetic field, mass and the radius of the neutron star respectively 
\cite{Toropina_11}. 
The mass accretion rate is given by 
\cite{Bondi_44} 
\begin{equation}\label{accretionrate}
    \dot M= \epsilon \pi \rho_{ej} R^{2}_{cap}v_{rel}=\epsilon \pi \rho_{ej}\frac{(2GM_{NS})^2}{v^{3}_{rel}}.
\end{equation}
Here, $0<\epsilon<1$, and depends on the medium in which the accretion process takes place. $\rho_{ej}(t)=\frac{3M_{ej}(t)}{4 \pi r_{ej}^{3}(t)}$ is the density of the ejected material which decreases with time. The supernova ejecta radius can be assumed to expand as $r_{ej}=a t^b$, where $a$ and $b$ are constants 
\cite{Chavalier_89}. 
Eq.(\ref{accretionrate}) can now be integrated to get the magnitude of the mass accreted by the neutron star. After some time, provided the accretion rate is high,in order that the effect of the magnetic field is negligible, the mass of the neutron star reaches the critical mass and further collapses unhindered. Neutron star stable configuations have been studied taking into account weak, strong, electromagnetic and gravitational interaction in the framework of general relativity 
\cite{Belvedere_12}. 
Tab.(\ref{tab2}) mentions the critical mass $M_{crit}$ and its corresponding radii for some parameterizations of neutron star models studied here. 
If we model the unhindered collapse of the neutron star after achieving the critical masses mentioned in Tab.(\ref{tab2}), by a spherical pressureless marginally bound cloud collapse, having the mass function of the form  of Eq.(\ref{msmf5}), then each model (i.e. $NL3$ 
\cite{Lalazissis_97}, 
$NL-SH$ 
\cite{Sharma_93}, 
$TM1$ 
\cite{Sugahara_93}, 
and $TM2$ 
\cite{Hirata_95}) 
ends up in a singularity which is hidden globally. This is because the event horizon forms before the singularity at $r=0$ in each case as seen using equations (\ref{tsr}), (\ref{ngde}) and (\ref{ngdeic}). However, the singularity is visible locally, as explained in the Appendix (\ref{AppendixA}).

\begin{table}
	\centering
	\begin{tabular}{lcccr} % four columns, alignment for each
		\hline
		Model  & $M_{crit}$ ($\times M_{\odot}$) &  $R$ ($km$) & $t_s(0)$ $(\times 10^{-5} s)$ & $t_{EH}(0)$ $(\times 10^{-5} s)$\\
		\hline
		NL3     & $2.67$  & $12.33$ & $3.42937$  &  $0.83442$  \\

   NL-SH       & $2.68$   & $12.54$ & $3.51078$ & $0.91288$  \\

   TM1    & $2.58$  & $12.31$ & $3.48018$ & $0.98919$   \\

   TM2     & $2.82$  &  $13.28$ & $3.72990$ & $1.00047$ \\

		\hline
	\end{tabular}
	\caption{Neutron star critical masses and the corresponding radii are obtained by fixing certain nuclear parameters, which includes the coupling constants and the meson masses \cite{Belvedere_12}. Each of these four models corresponds to one such fixed-parameter set. For each model, in the case of marginally bound unhindered gravitational collapse, the event horizon forms before the formation of the singularity at $r=0$. Hence the singularity thus formed is hidden globally.}
	\label{tab2}
\end{table}
\begin{figure}\label{fig4}
\centering
\includegraphics[scale=0.5]{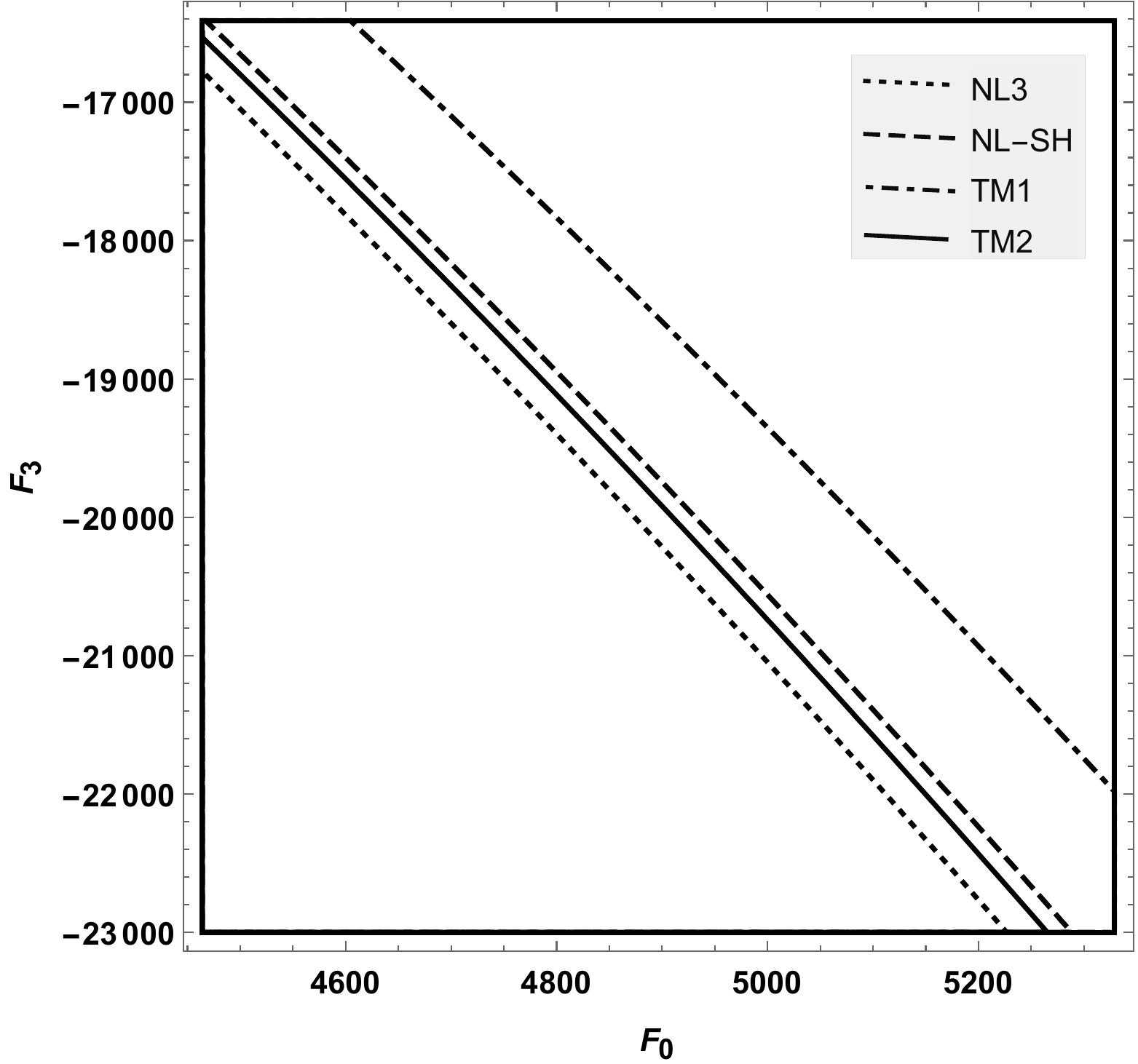}
\caption{Neutron star critical masses and the corresponding radii are obtained by fixing certain nuclear parameters which includes the coupling constants and the meson masses \cite{Belvedere_12}. Each of the four models: $NL3$, $NL-SH$, $TM1$, and $TM2$ corresponds to one such fixed parameter set. The Misner-Sharp mass function is chosen as $F(r,R)=F_0\left(\frac{r}{r_c}\right)^3+F_3\left(\frac{r}{r_c}\right)^6+F_R\left(\frac{R}{r_c}\right)^3$. We have considered a marginally bound collapse, i.e. $\mathcal{Y}=1$. The total initial mass, i.e. the critical mass of the collapsing cloud is given by $M_{crit}=(F_0+F_3+F_R)/2$. The magnitudes of $F_0$ and $F_3$ depicted in the plot are considerably smaller than that of $F_R$. Hence, $M_{crit}$ is approximately $F_R/2$. The region of parameter space $(F_0,F_3)$ below (above) the curve gives a globally visible (hidden) singularity at the end of the collapse of a neutron star after reaching the critical mass by accreting the supernova ejecta from its binary companion core progenitor.}
\end{figure}
%%%%%%%%%%%%%%%%%%%%%%%%%%%%%%%%%%%%%%%%%%%%%%%%%%%%%%%%%%%%%%%%%%%%%%%%%%%%%%%%%%%%%%%%%%%%%%%%%%%
One could argue that modeling the cloud formed due to matter accreting neutron star by zero pressure is not justified since neutron star has non-zero pressure, following a certain equation of state. To investigate the end state of the collapsing cloud formed after the neutron star accretes the supernova ejecta from its binary companion core progenitor, we rely on the work by 
\cite{Giambo_03, Giambo_06}. 
We give a brief overview of the formalism presented in these articles.

The general spacetime metric of a spherical collapsing cloud in the comoving coordinates is given by
\begin{equation}
    ds^2=-c^2\frac{\dot R^2}{\mathcal{H}}dt^2+\frac{R'^2}{\mathcal{Y}}dr^2+R^2d\Omega^2,
\end{equation}
where $R$, $\mathcal{Y}$ and $\mathcal{H}$ are functions of $t$ and $r$. The above metric can be rewritten in the transformed comoving area-radial coordinates $(r,R)$ as
\begin{equation}
     ds^2=-Adr^2-2BdRdr-\frac{1}{\mathcal{H}}dR^2+R^2d\Omega^2,
\end{equation}
where $A$ and $B$ are functions of $r$ and $R$. $\mathcal{H}$ can also be written as a function of $r$ and $R$ as
\begin{equation}
    \mathcal{H}(r,R)=\frac{F(r,R)}{R}+\mathcal{Y}(r,R)^2-1.
\end{equation}
The transformed metric is advantageous to obtain information about the global causal structure of the singularity.
A quantity $\Delta(r,R)$ is defined as
\begin{equation}
\Delta=B^2-\frac{A}{\mathcal{H}}= \frac{R'^2}{\mathcal{Y} \mathcal{H}}.
\end{equation}
Using the Einsteins' field equation, one can get an integral expression of $\sqrt{\Delta}$ as
\begin{equation}
    \sqrt{\Delta}=\int_R^r \frac{1}{\sqrt{\mathcal{Y}(r,R)}}\frac{\partial}{\partial r}\left(\frac{1}{\mathcal{H}(r,R)}\right)dR+\frac{1}{\sqrt{\mathcal{Y}(r,r)\mathcal{H}(r,r)}}.
\end{equation}
The formalism is restricted to only those cases where the energy density $\rho$ and matter density $\rho_m$ are related as 
\begin{equation}
    \rho=\omega \rho_m c^2,
\end{equation}
where 
\begin{equation}
    \omega=\frac{E(r)}{2}\left(\frac{F_r(r,R)}{\sqrt{\mathcal{Y}(r,R)}}+\frac{F_R(r,R)R'(r,R)}{\sqrt{\mathcal{Y}(r,R)}} \right).
\end{equation}
The energy density, the radial pressure, and the tangential pressure are respectively given by
\begin{equation}
\rho=\frac{c^2}{8\pi R^2}\left(\frac{F,_r}{R'}+F,_R\right),
\end{equation}
\begin{equation}\label{prgiambo}
p_r=-\frac{c^2 F,_R}{8\pi R^2},
\end{equation}
and
\begin{equation}
p_t=-\frac{c^2}{16\pi R R'}\left(F,_{rR}-\sqrt{\mathcal{Y}} F,_r\frac{\partial}{\partial R}\left(\sqrt{\mathcal{Y}}\right)+F,_{RR}R'\right).
\end{equation}
In a collapsing cloud, provided: $(1)$ the weak energy condition hold, $(2)$ the regularity conditions hold, $(3)$ the shell crossing singularity do not occur, and $(4)$ the shell focusing singularity form in a finite comoving time, the singularity formed as an end state of the gravitational collapse is globally visible if $n=3$ and $\frac{\zeta}{\alpha}>\frac{26+15\sqrt{3}}{2}$, where $\zeta$ and $n$ arise in the Taylor expansion of $\sqrt{\Delta}(r,0)$ around $r=0$ as
\begin{equation}
    \sqrt{\Delta}(r,0)=\zeta r^{n-1}+.....,
\end{equation}
and $2\alpha$ is the coefficient of $r^3$ in the expression of the Misner-Sharp mass function $F$ in the $(r,R)$ coordinates. One can choose a suitable $\mathcal{Y}$ and $F$, due to two degrees of freedom available to us, such that the end state of a non-zero pressured collapsing cloud, satisfying the weak energy condition and the regularity condition, is globally visible. Let us consider the Misner-Sharp mass function of the form
\begin{equation}
F(r,R)=F_0\left(\frac{r}{r_c}\right)^3+F_3\left(\frac{r}{r_c}\right)^6+F_R\left(\frac{R}{r_c}\right)^3.
\end{equation}
We set $\mathcal{Y}=1$, which corresponds to an acceleration-free marginally bound case. The total mass of the collapsing cloud is $(F_0+F_3+F_R)/2$.

It should be noted that for the collapsing fluid to satisfy the weak energy condition, one of the necessary criteria is to have the mass function to be a monotonically increasing function of the areal coordinate $R$ in the $(r,R)$ frame. Hence, one realizes that the formalism discussed in 
\cite{Giambo_03, Giambo_06} 
is restricted to studying the end state of only those collapsing fluid with negative radial pressure, as seen in Eq.(\ref{prgiambo}). It can be seen that for suitable values taken by the components of the Misner-Sharp mass function, one gets a globally visible end-state singularity.

\section{Concluding remarks}\label{5.3}
Some concluding points and concerns are discussed below:
\begin{enumerate}

\item A non-zero measured set of initial parameters, which relate to the total mass and the initial mean density of the collapsing cloud, is obtained, which leads to a globally visible singularity as its end state. The Misner-Sharp mass function is chosen such that the singularity formed is physically strong in the sense of Tipler. Hence, it can be concluded that such globally visible singularity is stable under small perturbations in the initial data ($\frac{F_0}{4}$ and $\Bar{\rho_0}$) of the collapsing spherical cloud. 

\item Globally visible singularity is possible to achieve in principle; however, here, we have discussed the possibility of its existence in an astrophysical scenario by looking into the possibility of occurrence of suitable mass and size required for its existence.  The singularity at the center of the M87 galaxy may be globally visible if the initial radius of the collapsing cloud forming it is more than $27.05$ times its Schwarzschild radius. Similarly, the collapsing cloud's initial radius forming the SgrA* singularity should be more than $26.42$ times its Schwarzschild radius to end up being globally visible. Here, we have modeled the collapsing cloud to be marginally bound, spherically symmetric, and pressureless. The outcome of the causal structure of the singularity may vary if these assumptions are dropped.

\item If a density contrast $\frac{\delta \rho}{\rho}$ of order one is achieved in the matter-dominated era just after the time of matter-radiation equality, the configuration detaches from the background universe and starts contracting due to gravity. A configuration having a total mass of the order less than $10^{45}$ kg forms a physically strong (in the sense of Tipler) singularity, which can be visible globally. It should be noted that we have not taken into account the process of virialization, which is an essential phenomenon in cosmology, and which could oppose the unhindered gravitational collapse of the configuration during the primordial time, causing it to attain an equilibrium, thereby avoiding the formation of the singularity at all.

\item A neutron star, which is a part of the binary system, can reach a critical mass by accreting the supernova ejecta of the companion exploding star, after which it can collapse unhindered. We consider four such models having different critical masses and their corresponding radii. Each of these models is obtained by fixing six nuclear parameters, as discussed in 
\cite{Belvedere_12}. 
The final state of such accreting neutron star is found to be a singularity that is globally hidden for all these models. One may claim that these act as positive evidence for the hypothesis of weak cosmic censorship. However, a loophole in this argument, which is of concern, is that the collapse formalism used for our purpose assumes zero pressure; however, neutron stars may have non-zero pressure satisfying the polytropic equation of state. To address this problem, we have incorporated the formalism proposed by Giambo \textit{et. al.} 
\cite{Giambo_03, Giambo_06}. 
However, the formalism works only for negative radial pressure if one has to satisfy the weak energy condition. One could obtain a suitable mass function $F(r,R)$ and the function $\mathcal{Y}(r,R)$ for the collapse of supernova ejecta accreting neutron star after achieving the critical mass and obtain a globally naked singularity as the end state of the collapse, as seen in Fig.(4). This end state is stable under small perturbation in the initial data (here $F_0$ and $F_3$).  The formalism developed in 
\cite{Giambo_03, Giambo_06} 
works only for the subset of configurations that have negative radial pressure. The formalism of finding out the global causal structure of the singularity formed due to gravitational collapse is not developed enough to incorporate arbitrary pressures and equation of state as demanded by the neutron star's complicated structure. A simplified model giving rise to both the possibilities: hidden and visible, globally, indicates that even a more realistic model may give rise to a globally visible singularity.

\item An external observer encountering an escaping singular null geodesic may find traces of quantum gravity encoded in it, making such singularities more tempting to investigate. We have shown here that the formation of a globally visible singularity may also arise in the astrophysical setup and is not merely a mathematical artifact. One could therefore interpret that such singularities, after all, may not be so elusive.  

\end{enumerate}
 
% Chapter Template

\chapter{Globally visible singularity in modified gravity theories} % Main chapter title

\label{Chapter6} % Change X to a consecutive number; for referencing this chapter elsewhere, use \ref{ChapterX}

\lhead{\emph{Chapter 6.}} % Change X to a consecutive number; this is for the header on each page - perhaps a shortened title

%----------------------------------------------------------------------------------------
%	SECTION 1
%----------------------------------------------------------------------------------------
As discussed in the Introduction, in the strong gravity regime, higher-order curvature terms in the action whose corresponding equation of motion is the field equation, if present, will dominate, and hence it may become essential to incorporate the modified action while investigating the neighborhood of the singularity formed due to gravitational collapse. Whether higher-order terms (or any change in the Lagrangian) has any effect on the causal property of the singularity is our concern which we address in this chapter. 

For a naked singularity to be visible for infinite time, the singularity should be a nodal point. A congruence of infinite null geodesics should escape the singularity for it to be a nodal point. We say that a singularity is a nodal point if more than one null geodesic solution escapes from the singularity. It is desirable to have infinite null geodesic solutions passing through the singularity, each differently redshifted, with the event horizon being infinitely redshifted.   We also address in this chapter, whether the first central singularity is a nodal point. 

The chapter is organized as follows: In sec.(\ref{6.1}), we give a brief overview of the Lemaitre-Tolman-Bondi (LTB) spacetime metric
\cite{Lemaitre_33, Tolman_34, Bondi_47} as a solution of the modified gravity, where the Lagrangian is $f(R)=R+\alpha R^2$. Here $R$ is the Ricci scalar, and $\alpha$ is a positive constant. We ensure that the matter field governed by the LTB metric satisfies the strong, weak, null, and dominant energy conditions. The first central singularity is then investigated for its causal property. Comparisons are made between the visible singularity formed due to two different matter fields, i.e., dust, and viscous fluid with heat flow, respectively in GR and $f(R)$ ), governed by the same LTB metric. However, the matching surface of the interior LTB metric and the exterior metric is different in different frameworks of gravity theory. In sec.(\ref{6.2}), we discuss the junction condition on the matching surface of the interior and the exterior spacetime in this modified gravity. In sec.(\ref{6.3}), we give a heuristic method to show that globally naked singularity formed is a nodal point by considering the above class of theory of gravity, of which general relativity is a particular case. Finally, we end the chapter with the results and the conclusions drawn after that, in sec.(\ref{6.4}).

\section{LTB spacetime in modified gravity}\label{6.1}
In general relativity, the collapse of a spherically symmetric dust cloud is governed by the Laimetre-Tolman-Bondi metric as follows:
\begin{equation}\label{LTBmg}
    ds^2=-dt^2+\frac{A'^2}{1+b}dr^2+A^2d\Omega^2, 
\end{equation}
in the comoving coordinate $t$ and $r$. Here, $A(t,r)$ is the physical radius of the collapsing cloud and is a monotone decreasing function of $t$, i.e., $\dot A<0$, $b(r)$ is called the velocity function and incorporates the information about the initial velocity of the collapsing cloud. The superscripts prime and dot, as usual, denote the partial derivative with respect to radial and time coordinate, respectively. We now rewrite some equations of LTB spacetime mentioned in sec.(\ref{3.1}) to avoid confusion caused due to usage of different notations.

Using the Einsteins' field equation, one can get the expression of the density as
\begin{equation} \label{fr1efegr}
\rho=\frac{S'}{A^2A'},
\end{equation}
where 
\begin{equation} \label{frS}
  S=A\left(\dot A^2-b \right).
\end{equation}
$S$ is the Misner-Sharp mass function
\cite{Misner_1969}. 
$S/2$ gives the mass of the cloud inside the shell of radial coordinate $r$ at time $t$. Since the pressure inside the cloud is zero, the second Einsteins equation gives us
\begin{equation} \label{fr2efegr}
    -\frac{\dot S}{A^2\dot A}=0,
\end{equation}
from which we can conclude that $S$ is independent of $t$. Hereafter, we will consider the case for which $b=0$. This corresponds to a marginally bound collapse. The mathematical difficulties by taking such case are drastically reduced, although the case of non-marginally bound collapse ($b\neq 0$) can give results which are not very different qualitatively 
\cite{Mosani_20(2)}. 
On integrating Eq.(\ref{frS}), we obtain 
\begin{equation}\label{frt-ts}
    t-t_s(r)=-\frac{2}{3}\frac{A^{\frac{3}{2}}}{\sqrt{S}}.
\end{equation}
Here $t_s(r)$ is the singularity curve. Rescaling the physical radius using the coordinate freedom such that initially $A(0,r)=r$, one express the singularity curve as 
\begin{equation}\label{frts}
    t_s(r)=\frac{2}{3}\frac{r^{\frac{3}{2}}}{\sqrt{S}}.
\end{equation}

The evolution of the apparent horizon, represented by the apparent horizon curve is obtained as
\begin{equation}\label{frtah}
    t_{AH}(r)=\frac{2}{3}\frac{r^{\frac{3}{2}}}{\sqrt{S}}-\frac{2}{3}S
\end{equation}
for the marginally bound collapse.
The dynamics of the EH is the solution of the null geodesic differential equation
\begin{equation}\label{frdeteh}
 \frac{dt_{EH}(r)}{dr}=A',   
\end{equation}
satisfying the condition 
\begin{equation}\label{frteh}
t_{EH}(r_c)=\frac{2}{3}\frac{{r_c}^{\frac{3}{2}}}{\sqrt{S(r_c)}}-\frac{2}{3}S(r_c),  
\end{equation}
which is obtained from Eq.(\ref{frtah}) (Recall that at the boundary of the collapsing cloud, the apparent horizon and the event horizon coincide. Refer sec.(\ref{3.1})). The necessary condition for the singularity to be globally visible is that at $r=0$, $t_{EH}=t_s$. 

For the theory of gravity where the generalized Lagrangian in the Einstein-Hilbert action is $f(R)=R+\alpha R^2$, the  collapsing cloud governed by the  LTB metric, as shown in Eq.(\ref{LTBmg}) is a viscous fluid with heat flow, unlike in GR, as seen in the equation:
\begin{equation}
    T^{\mu \nu}= \rho U^{\mu}U^{\nu}+ph^{\mu \nu}+2q^{(\mu}n^{\nu)}-\Pi \left(n^{\mu}n^{\nu}-\frac{1}{3}h^{\mu \nu}\right).
\end{equation}
Here,
\begin{itemize}
    \item $h_{\mu \nu}$ is the transverse metric (refer Eq.(\ref{transversemetric})), expressed as
\begin{equation}
    h_{\mu \nu}=U_{\mu}U_{\nu}+g_{\mu \nu},
\end{equation}

\item $n^{\mu}$ is a spatial unit vector in the radial direction satisfying 
\begin{equation}
n_{\mu}n^{\mu}=1 \hspace{1cm} \textrm{and} \hspace{1cm} n_{\mu}U^{\mu}=0.
\end{equation}
\item $q^{\mu}$ is the heat flux vector and is spacelike, i.e. 
\begin{equation}
    q_{\mu}U^{\mu}=0.
\end{equation}
It describes the heat conduction such that $q_{\mu}n^{\mu}$ is the heat, crossing a unit surface which is perpendicular to $n^{\mu}$, per unit time. 
\item $p$ is the effective pressure given by
\begin{equation}
    p=\frac{p_r+2p_t}{3},
\end{equation}
where $p_r$ and $p_t$ are the radial and tangential components of the pressure inside the collapsing cloud.
\item $\Pi$ measures the anisotropy in the pressure given by
\begin{equation}
    \Pi=p_t-p_r.
\end{equation}
\end{itemize}
%%%%%%%%%%%%%%%%%%%%%%%%%%%%%%%%%%%%%%%%%%%%%%%%%%%%%%%%%%%%%%%%%%%%%%%%%%%%%%%%%%%%%%%%%%%%%%%%%%%%
\begin{figure}
{\includegraphics[scale=0.65]{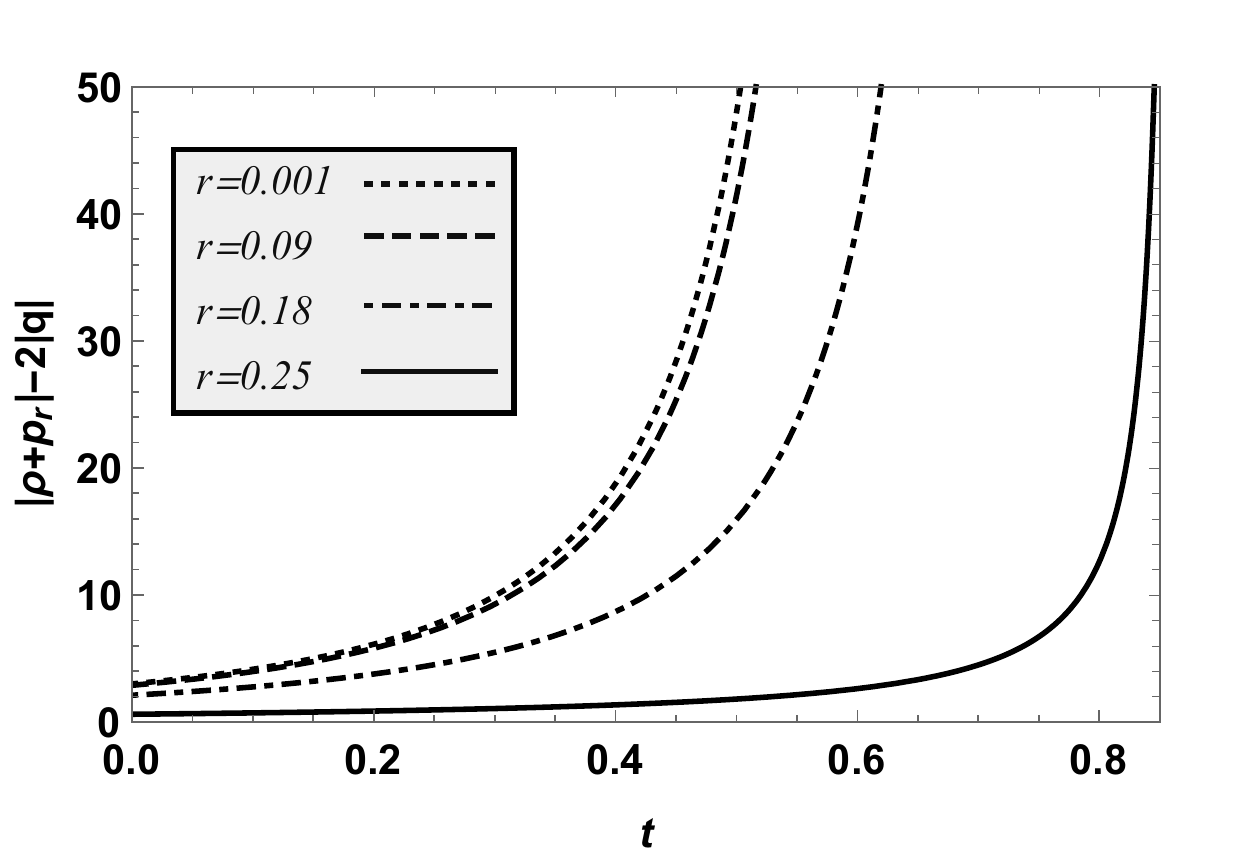}\includegraphics[scale=0.65]{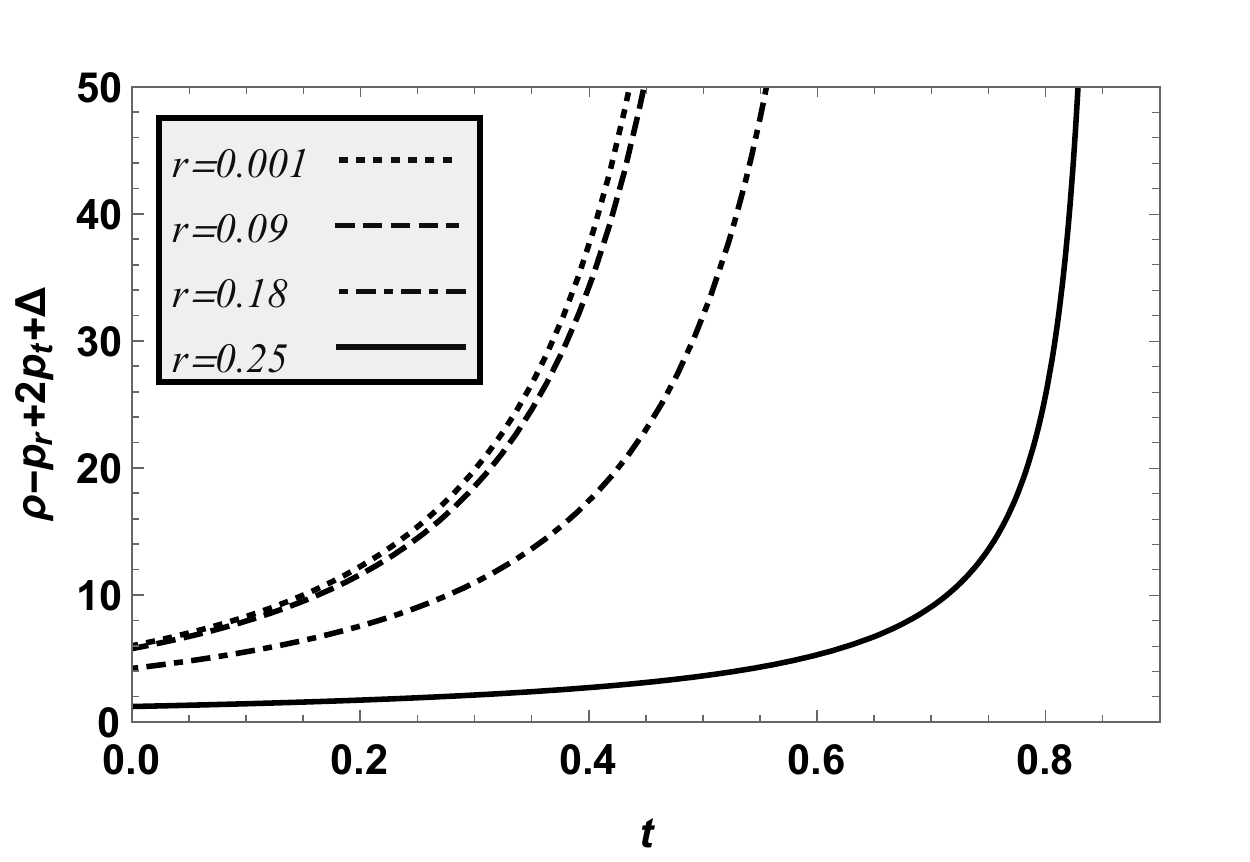}
\includegraphics[scale=0.65]{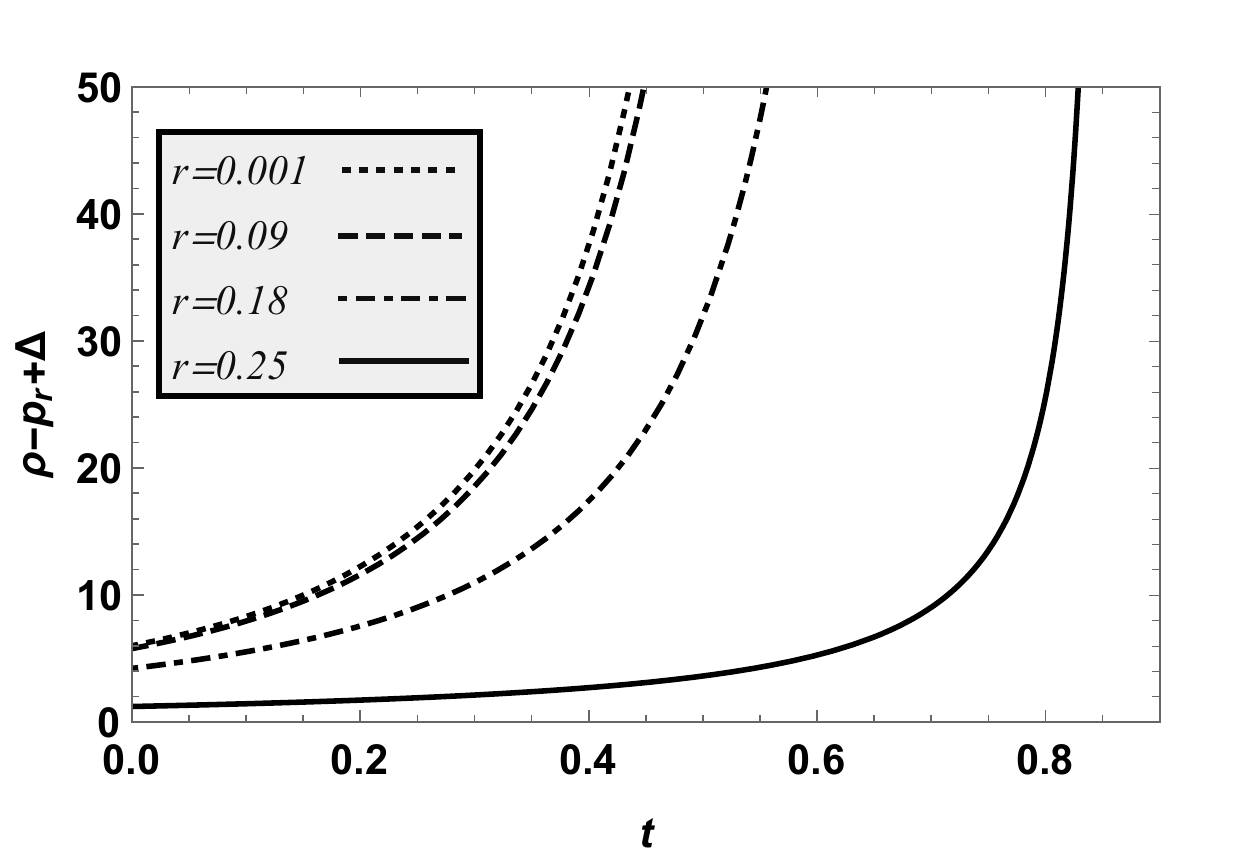}\includegraphics[scale=0.65]{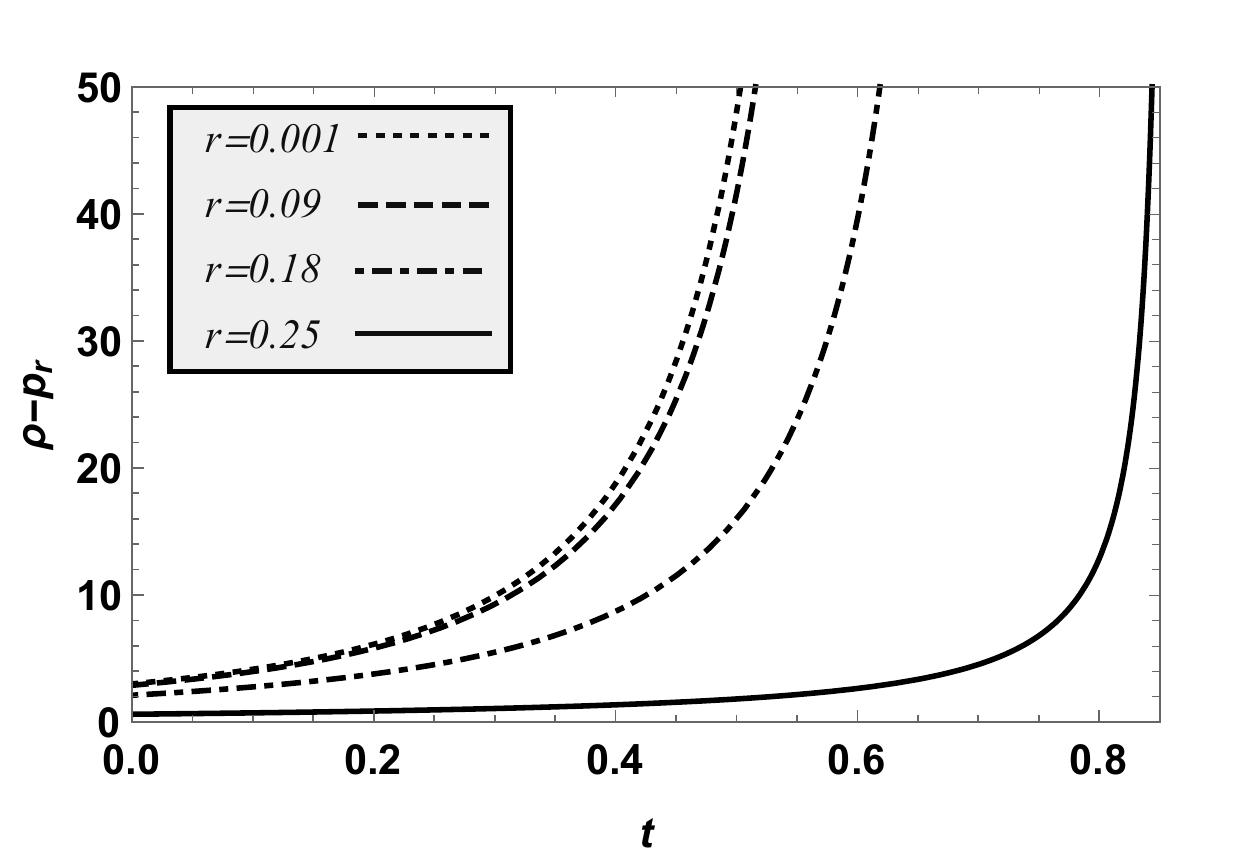}
\includegraphics[scale=0.65]{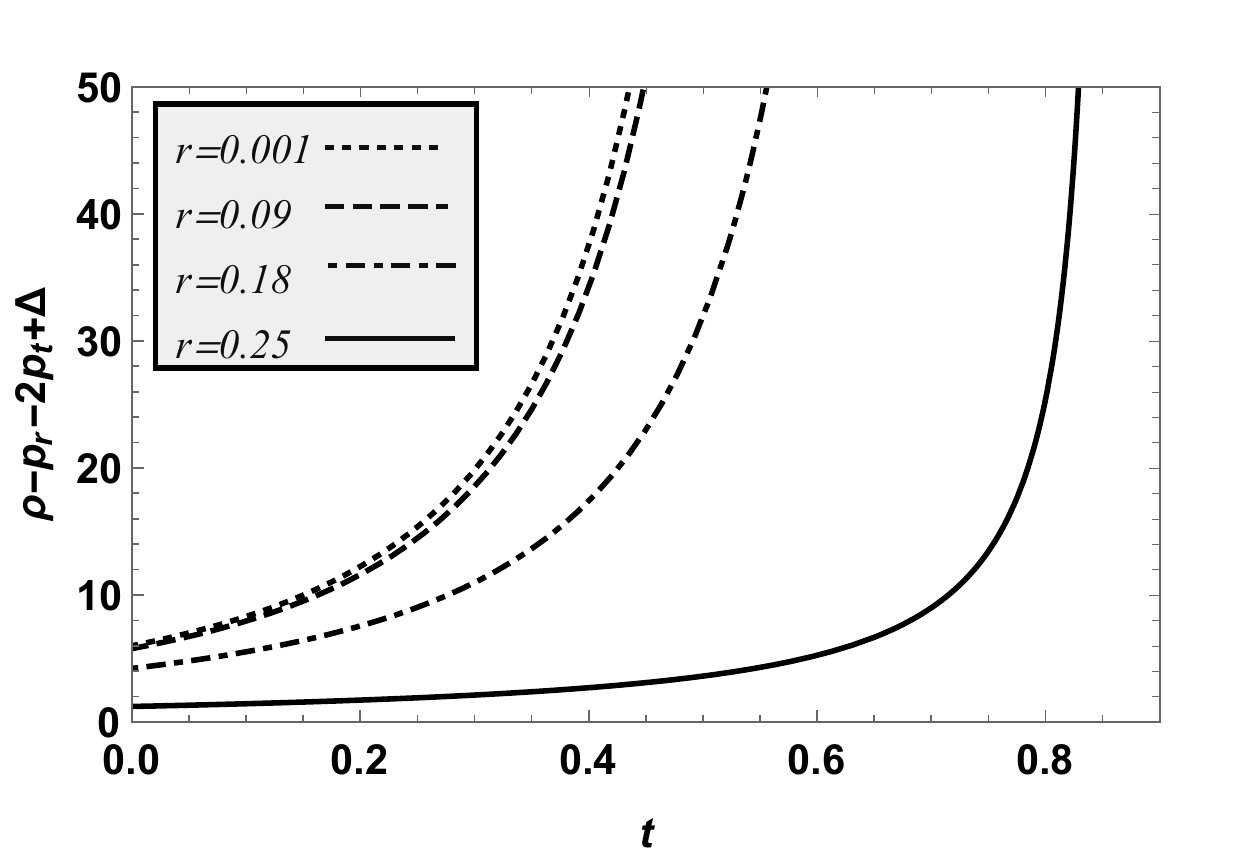}\includegraphics[scale=0.65]{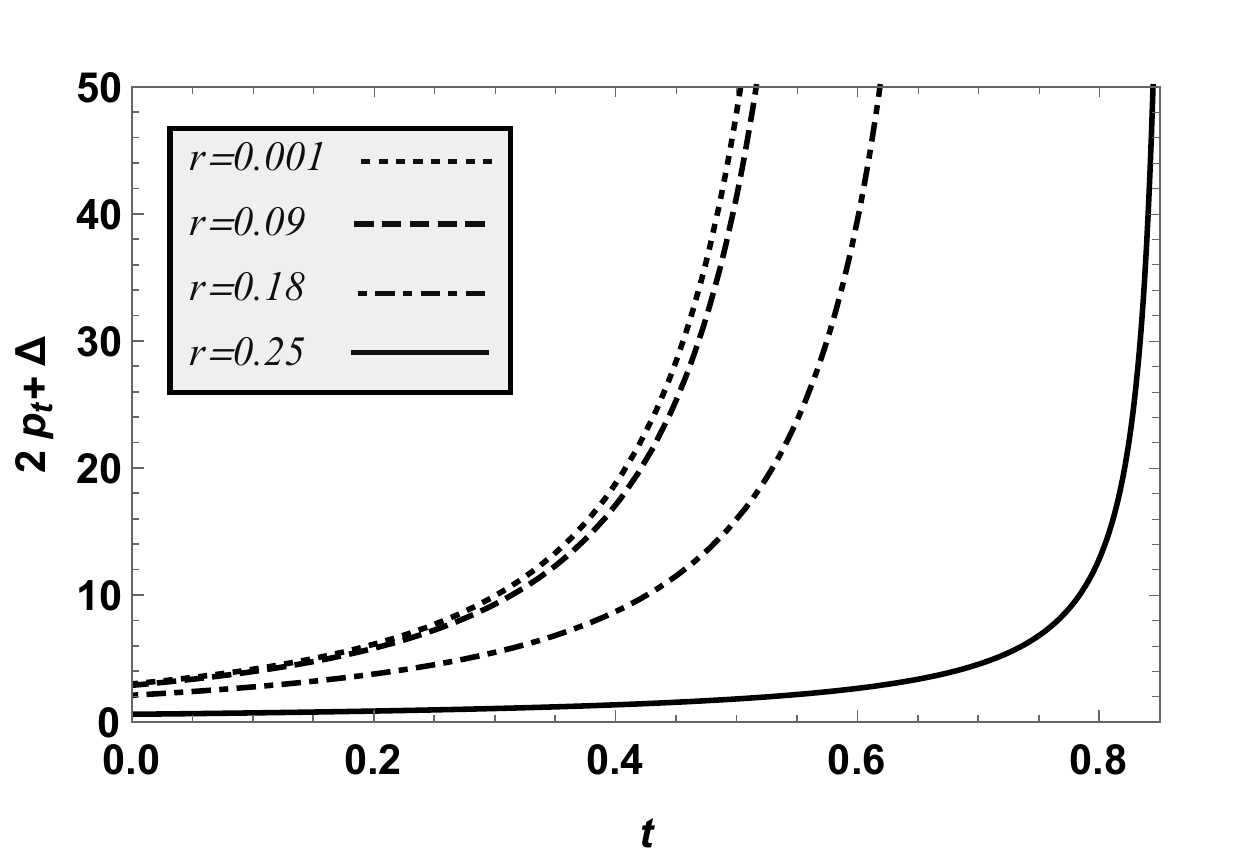}}
\caption{Various energy conditions of the collapsing matter cloud for different comoving radius $r$, throughout the collapse, is depicted here in the framework of $f(R)=R+\alpha R^2$ gravity with $\alpha=10^{-6}$. The initial data is taken as $S=r^3-25.5 r^6$. It can be seen from here that all the energy conditions (the inequalities in Eq.(\ref{frec1}-\ref{frec6})) are satisfied.}
\label{figfrenergycondition}
\end{figure}

%%%%%%%%%%%%%%%%%%%%%%%%%%%%%%%%%%%%%%%%%%%%%%%%%%%%%%%%%%%%%%%%%%%%%%%%%%%%%%%%%%%%%%%%%%%%%%%%%%%%
\begin{figure} 
{\includegraphics[scale=0.51]{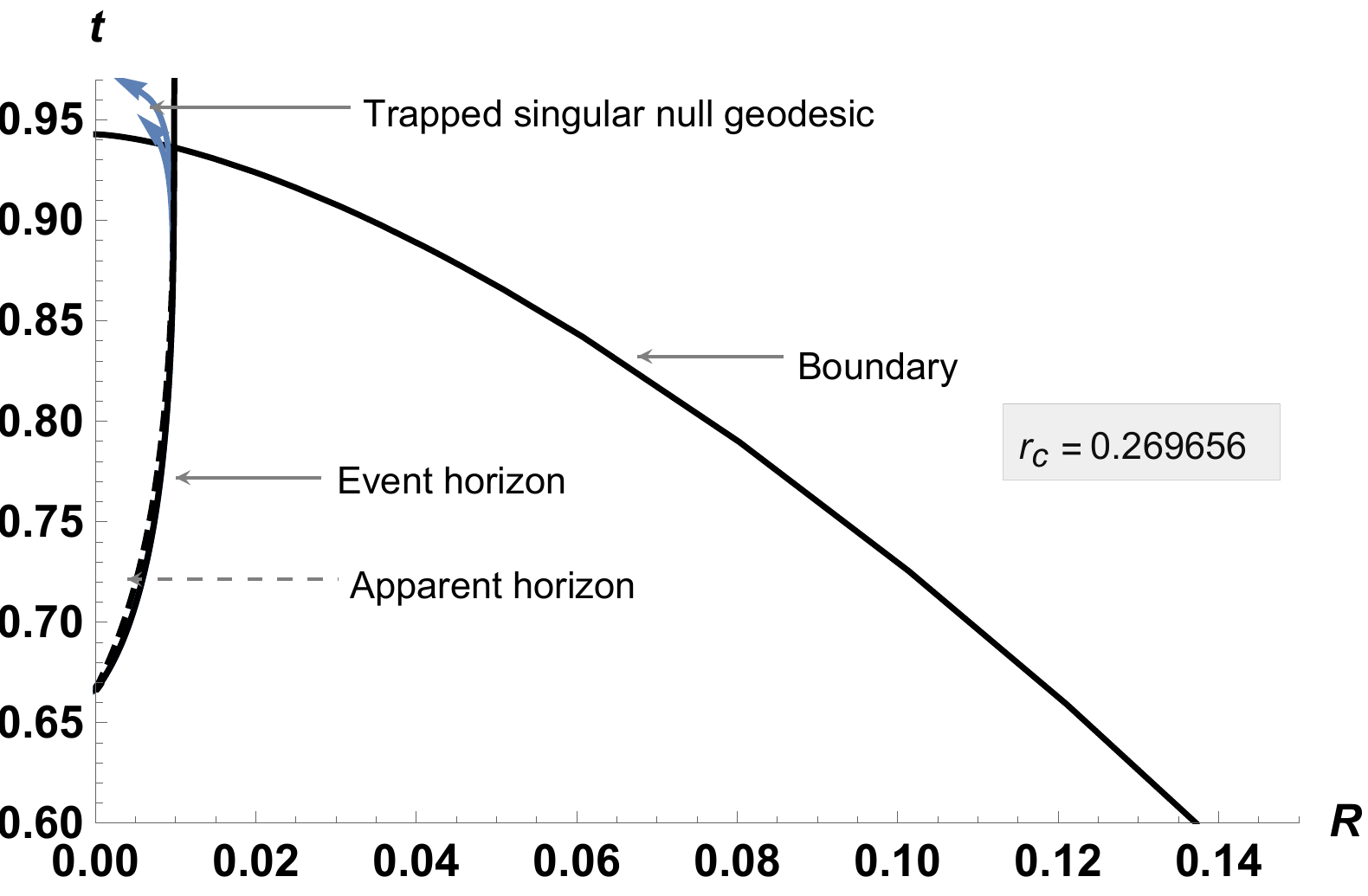}}\includegraphics[scale=0.51]{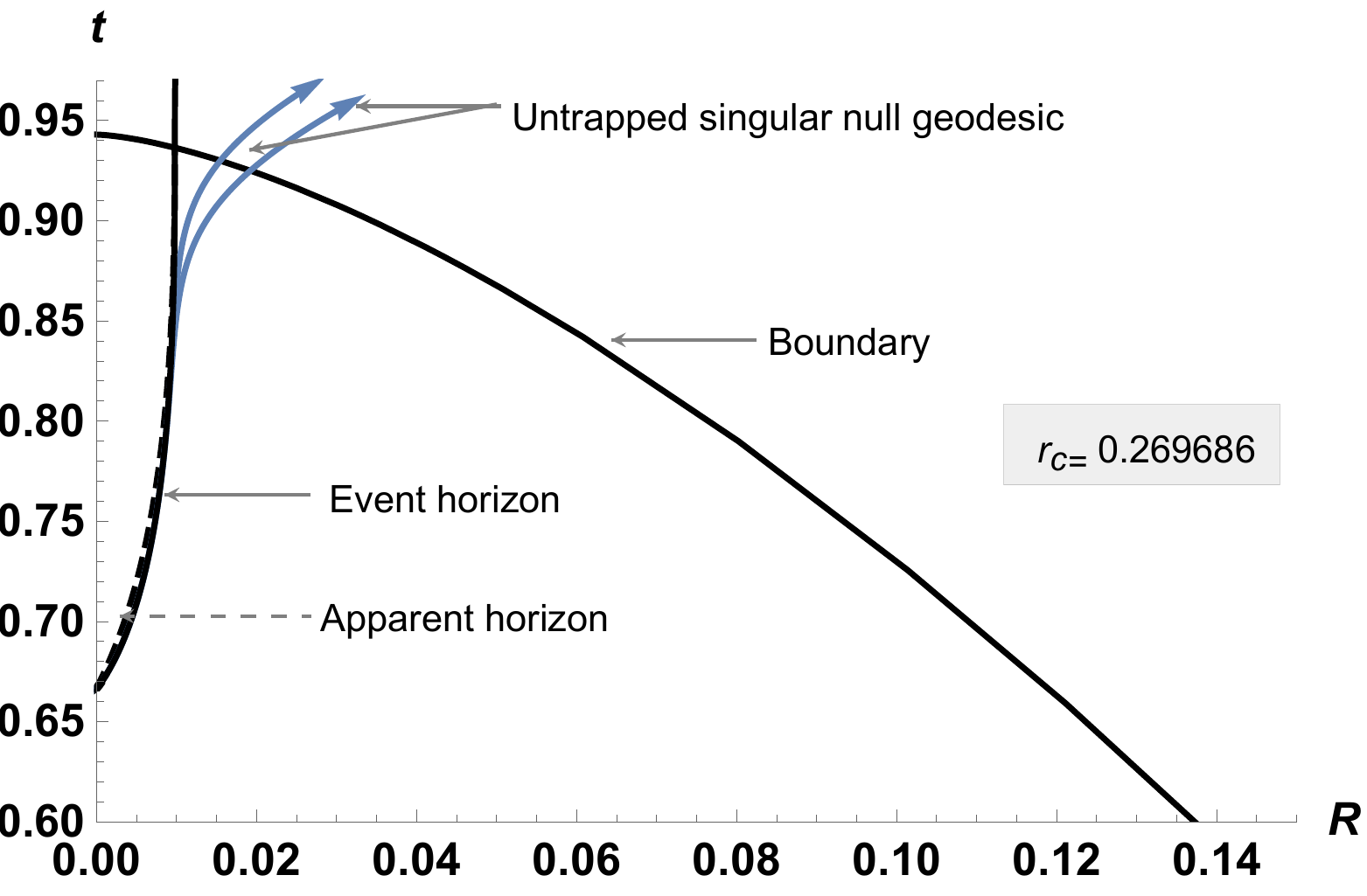}
\caption{The space-time plot depicting the causal structure of the singularity formed due to marginally bound ($b=0$) collapsing spherical matter cloud with the initial data $S=r^3-25.5r^6$. (a): In GR, the density Eq.(\ref{efe0}) vanishes at $r_c= 0.2696559088937193$. The event horizon forms before the formation of the first singularity, hence the singularity is only locally visible. (b): In $f(R)=R+\alpha R^2$ gravity theory, for $\alpha=10^{-6}$, the density Eq.(\ref{rhofr}) vanishes at $r_c=0.26968557639843954$. The event horizon forms together with the formation of the first singularity, hence the singularity is globally visible.}
\label{frspacetimeplotgrfr}
\end{figure}
%%%%%%%%%%%%%%%%%%%%%%%%%%%%%%%%%%%%%%%%%%%%%%%%%%%%%%%%%%%%%%%%%%%%%%%%%%%%%%%%%%%%%%%%%%%%%%%%%%%%
%%%%%%%%%%%%%%%%%%%%%%%%%%%%%%%%%%%%%%%%%%%%%%%%%%%%%%%%%%%%%%%%%%%%%%%%%%%%%%%%%%%%%%%%%%%%%%%%%%%%
%%%%%%%%%%%%%%%%%%%%%%%%%%%%%%%%%%%%%%%%%%%%%
%%%%%%%%%%%%%%%%%%%%%%%%%%%%%%%%%%%%%%%%%%%%%%%%%%%%%%%%%%%%%%%%%%%%%%%%%%%%%%%%%%%%%%%%%%%%%%%%%%%%
\begin{figure}
\centering
\includegraphics[scale=0.35]{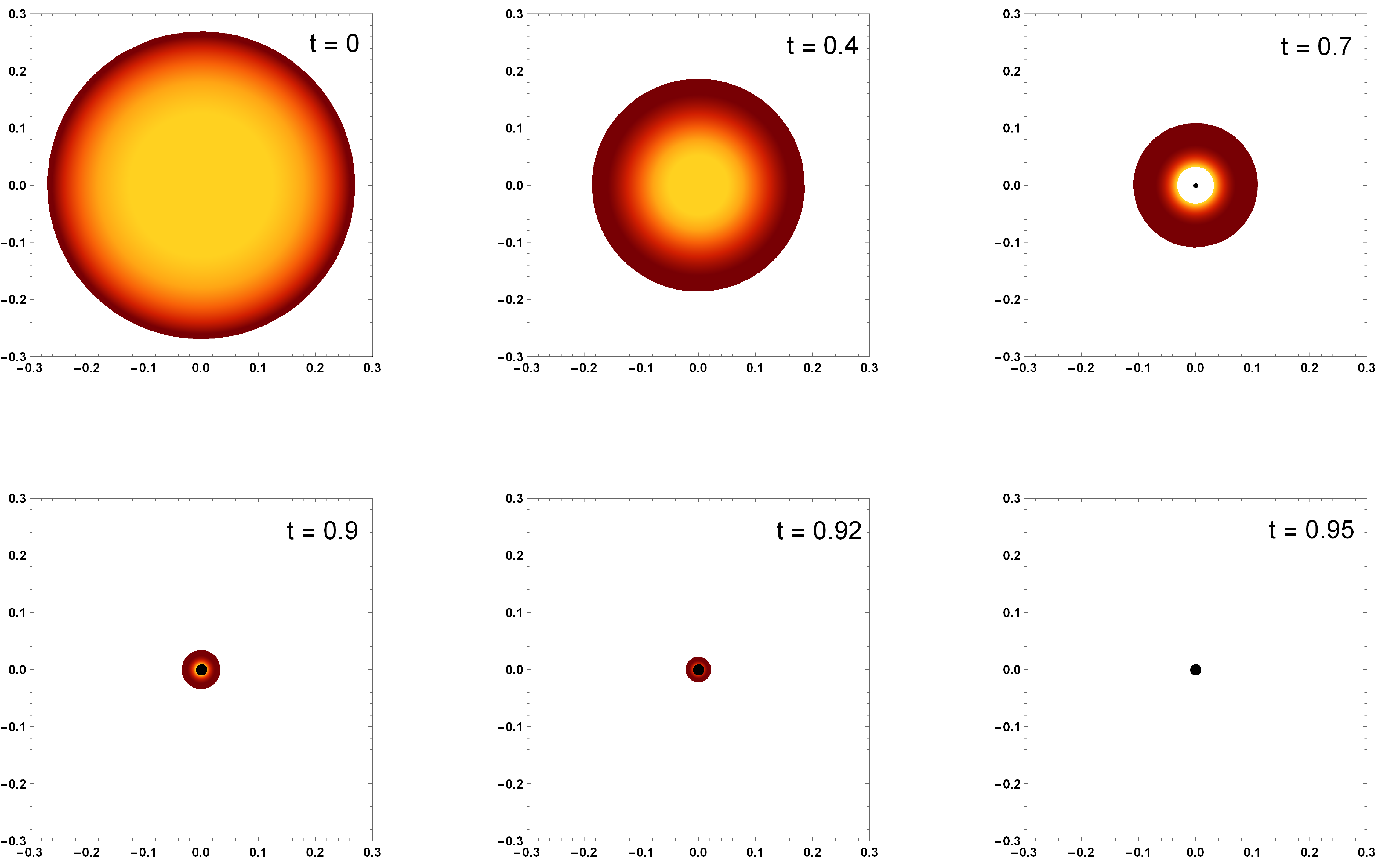}
\caption{the evolution of the density of the marginally bound collapsing cloud governed by LTB metric and made up of dust with the density Eq.(\ref{fr1efegr}) and the global causal structure of the first central singularity is depicted here. $S=r^3-25.5r^6$. The solid black disk represents the event horizon which increases in size with time and then achieves a fixed physical radius $S(r_c)$. The null geodesic wavefronts  are trapped by trapped surfaces, hence unable to escape from the singular region.}
\label{frdensityplotgr}
\end{figure}
%%%%%%%%%%%%%%%%%%%%%%%%%%%%%%%%%%%%%%%%%%%%%%%%%%%%%%%%%%%%%%%%%%%%%%%%%%%%%%%%%%%%%%%%%%%%%%%%%%%%
\begin{figure}
\centering
\includegraphics[scale=0.35]{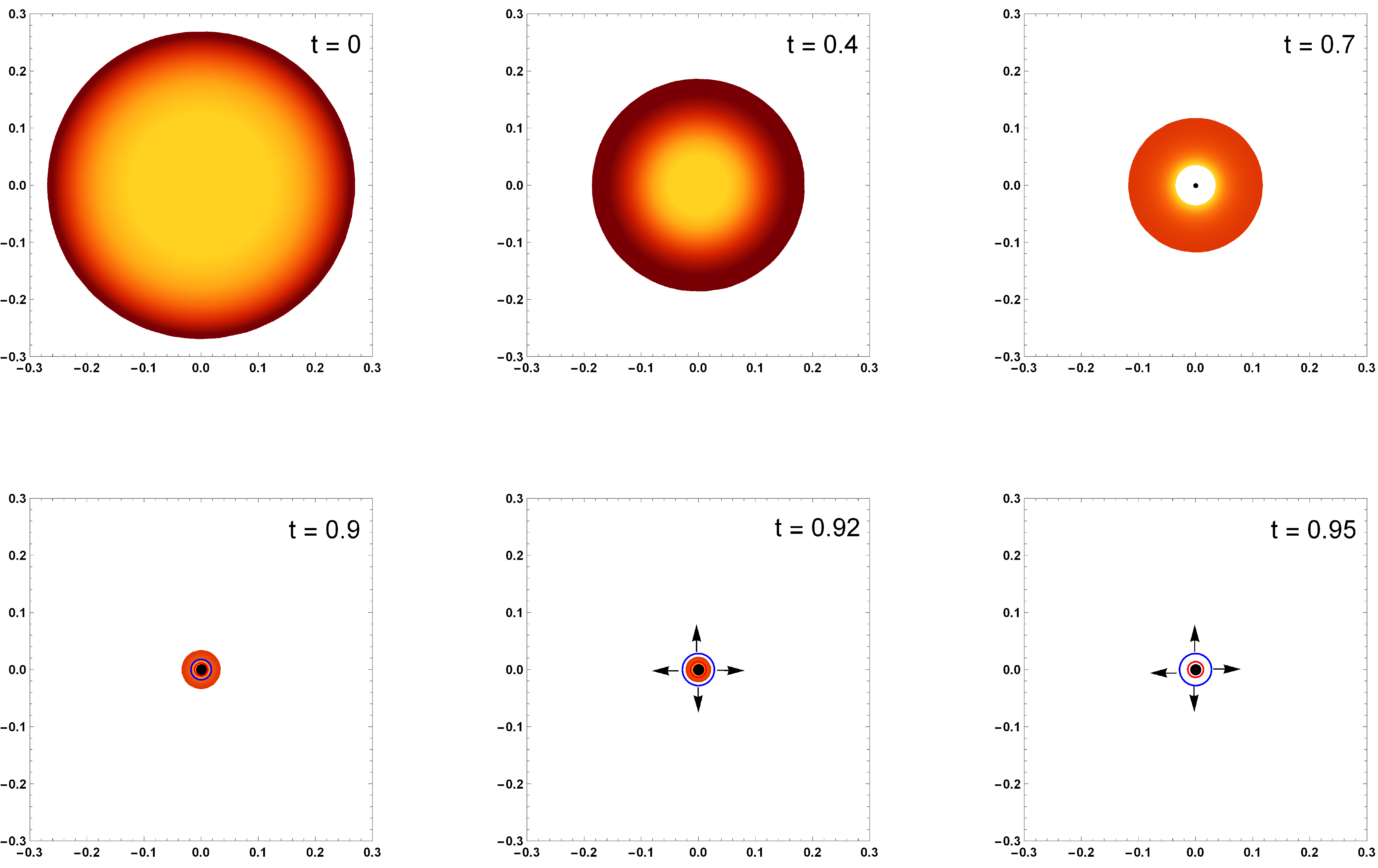}
\caption{The evolution of the density of the marginally bound collapsing cloud governed by LTB metric and made up of matter field with the profiles expressed in Eq.(\ref{rhofr}-\ref{qfr})  (which is in the framework of $f(R)=R+\alpha R^2$ gravity, with $\alpha=10^{-6}$) and the global causal structure of the first central singularity is depicted here. $S=r^3-25.5r^6$. The solid black disk represents the event horizon which increases in size with time and then achieves a fixed physical radius $S(r_c)$. The null geodesic wavefronts escaping from the singular region are depicted by red and blue concentric circles, which increase in size with time.}
\label{frdensityplotfr}
\end{figure}
%%%%%%%%%%%%%%%%%%%%%%%%%%%%%%%%%%%%%%%%%%%%%%%%%%%%%%%%%%%%%%%%%%%%%%%%%%%%%%%%%%%%%%%%%%%%%%%%%%%%
The energy density, radial and tangential pressures, and the heat flux of the marginally bound collapsing cloud governed by the LTB metric, Eq.(\ref{LTBmg}) with $b=0$, are respectively as follows:
\begin{equation}\label{rhofr}
\rho=\frac{S'}{A^2A'} +2\alpha \left(\frac{R \dot A^2}{A^2}+\frac{2R \dot A \dot A'}{A A'}-\frac{R^2}{4}+\frac{2\dot R \dot A}{A}+\frac{\dot R \dot A'}{A'}  -\frac{2R'}{A A'}+\frac{R' A''}{A'^{3}}=\frac{R''}{A'^{2}}\right),
\end{equation}
\begin{equation}\label{prfr}
    p_r=-\frac{\dot S}{A^2 \dot A}+2\alpha\left(-\frac{R\dot A^2 }{A^2}-\frac{2R\ddot A}{A}+\frac{R^2}{4}-\frac{2\dot R \dot A}{A}-\ddot R+\frac{R'}{A A'}\right),
\end{equation}
\begin{equation}\label{ptfr}
\begin{split}
      p_t=  &-\frac{\ddot A}{A}-\frac{\dot A \dot A'}{A A'}-\frac{\ddot A'}{A'}+2\alpha
    \Bigg (-\frac{R \ddot A}{A}-\frac{R \ddot A'}{A'}-\frac{R \dot A \dot A'}{A A'}+\frac{R^2}{4}-\frac{\dot R \dot A}{A}-\frac{\dot R \dot A'}{A'}-\ddot R \\
    & +\frac{R'}{A A'} -\frac{R' A''}{A'^{3}}+\frac{R''}{A'^{2}}\Bigg ),
\end{split}
\end{equation}
\begin{equation}\label{qfr}
    q=\frac{2\alpha}{A'}\left(\dot R'-\frac{R' \dot A'}{A'}\right).
\end{equation}
The Ricci scalar is given in terms of the metric components and their derivatives as
\begin{equation}
    R=2\left(\frac{\ddot A'}{A'}+\frac{\dot A^2}{A^2}+\frac{2\dot A \dot A'}{A A'}+\frac{2\ddot A}{A}\right).
\end{equation}

The function $S$ in the Eq.(\ref{rhofr}-\ref{prfr}), in terms of metric components is expressed as in Eq.(\ref{frS}) with $b=0$. However, it is necessary to note that unlike in GR, $S$ no more physically signifies the mass inside a collapsing shell of radial coordinate $r$ at time $t$. It can now be considered just an arbitrary function of the metric components. 

The collapsing matter field should satisfy all the energy conditions throughout the collapse, for which the following inequalities should be satisfied
\cite{Kolassis_88}:
\begin{equation}\label{frec1}
\vert \rho+p_r \vert-2\vert q \vert \geq 0,
\end{equation}
\begin{equation}\label{frec2}
    \rho-p_r+2p_t +\Delta \geq 0,
\end{equation}
\begin{equation}\label{frec3}
    \rho-p_r +\Delta \geq 0,
\end{equation}
\begin{equation}\label{frec4}
    \rho-p_r \geq 0,
\end{equation}
\begin{equation}\label{frec5}
    \rho-p_r-2p_t +\Delta \geq 0,
\end{equation}
\begin{equation}\label{frec6}
    2p_t +\Delta \geq 0,
\end{equation}
where $\Delta=\sqrt{\left(\rho+p_r \right)^{2}-4 q^2}$.
In Fig.(\ref{figfrenergycondition}), we have depicted the satisfaction of all the above inequalities, for the matter field having density, pressures, and heat flux as in Eq.(\ref{rhofr}- \ref{qfr}), and for certain fixed parametric values, as mentioned in the caption. It should be noted that the energy conditions are also satisfied for all nearby parametric values. Hence, there exists a non-zero measured set of parameters in the density profile and a non-zero range of $\alpha$ such that the energy conditions will not be violated for a small perturbation in these parametric values.

\section{Matching condition in f(R) gravity}\label{6.2}

For smooth matching of the spacetime governing the region of the collapsing cloud ($\mathcal{V}^{-}$)  with that of the exterior surrounding spacetime ($\mathcal{V}^{+}$), the first and second fundamental form induced by these two regions on the hypersurface $\Sigma$ which partitions these two regions should match
\cite{Darmois_27, Israel_84}. 
It can be shown that in GR, the interior LTB spacetime can be matched with the exterior static, spherically symmetric, asymptotically flat vacuum spacetime, which is the Schwarzschild spacetime. However, in the framework of $f(R)$ gravity, for smooth matching of the spacetimes of these two regions, apart from the above-mentioned two junction conditions, the continuity of the  Ricci scalar and its radial derivative at the boundary is also required. It was shown by Nzioki \textit{et. al} 
\cite{Nzioki_10} 
that for a class of $f(R)$ gravity model, which includes the Starobinsky one, the Schwarzchild solution is the only static, spherically symmetric, asymptotically flat vacuum spacetime with vanishing Ricci scalar, thereby extending the Birkhoff's theorem in $f(R)$ gravity.

Goswami \textit{et. al.} \cite{Goswami_14} in 2014 showed that smooth matching of the Ricci scalar and its radial derivative  constrains the previously free function $S(r)$, thereby fine-tuning it and making it unstable under matter perturbation. The argument goes as follows: The Ricci scalar is expressed in terms the arbitrary function $S$ as
\begin{equation}\label{R1}
    R=-\frac{3M+rM'}{ v^2\left(rv'+v \right)},
\end{equation}
where $M$ is related to $S$ as $M=\frac{S}{r^3}$, and $v=v(t,r)=\frac{A}{r}$ is called the scaling function. The scaling function can be thought of as the redefined time coordinate such that at the initiation of the collapse, $v(0,r)=1$ for all the shells, and $v(t,r)=0$ when the shell of comoving radius $r$ collapses to a singularity. Now, in order for the smooth matching of the Ricci scalar and its radial derivative at the junction connecting two regions $\mathcal{V}^{-}$ (LTB spacetime), and $\mathcal{V}^{-}$ (Schwarszchild spacetime), $R$ and $R'$ for the interior spacetime should vanish at the boundary $r_c$. Hence $R$ should have the form
\begin{equation}\label{R2}
    R=(r_c-r)^{2}g(t,r).
\end{equation}
Equating Eq.(\ref{R1}) and Eq.(\ref{R2}), we obtain
\begin{equation}\label{odem}
    rM'+3M=j(r) (r_c-r)^2,
\end{equation}
where
\begin{equation}
    j(r)=v^2(rv'+v)g(t,r).
\end{equation}
The functional form of $v$ and $g(t,r)$ is determined once the functional form of $M$ is determined. Hence we can say that $j=j(M,r)$. The Eq.(\ref{odem}) then becomes a first-order ordinary linear differential equation, which can only be satisfied by a class of functions $M$. This is how the additional matching condition constrains the function $S$ in $R+\alpha R^2$ gravity, which was free to choose in GR. 

\section{Globally visible nodal singularity in f(R) gravity.}\label{6.3}

The collapse formalism of the LTB metric in GR discussed in the previous section, i.e. Eq.(\ref{frS}) and Eq.(\ref{frt-ts}-\ref{frteh}), is same for the collapsing matter field governed by identical LTB metric, in $f(R)$ gravity for a time-independent function $S$ and vanishing $b$. However, the boundary $r_c$ of the collapsing cloud, which we define as the comoving radius where $\rho$ vanishes, will be different in $f(R)$ gravity. This is because the density profile Eq.(\ref{rhofr}) is different from Eq.(\ref{fr1efegr}). This causes a change in the evolution of the event horizon, thereby affecting the global causal structure of the singularity. The difference can be clearly seen in Fig.(\ref{frspacetimeplotgrfr}). In this figure, the geometry governing the collapse of two different matter fields in two different theories of gravity is the same. By this, we mean that apart from both matter fields being governed by the LTB metric, the initial data $(S,b)$ is also the same in both cases. However, since the boundaries of these two collapsing clouds are different, the initial condition Eq.(\ref{frteh}), which needs to be satisfied by the solution of the differential Eq.(\ref{frdeteh}) for it to represent the dynamics of the event horizon, is changed. Hence the previously locally visible singularity in GR is now globally visible in $f(R)$ gravity. The evolution of the density of the matter field along with the trapped (in GR) and escaped (in $f(R)$) null geodesics are depicted in Fig.(\ref{frdensityplotgr}) and Fig.(\ref{frdensityplotfr}), respectively. For a fixed functional form of $S$ as mentioned in the captions, for $\alpha=10^{-6}$, one gets a globally visible singularity. However, this is not the only value. One can show that for any greater value of  $\alpha$, globally visible singularity is achieved. This means that one can trace infinite event horizons, each corresponding to one value of $\alpha$, which are solutions of the differential equation (\ref{frdeteh}) and starting from $(t_s(0),0)$ in the $(t,r)$ plane. This is only possible if $(t_s(0),0)$ is a nodal point.

In order to check if the first singularity is a nodal point, consider two different frameworks of gravity, both of which are Starobinsky type, but with different value of scalar multiples non-minimally coupled with the quadratic curvature term in the Lagrangian. Let us call them $\alpha_1$ and $\alpha_2$, with
\begin{equation}
\alpha_2>\alpha_1     
\end{equation}
($\alpha_1$ can also be zero, which corresponds to GR). The evolution of the two distinct event horizons, each corresponding to distinct values of $\alpha$, are dictated by differential Eq.(\ref{frdeteh})
respectively satisfying 
\begin{equation}
    t_{EH}(r_{1})= \frac{2}{3}\frac{r_{1}^{\frac{3}{2}}}{\sqrt{S(r_1)}}-\frac{2}{3}S(r_1), 
\end{equation}
and
\begin{equation}
    t_{EH}(r_{2})= \frac{2}{3}\frac{r_{2}^{\frac{3}{2}}}{\sqrt{S(r_2)}}-\frac{2}{3}S(r_2).
\end{equation}
Here $r_1$ and $r_2$ are the largest comoving radius of the collapsing cloud corresponding to $\alpha_1$ and $\alpha_2$ respectively. Let us choose $\alpha_1$ such that for a given fixed functional form of $S$, the first singularity is globally visible. One can therefore see that for $\alpha=\alpha_1$ at $(t_{s}(0),0)$ in the $(t,r)$ plane, 
\begin{equation*}
    \frac{dt_{EH}(r)}{dr}
\end{equation*}
is not continuous. This is because for a small change in $t_{EH}$ from $t_{s}(0)$ to some $t_f$ where 
\begin{equation}
    t_f>t_s(0),
\end{equation}
there is zero change in $r$, since all null geodesics at $r=0$ are trapped after the time $t_s(0)$. Hence,  $\frac{dt_{EH}}{dr}$ is infinite at $r=0$.

Now, the uniqueness theorem of the first order linear differential equation says that if 
\begin{equation}
    g(x,y)\hspace{1cm} \textrm{and} \hspace{1cm} \frac{\partial g(x,y)}{\partial y}
\end{equation}
are continuous in the neighborhood around $x=0$, then the solution (in a possibly smaller neighborhood around $x=0$) of the differential equation (with initial condition) given by
\begin{equation}
    y'=g(x,y), \hspace{1cm} y(x_0)=y_0 
\end{equation}
is unique. However, this uniqueness theorem is not applicable in our case because of the discontinuity of $A'(t,r) (=\frac{dt_{EH}}{dr})$ at $(t_s(0),0)$. Therefore, one can have more than one solution of the differential Eq.(\ref{frdeteh}), and passing through $(t_s(0),0)$, making it a nodal point. 

One can check numerically that the singularity is globally visible for any $\alpha=\alpha_2>\alpha_1$, if it is globally visible for $\alpha=\alpha_1$. This supports the claim that the first central singularity is indeed a nodal point.

Let us now fix the framework of gravity. For the singularity to be visible by an asymptotic observer for infinite time, the central singularity should emit a congruence of infinite null geodesics, each redshifted by a different amount and the event horizon being the most redshifted (infinitely) null geodesic. This can happen because we have shown that $(t_s(0),0)$ is a nodal point.

\section{Concluding remarks}\label{6.4}
The concluding remarks are as follows:
\begin{enumerate}
\item In order to determine the global causal property of the singularity formed due to a collapsing spherically symmetric matter cloud, only knowing the spacetime metric governing the collapsing matter field is not sufficient. One also has to have the information of the extent to which the spacetime is governed by a given metric. In other words, one also has to have the information of the largest comoving radius $r_c$, which is also the initial size of the collapsing cloud. This boundary of the cloud affects the evolution of the event horizon in that it provides an initial condition to the differential equation whose solution satisfying this initial condition represents the dynamics of the event horizon. 

\item To show this, we considered the same spacetime metric (marginally bound LTB using up the remaining one-degree freedom by fixing the functional form of $S$) governing two different matter fields respectively collapsing unhindered in two different theories of gravity. The LTB metric corresponding to dust in GR corresponds to imperfect viscous fluid in  $f(R)=R+\alpha R^2$ gravity. Since the density profiles of the two clouds are different, their boundaries (which are determined by vanishing density) are also different. For this fixed metric with no remaining functional freedom of choice,  the event horizon, therefore, forms before the formation of the first singularity in GR, but forms together with the formation of the first singularity in $f(R)$ gravity, thereby making the singularity locally visible in GR but globally visible in $f(R)$ gravity. 

\item It should be noted that when we say: ``same spacetime metric in two different theories of gravity" or ``same geometry in two different theories of gravity," we don't mean that the two spacetimes are isomorphic to each other. They are not isomorphic because of the difference in the matching surfaces in both cases. This difference is because we have defined the boundary of the collapsing cloud such that it has the physical radius corresponding to that comoving radius where the density vanishes. These comoving radii are different because the density profiles of the matter fields are different for different theories of gravity.   

\item In scenarios where the collapsing cloud is such that its density does not vanish but has some known value $\rho_c$ at the boundary, the outermost comoving radius $r_c$ in $f(R)$ gravity will still be different from that in GR for the LTB cloud. This difference in $r_c$ causes the event horizons to evolve differently in different gravity theories, possibly affecting the global causal structure of the first central singularity. This is similar to the case of vanishing density at the boundary, which we have considered.  

\item The local causal structure of the singularity is, however, only determined by the behavior of the apparent horizon, which is the boundary of all trapped surfaces, and whose dynamics are completely determined once the governing spacetime metric is known.  

\item Matching the interior collapsing spacetime with the exterior spacetime in the framework of $f(R)$ theories of gravity impose a restriction on the otherwise free function $S$. The spacetime is singular at the matching surface if the junction conditions are violated. We have, however, chosen a specific form of the function $S$, which is $S=r^3-25.5r^6$, as an example to show the difference in the global causal structure of the singularity in GR and in $f(R)$ gravity. Whether or not this specific functional form maintains the continuity of the Ricci scalar and its radial derivative has not been investigated. However, even if there is a jump in the curvature term at the boundary, one could physically interpret this violation of the junction condition such that there exists surface stress-energy term on the matching hypersurface and should not be considered unphysical. 

\item For an asymptotic observer to be able to observe the singularity, apart from the event horizon to form with the formation of the first singularity, the singularity should also be a nodal point. Here we have argued that because of the discontinuity of the function $A'(t,r)$ at $(t_s(0),0)$, the uniqueness theorem of the first-order differential equation does not hold. Hence, there can exist more than one solutions of null geodesic equation starting from $(t_s(0),0)$. We have argued that there indeed exists more than one outgoing singular null geodesics, using the apparent property that the global causal structure of the singularity is stable under small perturbation in the value of $\alpha$, which can be verified numerically.

\item It should be noted that studying the global causal structure of the first central singularity formed due to a collapsing cloud requires the explicit expression of the physical radius in terms of $t$ and $r$. This can be easily obtained in GR in the case of dust collapse using Eq.(\ref{frt-ts}) and Eq.(\ref{frts}) to obtain
\begin{equation}
    A(t,r)=\left(r^{\frac{3}{2}}-\frac{3}{2}\sqrt{S}t\right)^{\frac{2}{3}}.
\end{equation}
However, in GR, in the case of the cloud having non-zero pressure, such explicit expression of $A(t,r)$ is difficult to obtain since integrating the analogous equation of Eq.(\ref{frS}) is not so straightforward. One way to interpret the resulting global visibility, which we show in $f(R)$ is that the LTB metric in $f(R)$ theory governs a collapsing cloud having some pressure as seen in  Eq.(\ref{prfr}) and Eq.(\ref{ptfr}). This global visibility seems to be generic in nature as far as small perturbations in the initial data $(S,b)$ is concerned. Now, it seems fairly reasonable to assume that even in GR, one should get a non-zero measured set of initial data for which the end state of a ``pressured" collapsing cloud ends up in a globally visible singularity.

\end{enumerate} 
% Chapter Template

\chapter{Conclusions and future scope} % Main chapter title

\label{Chapter7} % Change X to a consecutive number; for referencing this chapter elsewhere, use \ref{ChapterX}

\lhead{\emph{Chapter 7.}} % Change X to a consecutive number; this is for the header on each page - perhaps a shortened title

The thesis is focused on the existence of naked singularities as the end state of the gravitational collapse. Merely showing the presence of outgoing singular null geodesics is not sufficient to counter the cosmic censorship hypothesis. As discussed in the Introduction, properties such as stability, genericity, and strength of the singularity are necessary to address. Simplifying assumptions need to be dropped as much as possible to understand a more general scenario.  Additionally, suitable initial data giving rise to its existence needs to be checked in the astrophysical setup. Moreover, the results should also arise in the framework of other theories of gravity. 

We have considered these issues and addressed them here up to a certain extent. We argued in support of the violation of strong cosmic censorship by depicting a locally naked singularity formed due to gravitational collapse of non-zero pressured cloud, which is \textbf{strong} in the sense that the volume element formed due to the wedge product of independent Jacobi fields of the nonspacelike geodesic congruence vanish as the congruence approaches the singularity. Additionally, we found that the causal property of the singularity is \textbf{stable} under small perturbation in the subset of total initial data giving rise to the same outcome. 

Later, we dropped the \textbf{simplifying assumption} of the marginally bound case while studying the gravitational collapse of the LTB cloud, wherein we depicted the existence of a globally visible singularity that is strong in the sense mentioned above. We further dropped the simplifying assumption of spherical symmetry of a dust inhomogeneous cloud undergoing gravitational collapse and showed the formation of strong globally visible singularity having an interesting property of being off-centric, more specifically the singularity being close to the boundary of the collapsing cloud. The implication of the off-centric nature of the singularity is that the outgoing singular null geodesic is received by an external observer in a less distorted form, thereby preserving the information of the quantum gravity region. Here, we have followed Misner's interpretation of the singularity theorem, as mentioned in the Introduction. 

We then investigated the existence of a globally visible singularity in the \textbf{astrophysical setup} by deriving the criteria required by the collapsing cloud to form a globally visible singularity at the center of the milky way galaxy and the M87 galaxy. We also derived requirements to form a primordial visible singularity just after the time of matter-radiation equality. Further, we considered the scenario of the collapse of the neutron star after reaching a critical mass by accreting the supernova ejecta of the binary companion core progenitor and found that such singularity will always be globally hidden. All the while, we modeled the collapsing cloud by LTB spacetime. 

Finally, we addressed the problem of the existence of a globally visible singularity in an \textbf{alternative gravity theory} having the Lagrangian $R+\alpha R^2$. We gave a heuristic method to prove that such singularity is a nodal point, thereby making such singularity visible to an asymptotic observer for infinite time. This property of singularity is independent of the theory of gravity. In fact, we considered a class of theory of gravity to reach such a conclusion by identifying one outgoing singular null geodesic with one value of $\alpha$.

The fundamental philosophy of physics involves validating a theory describing the physical system by observations and measurements. As far as the future scope is concerned, one of the immediate concerns is an investigation of the observational features of the naked singularities, thereby bridging the gap between theoreticians and astronomers. In a recent article
\cite{Shaikh_18},
the authors found that the presence of a shadow does not necessarily imply the existence of a black hole. Even naked singularities can cast a shadow. A natural quest then one can follow is to find the observational signatures that distinguish blackholes from naked singularities. 

\textit{On a final note...} 

One of the properties on which the classical laws of nature are premised on is \textit{predictability}. If we know the complete initial data, we can predict the entire future and trace the entire past of the system under consideration, in principle. For example, according to Newton's law of motion, given a particle's initial position and initial velocity, one can, in principle, predict its entire future trajectory. On a classical level, hence, Physics goes hand in hand with the power of predictability, unlike the laws of quantum physics, in which, due to Heisenberg's uncertainty principle, nothing can be predicted with certainty. However, even on a classical level, the existence of a naked singularity implies an absence of the Cauchy surface, breaking down the future's deterministic nature. Of course, one still has to look into several aspects of its existence in our universe before arriving at a concrete conclusion. Till then, we hope that one day we will accomplish our goal of completely understanding this strange object that challenges the doctrine of predeterminism. $.\hspace{11cm}$\textit{...the end may be the beginning} 

%----------------------------------------------------------------------------------------
%	THESIS CONTENT - APPENDICES
%----------------------------------------------------------------------------------------
\addtocontents{toc}{\vspace{2em}} % Add a gap in the Contents, for aesthetics
\appendix % Cue to tell LaTeX that the following 'chapters' are Appendices

% Include the appendices of the thesis as separate files from the Appendices folder
% Uncomment the lines as you write the Appendices
% Appendix A

% Appendix Template

\chapter{Locally visible singularity in dust collapse} % Main appendix title

\label{AppendixA} % Change X to a consecutive letter; for referencing this appendix elsewhere, use \ref{AppendixX}

\lhead{Appendix A. \emph{Locally visible singularity in dust collapse}} % Change X to a consecutive letter; this is for the header on each page - perhaps a shortened title

Here, we discuss the local visibility of the singularity formed due to a marginally as well as non- marginally bound collapse of the dust sphere. The Taylor expansion of the time curve Eq.(\ref{timecurve1} around $r=0$ is expressed as
\begin{equation}
     t(r,v)=t(0,v)+r\chi_1(v)+r^2\chi_2(v)+r^3\chi_3(v)+O(r^4),
\end{equation}
where
\begin{equation}\label{Achii}
    \chi_i(v)=\frac{1}{i!}\frac{d^i t}{dr^i}\bigg |_{r=0}.
\end{equation}
Here $v$ is the scaling function and is related to the physical radius and the comoving radius as 
\begin{equation}
    R(t,r)=r v(t,r).
\end{equation}
Coordinate freedom gives us the option of rescaling the physical radius as
\begin{equation}
    R(t_i,r)=r,
\end{equation}
where $t_i$ is the initial time of the collapse. This is equivalent to having $v(t_i,r)=1$. First three $\chi_i$'s can be calculated by integrating Eq.(\ref{friedmann}), and is expressed as 
\citep{Joshi_12, Mosani_20}:
\begin{equation}\label{Achi1}
    \chi_1(v)=-\frac{1}{2}\int^{1}_{v} \frac{\frac{G \mathcal {F}_1}{ v}+f_1c^2}{\left(\frac{G \mathcal{F}_0}{ v}+f_0c^2\right)^{\frac{3}{2}}}dv, 
\end{equation}
\begin{equation}\label{Achi2}
    \chi_2(v)=\int^{1}_{v}\left[\frac{3}{8}\frac{\left(G\frac{\mathcal{F}_1}{v}+f_1 c^2\right)^2}{\left(G\frac{\mathcal{F}_0}{v}+f_{0}c^2\right)^{\frac{5}{2}}}-\frac{1}{2}\frac{\frac{G\mathcal{F}_2}{v}+f_{2}c^2}{\left(\frac{G\mathcal{F}_0}{v}+f_{0}c^2\right)^{\frac{3}{2}}}\right] dv
\end{equation}
and

\begin{equation}\label{Achi3}
\begin{split}
 \chi_3(v) & =  \int_v^1 \frac{f_{1}}{\left(\frac{G \mathcal{F}_0}{v}+f_{0}\right)^{\frac{3}{2}}}\left(-\frac{5}{16}\left(\frac{f_{1}}{\frac{G\mathcal{F}_0}{v}+f_{0}c^2}\right)^2+\frac{3}{4}\left(\frac{\frac{G \mathcal{F}_2}{v}+f_{2}c^2}{\frac{G\mathcal{F}_0}{v}+f_{0}c^2}\right)\right)  \\
     & -\frac{1}{2}\frac{\left(\frac{G\mathcal{F}_3}{v}+f_{3}c^2\right)}{\left(G\frac{\mathcal{F}_0}{v}+f_{0}c^2\right)^{\frac{3}{2}}}dv.
     \end{split}
\end{equation}
Here, $f_i$ and $\mathcal{F}_i$ are such that
\begin{equation}
F= \sum_{i=0}^{\infty} \mathcal{F}_i r^{i+3} 
\end{equation}
and
\begin{equation}
    f=\sum_{j=0}^{\infty}f_j r^{j+2}.
\end{equation}
near $r=0$. Here, $F$ and $f$ are the Misner-Sharp mass function
\cite{Misner_1969}, and the velocity function respectively. It was shown in 
\cite{Mosani_20} 
that for a strong singularity to be locally naked, we should have $\chi_1=\chi_2=0$ and $\chi_3>0$ at $v=0$. This condition is always satisfied for the case of marginally bound collapse with the collapsing fluid having mass function as in Eq.(\ref{strongmassfunction}) with $F_3<0$, as is apparent from Eq.(\ref{Achi1}-\ref{Achi3}). The condition may or may not satisfy for a non zero velocity function. However, for the velocity function $f=10^{-5}\frac{G F}{c^2r}$, as chosen in Fig.(1b), the above mentioned condition is satisfied. Hence, all the singularities arising in chapter (\ref{Chapter4}) are at least locally naked. 

\addtocontents{toc}{\vspace{2em}} % Add a gap in the Contents, for aesthetics
\backmatter

%----------------------------------------------------------------------------------------
%	BIBLIOGRAPHY
%----------------------------------------------------------------------------------------
%\printbibliography
%\label{Bibliography}
%

\lhead{\emph{Bibliography}} % Change the page header to say "Bibliography"
\bibliographystyle{IEEEtran} % Use the "unsrtnat" BibTeX style for formatting the Bibliography

\bibliography{Bibliography} % The references (bibliography) information are stored in the file named "Bibliography.bib"
\chapter{List of publications}
\begin{itemize}
   \item \textbf{On the visibility of singularities in general relativity and modified gravity theories}\\
   Karim Mosani, Dipanjan Dey, Pankaj S. Joshi, Gauranga C. Samanta, Harikrishnan Menon, and Vaishnavi Patel\\
   \href{https://arxiv.org/abs/2106.01773}{arXiv:2106.01773}.

   \item \textbf{Globally visible singularity in an astrophysical setup}\\
   Karim Mosani, Dipanjan Dey and Pankaj S. Joshi \\
   \href{https://academic.oup.com/mnras/article-abstract/504/4/4743/6253197}{Monthly Notices of the Royal Astronomical Society, \textbf{504}, 4, 4743 (2021)}.
   %\\ \href{https://arxiv.org/abs/2103.07190}{arXiv:2103.07190}.

\item  \textbf{Causal structure of singularity in non-spherical gravitational collapse}\\
Dipanjan Dey, Pankaj S. Joshi, Karim Mosani, and Vitalii Vertogradov, \\ \href{https://doi.org/10.1140/epjc/s10052-022-10401-1}{European Physical Journal C, \textbf{82}, 431 (2022)}.

\item \textbf{Global visibility of a strong curvature singularity in non-marginally bound dust collapse}\\
Karim Mosani, Dipanjan Dey and Pankaj S. Joshi \\  \href{https://journals.aps.org/prd/abstract/10.1103/PhysRevD.102.044037}{Phys. Rev. D \textbf{102}, 044037 (2020)}.

\item \textbf{Strong curvature naked singularities in spherically symmetric perfect fluid collapse}\\
Karim Mosani, Dipanjan Dey and Pankaj S. Joshi\\
\href{https://journals.aps.org/prd/abstract/10.1103/PhysRevD.101.044052}{Phys. Rev. D \textbf{101}, 044052 (2020)}.

%\item \textbf{Karim Mosani}, Dipanjan Dey and Pankaj S. Joshi, On viability of isentropic perfect fluid collapse with a linear equation of state, \textit{submitted}, \href{https://arxiv.org/abs/1910.13678}{arXiv:1910.1368}.
\item \textbf{Dynamics of General Barotropic Stellar Fluid in the Framework of $R+2 \alpha T$ Gravity}\\
Karim Mosani and Gauranga C. Samanta\\  \href{https://www.worldscientific.com/doi/abs/10.1142/S0217732320500893}{Mod. Phys. Lett. A, \textbf{33}, 1, 2050089 (2020)}.

\item \textbf{Gravitational Collapse in General Relativity and in $R^2$ Gravity : A Comparative Study}\\
Artyom V. Astashenok, Karim Mosani, Sergey D. Odintsov and Gauranga C. Samanta\\ \href{https://www.worldscientific.com/doi/abs/10.1142/S021988781950035X}{Int. J. Geom. Methods Mod. Phys, \textbf{16}, 03, 1950035 (2019)}.
\end{itemize}
\chapter{Brief Biodata of the Candidate}

Karim Mosani joined the Department of Mathematics, BITS Pilani, K. K. Birla Goa Campus as a Ph.D. student in January 2017. 

He holds a Bachelors's degree in Science (2014) with Majors in Mathematics and Minors in Physics and Statistics from the University of Mumbai. He also has a Master's degree in Mathematics (2016) from the University of Mumbai. 

His Ph.D. research is supported by Junior Research fellowship (initial two years) and Senior Research Fellowship (subsequent three years) granted by the Council of Scientific and Industrial Research (CSIR, India, Ref: 09/919(0031)/2017-EMR-1). 

He has participated in and presented his work at various conferences and workshops: Yukawa Institute of Theoretical Physics, Kyoto University, Kyoto (2019), Indian Institute of Science Education and Research, Mohali (2019), International Center for Cosmology, Anand (2019), Herzen State Pedagogcal University of Russia (online invited seminar 2021), Birla Institute of Technology and Science - Pilani, Hyderabad (online 2021).

\chapter{Brief Biodata of the Guide}

Prof. Gauranga C. Samanta has been an Associate Professor in the PG department of Mathematics at Fakir Mohan University, Odisha, since June 2020. Before joining the present institution, he held the position of Assistant Professor at the Dept. of Mathematics in BITS Pilani, Goa Campus. He also has the post of Visiting Associate at the Inter-University Centre for Astronomy and Astrophysics (IUCAA), Pune, since 2020. 

He holds a Ph.D. degree from Sambalpur University, Odisha, in 2011. His thesis title was: Some Cosmological Models in Kaluza Klein Spacetime. 

He was the Principle Investigator of the project titled: Multidimensional dark energy cosmological models in modified theories of gravitation. It was funded by the UGC-Startup grant, Govt. of India from Dec 2013 to Dec 2015. He was also the Principle Investigator of the project titled: Study on dark energy, dark matter, and cosmic acceleration of the universe in modified gravity. It was funded by CSIR, Govt. of India, from May 2017 to May 2020. He has worked extensively in the field of theoretical cosmology, and has sixty two publications to his name.
\end{document}